\documentclass[aos]{imsart}

\RequirePackage{amsthm,amsmath,amsfonts,amssymb}
\RequirePackage[numbers]{natbib}
\RequirePackage[colorlinks,citecolor=blue,urlcolor=blue]{hyperref}
\RequirePackage{graphicx} 

\usepackage{breqn}
\usepackage{lmodern}
\usepackage{enumerate, lineno}
\usepackage{subcaption}
\usepackage{wrapfig}
\usepackage{tikz-network}
\usepackage{float}
\usepackage{tikz-cd}
\usepackage{tikz}
\usetikzlibrary {arrows.meta}
\usetikzlibrary {patterns, patterns.meta} 
\usetikzlibrary{shapes.geometric}

\startlocaldefs

\newcommand{\nrm}[2]{\left\|#1\right\|_{#2}}

\newcommand{\lrp}[1]{\left( #1 \right)}
\newcommand{\brcs}[1]{\left\{ #1 \right\}}
\newcommand{\bxs}[1]{\left[ #1 \right]}
\newcommand{\vrt}[1]{\left| #1 \right|}

\newtheorem{definition}{Definition}
\newtheorem{theorem}{Theorem}[section]
\newtheorem{lemma}{Lemma}[section]
\newtheorem{corollary}{Corollary}[section]
\newtheorem{proposition}{Proposition}[section]
\newtheorem{remark}{Remark}

\newtheorem{claim}{Claim}[subsection]

\newcommand{\ba}{\mathbf{a}}
\newcommand{\bb}{\mathbf{b}}
\newcommand{\bc}{\mathbf{c}}
\newcommand{\bd}{\mathbf{d}}
\newcommand{\be}{\mathbf{e}}

\newcommand{\bh}{\mathbf{h}}

\newcommand{\bl}{\mathbf{l}}
\newcommand{\bmm}{\mathbf{m}}

\newcommand{\bo}{\mathbf{o}}

\newcommand{\bq}{\mathbf{q}}
\newcommand{\br}{\mathbf{r}}

\newcommand{\bt}{\mathbf{t}}
\newcommand{\bu}{\mathbf{u}}
\newcommand{\bv}{\mathbf{v}}
\newcommand{\bw}{\mathbf{w}}
\newcommand{\bx}{\mathbf{x}}
\newcommand{\by}{\mathbf{y}}
\newcommand{\bz}{\mathbf{z}}

\newcommand{\bA}{\mathbf{A}}
\newcommand{\bB}{\mathbf{B}}
\newcommand{\bC}{\mathbf{C}}
\newcommand{\bD}{\mathbf{D}}
\newcommand{\bE}{\mathbf{E}}

\newcommand{\bG}{\mathbf{G}}
\newcommand{\bH}{\mathbf{H}}
\newcommand{\bI}{\mathbf{I}}
\newcommand{\bJ}{\mathbf{J}}

\newcommand{\bL}{\mathbf{L}}
\newcommand{\bM}{\mathbf{M}}

\newcommand{\bO}{\mathbf{O}}
\newcommand{\bP}{\mathbf{P}}
\newcommand{\bQ}{\mathbf{Q}}
\newcommand{\bR}{\mathbf{R}}
\newcommand{\bS}{\mathbf{S}}

\newcommand{\bU}{\mathbf{U}}
\newcommand{\bV}{\mathbf{V}}
\newcommand{\bW}{\mathbf{W}}
\newcommand{\bX}{\mathbf{X}}

\newcommand{\bZ}{\mathbf{Z}}

\newcommand{\cB}{\mathcal{B}}
\newcommand{\cC}{\mathcal{C}}
\newcommand{\cD}{\mathcal{D}}

\newcommand{\bcE}{\mathbf{\cal E}}
\newcommand{\cF}{\mathcal{F}}
\newcommand{\cL}{\mathcal{L}}
\newcommand{\cM}{\mathcal{M}}
\newcommand{\cN}{\mathcal{N}}

\newcommand{\cO}{\mathcal{O}}

\newcommand{\cW}{\mathcal{W}}
\newcommand{\cX}{\mathcal{X}}

\newcommand{\bon}{\mathbf{1}}
\newcommand{\E}{\mathbb{E}}

\newcommand{\R}{\mathbb{R}}
\newcommand{\pr}{\mathbb{P}}
\newcommand{\tti}{2,\infty}
\newcommand{\T}{^{\intercal}}

\newcommand{\vecop}{\operatorname{vec}}
\newcommand{\tr}{\operatorname{tr}}
\endlocaldefs

\begin{document}

\begin{frontmatter}
\title{Embedding Network Autoregression for time series analysis and causal peer effect inference}
\runtitle{Embedding Network Autoregression}

\begin{aug}
\author[A]{\fnms{Jae Ho}~\snm{Chang}\ead[label=e1]{chang.2090@osu.edu}}
\and
\author[A]{\fnms{Subhadeep}~\snm{Paul}\ead[label=e2]{paul.963@osu.edu}}
\address[A]{Department of Statistics,
The Ohio State University \printead[presep={ ,\ }]{e1,e2}}
\end{aug}
\begin{abstract}
We propose an Embedding Network Autoregressive Model for multivariate networked longitudinal data. We assume the network is generated from a latent variable model, and these unobserved variables are included in a structural peer effect model or a time series network autoregressive model as additive effects. This approach takes a unified view of two related yet fundamentally different problems: (1) modeling and predicting multivariate networked time series data and (2) causal peer influence estimation in the presence of homophily from finite time longitudinal data. Our estimation strategy comprises estimating latent variables from the observed network
followed by least squares estimation of the network autoregressive model.  We show that the estimated momentum and peer effect parameters are consistent and asymptotically normally distributed in setups with a growing number of network vertices (N) while considering both a growing number of time points T (for the time series problem) and finite T cases (for the peer effect problem). We allow the number of latent vectors K to grow at appropriate rates, which improves upon existing rates when such results are available for related models. Our theoretical results encompass cases both when the network is modeled with the random dot product graph model (ENAR) and a more general latent space model with both additive and multiplicative effects (AMNAR). We also develop a selection criterion when K is unknown that provably does not under-select and show that the theoretical guarantees hold with the selected number for K as well. Interestingly, even though we propose a unified model, our theoretical results find that different growth rates and restrictions on the latent vectors are needed to induce omitted variable bias in the peer effect problem and to ensure consistent estimation in the time series problem.
\end{abstract}

\begin{keyword}[class=MSC]
\kwd[Primary ]{00X00}
\kwd{00X00}
\kwd[; secondary ]{00X00}
\end{keyword}

\begin{keyword}
\kwd{Network time series}
\kwd{social influence}
\kwd{peer effect}
\kwd{social network}
\kwd{latent homophily}
\kwd{network embedding}
\end{keyword}

\end{frontmatter}

\section{Introduction}

A network of relationships and longitudinal node-level responses commonly appear in research problems in multiple domains, including social sciences, economics, public health, and biomedical sciences. We consider two key statistical problems associated with such data that have been widely investigated in the literature. The first is to causally estimate peer effects or social influence propagating through an observed network when the node level outcome of interest is measured in at least two but finite time points \citep{shalizi2011homophily,vanderweele2011sensitivity,goldsmith2013social,mcfowland2021estimating,nath2022identifying}. The second problem is to model and predict a high dimensional time series when network information is also observed \citep{zhu2017network,zhu2019network,knight2019generalised,zhu2020grouped,chen2023community}. The goals in these two problems are generally different and have been investigated separately in the literature. We take a unified view of these two problems and propose to include latent homophily variables in both of these problems. In the case of estimating peer influence, this approach aids causal identification by controlling for latent confounding. On the other hand, in the case of network time series, this approach leads to an improved prediction and model estimation.

It will be convenient to formally introduce the network autoregressive model (NAR or NAM) model, which has been used in the literature for both problems, to facilitate further discussion on its interpretation. We assume that we have measurements $y_{it}$ for $i =1, \ldots, N,$ and $t=0, \ldots, T-1$ with $N, T \in \mathbb{N}$ for an univariate outcome measured at $N$ vertices of a network over $T$ time periods. Let $\ell_{ij}$ denote the $(i,j)$-th entry of the normalized (symmetric) adjacency matrix.
 Then, our measurements $y_{it}$ are assumed to be generated via
 \begin{equation}
y_{i,t+1}=\alpha y_{it}+\theta\sum_{j\ne i}\ell_{ij}y_{jt} +{\bz_{it}}\T\gamma+\epsilon_{i,t+1},
\label{NAR}
 \end{equation}
where $\bz_{it}$ is a vector of (possibly time-varying) covariates and $\epsilon_{it}$ is the error term.  
We differentiate between the utility and the interpretation of this model in terms of whether $T$ is finite or growing, with the former being useful as a linear structural model for peer influence estimation and the latter being useful for multivariate time series modeling.  The parameter  $\theta$ has been termed as the ``peer effect'' parameter in the literature on both problems, while the parameter $\alpha$ is referred to as the ``momentum effect.''

The above NAR or NAM model with longitudinally measured outcomes in Equation \ref{NAR} has been widely employed for the causal identification of peer influence \citep{christakis2007spread,shalizi2011homophily,vanderweele2012and,christakis2013social,o2014estimating,mcfowland2021estimating,nath2022identifying}. Typical restrictions on the model would be an assumption of exogeneity of error term $\E[\epsilon_{i,t+1}|\ell_{ij},\bz_{it},y_{it}]=0,$ along with the regressors not being linearly dependent.  However, several authors have noted issues with identifying peer effect parameter $\theta$ from observational data with such models, including confounding due to latent homophily in peer selection and other unobserved omitted variables \citep{shalizi2011homophily,goldsmith2013social,o2014estimating,an2022causal}. The authors in \cite{mcfowland2021estimating,nath2022identifying} suggest augmenting the linear peer effects model with additive latent variables, which are responsible for network formation as a way of controlling for latent homophily. They proceed to show that asymptotic identification of the peer influence parameter is possible with this approach. However, consistency, limiting distribution, and the rate of convergence of the estimator of the peer effect parameter remain unsolved problems. 

The NAR model in Equation \ref{NAR} has also been used in the context of modeling and predicting multivariate high-dimensional network-linked time series in a line of work including \cite{zhu2017network, zhu2019network,knight2019generalised,zhu2020grouped,chen2023community}. Those papers then investigate the stationarity of the model along with consistency and asymptotic normality of the parameter estimates in an asymptotic setup where $T \to \infty$ under the assumptions $\epsilon_{it}\overset{iid}{\sim}\cN(0,\sigma^2)$. While several extensions of the baseline NAR model in Equation \ref{NAR} have been proposed, additional latent information from the network, in addition to the peer effects, has not been incorporated into the model. Therefore, the benefits of including latent variables in improving predictive performance and the conditions under which consistent estimation is possible are also currently unknown.

Our proposal in this paper is to augment the NAR model (Equation \ref{NAR}) with latent variables that are related to both the outcome and the formation of the network. 
Accordingly,  we assume that the network adjacency matrix is generated from either the Random Dot Product Graph (RDPG) model \cite{athreya2017statistical}, or the additive and multiplicative effects latent space network (LSN) model \cite{hoff2021additive,ma2020universal,li2023statistical}. The RDPG model is a general latent variable model for network data whose special cases include several popular latent variable models, namely, the Stochastic Block Model (SBM), Degree Corrected SBM (DCSBM), Mixed Membership SBM (MMSBM), and  DCMMSBM \citep{athreya2017statistical,rubin2022statistical}. The LSN model is more general since it includes both additive and multiplicative effects and connects latent variables with probabilities through non-linear link functions \citep{ma2020universal,hoff2021additive,li2023statistical}. 

Since most of the paper will use the RPDG model, we describe our approach using this model. We assume network adjacency matrix $\bA$ is generated from a $K$ dimensional RDPG model (defined later) with sparsity parameter $\rho_N$ and latent position parameter matrix $\bL \in \mathbb{R}^{N\times K}$ 
such that the probability of connections $\bP=\rho_N\bL\bL\T$. 
We let $\bU \in \mathbb{R}^{N\times K}$ be the matrix which contains latent homophily vectors for the vertices. This matrix could be $\bU_\bP$, the orthogonal eigenvectors of $\bP$ for $K$ largest magnitude eigenvalues, or the matrix of corresponding spectral embedding vectors $\bU_\bP\bS_\bP^{1/2}$, where $\bS_\bP$ is the $K \times K$ diagonal matrix containing the $K$ largest magnitude eigenvalues. Then, our Embedding Network Autoregression (ENAR) model augments the NAR model with these $K$ dimensional unknown (latent) vectors. Formally, the ENAR model assumes that the measurements $y_{it}$  for $i=1,\dots,N$, are assumed to be generated via
\begin{equation}
 y_{i,t+1}=\alpha y_{it}+\theta\sum_{j\ne i}\ell_{ij}y_{jt}+{\bu_i}\T\beta+{\bz_{it}}\T\gamma+\epsilon_{it}.
 \label{ENAR}
\end{equation}
However, the vectors of latent variables $\bu_i$s are not observed. Therefore, we propose to estimate the latent variables from the observed network either through spectral embedding (for RDPG) or maximum likelihood estimation (for LSN) and replace $\bu_i$ with its estimated version $\hat{\bu}_i$ in Equation \ref{ENAR}. 
However, a natural concern is whether the true peer influence parameter $\theta$ can be estimated consistently and the asymptotic variance can be characterized to enable inference when the $\bu_i$ is replaced with estimated $\hat{\bu}_i$. \cite{mcfowland2021estimating} considered this issue as trading off omitted variable bias with measurement error bias and remarked that it would only succeed if the measurement error bias is low.  

In this article, we show that the answer is affirmative and develop a theory for the consistency and asymptotic normality of the peer influence parameter. In addition, we propose to include estimated latent variables in the context of time series modeling as well to enable accurate inference on the parameters of the model and improve predictive performance. For the network time series problem, we develop theoretical results under an asymptotic setup where the number of nodes $N$, time $T$, and the dimension of the latent space $K$ grow.  We explore several questions that naturally arise in modeling two distinct phenomena with a similar model. These questions include differences in modeling latent homophily with eigenvectors $\bU_\bP$ and with spectral embedding vectors $\bU_\bP\bS_\bP^{1/2}$. We discuss asymptotic biases of ignoring latent homophily vectors with both choices and develop theoretical results for the peer effect problem for both choices.  We also delve into the challenging problem of obtaining results when the number of latent variables $K$ is unknown, which is often the case in practice but has not been explored in the literature before. We develop a selection criterion for selecting $K$ and derive theoretical results with the selected $K$.  We further study the asymptotic properties of the estimators when the general latent space model is used to model the network. Together, these results greatly expand the scope of the methodology and theory for both the peer effect problem and the time series problem.

\subsection{Connection to related literature} Our work is connected to several recent works encompassing network time series, network peer effect estimation, and network regression. In the network time series literature, our approach is related to the CNAR model of \cite{chen2023community}. We compare our results theoretically and in simulation extensively with the CNAR model and show the theoretical advantage of our approach as well as superior or comparable prediction performance in simulation. In the peer effect literature, our work is related to \cite{mcfowland2021estimating,nath2022identifying}. However, the theoretical results for the peer effect problem in this paper go well beyond what was considered in those references and delve into consistency and asymptotic normality of peer effect parameters, selection of latent dimension, and differences between modeling choices. For both problems, no earlier work addresses the problem of selecting the number of latent dimensions or communities. Our work is also related to network regression  \cite{fosdick2015testing,he2019multiplicative,le2022linear,lunde2023conformal}. For this problem, our corollaries develop consistency and asymptotic normality for growing embedding dimensions and our model selection procedure provides theoretical guarantees on selecting the number of embedding vectors. A paper related to ours is the random graph autoregressive model of \cite{wu2023random}. However, the theoretical results in \cite{wu2023random} are only developed for $T \to \infty$, and the method itself does not work unless we have data available in at least a few time points, which is not appropriate for the peer effect problem where we often will have data on only 2 or 3-time points.

\subsection{Main results and contributions} 
We summarize our theoretical results for both of these statistical goals. We always operate under a large $N$ asymptotic setup i.e., assume $N \to \infty$. We define the set of parameters as $\mu^H\triangleq(\beta\T\bH,\alpha,\theta,\gamma\T)\T$ with $\bH\in\cO_K$,  is a $K \times K$ matrix with orthonormal columns. It is well known that the multiplicative latent variables can only be estimated from a network up to an ambiguity of such a matrix from the class $\cO_K$ \citep{hoff2002latent,athreya2017statistical}. Therefore, the parameter $\beta$ can only be recovered up to the ambiguity of $\bH\T\beta$.

\subsubsection{Network time series results:} In the case of modeling time series, we assume $T \to \infty$. The true data-generating model is augmented with eigenvectors $\bU_\bP$ from the RDPG model. We show that under certain regularity conditions on the expected density of the network $N\rho_N$  (or equivalently on the eigengap $s_K(\bP)$) namely, $s_K(\bP) \asymp N\rho_N  =\omega(T+\log^4N)$, the estimated parameter vector $\hat{\mu}$, suitably normalized, converges to a multivariate normal distribution with finite variance around the true parameter vector $\mu^H$ as long as $K^2 = o( N)$ and $\log K=o( T)$. The rates of convergence are different for different sets of parameters, namely, it is $\sqrt{T}$ for the parameters $\bH\T\beta$, while it is $\sqrt{NT}$ for the remaining model parameters (Theorems \ref{thm-enar-ev-clt-NT} and \ref{thm-enar-ev-clt-NTK}). To compare with existing results in related literature, the asymptotic growth rates necessary for the CNAR model in \cite{chen2023community} (which albeit is a different but related model) is $K^4 = o( N)$ and $K^2 \log K =o( T)$ and the eigengap $s_K(\bP) = \omega( K \sqrt{NT\rho_N})$. Therefore, our results on the ENAR model require strictly weaker assumptions on the eigengap and the necessary sample size. In the simulation, we see that ENAR can consistently estimate the latent effects and network effects, yet it achieves comparable or better prediction performance than CNAR and NAR.

We further show that when the model is instead augmented with spectral embeddings $\bU_{\bP} \bS_{\bP}^{1/2}$, the error of approximation from estimated spectral embedding is often too large for the time series problem and leads to inconsistent estimation (Claim \ref{clm-enar-se-err} of the proof of Theorem \ref{thm-enar-se-clt-N}).

\subsubsection{Peer effect inference results} Second, for the case of finite $T$ (with $T \geq 2$), we are interested in the accurate inference of the peer influence parameter $\theta$. We start with results under the ENAR model that uses eigenvectors $\bU_{\bP}$ to model the effects of latent homophily. Our result shows if  $N \rho_N =\omega(\log ^4 N)$ and $N = \omega(K^2)$, while  $\hat{\beta}_{ev} - \bH_{ev}\T\beta_{ev} = O_\pr(1)$, the rest of the parameters including the peer influence parameter is estimated with a $\sqrt{N}$ convergence to a multivariate normal distribution with finite variance (Theorems \ref{thm-enar-clt-N} and \ref{thm-enar-clt-NK}). 

We find that for the peer effect problem, the more interesting case is when the true data generating model contains the spectral embedding vectors $\bU_{\bP}\bS_{\bP}^{1/2}$ as latent homophily as opposed to eigenvectors. In Theorem \ref{thm-enar-se-underK-consist-N}, we show that \textit{ignoring latent homophily in terms of spectral embedding vectors leads to inconsistent estimation of peer effect} in the case of dense networks, and less than $\sqrt{N}$ rate of consistency even for the sparse case, as $\sqrt{N}(\hat\theta_{se,-}-\theta_{se})=O_\pr(\sqrt{N}\rho_N)$. On the other hand, our method that fits least squares with estimated spectral embeddings $\hat{\bU}_{\bA}\hat{\bS}_{\bA}^{1/2}$ leads to consistent and asymptotically normal estimators of peer effect for fixed $K$ (Theorem \ref{thm-enar-se-clt-N}) and growing $K$ (Theorem \ref{thm-enar-se-clt-NK}) regimes. For the growing $K$ case, under conditions $N \rho_N=\omega( \log ^4 N)$ and $N =\omega(K^2 \log K)$ we have $\frac{\sqrt{N} \log K}{\sqrt{K}}(\hat\theta_{se}-\theta_{se}) = O_\pr(1)$. Therefore, we have $\sqrt{N}$ consistency for the fixed $K$ case and a slightly slower rate of convergence for the growing $K$ case depending upon the growth rate of $K$. The theoretical guarantees also hold even when $K$ is unknown and is either correctly selected or over-selected (Theorem \ref{thm-enar-se-overK-clt-N}). More details on theoretical guarantees for the selection of $K$ are in the following paragraph.

These results can be compared with the results in \cite{mcfowland2021estimating,nath2022identifying} where the authors showed asymptotic unbiasedness of the peer influence parameter under the SBM and the RDPG models respectively. In contrast, our results hold for RDPG models with growing dimensions of the latent space $K$ and generalize and supplement those results to include consistency and asymptotic normality.  

 The theoretical results for both problems are succinctly summarized in Table \ref{tab:theory}.

\subsubsection{Selection of \texorpdfstring{$K$}{K} and results with model misspecification} We address the challenging problem of selecting the number of latent vectors to include in the ENAR model (or equivalently the dimension of the latent homophily vectors). While this is an important problem for practitioners, previous works in both network time series literature (\cite{chen2023community}) and Peer effect literature \cite{mcfowland2021estimating,nath2022identifying} did not address this problem. We develop a criterion for the selection of the number of latent vectors that involve a modified residual sum of squares and a penalty term for model complexity. We prove in Theorem \ref{thm-enar-overK} that the procedure does not underselect the number of latent vectors $K$ both when the eigenvectors are used or when the spectral embedding vectors are used, i.e., $\pr\lrp{\hat K\ge K}\to 1$. We complement this result with results that show the same level of theoretical guarantees on the estimators described above are available when the selected $K$ is higher than the true $K$, and consequently, we include some additional latent vectors in the model (Theorems \ref{thm-enar-ev-overK-clt-NT} and \ref{thm-enar-se-overK-clt-N}).

\begin{table}[h]
    \centering
    \begin{tabular}{p{3cm}|p{1.5cm}|p{4cm}|p{1.5cm}|p{2cm}}
       \multicolumn{5}{c}{\textbf{Time series problem} ($T \to \infty$, $N \to \infty$)} \\ &&&&\\ 
       Model & Rate of consistency & Growing $K$ conditions & Consistency with selection of $K$ & Effect of omitting \\ \hline &&&&\\
       eigenvectors $\bU_{\bP}$ & $\sqrt{NT}$ (Thm \ref{thm-enar-ev-clt-NT}) & $N=\omega(K^2)$, $T=\omega(\log K)$,$ N\rho_N=\omega(T+\log^4N)$ 
 (Thm \ref{thm-enar-ev-clt-NTK})& Yes (Thm \ref{thm-enar-overK} and Thm \ref{thm-enar-ev-overK-clt-NT}) & Higher rate of bias (Thm \ref{thm-enar-ev-underK-consist-NT}) \\
       &&&&\\
 \hline \\
       \multicolumn{5}{c}{\textbf{Peer effect problem}  ($N \to \infty$, finite $T$)} \\ \\
        Model & Rate of consistency & Growing $K$ conditions & Consistency with selection of $K$ & Effect of omitting \\ \hline
        eigenvectors $\bU_{\bP}$ & $\sqrt{N}$ (Thm \ref{thm-enar-clt-N}) & $N=\omega(K^2), N\rho_N=\omega(\log^4 N)$ for $\sqrt{N}$ consistency (Thm \ref{thm-enar-clt-NK}) & Yes (Thm \ref{thm-enar-overK} and Thm \ref{thm-enar-ev-overK-clt-NT}) & Consistent, but higher bias (Thm \ref{thm-enar-ev-underK-consist-NT})\\ &&&& \\
        embedding $\bU_{\bP}\bS_{\bP}^{1/2}$& $\sqrt{N}$ (Thm \ref{thm-enar-se-clt-N}) & $N\rho_N=\omega\lrp{\log^4N}$ and $N=\omega(K^2 \log K)$ for $\frac{\sqrt{N}\log K}{\sqrt{K}}$ consistency (Thm \ref{thm-enar-se-clt-NK}) & Yes (Thm \ref{thm-enar-overK} and \ref{thm-enar-se-overK-clt-N}) & Inconsistent (Thm \ref{thm-enar-se-underK-consist-N})\\ 
        &&&&
       \\
        \hline
    \end{tabular}
    \caption{Table for rates of consistency of peer effect ($\theta$) and momentum effect parameters ($\alpha$) under various model and asymptotic setup}
    \label{tab:theory}
\end{table}

\subsubsection{Omitted variable bias when latent vectors are ignored} In Theorems \ref{thm-enar-ev-underK-consist-NT} and \ref{thm-enar-se-underK-consist-N}, we study the omitted variable bias due to under-selecting the latent vectors and as a special case the situation of not including the latent vectors at all in the fitted model. First, to study this situation, we derive new results on the concentration of eigenvectors and spectral embedding vectors when the dimension is under-selected in Theorem \ref{thm-rdpg-cnt2}. In Theorem \ref{thm-enar-ev-underK-consist-NT} we see that $\sqrt{NT}(\hat\theta_{ev,-}-\theta_{ev})=O_\pr\lrp{\sqrt{\frac{T}{N}}}$, while in Theorem \ref{thm-enar-se-underK-consist-N} we see that $\hat\theta_{se,-}-\theta_{se}=O_\pr\lrp{\rho_N}$ when the latent vectors are omitted. This tells us that the model with eigenvectors is unable to induce effective confounding and the omitted variable bias asymptotically vanishes. On the other hand, the spectral embedding vectors introduce significant omitted variable bias and are therefore well suited as a model for the peer effect problem. However, in both cases, ignoring the latent vectors leads to higher bias as can be seen by comparing rate of convergence of the bias terms between those theorems and the corresponding theorems with latent vectors.

\subsubsection{Difference between eigenvectors and spectral embedding vectors}

We consider the conceptual and theoretical differences between using spectral embedding vectors and eigenvectors from the network in the outcome model. As can be seen from Table \ref{tab:theory}, for the peer effect problem, the true data-generating model should include latent positions or embedding vectors and not eigenvectors to induce the desired confounding. Instead, if the true model contains eigenvectors then the peer effect parameter is asymptotically identified as the omitted variable bias asymptotically goes away and one does not need to do anything to remove the confounding. On the other hand, with embedding vectors, the peer effect parameter is inconsistent due to the omitted variable bias and trading off the omitted variable bias with measurement error bias is fruitful since that leads to consistent estimation of peer effect. 

Almost a reverse situation happens in terms of modeling choice for the time series problem. We recommend the model for this problem should include eigenvectors and not latent positions or spectral embedding vectors. This is because when we use estimated spectral embedding vectors in place of the true embedding vectors, the measurement error compounds due to large $T$, and the parameter estimates become inconsistent. However, embedding vectors can be used in the model with a certain scaling factor that depends on $N$ and $T$, as we do for the AMNAR model with latent space position vectors described below.

\subsubsection{Latent space model (LSN) results} We also consider the additive and multiplicative effects of latent space model (LSN) \citep{hoff2021additive,ma2020universal,li2023statistical} for modeling the network data. We incorporate these latent variables from the network model with appropriate scaling into a model that we call the additive and multiplicative NAR (AMNAR) model. For estimation, the latent variables are estimated from a maximum likelihood estimator of \citep{ma2020universal}. We study the consistency and asymptotic normality properties of the resulting estimators for the time series problem (Theorem \ref{thm-amnar-clt-NT}) and the peer effect problem (Theorem \ref{thm-amnar-clt-N}).

\section{Embedding Network Autoregressive Model}
In this section, we describe the proposed Embedding Network Autoregressive (ENAR) model. We start by defining our notations and then describe the NAR and RDPG models, which the ENAR model builds upon. Then, we describe our model with motivations from both multivariate networked time series analysis and causal peer influence estimation literature. Finally, we also propose an extension of ENAR based on the additive and multiplicative latent variables from the LSN model (AMNAR).

Let $a$, $\ba$, and $\bA$ be generic notations for scalars, vectors, and matrices, respectively. Let
$\mathbb{I}_A$ denotes an indicator with a support set $A$.
For a matrix $\bA\in\R^{n\times m}$, write its element at $(i,j)$-th entry as $a_{ij}$. 
Let $\cO_{N,K}:=\brcs{\bH\in\R^{N\times K};\bH\T\bH=\bI_K}$ be a collection of $N\times K$ matrices with real orthonormal columns.
Write an $m$-dimensional vector and $m\times m$ matrix with zeros as $\bo_m$ and $\bO_m$, and let
$\bI_m$ denotes an $m\times m$ identity matrix. 
Let $S^{p-1}=\brcs{\bh\in\R^p;\nrm{\bh}{}=1}$ and $B^{p-1}=\brcs{\bh\in\R^p;\nrm{\bh}{}\le1}$ denote $p$-dimensional unit sphere and ball, respectively.

For an $n$-row square matrix $\bX$, let $\tr\lrp{\bX}\triangleq\sum_{i=1}^nx_{ii}$ denote its trace and $s_i(\bX)$ denote the $i$-th largest magnitude eigenvalue of $\bX$ such that $|s_1(\bX)|\ge\dots\ge|s_n(\bX)|$.
The $i$-th leading eigenvector of $\bX$  will mean the eigenvector corresponding to this $i$th leading eigenvalue.
Following are some matrix norms:
$\ell_{\infty}$-norm $\nrm{\bX}{\infty} \triangleq  \max_i \; \sum_{j} \lvert x_{ij} \rvert$,
$\ell_2$-norm $\nrm{\bX}{}\triangleq\sup_{\bh\in S^{n-1}}\nrm{\bX\bh}{}$, 
max-norm $\nrm{\bX}{\max}\triangleq\max_{i,j}|x_{ij}|$,
and Frobenius norm $\nrm{\bX}{F}\triangleq\tr\lrp{\bX^\top\bX}^{1/2}$.

For $a_n,b_n\geq0$, we write $a_n\gtrsim b_n$ if there exists $c>0$ independent with $n$ such that $a_n\geq cb_n$ for all $n$. 
Also, for all $e>0$, write $a_n=\omega(b_n)$ if  there exists $n_0>0$ such that $a_n> eb_n$ for any $n>n_0$ and write $a_n=\Omega(b_n)$ if there exists $n_0>0$ such that $a_n\ge eb_n$ for any $n>n_0$.
The notation $f_n=\Theta(g_n)$ means that both $f_n=O(g_n)$ and $g_n=O(f_n)$ hold.
We say $X_n=O(Y_n)$ whp. (with high probability) if for any $c>0$ there exist $C=C(c)>0, n_0>0$ such that $\pr\lrp{\vrt{X_n}>C\vrt{Y_n}}<n^{-c}$ for all $n>n_0$. 
Also, we write $X_n=O(Y_n)$ as. if $\vrt{X_n}\lesssim\vrt{Y_n}$ almost surely, and write $X_n=O(1)$ as. if there exists $M>0$ such that $\vrt{X_n}\leq M$ for all $n$ almost surely.
Finally, we denote a weak convergence of a sequence of random variables by $X_n\Rightarrow X$.

\subsection{Network Vector Autoregression}
As stated in the introduction, we assume a statistical problem where we have longitudinal measurements $y_{it}$ on a univariate outcome over $N$ subjects at $T$ time points. We let
$i=1,...,N$ and $t=0,...,T-1 $ with $(N,T\in\mathbb{N})$. We further assume that these $N$ individuals are connected in a network with an (undirected) adjacency matrix $\bA$, which is also observed. We define its normalized (symmetric) Laplacian matrix as $\cL_\bA=\cD^{-1/2}\bA\cD^{-1/2}$ where $\cD$ is a diagonal matrix containing its degrees.
For each unit $i$, we further have measurements on $p$ dimensional covariates $\bz_{it}$, where the subscript $t$ indicates that the covariates may vary over time. Recall the network vector autoregressive model \citep{zhu2017network,mcfowland2021estimating} in Equation \ref{NAR}, $
y_{i,t+1}=\alpha y_{it}+\theta\sum_{j\ne i}\ell_{ij}y_{jt} +{\bz_{it}}\T\gamma+\epsilon_{i,t+1}.$
We assume that $\bz_{it}\in\R^p$ are i.i.d. Gaussian random vectors with independent coordinates with zero mean and finite fourth-order-moments, and $\Sigma_z$ denotes their common covariance. This assumption is similar to one described in \cite{chen2023community}. 
For the model errors, we assume $\epsilon_{it}\overset{iid}{\sim}\cN(0,\sigma^2)$ for $\sigma>0$. Among the parameters, $\alpha \in \R$ denotes the momentum effect, $\theta \in \R$ denotes the peer influence effect, and $\gamma \in \R^p$ denotes the time-invariant covariate effects \citep{zhu2017network}.

Let $\bcE_t \triangleq \lrp{\epsilon_{1t},\ldots,\epsilon_{Nt}}\T \in \R^N$, and $\by_t \triangleq (y_{1t},\ldots,y_{Nt})\T \in \R^N$ be the
vectorized forms of the error term and the response obtained by stacking the corresponding terms for the $N$ individuals. Similarly, let $\bZ_t \triangleq [\bz_{1t},\ldots,\bz_{Nt}]\T \in \R^{N \times p}$ be the matrix of covariates whose $i$th row is $\bz_{it}$, the covariate for the $i$th subject. Then the above model can be expressed in the vector and matrix notations as $
    \by_{t+1} = \alpha \by_t + \theta \cL_\bA \by_t + \bZ_t \gamma + \bcE_{t+1}$.

\subsection{Random Dot Product Graph}

Latent position random graph models assume that a network is created by random edges independently sampled with probabilities that are functions of distance kernels between latent positions of vertices in some underlying latent space. In a $K$-dimensional random dot product graph (RDPG), this distance kernel is the dot product of two $K$-dimensional latent vectors.  
We outline the RDPG model below.

\begin{definition}
Let $\cX$ be a subset of $\R^K$ such that ${\bl_1}\T\bl_2\in[0,1]$ for all $\bl_1,\bl_2\in\cX$. 
Let $K\leq N$ and $\rho_N$ be a sequence such that $\rho_N\in(0,1]$ for all $N$.
Then, $\bA$ is said to follow a random dot product graph with latent positions $\bL=\bxs{\bl_1,...,\bl_N}\T\in\cX^N$ and sparsity factor $\rho_N$, denoted by $\bA\sim\operatorname{RDPG}({\rho_N}, \bL)$, if $a_{ij}\overset{ind.}{\sim}\operatorname{Bernoulli}(\rho_N{\bl_i}\T\bl_j)\mathbb{I}_{\brcs{i\ne j}}$ and $a_{ji}\triangleq a_{ij}$ for $\quad 1\leq i\le j\leq N$.
\end{definition}

Note that the indicator $\mathbb{I}_{\brcs{i\ne j}}$ in the above definition ensures that $\bA$ is hollow without self-loops. 
Using matrix notations, we have $\E[\bA] = \bP$ for $\bP\triangleq\rho_N\bL\bL\T$.
The role of $\rho_N$ is controlling the sparsity of the network. 
For example, the expected degrees are $\sum_{j=1}^Np_{ij}\in\bxs{0,N\rho_N}$ for every $i=1,...,N$. Hence when $\rho_N=1$ $\forall N$, the resulting graph is dense in the sense that the expected number of edges $\sum_{i<j}p_{ij}\sim N^2$.
If $\rho_N\rightarrow0$ as $N\rightarrow\infty$, the graph becomes sparse in the sense that $\sum_{i<j}p_{ij}=o(N^2)$ \citep{Xie2019EfficientEF}.

Under our asymptotic framework where the network size $N$ grows, it's reasonable to expect that the dimension of the latent space, denoted as $K$, will also increase. 
Under moderate sparsity assumption, the difference in eigenvectors of $\bP$ and $\bA$ remains bounded with high probability \citep{lei2015consistency}. 
The estimation of latent positions for random graphs has received extensive attention in the literature \citep{tang2018limit,cape2019signal,Xie2019EfficientEF,rubin2022statistical}, and the results often involve an assumption on the minimum growth rate of $\rho_n$.

\subsection{Embedding Network Autoregression}
Now we define our model which augments the NAR model with latent variables that are common for both the model of the univariate responses and the model for the network. 
Accordingly, in this section, we further assume the network is generated from an RDPG model, $\bA\sim\operatorname{RDPG}({\rho_N}, \bL)$.
We assume that every vertex in $\bA$ is connected to at least one vertex.

The eigenvalue decomposition of $\bP$ can be expressed as $\bP=\bU_\bP\bS_\bP{\bU_\bP}\T$, where $\bU_\bP\in\R^{N\times K}$ contains orthonormal eigenvectors for K largest magnitude eigenvalues, and $\bS_\bP$ is a diagonal matrix containing corresponding eigenvalues $s_1(\bP)\ge\dots\ge s_K(\bP)>0$. 
Write $\bU_{ev}\triangleq\bU_\bP$ and $\bU_{se}\triangleq\bU_\bP{\bS_\bP}^{1/2}$ so that each row of $\bU\in\brcs{\bU_{ev},\bU_{se}}$, denoted by $\bu_i$, contains information about the associated latent variables of each vertex.
Then, we define the ENAR model as a set of two models as follows,
\begin{align}
    y_{i,t+1} & =\alpha y_{it}+\theta\sum_{j\ne i}\ell_{ij}y_{jt}+{\bu_i}\T\beta+{\bz_{it}}\T\gamma+\epsilon_{it}, \\
    a_{ij} & \overset{ind.}{\sim}\operatorname{Bernoulli}(\rho_N{\bl_i}\T\bl_j)\mathbb{I}_{\brcs{i\ne j}}. \nonumber
    \label{eq-enar}
\end{align}
 
 As before, we can write the model in vector and matrix notation as follows,
\begin{align}
\by_{t+1}=\alpha\by_{t}+\theta\cL_\bA\by_{t}+\bU\beta+\bZ_{t}\gamma+\bcE_{t+1},
\end{align}
where now the parameter $\beta$ denotes the global effect of latent variables, and $\bU$ is either the matrix of eigenvectors $\bU_{ev}$ or the matrix of spectral embedding vectors $\bU_{se}$. 
 Finally, assume that $\bZ_{t}$ is independent with $\brcs{\bcE_{t+1},\bcE_t,\ldots}$ and $\brcs{\by_t,\by_{t-1},\ldots}$ for each $t$ and $\bA$ is generated independently with the rest of random components across all $t$.

\subsection{Finite time model for peer influence}
\label{sec-enar-n-enr}

The ENAR model is also motivated by the problem of estimating causal peer influence adjusting for latent homophily in the settings of longitudinal data but perhaps with finite time points (e.g., $T=2$). The model is identical to the ENAR model described in the earlier section, except we do not have a time series, but only a finite number of time points, and the asymptotic setup is for $N \to \infty$.

The problem can be illustrated with a causal diagram similar to \cite{mcfowland2021estimating,nath2022identifying,o2014estimating}. In the causal diagram \cite{pearl2009causality} in Figure \ref{fig:DAG}, the observed variables are represented by rectangles while the unobserved or latent variables are represented by circles. Intuitively, the causal peer influence is the causal effect of the outcome of a ``peer'' who is linked in the network on the outcome of an individual.
The problem of estimating causal peer influence is then estimating the causal effect corresponding to the path $
Y_j^{t} \rightarrow Y_i^{t+1},$
conditioning on the observed network links $a_{ij}$s. 

\usetikzlibrary{positioning}

\tikzset{
    myboxcircle/.style={circle,draw=black,align=center},
}
\tikzset{
    myboxrounded/.style={rounded rectangle,draw=black,align=center},
}
\tikzset{
    myboxrectangle/.style={rectangle,draw=black,align=center},
}

\tikzset{
        > = stealth,
        every node/.append style = {
            text = black
        },
        every path/.append style = {
            arrows = ->,
            line width = .3mm
        },
        hidden/.style = {
            draw = black,
            shape = circle,
            inner sep = 3pt
        }
        ,
        hidden1/.style = {
            draw = black,
            shape = rectangle   
        }
    }
    
\begin{wrapfigure}{l}{0.6 \textwidth}
    \caption{Causal diagram for peer effects}
    \label{fig:DAG}
\begin{center}
    \tikz{
    \node[hidden1][fill=cyan] (a) {$Y_{i}^{t+1}$};
    \node[hidden1] (x) [below left = of a] {$Y_{i}^{t}$};
        \node[hidden1] (x1) [below left = of x] {$Y_{i}^{t-1}$};
    \node[hidden1] (c) [below right = of x] {$a_{ij}$};
    \node[hidden1][fill=green] (y) [below right = of a] {$Y_{j}^{t}$};
     \node[hidden1] (y1) [below right = of y] {$Y_{j}^{t-1}$};
        \node[hidden] (e) [below  left = of c] {\textcolor{red}{$\bu_{i}$}};
                \node[hidden] (f) [below right = of c] {\textcolor{red}{$\bu_{j}$}};
    \node[hidden1] (b) [ below left = of e] {$\bz_{i}$};
        \node[hidden1] (d) [ below right = of f] {$\bz_{j}$}; 

    \path (x) edge (a);
    \path (y) [fill=blue, draw=blue] edge (a);
    \path (e) edge (a);
    \path (e) edge (x);
    \path (e) edge (x1);
    \path (f) edge (y);
    \path (f) edge (y1);
    \path (b) edge (a);
    \path (b) edge (x);
    \path (b) edge (x1);
     \path (d) edge (y);
     \path (d) edge (a);
     \path (d) edge (y1);
    \path (e) edge (c);
    \path (f) edge (c);
    \path (x1) edge (x);
    \path (x1) edge (y);
        \path (y1) edge (x);
    \path (y1) edge (y);
}
\end{center}
\end{wrapfigure}

However,  as the causal diagram shows, there are already several backdoor paths open. Moreover, conditioning on $a_{ij}$ opens additional backdoor paths involving $\bu_i$ and $\bu_j$ since $a_{ij}$ is a collider variable in those paths. Intuitively, we have the same latent variables that are responsible for individuals' network selections (homophily) and also affect the outcomes. It is our conditioning on the observed network that opens those backdoor paths involving the latent variables that lead to confounding. Below, we enumerate all the backdoor paths as follows. These are,
(1) $Y_j^{t}   \leftarrow Y_i^{t-1} \rightarrow Y_i^t  \rightarrow Y_i^{t+1}$, (2) $ 
 Y_j^{t}   \leftarrow \bu_j \rightarrow a_{ij}  \rightarrow Y_i^{t+1}$, (3) $
 Y_j^{t}   \leftarrow \bu_j \rightarrow a_{ij} \leftarrow \bu_i \rightarrow Y_i^{t+1} $, (4) $
  Y_j^{t}   \leftarrow \bu_j \rightarrow a_{ij} \leftarrow \bu_i \rightarrow Y_i^t \rightarrow Y_i^{t+1}$, (5) $
    Y_j^{t}  \leftarrow \bu_j \rightarrow a_{ij} \leftarrow \bu_i \rightarrow Y_i^{t-l} \ldots \rightarrow Y_i^{t-1} \rightarrow Y_i^t \rightarrow Y_i^{t+1},$ and several paths that involve $\bz_i$ and $\bz_j$.
The first backdoor path can be closed by conditioning on $Y_i^t$, while the backdoor paths involving $\bz_i$ and $\bz_j$ can be closed by conditioning on those observed covariates. However, as we can see from the other open backdoor paths, we need to condition on $\bu_i$ and $\bu_j$ to close all of those backdoor paths. Therefore the linear structural equation model that we want to estimate is
\[\by_{t+1}=\alpha\by_{t}+\theta\cL_\bA\by_{t}+\bU\beta+\bZ_t\gamma+\bcE_{t+1}.\]

Here $\theta$ is our target peer influence parameter that we want to estimate consistently. We emphasize that $\bU$ is a latent variable that is not observed. We theoretically show that we \textit{can} estimate the structural peer effect parameter $\theta$ consistently under our modeling assumptions with the methodology described below.

We also consider a regression model for data observed at a single time point ($T=1$). Here our goal is to model a response observed only once as a function of several covariates or predictors while controlling for latent homophily variables that may be correlated with the covariates whose effects we want to estimate. Accordingly, we propose the Embedding Network Regression (ENR) model as,
$\by=\alpha\bon_N+\bU\beta+\bZ\gamma+\bcE$. 
This model has appeared in various forms previously in the literature \cite{fosdick2015testing,he2019multiplicative,le2022linear}. However, we are not aware of a study of the theoretical properties of estimators of this specific model. In addition, this model is a network analog of the popular spatial confounding regression model used in spatial data analysis \cite{guan2023spectral}.

\subsection{Additive and multiplicative latent variables from latent space}\label{sec-lsn}

Finally, we consider an additive and multiplicative latent space model for the network data and propose to include both the additive and multiplicative latent variables in the network autoregressive model. 
Let $\bq_i,v_i$ be $K$-dimensional real vector and real scalars representing the multiplicative homophily and additive degree (or activity) parameter of node $i$, respectively.
Let $\sigma:\R\to\R$ be a known smooth and increasing link function that maps latent variables.
Then, a latent space network can be generated via the density
$f_{ij}\triangleq f\lrp{\cdot;\sigma\lrp{{\bq_i}\T\bq_j+v_i+v_j}}$ \citep{li2023statistical,ma2020universal}. 
Let $\bx_i\triangleq\lrp{\bq_i\T,v_i}\T$ denote a parameter vector containing all latent variables associated with node $i$. 
Collecting all latent variables for the network, we obtain $\bQ\triangleq\bxs{\bq_1,...,\bq_N}\T$ and $\bv\triangleq\lrp{v_1,...,v_N}\T$ hence $\bX\triangleq\bxs{\bQ\,\vert\,\bv}\in\R^{N\times\lrp{K+1}}$.
Here, we treat $\bX$ as fixed parameters.
If we let $\chi\triangleq\bQ\bQ\T+\bv{\bon_N}\T+\bon_N\bv\T\in\R^{N\times N}$ denote a latent variable matrix, then we can define the network connectivity $\bP$ as $p_{ij}\triangleq\sigma(\chi_{ij})$.
Then, the log-likelihood of $\chi$ becomes
\[f(\bA;\chi)=\sum_{i=1}^n\sum_{j>i}\log f\lrp{a_{ij};p_{ij}}=\sum_{i}\sum_{j>i}\log f\lrp{a_{ij};\sigma\lrp{\chi_{ij}}}.\]

Equipped with this network model, we define the additive and multiplicative effect network autoregressive model (AMNAR) as follows.
 Assume that a hollow and undirected graph is generated by $a_{ij}\overset{ind}{\sim}f_{ij}$.
 Define global effects of latent variables and degree parameters on responses as $\beta_1\in\R^K$ and $\beta_2\in\R$.
Then, our measurements $y_{it}$ are now assumed to be generated via $y_{i,t+1}=\alpha y_{it}+\theta\sum_{j\ne i}\ell_{ij}y_{jt}+\lrp{{\tilde\rho_{NT}}^{1/2}\bq_i}\T\beta_1+\lrp{{\tilde\rho_{NT}}^{1/2}v_i}\beta_2+{\bz_{it}}\T\gamma+\epsilon_{it}$, 
 for $i=1,...,N$ and $t=0,...,T-1 (N,T\in\mathbb{N})$. 
The multiplier $\tilde\rho_{NT}\in(0,1]$ is a function of $N$ and $T$ and controls the growth rate of latent variables in the AMNAR model.
The above model can be expressed with matrices and vectors as (with $\beta\triangleq\lrp{\beta_1\T,\beta_2}\T\in\R^{K+1}$)
\begin{align}\label{eq-lnar}    
    \by_{t+1}=\alpha\by_{t}+\theta\cL_\bA\by_{t}+{\tilde\rho_{NT}}^{1/2}\bX\beta+\bZ_{t}\gamma+\bcE_{t+1}.
\end{align} 

Our proposals for ENAR and AMNAR models are related to but different from the Community NAR model of \cite{chen2023community}. Similar to CNAR our frameworks use latent variables which are part of network formation, but the use of those factors is quite distinct. As \cite{chen2023community} noted, the CNAR outcome model is different from the peer effect NAR model since it contains the term $\bU\cB\bU\T$ in place of the observed network $\cL_\bA$, where $\cB$ is a matrix of unknown parameters. Our framework allows for the interpretation of peer influence given the observed network, and consistent estimation of peer effects is an important goal for us. In addition, our framework allows us to include more general latent effects than multiplicative factors.

\section{Estimation}

\subsection{Strict stationarity}
\label{sec-station}
We start describing our estimation methodology with a discussion on the stationary distribution of $\by_t$ for ENAR.
Given our aim to establish its asymptotic distribution under both finite and diverging $T$ and growing network size $N$, we first derive a stationary solution for $\by_t$.
We denote $\pr^*\triangleq\pr\lrp{\cdot|\bA}$, $\E^*\triangleq\E\lrp{\cdot|\bA}$, and $\operatorname{Cov}^*\triangleq\operatorname{Cov}\lrp{\cdot,\cdot|\bA}$, as the conditional probability, expectation, and covariance respectively conditioning on $\bA$. Using notations similar to \cite{zhu2017network}, define $\bG\triangleq\alpha\bI_N+\theta\cL_\bA$ and $\tilde\bcE_{t+1}\triangleq\bZ_{t}\gamma+\bcE_{t+1}$.
Then, we can rewrite the ENAR model equivalently as 
\begin{equation}      
    \by_{t+1}=\bU\beta+\bG\by_{t}+\tilde\bcE_{t+1}.
    \label{eq-enar_vec}
\end{equation}

When $N$ is fixed, the following results hold.
\begin{theorem}\label{thm-enar-station}
    If $|\alpha|+|\theta|<1$, then there is a unique strictly stationary solution to the ENAR model \ref{eq-enar_vec} with a finite first moment, and the solution is given by,
    \begin{align}\label{eq-enar-stationsol}
        \by_t=(\bI_N-\bG)^{-1}\bU\beta+\sum_{j=0}^\infty\bG^j\tilde\bcE_{t-j}
    \end{align}
    where the latent variable $\bU$ is either $\bU_{ev}$ or $\bU_{se}$.
\end{theorem}
\begin{lemma}\label{lmm-enar-meancov}
    Define $\Gamma(h)\triangleq\operatorname{Cov}^*(\by_t,\by_{t-h})$ for all $t$. 
    Upon the conditions in Theorem \ref{thm-enar-station} and conditional on $\bA$, \ref{eq-enar-stationsol} follows a normal distribution with the mean 
    $\varphi\triangleq(\bI_N-\bG)^{-1}\bU\beta$
    and $\vecop\Gamma(0)=(\bI_{N^2}-\bG\otimes\bG)^{-1}\vecop\brcs{\lrp{\sigma^2+\gamma\T\Sigma_z\gamma}\bI_N}$ and $\Gamma(h)=\begin{cases}
    \bG^h\Gamma(0)       & ,h>0 \\
    \Gamma(0)\bG^{-h} & ,h<0. 
    \end{cases}$.
\end{lemma}
The proof of Theorem \ref{thm-enar-station}, along with all other theorems and lemmas, is contained in the Appendix. These results closely resemble the stationarity results presented in \cite{zhu2017network} for the model without the latent effects. 
In addition, we now need to distinguish the mean $\E(\by_t) = \varphi$ between the cases $\bU = \bU_{ev}$ and $\bU = \bU_{se}$ by denoting them as $\varphi_{ev}$ and $\varphi_{se}$, respectively, since they correspond to different means under different modeling choices.
For simplicity, we will denote $\Gamma(0)$ by $\Gamma$ henceforth. Next, we note that $\by_t$ is asymptotically stationary in the sense of Definition \ref{def-sstation} in the Appendix when the network size $N$ grows. The following theorem is also proved in the Appendix.
\begin{theorem}\label{thm-enar-station2}
Upon the conditions in Theorem \ref{thm-enar-station} with $N\rightarrow\infty$, \ref{eq-enar-stationsol} is a unique strictly stationary solution with a finite first moment for either latent variable $\bU_{ev}$ or $\bU_{se}$. i.e., $\max_{1\leq i<\infty}\E\vrt{y_{it}}<\infty$.
\end{theorem}

\subsection{Spectral Embedding under RDPG}
The ENAR model embeds the latent variables using the population spectra of an expected adjacency matrix--which represents the connection probability between entities.
One challenge in estimating the parameters of the ENAR model is that these latent vectors $\bU$ are unobservable.
To address this, we can utilize the asymptotic properties governing the differences between population spectra and sample spectra.
According to a line of works in RDPG literature \citep{athreya2017statistical,tang2018limit,Xie2019EfficientEF,rubin2022statistical}, if $\bA$ follows $\operatorname{RDPG}({\rho_N}, \bL)$ then we have $\nrm{\bA-\bP}{}=O(\sqrt{N\rho_N})~whp.$ provided that $N\rho_N=\omega(\log^4N)$.
Also, \cite{Xie2019EfficientEF} noted that provided the existence of $\lim_{N\to\infty}\frac{1}{N}\bL\T\bL$ as a positive definite matrix, we have $|s_i(\bA)|=\Theta(N\rho_N)~whp.$ for all $i=1,\dots,N$ by Weyl's inequality and $s_i(\bP)=\Theta(N\rho_N)$ for all $i=1,\dots,K$ deterministically. These results are comparable to the rate in \cite{cape2019signal} for ``signal-plus-noise'' models where the authors assume that the smallest non-zero eigenvalue of $\bP$ is of $\Omega (N\rho_N)$ for eigenvector deviation results.

There are many concentration results available for the subspace approximation problem, and we highlight the works devoted to the analysis of entrywise approximation of adjacency spectral embedding (ASE) in the case of RDPG \citep{tang2017semiparametric, athreya2017statistical, tang2018limit, cape2019two, Xie2019EfficientEF, rubin2022statistical, xie2024entrywise}.
Specifically, we employ the construction of the RDPG model with fixed latent positions.
\begin{definition}[\citealp{Xie2019EfficientEF}]
    The adjacency spectral embedding (ASE) of $\bA$ into $\R^K$ is given by $\bU_{\bA}|\bS_{\bA}|^{1/2}$ for $(|\bS_\bA|)_{ij}\triangleq|(\bS_\bA)_{ij}|$ where $\bS_{\bA}\in\R^{K\times K}$ is the diagonal matrix with the $K$ largest magnitude eigenvalues of $\bA$ and $\bU_\bA\in\cO_{N,K}$ is the corresponding eigenvectors of $\bA$.
\end{definition}
We prove the following concentration result for ASE in terms of the spectral norm instead of the more commonly used Frobenius norm by compiling known proof techniques from the literature while tracking the order in terms of $K$.
\begin{proposition}\label{prp-rdpg-cnt}
    Assume that $\bA\sim\operatorname{RDPG}(\rho_N,\bL)$ and there exists constants $c,N_0>0$ such that $cN\rho_N\le s_K(\bP)$ for all $N>N_0$.
    Then, if $N\rho_N=\omega\lrp{\log^4N}$, there exists a sequence of matrices $\bH_{ev},\bH_{se}\in\cO_K$ such that
    \[\nrm{\bU_\bA-\bU_\bP\bH_{ev}}{}=O\lrp{\frac{1}{\sqrt{N\rho_N}}}~whp.\]
    and
    \[\nrm{\bU_\bA|\bS_\bA|^{1/2}-\bU_\bP{\bS_\bP}^{1/2}\bH_{se}}{}=O\lrp{1+\sqrt{\frac{K\log^2N}{N\rho_N}}}~whp.\]
\end{proposition}
Proof of Proposition \ref{prp-rdpg-cnt} is contained in Section \ref{sec-prf-ase-cncnt}.
Therefore, we propose to use $\hat\bU_{ev}\triangleq\bU_\bA$ and $\hat\bU_{se}\triangleq\bU_\bA|\bS_\bA|^{1/2}$ as the approximations of unknown latent variables modeled by eigenvector and embedding, respectively.

\subsection{Least Squares Estimation of Parameters}
From the equation of the model \ref{eq-enar}, we obtain the linear regression representation
$ y_{i,t+1} = \bw_{i,t}\T\mu + \epsilon_{i,t+1}$,
where $\bw_{it}\T\triangleq(\bu_i\T,y_{it},\ell_{i\cdot}\T\by_{t},\bz_{it}\T)$ and our parameters of interest, $\mu\triangleq(\beta\T,\alpha,\theta,\gamma\T)\T\in\R^{K+p+2}$. Thus, the auto-regression of the networked measurements at time $t+1$ can be written as
\[ \by_{t+1} = \bW_{t}\mu + \bcE_{t+1} \]
where $\bW_t=[\bw_{1t},...,\bw_{Nt}]\T$. We can further collect the entire time series as $\by\triangleq({\by_{1}}\T,...,{\by_{T}}\T)\T$, $\bW\triangleq\bxs{{\bW_0}\T,...,{\bW_{T-1}}\T}\T$, and $\bcE\triangleq\lrp{{\bcE_1}\T,...,{\bcE_T}\T}\T$, thereby obtaining the representation $\by=\bW\mu+\bcE$ in $\R^{NT}$. Note that $\bW$ contains the population latent variables $\bU$ from the network model. Therefore, it is interpreted as the population design matrix for the ENAR model \citep{chen2023community}. Utilizing the estimated latent variables $\hat\bU$, we can obtain the approximated version $\hat\bW_t=[\hat\bw_{1t},...,\hat\bw_{Nt}]\T$ for ${\hat\bw_{it}}\T\triangleq({\hat\bu_i}\T,y_{it},{\ell_{i\cdot}}\T\by_{t},{\bz_{it}}\T)$ and $\hat\bW$ accordingly.

This expression naturally motivates the least squares estimation of $\mu$. Therefore, we target the least squares estimator 
\[ \hat\mu = \lrp{\hat\bW\T\hat\bW}^{-1}\hat\bW\T\by = \lrp{{\hat\beta}\T,\hat\alpha,\hat\theta,{\hat\gamma}\T}\T \]
and study its asymptotic properties as an estimator of $\mu$. We establish the asymptotic distribution of the estimator through non-asymptotic concentration results and martingale central limit theorem \citep{hall2014martingale,zhu2017network,chen2023community}.
To avoid notation abuse, we write $\hat\mu_{ev}$ and $\hat\mu_{se}$ to denote the least square estimators with $\hat\bU=\hat\bU_{ev}$ and $\hat\bU=\hat\bU_{se}$, respectively.

\subsection{Estimation with LSN model}
\label{sec-amnar-estimation}
The discussion of model stationarity for AMNAR is very similar to that of the ENAR model, and their proofs are contained in the Appendix \ref{lnar-station-prf}.

Our main interest is to estimate the parameters of AMNAR consistently, but similarly to ENAR, the latent variables $\bX$ are not observable.
We tackle this issue by introducing the maximum likelihood estimation with Lagrange adjustment given in \cite{li2023statistical} where the authors found its asymptotic properties as well.
Let $\hat\bX\triangleq\arg\max_{\bX\in\Xi_K}\log f\lrp{\chi;\bA}$ denote the maximum likelihood estimator over the constrained parameter space $\Xi_K\triangleq\brcs{\bX\in\R^{N\times(K+1)} \,;\,{\bQ}\T\bon_N = \bo_K, \,{\bQ}\T{\bQ} \text{ is diagonal, } \nrm{\bX}{\tti} = O(1)\text{ as }N\rightarrow \infty}.$
The following proposition is analogous to Theorem 3.3 of \cite{li2023statistical} with slightly modified assumptions stated in the Appendix section \ref{sec-amnar-app}.
\begin{proposition}
    \label{prp-lsn-concent}
    Under assumptions on latent space model given in section \ref{sec-amnar-app} in Appendix, we have $\nrm{\hat\bX-\bX}{F}=O_\pr(1)$.
\end{proposition}

Following the same estimation strategy as in the ENAR model, we begin with the expression
$ y_{i,t+1} = {\bmm_{it}}\T\mu + \epsilon_{i,t+1}$,
where ${\bmm_{it}}\T\triangleq\lrp{{\tilde\rho_{NT}}^{1/2}{\bx_i}\T,y_{it},{\ell_{i\cdot}}\T\by_{t},{\bz_{it}}\T}$ and our parameters of interest, $\mu_m\triangleq(\beta_m\T,\alpha_m,\theta_m,\gamma_m\T)\T\in\R^{K+p+3}$. 
Thus, the autoregression of the networked measurements at time $t+1$ can be written as
\[ \by_{t+1} = \bM_{t}\mu_m + \bcE_{t+1}\in\R^N\]
and obtain the representation for the total observed data as $\by=\bM\mu_m+\bcE$ in $\R^{NT}$. Note that $\bM$ contains the latent variables $\bX$. 
Therefore, the true design matrix $\bM$ can be approximated by utilizing Proposition \ref{prp-lsn-concent} as $\hat\bM_t=[\hat\bmm_{1t},...,\hat\bmm_{Nt}]\T$ for ${\hat\bmm_{it}}\T\triangleq({\tilde\rho_{NT}}^{1/2}{\hat\bx_i}\T,y_{it},{\ell_{i\cdot}}\T\by_{t},{\bz_{it}}\T)$ and $\hat\bM$ accordingly. 
Then, the least squares estimator for $\mu_m$ can be easily found as $\hat\mu_m=\lrp{{\hat\bM}\T\hat\bM}^{-1}{\hat\bM}\T\by = \lrp{{\hat\beta_m}\T,\hat\alpha_m,\hat\theta_m,{\hat\gamma_m}\T}\T$ which minimizes the residual sum of squares.

\section{Large Sample Results}
\label{sec-enar-consist}

We next develop a theory on consistency and asymptotic normality of the least squares estimator of the ENAR model under an asymptotic setup where we always assume $N \to \infty$, but consider both finite $T$ case and the case of $T \to \infty$ separately. We start with the results for the ENAR model with eigenvectors from the RDPG network model and then discuss the results for the case spectral embedding vectors and the latent position vectors from the LSN model.
In this section, we write  $\mu_f^H\triangleq(\beta_f\T\bH_f,\alpha_f,\theta_f,\gamma_f\T)\T$, where $\bH_f \in \cO_K$ is an arbitrary matrix with orthonormal columns from Proposition \ref{prp-rdpg-cnt} and $f\in\brcs{ev,se}$.
We differentiate between $\mu_{ev}$ and $\mu_{se}$ as they represent different modeling choices.

\subsection{Growing \texorpdfstring{$T$}{T} results}

The first two theorems stated below show the asymptotic normality of the least squares estimator when $\bU=\bU_{ev}$ and both $N$ and $T$ grow under two cases, first when the number of latent dimensions $K$ is fixed and then second when $K$ also grows along with $N$ and possibly also $T$. 
\begin{theorem} 
    \label{thm-enar-ev-clt-NT} Assume that $|\alpha_{ev}|+|\theta_{ev}|<1$ and $N\rho_N =\omega(T+\log^4N)$.
    \[
        \Sigma_{ev}\triangleq\bxs{\begin{matrix}
        \bI_K & (0) & (0) \\
        & \Sigma_{\alpha\theta} & (0)\\
        & & \Sigma_z
        \end{matrix}},\quad
        \Sigma_{\alpha\theta}\triangleq\bxs{\begin{matrix}
            \tau_2 & \tau_{23}\\
            & \tau_3
        \end{matrix}}
    \]
    with constants
    \[\tau_2\triangleq\lim_{N\rightarrow\infty}\frac{1}{N}\E\tr\lrp{\Gamma}, \tau_{23}\triangleq\lim_{N\rightarrow\infty}\frac{1}{N}\E\tr\lrp{\Gamma\cL_\bA},
    \tau_3\triangleq\lim_{N\rightarrow\infty}\frac{1}{N}\E\tr\lrp{\Gamma{\cL_\bA}^2}.\]
    Then, for $\bD_{ev}\triangleq\sqrt{T}\operatorname{diag}\lrp{\bI_{K},\sqrt{N}\bI_{p+2}}$, we have, as $N,T\rightarrow\infty$,
    \begin{align*}
        \bD_{ev}\lrp{\hat\mu_{ev}-\mu_{ev}^H}\Rightarrow\cN\lrp{\bo_{K+p+2},\sigma^2{\Sigma_{ev}}^{-1}}.
    \end{align*}
\end{theorem}
The condition $|\alpha_{ev}|+|\theta_{ev}|<1$ is required to ensure that $\by_t$ has the stationary distribution discussed in section \ref{sec-station} as $N,T\rightarrow\infty$.
It is noteworthy that to consistently estimate the parameters, the population network density $N\rho_N$ should grow faster than $T$ and $\log^4N$ in Theorem \ref{thm-enar-ev-clt-NT}. 
The growth rate needed in Theorem \ref{thm-enar-ev-clt-NT} matches the rate in \cite{cape2019signal} as long as $T = O(\log^4N)$. If $T$ grows faster then, the Theorem \ref{thm-enar-ev-clt-NT} requires a better concentration of $\hat{\bU}$ for ENAR estimation to remain accurate. The convergence rate of $\sqrt{T}$ for the latent position effects $\hat\beta_{ev}$ is slower than $\sqrt{NT}$ for the rest, which matches the rates obtained in \cite{chen2023community} for their CNAR model.

Next, we consider the case where $K$ grows along with $N$ and $T$ as well.

\begin{theorem}
    \label{thm-enar-ev-clt-NTK}
    Assume that $|\alpha_{ev}|+|\theta_{ev}|<1$ and $\nrm{\beta}{}=O\lrp{1}$ as $K\rightarrow\infty$.
    Furthermore, assume that $N\rho_N=\omega\lrp{T+\log^4N}$, $N=\omega (K^2)$, and $T =\omega(\log K)$.
    For a positive integer $m$, suppose we have an $m\times\lrp{K+p+2}$ matrix $\bA_K$ such that $\nrm{\bA_K}{}=O(1)$ as $K\rightarrow\infty$. If we define  $\bV_{ev}\triangleq\lim_{K\rightarrow\infty}\bA_K{\Sigma_{ev}}^{-1}{\bA_K}\T\in\R^{m\times m}$, then we have, as $N,T,K\rightarrow\infty$,
    \[\bA_K\bD_{ev}\lrp{\hat\mu_{ev}-\mu_{ev}^H}\Rightarrow\cN\lrp{\bo_m,\sigma^2\bV_{ev}}.\] 
\end{theorem}

The condition on the population eigen gap $s_K(\bP)=\Theta(N\rho_N)$ is that  $N\rho_N =\omega\lrp{T+\log^4N}$, which improves upon the rate $s_K(\bP)=\omega\lrp{K\sqrt{TN\rho_N}}$, i.e., $N\rho_N=\omega\lrp{K^2T}$ presented in the context of CNAR model in \cite{chen2023community} by a factor of $K^2$ which is a meaningful difference when $K$ is large. Further, to consistently estimate the parameters, we only require $N = \omega(K^2)$ and $T =\omega(\log K)$. This can be compared to the required rates of $N =\omega (K^4)$ and $T =\omega (K^2\log K)$ in \cite{chen2023community}. Hence, consistent estimation in the CNAR model \cite{chen2023community} requires $O(K^2)$ times more sample size both in terms of $N$ and $T$. Therefore, this represents a substantial relaxation of conditions. These reductions are largely due to the ENAR model having $\bU$ entering the model in an additive form as opposed to a multiplicative form in the CNAR model, and consequently, requiring us to estimate fewer parameters attached to latent variables.

\subsection{Finite \texorpdfstring{$T$}{T} results}
Now, we prove asymptotic results for the model in the finite $T$ case which is appropriate for the problem of causal peer influence estimation. This finite $T$ case was not studied in earlier works of \cite{zhu2017network,chen2023community} on network autoregressive models. Although motivated by the peer effect literature, the recent work of \cite{wu2023random} does not study the finite $T$ case either. 
Note that we no longer require $\vrt{\alpha}+\vrt{\theta}<1$ in the finite $T$ case.
Distinctively from the Theorem \ref{thm-enar-ev-clt-NT} and \ref{thm-enar-ev-clt-NTK}, the consistency of $\hat\mu$ given finite $T$ is different as the estimation error for the latent position effects, $\hat\beta_{ev}-{\bH_{ev}}\T\beta_{ev}$, will only be bounded in probability. However, our result still shows the peer influence parameter $\theta_{ev}$ along with other parameters, namely, $\alpha_{ev}$ and $\gamma_{ev}$, which we are typically interested in inferring are all $\sqrt{N}$ consistent.
\begin{theorem}\label{thm-enar-clt-N}
    Assume that $N\rho_N =\omega(\log^4N)$.
    Partition the parameter vector as $\mu_{ev}=\lrp{\beta_{ev}\T,{\mu_{ev,-\beta}}\T}\T$ and $\hat\mu_{ev}$ accordingly as well.
    Let $\Sigma_{ev,-u}\triangleq\begin{bmatrix}
        \Sigma_{\alpha\theta} & (0)\\
        & \Sigma_z
    \end{bmatrix}$.
    Then, we have $\hat\beta_{ev}-{\bH_{ev}}\T\beta_{ev}=O_\pr(1)$ and
    \begin{align}
        \sqrt{N}\lrp{\hat\mu_{ev,-\beta}-\mu^H_{ev,-\beta}}&\Rightarrow\cN\lrp{\bo_{p+2},\frac{\sigma^2}{T}{\Sigma_{ev,-u}}^{-1}}, \quad \text{ as } N\rightarrow\infty.
    \end{align}
\end{theorem}
The result for growing $K$ is as follows.
\begin{theorem}\label{thm-enar-clt-NK}
In Theorem \ref{thm-enar-clt-N}, further assume that $\nrm{\beta}{}=O\lrp{1}$ as $K\rightarrow\infty$, $N\rho_N =\omega(\log^4N)$, and $N =\omega (K^2)$.
For a positive integer $m$, suppose we have an $m\times\lrp{K+p+2}$ matrix $\bA_K$ such that $\nrm{\bA_k}{}=O(1)$ as $K\rightarrow\infty$.
Then, $\bA_K\lrp{\hat\beta_{ev}-{\bH_{ev}}\T\beta_{ev}}=O_\pr(1)$ and we have
\[\sqrt{N}\lrp{\hat\mu_{ev,-\beta}-\mu^H_{ev,-\beta}}\Rightarrow\cN\lrp{\bo_{p+2},\frac{\sigma^2}{T}{\Sigma_{ev,-u}}^{-1}}\]
as $N,K\rightarrow\infty$.
\end{theorem}
As we limit $T$ to be finite, $N\rho_N$ is allowed to grow at less restrictive rates compared to the case where $T$ diverges. However, we still require the number of dimensions to grow at the same rate with the network size $N$ as described in Theorem \ref{thm-enar-ev-clt-NTK}. We discuss the peer effect problem more in the next section with spectral embeddings.

Our next two results are related to the Embedding Network Regression (ENR) model. As aforementioned, ENR is a special case of ENAR with $T=1<\infty$. Specifically, it can be derived from ENAR with finite $T$: let $\by_0\triangleq\bon_N$, $\bZ\triangleq\bZ_0$, and assume that there is no peer influence effect, i.e., $\theta=0$. 

Without loss of generality, we may omit the grand mean effect, $\alpha$.
Therefore, we have reduced data model as ${\bw_{i}^R}\T\triangleq(\bu_{ev,i}\T,{\bz_{i}}\T)$ and $\mu^R\triangleq(\beta_{ev}^R\,\T,\gamma_{ev}^R\,\T)\T\in\R^{K+p}$ hence giving the representation $\by_1=\bW^R\mu_{ev}^R+\bcE_1$ for $\bcE_1\sim\cN(\bo_N,\sigma^2\bI_N)$. With a usual least square estimator $\hat\mu_{ev}^R\triangleq\lrp{{\bW^R}\T\bW^R}^{-1}{\bW^R}\T\by_1$, the following two results are corollaries that follow from Theorems \ref{thm-enar-clt-N} and \ref{thm-enar-clt-NK} respectively.

\begin{corollary}[Fixed $K$]
    Assume that $N\rho_N=\omega(\log^4N)$.
    Then, we have $\hat\beta_{ev}^R-{\bH_{ev}}\T\beta_{ev}^R=O_\pr(1)$ and $
    \sqrt{N}\lrp{\hat\gamma_{ev}^R-\gamma_{ev}^R}\Rightarrow\cN\lrp{\bo_{p},\sigma^2{\Sigma_z}^{-1}}$, 
    as $N\rightarrow\infty$.
\end{corollary}
\begin{corollary}[Growing $K$]
    \label{cor-enar-ev-clt-NK}
    Further assume that $\nrm{\beta_{ev}^R}{}=O\lrp{1}$ as $K\rightarrow\infty$, $N\rho_N=\omega\lrp{\log^4N}$, and $N =\omega (K^2)$.
    For a positive integer $m$, suppose we have an $m\times\lrp{K+p+2}$ matrix $\bA_K$ such that $\nrm{\bA_k}{}=O(1)$ as $K\rightarrow\infty$.
    Then, we have $\bA_K\lrp{\hat\beta^R_{ev}-{\bH_{ev}^R}\,\T\beta_{ev}^R}=O_\pr(1)$ and $\sqrt{N}\lrp{\hat\gamma^R_{ev}-\gamma_{ev}^R}\Rightarrow\cN\lrp{\bo_{p},\sigma^2{\Sigma_z}^{-1}}$
    as $N,K\rightarrow\infty$.
\end{corollary}

\subsection{Results with spectral embeddings for the finite $T$ peer effect problem} The previous section comprehensively dealt with consistency and asymptotic normality for the ENAR model with eigenvectors for both the problems of network time series and peer effect. However, as will be shown later in Section \ref{omitbias}, the model may not be appropriate to induce the confounding necessary for the peer effect problem. Intuitively, because eigenvectors are constrained to have columns whose $\ell_2$ norms are 1 by definition, the entries of the design matrix made with them are very small. In particular, asymptotically, the effect of ignoring the eigenvectors in the model on the peer effect parameter estimates (``omitted variable bias'') is negligible. Hence, this model is not appropriate for the peer effect problem where we expect non-trivial omitted variable bias due to the omission of latent variables. Therefore, we next develop a theory on consistency and asymptotic normality of the least squares estimator for the ENAR model with spectral embedding vectors. In this case, we consider the finite $T$ case, while $N \to \infty$.

The first two theorems stated below show the asymptotic normality of the least squares estimator when $N$ grows under two cases, first when the number of latent dimensions $K$ is fixed and then second when $K$ also grows along with $N$.
Recall from  Proposition \ref{prp-rdpg-cnt} that $s_K(\bP)$ denotes the smallest non-zero eigenvalue of $\bP$.
\begin{theorem} \label{thm-enar-se-clt-N} 
    Assume that $N\rho_N=\omega\lrp{\log^4N}$. Define
    \[\Sigma_{se}\triangleq\begin{bmatrix}
        \bE & \bC & (0) \\
        & \Sigma_{se,\alpha\theta}+\Sigma_{\alpha\theta} & (0) \\
        & & \Sigma_z
    \end{bmatrix},\quad
    \Sigma_{se,\alpha\theta}\triangleq\lim_{N\to\infty}\E\begin{bmatrix}
        \frac{1}{N}{\varphi_{se}}\T\varphi_{se} & \frac{1}{N}{\varphi_{se}}\T\cL_\bA\varphi_{se}\\
        & \frac{1}{N}{\varphi_{se}}\T{\cL_\bA}^2\varphi_{se}
    \end{bmatrix}\]
    where $\bE\triangleq\lim_{N\to\infty}\frac{1}{N\rho_N}\bS_\bP$, 
    $\bC\triangleq\bxs{\bc_{12},\bc_{13}}\in\R^{K\times2}$, and
    ${\bc_{12}}\T,{\bc_{13}}\T\in\R^K$ are vectors with $\ell_2$-norm of $\sqrt{\rho_\infty}$ multiplied by positive constants.
    Then, for $\bD_{se}\triangleq\sqrt{N}\operatorname{diag}\lrp{\sqrt{\rho_N}\bI_{K},\bI_{p+2}}$, we have, as $N\rightarrow\infty$,
    \begin{align*}
        \bD_{se}\lrp{\hat\mu_{se}-\mu_{se}^H}\Rightarrow\cN\lrp{\bb,\frac{\sigma^2}{T}{\Sigma_{se}}^{-1}}
    \end{align*}
    for $\bb\triangleq\lrp{{\bb_1}\T,{\bb_2}\T,{\bo_p}\T}\T$ for some fixed $\bb_1\in\R^K,\bb_2\in\R^2$ such that $\nrm{\bb_1}{}<c_1$ and $\nrm{\bb_2}{}<\sqrt{c_2\rho_\infty}$ for some constants $c_1,c_2>0$.
\end{theorem}
This theorem establishes consistency and asymptotic normality of the peer effect parameter. This result, therefore, expands the results of \cite{mcfowland2021estimating,nath2022identifying} on asymptotically unbiased estimation to $\sqrt{N}$ consistency in the sense that the estimation error as $O_\pr\lrp{N^{-1/2}}$. We note that there is a small $O\lrp{N^{-1/2}}$ asymptotic bias in the estimation of the peer effect parameter. As we will show in Section \ref{omitbias}, the omitted variable bias in this model is substantial, and peer effect parameter estimation is not possible in $\sqrt{N}$-consistent rate (and even maybe inconsistent) if one ignores the latent variable. Next, we consider the case where $K$ grows along with $N$.
\begin{theorem}
    \label{thm-enar-se-clt-NK}
    Assume that $\nrm{\beta}{}=O\lrp{1}$ as $K\rightarrow\infty$.
    Furthermore, assume that $N\rho_N=\omega\lrp{\log^4N}$ and $N=\omega\lrp{K^2\log K}$.
    For a positive integer $m$, suppose we have a matrix $\bA_K\triangleq\bxs{\bA_1\,|\,\bA_2}\in\R^{m\times\lrp{K+p+2}}$ for $\bA_1\in\R^{m\times (K+2)},\bA_2\in\R^{m\times p}$ such that $\nrm{\bA_1}{}=O\lrp{\frac{\log K}{\sqrt{K}}}$ and $\nrm{\bA_2}{}=O(1)$  as $K\rightarrow\infty$. 
    If we define  $\bV_{se}\triangleq\lim_{K\rightarrow\infty}\bA_K{\Sigma_{se}}^{-1}{\bA_K}\T\in\R^{m\times m}$, then we have, as $N,K\rightarrow\infty$,
    \[\bA_K\bD_{se}\lrp{\hat\mu_{se}-\mu_{se}^H}\Rightarrow\cN\lrp{\bo_m,\frac{\sigma^2}{T}\bV_{se}}.\]
\end{theorem}
Compared to Theorem \ref{thm-enar-clt-NK}, we require that $N$ grows faster by a $\log K$ factor so can attain the convergence of the asymptotic covariance (see Lemma \ref{lmm-enar-se-∑nasym}).
As evident from the order of $\nrm{\bA_1}{}$, now the convergence rates of $\hat\beta_{se}$ and $\hat\alpha_{se},\hat\theta_{se}$ are slightly slower by a factor of $\frac{\log K}{\sqrt{K}}$ from $\sqrt{N}$-consistency.

\subsection{Results for additive and multiplicative latent effects model}\label{sec-amnar-consist}
Before starting the asymptotic properties of $\hat\mu_m$, we first define its asymptotic precision matrix as 
\[\Omega_x\triangleq\lim_{N\to\infty}\begin{bmatrix}
    \frac{1}{N}\bQ\T\bQ & \frac{1}{N}\bQ\T\bv \\
    & \frac{\nrm{\bv}{}^2}{N}
\end{bmatrix},\Omega_{\alpha\theta}\triangleq\lim_{N\to\infty}\E\begin{bmatrix}
    \frac{\psi\T\psi}{N} & \frac{\psi\T\cL_\bA\psi}{N}\\
    & \frac{\psi\T{\cL_\bA}^2\psi}{N}
\end{bmatrix},  \Omega_m\triangleq\begin{bmatrix}
    \Omega_x & \Omega_{x,\alpha\theta} & (0)\\
    & \Omega_{\alpha\theta}+\Sigma_{\alpha\theta} & (0)\\
    & & \Sigma_z
\end{bmatrix}\]
for $\Omega_{x,\alpha\theta}\in\R^{(K+1)\times(p+2)}$ with $\Omega_{x,\alpha\theta}\triangleq\bxs{\bc_1,\bc_2}\in\R^{(K+1)\times2}$ such that $\bc_1,\bc_2\in\R^{K+1}$ are constant vectors with $\nrm{\bc_1}{}=c_1{\tilde\rho_\infty}^{1/2}$,$\nrm{\bc_2}{}=c_2{\tilde\rho_\infty}^{1/2}$ for some $c_1,c_2>0$ where $\tilde\rho_\infty\triangleq\lim_{N,T\to\infty}\tilde\rho_{NT}$.
Then, the following result holds for $\hat\mu_m$.
\begin{theorem} \label{thm-amnar-clt-NT}
    Assume that $|\alpha|+|\theta|<1$, $\tilde\rho_{NT}=o\lrp{T^{-1/2}}$, and the assumptions of Proposition \ref{prp-lsn-concent}.
    Then, for
    $\bD_a\triangleq\sqrt{NT}\operatorname{diag}\lrp{{\tilde\rho_{NT}}^{1/2}\bI_{K+1},\bI_{p+2}}$ and as $N,T\rightarrow\infty$, we have
    \[
    \bD_a(\hat\mu_m-\mu_m)\Rightarrow\cN\lrp{\bo_{K+p+3},\sigma^2{\Omega_m}^{-1}}
    \]
\end{theorem}
As Theorem \ref{thm-amnar-clt-NT} implies, we have $\sqrt{NT}$-consistency for $\alpha$ and $\theta$ under AMNAR model as well, and $\hat\beta_m$ converges at the rate of $\lrp{TN\tilde\rho_{NT}}^{1/2}$, which is slower than $\sqrt{N}\sqrt[4]{T}$ by the assumption $\tilde\rho_{NT}=o\lrp{T^{-1/2}}$.
The condition $\tilde\rho_{NT}=o\lrp{T^{-1/2}}$ guarantees that $\alpha,\theta$ are estimated with $\sqrt{NT}$ consistency for the network time series problem.
Since $\tilde\rho_\infty=0$ in this case, it can be seen from the form of $\Omega_{x,\alpha\theta}$, that  $\hat\beta_m$ and $\hat\alpha_m,\hat\theta_m$ are asymptotically uncorrelated under this setup.

However, under finite $T$ case, asymptotically non-negligible correlation can be induced by the scaling of $\lim_{N\to\infty}{\tilde\rho_{NT}}^{1/2}$, while paying the price of estimation error for $\alpha$ and $\theta$.
Such a trade-off has also been observed for the case of ENAR with $\bU_{se}$ as well in Theorem \ref{thm-enar-se-clt-N}.
Next, the asymptotic results for the model under finite time is given below.
\begin{theorem} \label{thm-amnar-clt-N}
    Assume the settings for Theorem \ref{thm-amnar-clt-NT} with finite $T$. Then define $\tilde\bD_a\triangleq\sqrt{N}\operatorname{diag}\lrp{{\tilde\rho_{NT}}^{1/2}\bI_{K+1},\bI_{p+2}}$.
    As $N\rightarrow\infty$, we have 
    \[\tilde\bD_a(\hat\mu_m-\mu_m)\Rightarrow\cN\lrp{\bb,
    \frac{\sigma^2}{T}{\Omega_m}^{-1}}\]
    for $\bb\triangleq\lrp{{\bb_1}\T,{\bb_2}\T,{\bo_p}\T}\T$ for some fixed $\bb_1\in\R^{K+1},\bb_2\in\R^2$ such that $\nrm{\bb_1}{}\le c_1{\tilde\rho_\infty}^{1/2}$ and $\nrm{\bb_2}{}\le c_2\tilde\rho_\infty$ for some constants $c_1,c_2>0$.
\end{theorem}
Therefore, the peer effect parameter in the AMNAR model is estimated at $\sqrt{N}$ rate of consistency and we do not need any assumption on the scaling factors $\tilde{\rho}_{NT}$. This result is consistent with the results we obtained for the peer effect problem with ENAR.

\section{Selection of number of latent vectors \texorpdfstring{$K$}{K} and results under misspecification}
Until now, we have assumed the number of latent vectors to include in the ENAR model $K$ is known. However, in practice, $K$ is unknown and must be obtained from data. In this section, we provide several results for the selection of $K$ and results for the cases of over-selection and under-selection. We develop a strategy for guaranteed \textit{over-estimating} of $K$, 
based on a new model selection criterion developed in this paper. 
The reason for focusing on guarantee for non-under-selection is that we show the consistency and asymptotic normality of some model parameters, including the peer effect parameter, holds under the model misspecification in terms of overestimating $K$. Throughout the section, we assume $K$ does not grow with $N$ and $T$.

\subsection{Over-selection via criterion}\label{sec-enar-IC}
Our goal in this section is to develop a criterion that can guarantee non-under-selection of $K$.
Since the true latent variables $\bU$ and the true latent dimension $K$ are not observable in practice, we frame this as a model selection problem. 
This problem can be addressed using a criterion, leading to the estimation of the number of latent variables as $\hat K$.
Then, we instead estimate $\hat\bU_{\hat K}$ which contains $\hat K$ latent variables derived from the observed network $\bA$.

Let $K$ be the true dimension of the RDPG model and consequently, the true number of latent variables included in the ENAR model.
Then, Consider the following criterion with a penalty function $p_{NT}$ whose order depends on $N,T$ for the dimension determination:
\[Cr_f(k)\triangleq\hat R_f^{(k)}+p_{NT}(k),~k\in\brcs{0,1,...,J_N},~J_N\in\mathbb{N}\]
where $\hat R_f^{(k)}\triangleq\frac{1}{T}{\by}\T\lrp{\bI_{NT}-\hat\Pi_f^{(k)}}\by$ for the projection $\hat\Pi_f^{(k)}$ onto the column space of $\bon_T\otimes\hat\bU_f^{(k)}$ where $\hat\bU_f^{(k)}\in\R^{N\times k}$ is one of eigenvectors $\hat\bU_{ev}^{(k)}$ or spectral embedding $\hat\bU_{se}^{(k)}$ with respect to $k$ largest magnitude eigenvalues of $\bA$. The term $\hat R_f^{(k)}$ is, therefore, a modified residual sum of squares (RSS), namely, the RSS one would obtain by regressing only on the latent variables. The penalty 
$p_{NT}(k)$ is a positive non-decreasing function in $k$.
Then, our choice of $k$ will be $\hat K\triangleq{\arg\min}_{k=0,1,...,J_N}Cr_f(k)$, where $J_N$ is a large number. We provide some guidance on selecting $J_N$ in the Remark \ref{selectJN}.
\begin{theorem}\label{thm-enar-overK}
    Assume that $K\le J_N$ and $T=\omega\lrp{\log K}$.
    Let $p_{NT}$ be a penalty such that $p_{NT}(J_N)=o(1)$ as $N,T\to\infty$. Then, we have
    \[\pr\bxs{\liminf\brcs{\hat K\ge K}}=1\]
    when $N,T,K\to\infty$.
\end{theorem}
\begin{remark}
\label{selectJN}
    (Practical guidance on model selection) 
    It is well known that $N\rho_N=\omega\lrp{\log^2N}$ guarantees $\nrm{\bA}{\infty}\to\nrm{\bP}{\infty}$ as. (see Proposition A.2 of \citealp{tang2017semiparametric} for example) where $\nrm{\bP}{\infty}=O\lrp{N\rho_N}$.
    Then, with an additional condition on $K$ as $K=o(N\rho_N)$, we can choose $J_N$ such that $J_N=O\lrp{\nrm{\bA}{\infty}}$ and $p_{NT}(k)=\frac{k}{N+T}$ for the search of $K$, for example. A second choice of penalty that may be more appropriate for larger $T$ is $p_{NT}(k)=\frac{k^2}{N+T}$. With this penalty, when $T =\omega(N)$, we may choose $J_N=\sqrt{N}$, while for $T=o(N)$ we may choose any $J_N=o(\sqrt{N})$.
\end{remark}
While we have assumed a finite time model for the case of spectral embedding when estimating parameters, we do not put such an asymptotic condition when selecting the number of latent variables $K$, and the above theorem holds for both $\bU_{ev}$ and $\bU_{se}$.

\subsection{Asymptotic results under misspecification of \texorpdfstring{$K$}{K}}
\label{omitbias}
Here, we discuss the theoretical results when we choose $K_o\ne K$ number of latent variables to embed for ENAR.
\subsubsection{When \texorpdfstring{$K_o>K$}{Ko > K}}
Suppose that we have embedded more latent variables than the true $K$ value in ENAR and obtained the over-specified model.
This is guaranteed with a high probability according to Theorem \ref{thm-enar-overK}.
Therefore, we obtain an augmented estimation of latent variables as $\hat\bU_{ K_o}\triangleq\bxs{\hat\bU|\hat\bU_{K:}}\in\R^{N\times K_o}$ with $\hat\bU_{K:}$ which contains $K_o-K$ auxiliary latent variables.
Then, we can pick its population counterpart which is partitioned as $\bU=\bxs{\bU|\bU_{K:}}$.
For the case of spectral embedding, $\bU_{se}=\bU_\bP{\bS_\bP}^{1/2}$ contains original $K$-dimensional latent variable information while $\bU_{se,K:}=\bO_{N\times(K_o-K)}$ because, for all $i>K$, $s_i(\bP)=0$ as the rank of the population network connectivity $\bP$ is $K$.
For eigenvectors we have $\bU_{ev}=\bxs{\bU_\bP|\bU_{\bP,K:}}$ where $\bU_{\bP,K:}\in\R^{N\times(K_o-K)}$ contains the eigenvectors corresponding to those trivial eigenvalues.

Now let us augment the estimate of the design matrix $\hat\bW$ as $\hat\bW_+\triangleq\bxs{\bon_T\otimes\hat\bU_{K:}\,|\,\hat\bW}\in\R^{NT\times( K_o+p+2)}$.
Also, denote the augmented true parameter vector by $\mu_+\in\R^{ K_o+p+2}$ which puts $K_o-K$ null latent effects on those auxiliary latent variables as the true data generating model is assumed to regress on $\bU\in\R^{N\times K}$ only.
Then, naturally we obtain a least square estimator $\hat\mu_+\triangleq\lrp{{\hat\bW_+}\T\hat\bW_+}^{-1}{\bW_+}\T\by$ which can be decomposed as
\[\hat\mu_+=\lrp{{\hat\bW_+}\T\hat\bW_+}^{-1}{\hat\bW_+}\T\lrp{\bW^H\mu^H+\bcE}.\]
To keep the notation consistent, we write $\hat\mu_{se,+},\hat\mu_{ev,+}$ to denote the estimators with spectral embedding and eigenvectors, respectively.
Now we state the central limit theorem under the case of over-selection of $K$.
\begin{theorem} \label{thm-enar-ev-overK-clt-NT}
    Assume that $N\rho_N=\omega\lrp{T+\log^4N}$. Define an augmented asymptotic precision covariance as $\Sigma_{ev,+}\triangleq\bxs{\begin{matrix}
        \bI_{K_o-K} & \bO_{(K_o-K)\times(K+p+2)}\\
        & \Sigma_{ev}
    \end{matrix}}$.
    Then, for an augmented convergence rates $\bD_{ev,+}\triangleq\sqrt{T}\operatorname{diag}\lrp{\bI_{K_o},\sqrt{N}\bI_{p+2}}$ and $\hat\Sigma_{ev,+}$ defined in \ref{eq-ev-hSw+}, we have, as $N,T\to\infty$,
    \begin{align*}
        \bD_{ev,+}\lrp{\hat\mu_{ev,+}-\mu_{ev,+}^H}\Rightarrow\cN\lrp{\bo_{K_o+p+2},\sigma^2{\Sigma_{ev,+}}^{-1}}.
    \end{align*}
\end{theorem}
\begin{theorem} \label{thm-enar-se-overK-clt-N}
    Assume that $N\rho_N=\omega\lrp{\log^4N}$. Define an augmented asymptotic precision covariance as $\Sigma_{se,+}\triangleq\bxs{\begin{matrix}
        \bO_{K_o-K} & \bO_{(K_o-K)\times(K+p+2)}\\
        & \Sigma_{se}
    \end{matrix}}$.
    Then, for an augmented convergence rates $\bD_{se,+}\triangleq\sqrt{N}\operatorname{diag}\lrp{\sqrt{\rho_N}\bI_{K_o},\bI_{p+2}}$ and $\hat\Sigma_{se,+}$ defined in \ref{eq-se-hSw+}, we have, as $N\rightarrow\infty$,
    \begin{align*}
        \hat\Sigma_{se,+}\bD_{se,+}\lrp{\hat\mu_{se,+}-\mu_{se,+}^H}\Rightarrow\cN\lrp{\bb,\frac{\sigma^2}{T}{\Sigma_{se,+}}}
    \end{align*}
    for $\bb\triangleq\lrp{{\bo_{K_o-K}}\T,{\bb_1}\T,{\bb_2}\T,{\bo_p}\T}\T$ for some fixed $\bb_1\in\R^K,\bb_2\in\R^2$ such that $\nrm{\bb_1}{}=c_1$ and $\nrm{\bb_2}{}=\sqrt{c_2\rho_\infty}$ for some constants $c_1,c_2>0$.
\end{theorem}
The above theorems show that for the fixed $K$ case, the same theoretical results on consistency and asymptotic normality of parameter estimates hold when $K$ is over-selected as well. The rates of convergence for all parameters, including the $\beta$ parameters corresponding to the latent factors that are truly in the model are the same as the case when $K$ is known. Note for Theorem \ref{thm-enar-ev-overK-clt-NT}, we let both $N,T \to \infty$ since that pertains to the network time series and peer effect models with eigenvectors, while for Theorem \ref{thm-enar-se-overK-clt-N}, we let only $N \to \infty$ and keep $T$ fixed since the model with spectral embeddings is intended to be used for the peer effect problem only. 

\subsubsection{When \texorpdfstring{$K_o<K$}{Ko < K}}
Now suppose we have selected the number of latent variables $ K_o\ge0$ smaller than the true latent dimension $K$.
Note that the ENAR with $K_o=0$ reduces to the NAR model in the time series context and to the model with omitting latent variables in the peer effect context.
The true latent position information can now be partitioned as $\bU=\bxs{\bU_-|\bU_{:K}}$ where $\bU_-\in\R^{N\times K_o}$ contains first $K_o$ columns of $\bU$ and $\bU_{:K}\in\R^{N\times (K-K_o)}$ contains the next $K-K_o$ vectors of $\bU$ which will be missed in the model fit. 
If $K_o=0$, then we will have $\bU_{:K}=\bU$.

To study the under-selection case, we need new results that quantify the deviation of the estimated $N \times K_0$ matrix of eigenvectors and spectral embedding vectors from the true $N \times K$ matrix of eigenvectors and spectral embedding vectors, respectively. The next theorem proves two concentration inequalities, providing such results that may be of independent interest to researchers in statistical inference in networks.

\begin{theorem}\label{thm-rdpg-cnt2}
    Assume the network $\bA$ is generated from a $K$ dimensional RDPG model, $\bA \sim \operatorname{RDPG}(\rho_N,\bL)$. Let $k\in\brcs{1,\dots,K-1}$ for $K>1$.
    Write a truncated eigenvector of $\bA$ as $\bU_{\bA,k}$ and truncated spectral embedding as $\bU_{\bA,k}|\bS_{\bA,k}|^{1/2}$ which contain $k$ leading eigenvectors (and eigenvalues). Then, provided that $N\rho_N=\omega\lrp{\log^4N}$, there exists a sequence of $\bH_{ev,k},\bH_{se,k}\in\cO_{K,k}$ such that
    \[\nrm{\bU_{\bA,k}-\bU_\bP\bH_{ev,k}}{}=O\lrp{1}~whp.\]
    \[\nrm{\bU_{\bA,k}|\bS_{\bA,k}|^{1/2}-\bU_\bP{\bS_\bP}^{1/2}\bH_{se,k}}{}=O\lrp{\sqrt{N\rho_N}}~whp.\]
\end{theorem}

Now denote the design matrix with $\hat\bU_-$ by $\hat\bW_-\in\R^{NT\times(K_o+p+2)}$ which contains an underly selected number of latent variables.
A least square estimator $\hat\mu_-$ can be obtained as $\lrp{{\hat\bW_-}\T\hat\bW_-}^{-1}{\hat\bW_-}\T\by$.
\begin{theorem}
    \label{thm-enar-ev-underK-consist-NT}
    Assume that $N\rho_N=\omega\lrp{\log^4N}$.
    Then, for $K_o\in\brcs{0,1,\dots,K-1}$ we have 
    \begin{align*}
        \sqrt{T}\lrp{\hat\beta_{ev,-}-{\bH_{ev,-}}\T\beta_{ev}}&=\bb_{ev,\beta}+\bt_1,\quad
        \sqrt{NT}\begin{bmatrix}
            \hat\alpha_{ev,-}-\alpha_{ev}\\
            \hat\theta_{ev,-}-\theta_{ev}
        \end{bmatrix}=\bb_{ev,\alpha\theta}+\bt_2,\\
        \sqrt{NT}\lrp{\hat\gamma_{ev,-}-\gamma_{ev}}&=\bt_3
    \end{align*}
    where ${\bh_1}\T\bb_{ev,\beta}=O\lrp{\sqrt{T}}~whp.,\quad{\bh_2}\T\bb_{ev,\alpha\theta}=O\lrp{\sqrt{T/N}}~whp.,{\bh_1}\T\bt_1=O_\pr(1),{\bh_2}\T\bt_2=O_\pr(1),{\bh_3}\T\bt_3=O_\pr(1)$,
    for any $\bh_1\in B^{K_o-1},\bh_2\in B^1,\bh_3\in B^{p-1}$ as $N,T\to\infty$.
\end{theorem}
In the above Theorem \ref{thm-enar-ev-underK-consist-NT}, the limit of suitably normalized estimators is written in terms of two terms. The terms denoted as $\bt_1, \bt_2, \bt_3$ are all $O_\pr(1)$ and converge to normal distributions as in Theorem \ref{thm-enar-ev-clt-NT}, while the terms denoted by $\bb_{ev,\beta},\bb_{ev,\alpha\theta}$ are asymptotic bias terms. We can compare these limits with the results in Theorem \ref{thm-enar-ev-clt-NT}, where the orders for the estimation errors of $\mu_{sv}^H$ corresponding to $\bb_{ev,\beta},\bb_{ev,\alpha\theta}$ were $O\lrp{\sqrt{\frac{T}{N\rho_N}}}~whp.$, and  $O\lrp{\sqrt{\frac{T}{N^2\rho_N}}}~whp.$ respectively. Therefore, in Theorem \ref{thm-enar-ev-underK-consist-NT}, even though $\bt_2$ will converge to a normal limiting distribution and we will have $\sqrt{NT}$ consistency for $\alpha, \theta$ parameters, the biases in those parameter estimates are of a higher order than in the correct or over-selected $K$ case.

\begin{theorem} \label{thm-enar-se-underK-consist-N}
    Assume that $N\rho_N=\omega\lrp{\log^4N}$. Then, for $K_o\in\brcs{0,1,\dots,K-1}$ we have
    \begin{align*}
        \sqrt{N\rho_N}\lrp{\hat\beta_{se,-}-{\bH_{se,-}}\T\beta_{se}}&=\bb_{se,\beta}+\bt_1,\quad
        \sqrt{N}\begin{bmatrix}
            \hat\alpha_{se,-}-\alpha_{se}\\
            \hat\theta_{se,-}-\theta_{se}
        \end{bmatrix}=\bb_{se,\alpha\theta}+\bt_2,\\
        \sqrt{N}\lrp{\hat\gamma_{se,-}-\gamma_{se}}&=\bb_{se,\gamma}+\bt_3
    \end{align*}
    where ${\bh_1}\T\bb_{se,\beta}=O\lrp{\sqrt{N\rho_N}}~whp.,\quad{\bh_2}\T\bb_{se,\alpha\theta}=O\lrp{\sqrt{N}\rho_N}~whp.,\quad{\bh_3}\T\bb_{se,\gamma}=O_\pr\lrp{\sqrt{\rho_N}},{\bh_1}\T\bt_1=O_\pr(1),{\bh_2}\T\bt_2=O_\pr(1),{\bh_3}\T\bt_3=O_\pr(1)$,
    for any $\bh_1\in B^{K_o-1},\bh_2\in B^1,\bh_3\in B^{p-1}$ as $N\to\infty$.
\end{theorem}
This result can also be compared with the result in Theorem \ref{thm-enar-se-clt-N}, where the orders of estimation error corresponding to the ``bias terms'' $\bb_{se,\beta},\bb_{se,\alpha\theta}$ were $O\lrp{1}~whp.$, and $O\lrp{\sqrt{\rho_N}}~whp.$ respectively. In comparison, the bias terms $\bb_{se,\beta},\bb_{se,\alpha\theta}$ in the above theorem are $O\lrp{\sqrt{N \rho_N}}~whp.$, and $O\lrp{\sqrt{N} \rho_N}~whp.$  respectively. Theorem \ref{thm-enar-se-underK-consist-N} shows that when the true data generating peer effect model contains spectral embedding vectors, ignoring those vectors in the estimation process leads to possibly inconsistent estimation. The estimation error for $\theta_{se}$ is $O_\pr(\rho_N+N^{-1/2})$, which increases as the network gets denser, and in the limit of dense networks, the estimator of the peer effect parameter becomes inconsistent.
In contrast, Theorem \ref{thm-enar-ev-underK-consist-NT} showed that for the model with eigenvectors when $K_0=0$, i.e, all the latent variables are ignored during estimation, the estimation error for $\theta_{ev}$ is $O_\pr\lrp{1/\sqrt{NT}}$. 
It appears that with eigenvectors, there is no detrimental effect on the estimation of peer effect due to omitting latent variables (higher asymptotic bias, but same rate of consistency). This is contrary to the assumptions from the Directed Acyclic Graph, which says that there is non-trivial confounding due to the latent variables. Since modeling choice should be made based on the DAG \citep{pearl2009causality}, the model with spectral embedding vectors is appropriate for the peer effect problem, while the model with eigenvectors is not. Both the theorems also quantify the estimation error when the number of latent vectors selected $K_0$ is less than the true number of latent vectors. 
The asymptotic orders for the parameter estimation errors when $0<K_o<K$ are the same as when $K_0=0$, as both theorems show.

\section{Simulations}
\label{sec-enar-sim}
In this section, we use Monte Carlo simulations to illustrate the finite sample performance of the ENAR model estimator with $\bU=\bU_{ev}$ and compare it with NAR and CNAR. We examine the sensitivity of the considered models under model misspecification in terms of estimations of model parameters and one-step-ahead prediction of $\by_{T+1}$.
In this regard, we consider the scenarios where $\brcs{\by_0,...,\by_T}$ and $\by_{T+1}$ follow each of ENAR, CNAR, and NAR.
In the case where we generate $\by_t$ with CNAR \citep{chen2023community}, we assumed that there is no latent variable structure in the model noise. 

We consider the DCSBM and DCMMSBM for the population distributions of $\bA$, and generate the networks using \texttt{fastRG} package in R \cite{rohe2018note}.
First, we used the matrix $2q\bI_K+q\bon_K{\bon_K}\T$ where $q=\frac{9}{40}$ to generate the $K \times K$ block matrix of connection probabilities.
As a result, the ratio of inter-community and between-community connectivity is 3. The maximum expected degree for each graph was set to be $N\rho_N$ where $\rho_N\triangleq N^{-1/2}$, ensuring that the graphs are sparse. The degree heterogeneity parameters associated with both DCSBM and DCMMSBM were sampled from a standard log-normal distribution. For DCMMSBM, the (mixed) block memberships were generated from a Dirichlet distribution with parameter vector $(1, \ldots, 1)$. For DCSBM, the block memberships were sampled from a categorical distribution with equal probabilities. 
\begin{figure}[h]
    \centering
    \begin{subfigure}{0.5\textwidth}
        \includegraphics[width=1\textwidth]{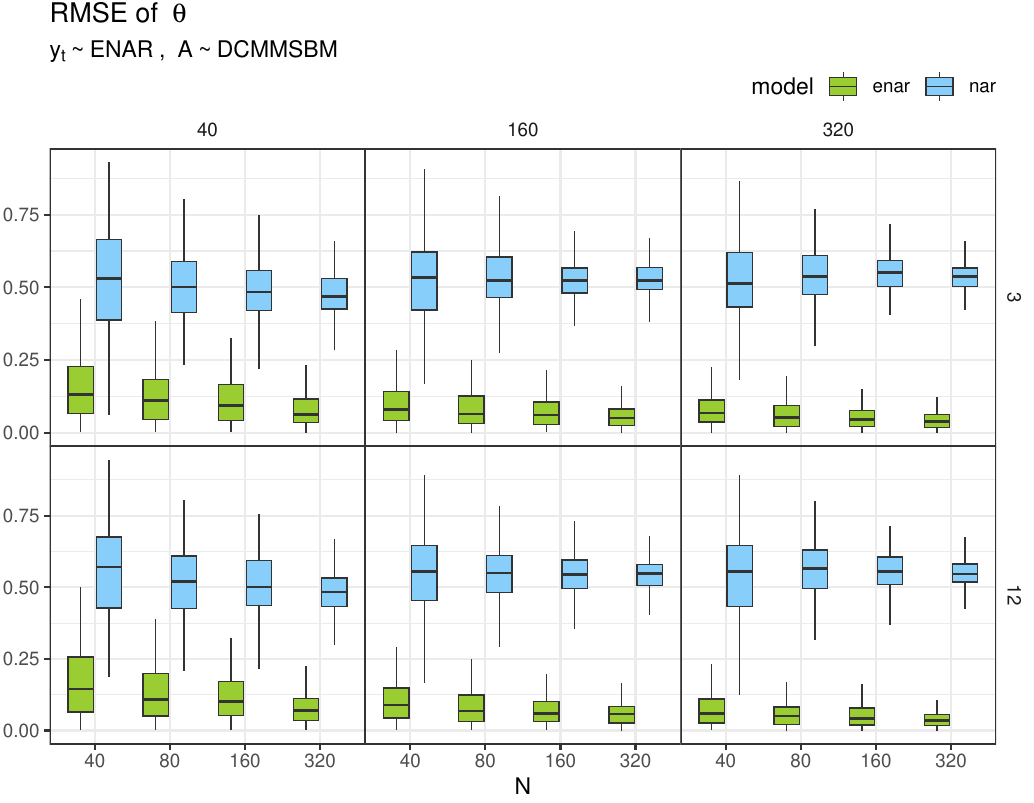}
    \end{subfigure}%
        \begin{subfigure}{0.5\textwidth}
        \includegraphics[width=1\textwidth]{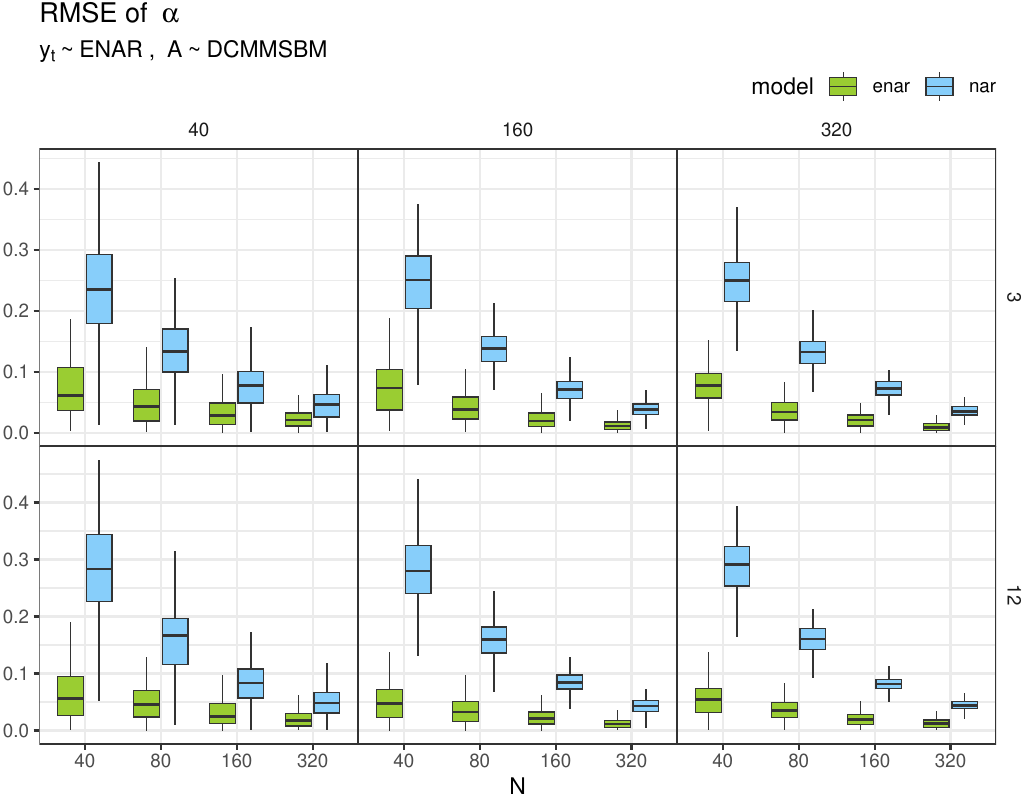}
    \end{subfigure}
    \caption{Boxplot of estimates of $\theta$ (left) and  $\alpha$ (right) from  ENAR and NAR model with increasing $N$ when data is generated from ENAR model with DCMMSBM. The rows corresponds to $K=3$ and $K=12$, and the columns corresponds to $T=40, 160, 320$.
    \vspace{-5pt}}
    \label{fig-enar_dcmmsbm}
\end{figure}
For model parametrization, we set the parameters $\beta^{\rm enar}$ associated with the latent effects as
$\lrp{1,-1/2,...,\lrp{-1}^{K-1}/K}\T\in\R^K$.
For the latent peer effect measured by the community structure of CNAR, we used $\cB_1\triangleq\frac{1}{10}\operatorname{diag}\lrp{\beta^{\rm enar}}$ where the definition of $\cB_1$ comes from \cite{chen2023community}.
For ENAR, and NAR we set $\alpha=\theta=\frac{1}{5}$, and set the covariate effects as $\gamma=\lrp{\frac{1}{3},-\frac{1}{6},0}\T$ for all of considered models.

To compare estimation performance, we computed relative root mean squared errors (RMSE) as $\frac{\nrm{\bB-\hat\bB}{}}{\nrm{\bB}{}}$ for an arbitrary matrix  $\bB$ and its estimate $\hat\bB$.
Also, we report one-step-ahead prediction errors as $\frac{\nrm{\bW_{T}\lrp{\hat\mu-\mu}}{}}{\nrm{\bW_{T}\mu}{}}$ i.e., root mean squared prediction errors (RMSP), as the systematic noise incurred by $\bcE_{T+1}$ is difficult to predict.
Throughout simulations we generate the covariates $\bz_{it}$ from $\cN\lrp{\bo_3,\operatorname{diag}\lrp{3,2,1}}$ and $\bcE_t$ from $\cN\lrp{\bo_N,0.25\,\bI_N}$.
To track the model performance as its dimension grows, we take $N\in\brcs{40, 80, 160, 320}$ and $K\in\brcs{3,12}$. 
We also consider finite $T$ case where $T=2$ and growing $T$ case where $T\in\brcs{40, 160, 320}$. In all cases, we run 200 replications.

\subsection{Generating data from ENAR model}
\begin{figure}
    \centering
         \begin{subfigure}{0.5\textwidth}
        \includegraphics[width=1\textwidth]{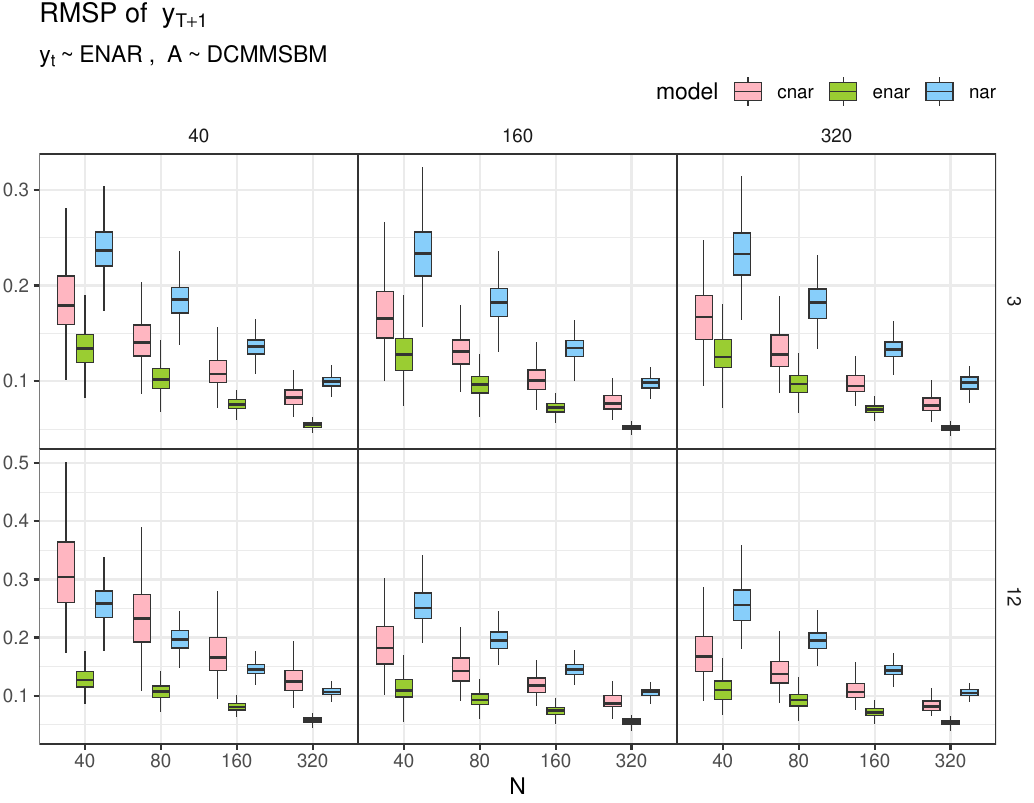}
    \end{subfigure}%
            \begin{subfigure}{0.5\textwidth}
        \includegraphics[width=1\textwidth]{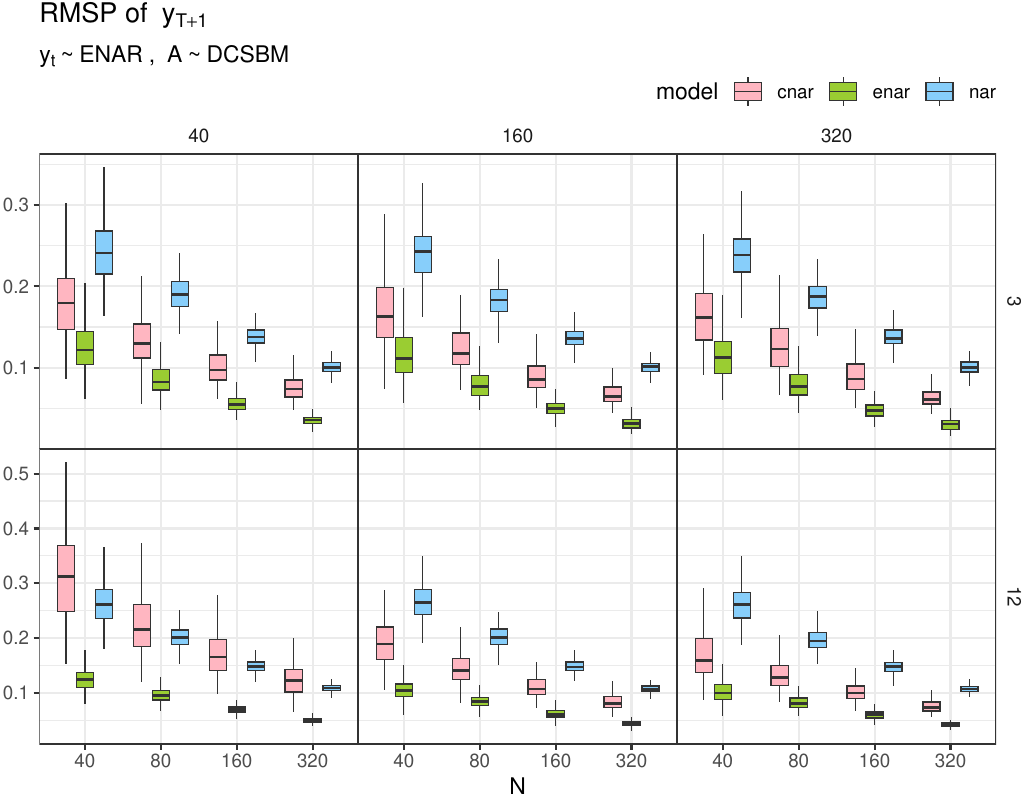}
    \end{subfigure}
    \caption{Boxplot of the prediction error from ENAR, CNAR, and NAR model when data is generated from ENAR model with DCMMSBM and DCSBM respectively.
     \vspace{-5pt}}
    \label{fig-enar_pred}
\end{figure}
The RMSE boxplots of peer influence and momentum effects are shown in Figures \ref{fig-enar_dcmmsbm} for data generated from the ENAR model with DCMMSBM. 
Each column and row of the facet grid corresponds to a different value of \( T \) and \( K \), respectively, while the figures within the grids are with increasing $N$. 
From Figure \ref{fig-enar_dcmmsbm}, we observe that when the true model is ENAR, the RMSE of its estimates of \( \alpha \) and \( \theta \) consistency decreases as \( N \) and \( T \) grows, which is the expected phenomenon from our asymptotic theories.
 
For estimates from the NAR model, while the RMSE for $\alpha$ parameter still decreases with increasing $N$, the RMSE for estimation of \( \theta \)  continues to remain high even when $N$ and $T$ increase. This is because since the NAR fit omits the latent variable effects, it incurs irreducible bias in parameter estimation,

\begin{figure}
    \centering

    \begin{subfigure}{0.5\textwidth}
        \includegraphics[width=1\textwidth]{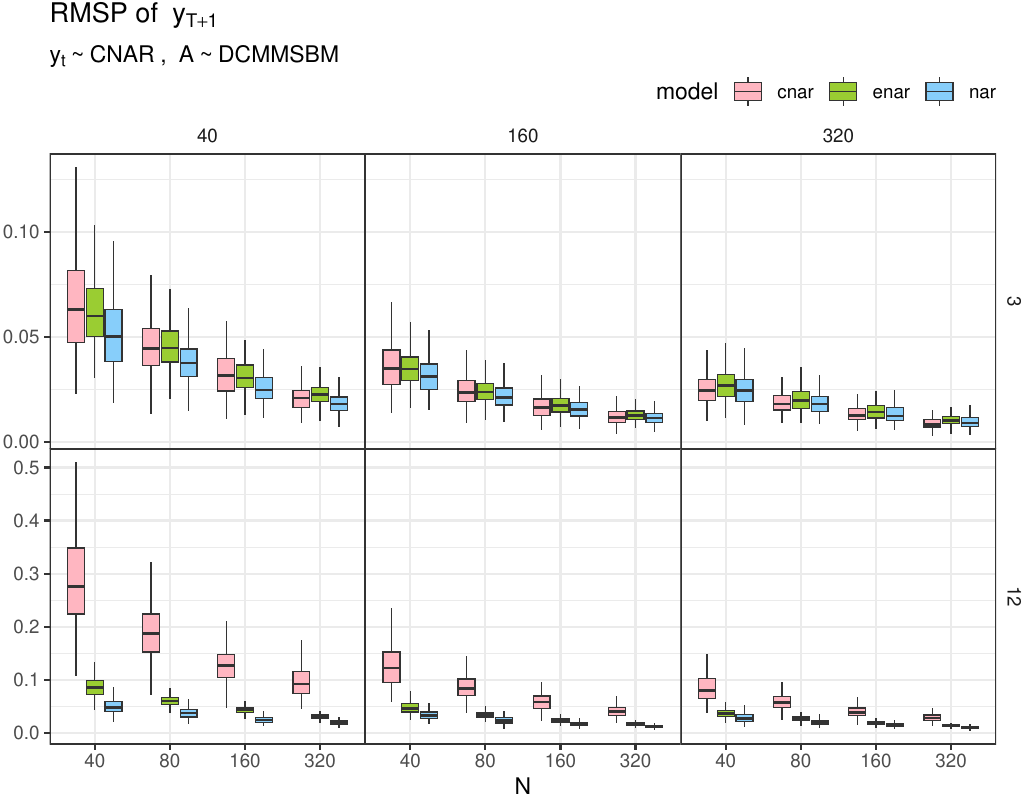}
    \end{subfigure}%
    \begin{subfigure}{0.5\textwidth}
        \includegraphics[width=1\textwidth]{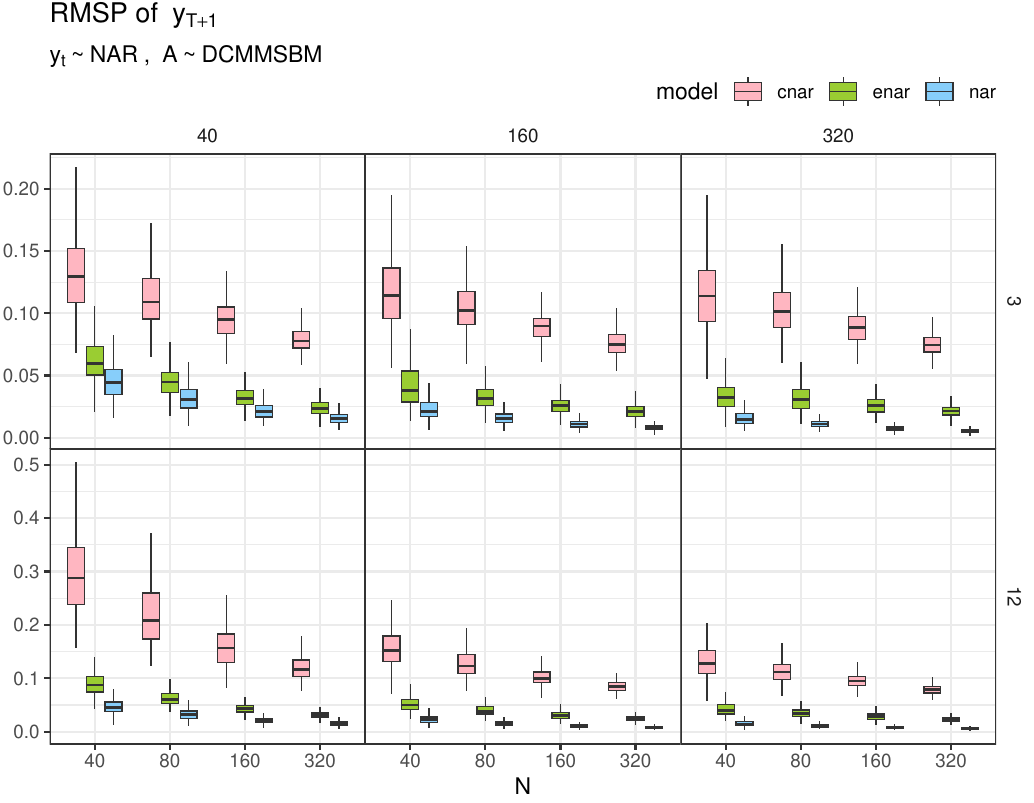}
    \end{subfigure}%
    \caption{Model misspecification: Boxplot of prediction error from CNAR, ENAR, and NAR model when data is generated from  CNAR and NAR models respectively.
     \vspace{-5pt}}
    \label{fig-pred_all}
\end{figure}

The predictions of ENAR, as shown in Figure \ref{fig-enar_pred} for graphs generated from DCMMSBM and DCSBM also improve with both \( N \) and \( T \). This figure shows that if ENAR is the true data-generating model, then omitting the latent variables from the fitted model (i.e., fitting the NAR model) not only leads to inaccurate parameter estimates but also to higher prediction error. The CNAR model also does not perform well for prediction when ENAR is the true data-generating model.

\subsection{Comparison Under Model misspecification}

Next, we generate $\mathbf{y}_t$ from NAR and CNAR while the underlying networks are generated from the DCMMSBM model. This corresponds to model misspecification for the ENAR model and we can compare the accuracy of prediction as well as parameter estimation in this setting. 
In Figure \ref{fig-pred_all} (left), we see a very good performance of both ENAR and NAR even when the data is generated from the CNAR model. This is especially true for smaller values of $N$ and $T$ (e.g., $N=T=40$). When $K$ is increased to 12, the performance of CNAR is worse than NAR and ENAR even when the data is generated from CNAR. This is because with increasing $K$ the performance of CNAR model estimators becomes worse. 

When we assumed the true model NAR, as expected, the predictive ability of NAR was the best overall for all settings of $N, T, K$ (Figure \ref{fig-pred_all} (right)). However, in each case the ENAR model came close in terms of predictive ability while the CNAR model produced large errors, especially when $K$ was larger and $N$ and $T$ were smaller. 

Finally,  we evaluate the estimates of the $\theta$ and $\alpha$ parameters under model misspecification when the data is generated from the NAR model. Then, we observe that the estimation of the peer influence effect becomes biased for ENAR especially when $T$ and $N$ are larger (Figure \ref{fig-nar_dcmmsbm} in Appendix). 
However, their estimation bias is significantly smaller than that of using NAR when data is generated from ENAR as shown in Figure \ref{fig-enar_dcmmsbm}.
Moreover, the ENAR model was able to consistently estimate $\alpha$ at comparable rates to NAR, showing robustness under model misspecification. In contrast, the NAR model produced large errors even for estimating $\alpha$ when ENAR was the true data generating model in Figure  \ref{fig-enar_dcmmsbm}.

\subsection{Finite \texorpdfstring{$T$}{T} case}

\begin{figure}
    \centering
    \begin{subfigure}{.245\textwidth}
        \includegraphics[width=1\textwidth]{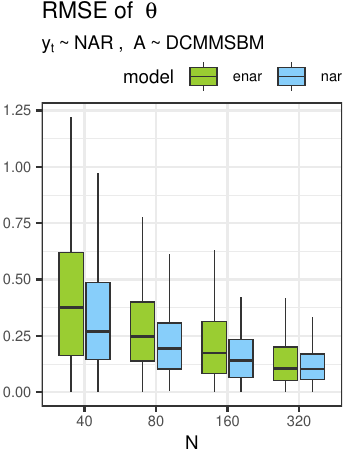}
    \end{subfigure}%
    \begin{subfigure}{.245\textwidth}
        \includegraphics[width=1\textwidth]{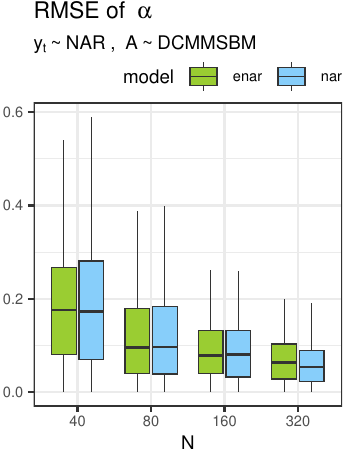}
    \end{subfigure}
    \begin{subfigure}{.245\textwidth}
        \includegraphics[width=1\textwidth]{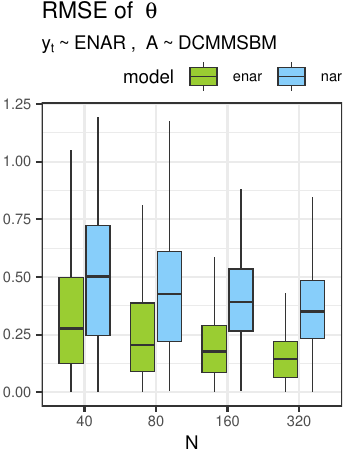}
    \end{subfigure}%
    \begin{subfigure}{.245\textwidth}
        \includegraphics[width=1\textwidth]{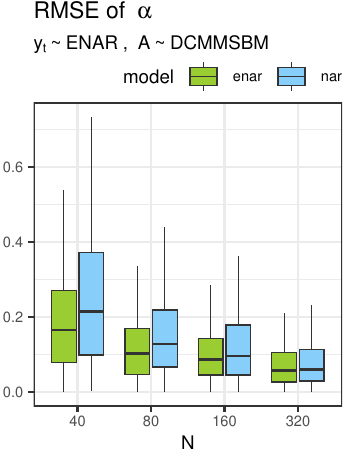}
    \end{subfigure}
    \caption{Boxplot of estimates of $\theta$ and $\alpha$ from ENAR, and NAR model when data is generated from ENAR and NAR model with DCMMSBM.
     \vspace{-5pt}}
    \label{fig-finiteT_dcmmsbm}
\end{figure}

\begin{wrapfigure}{l}{0.4\textwidth}
    \centering
    \begin{subfigure}{.4\textwidth}
        \centering
        \includegraphics[width=.485\textwidth]{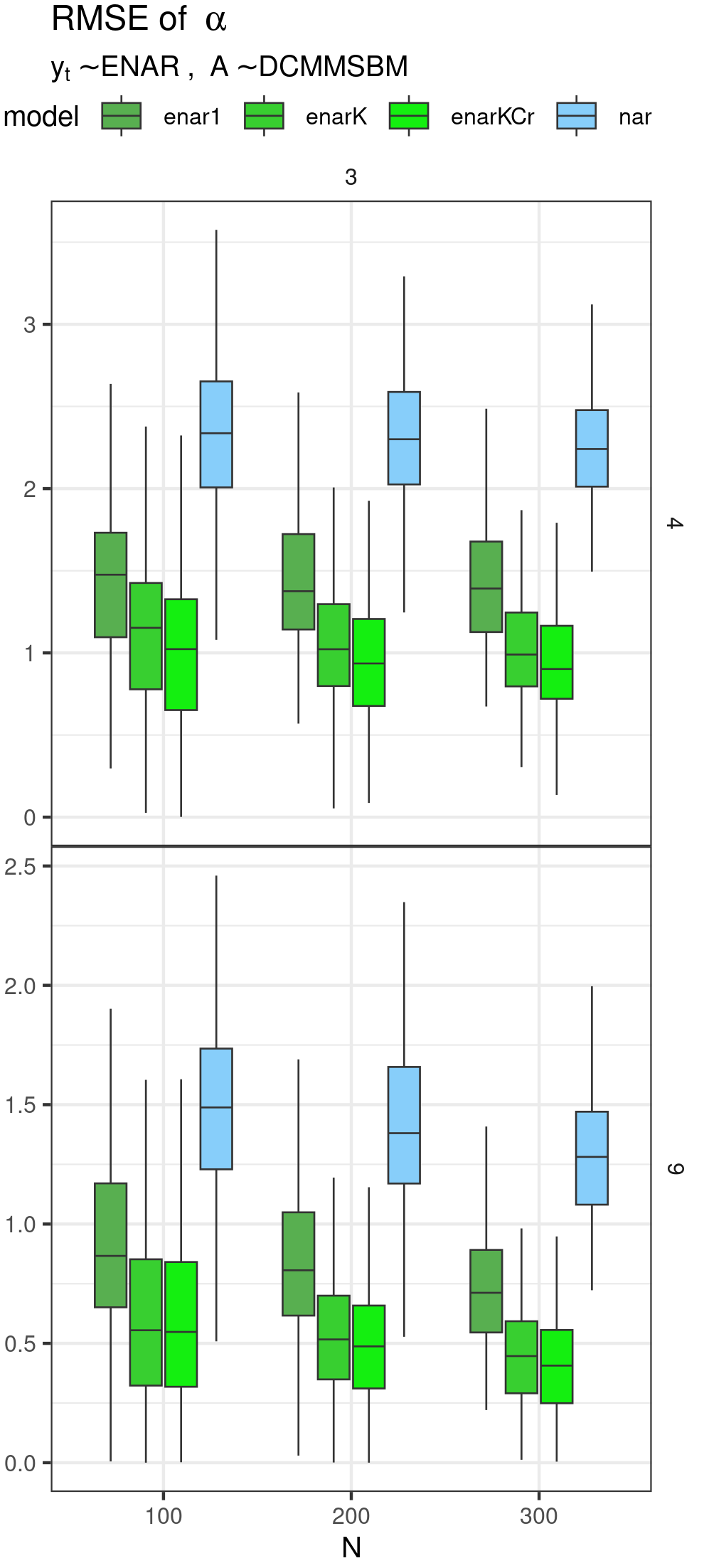}
        \includegraphics[width=.485\textwidth]{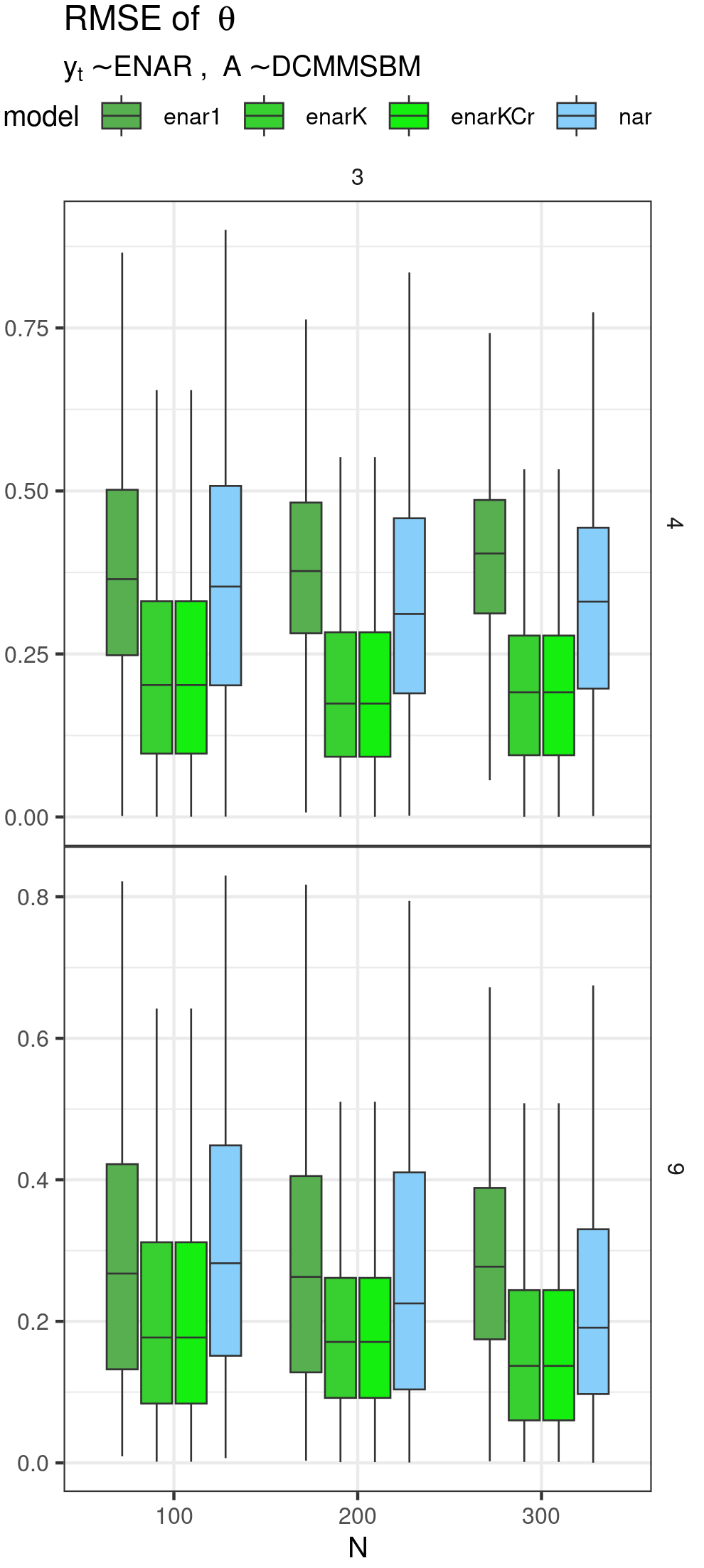}
    \end{subfigure}%
    \caption{ENAR with $\bU=\bU_{se}$\vspace{-5pt}}
    \label{fig-misK_dcmmsbm-se}
\end{wrapfigure}

Next, we investigate the performance of ENAR and NAR estimators in terms of the accuracy of parameter estimation in the fixed $T$ case. We set $T=2$, and $K=3$ and increase $N$.
In Figure \ref{fig-finiteT_dcmmsbm}, it is clear that ENAR can estimate $\alpha$ and $\theta$ well even under model misspecification of generating data from NAR. The estimation error for estimating both $\alpha$ and $\theta$ from ENAR are comparable to NAR and decreases with increasing $N$. In contrast, when the data is generated from ENAR, we see that the estimate of $\theta$ from NAR is biased and continues to show high error even when $N$ increases, while the estimation error decreases for ENAR.

Combined with our previous observations in growing $N$ and $T$ cases, ENAR shows robust estimation performance under various model misspecifications, while achieving better prediction and parameter estimation performance when the data is generated from ENAR in both growing and fixed model dimension cases.

\subsection{Misspecification of $K$ in ENAR}

\begin{figure}
    \begin{subfigure}{1\textwidth}
        \centering
        \includegraphics[width=.45\textwidth]{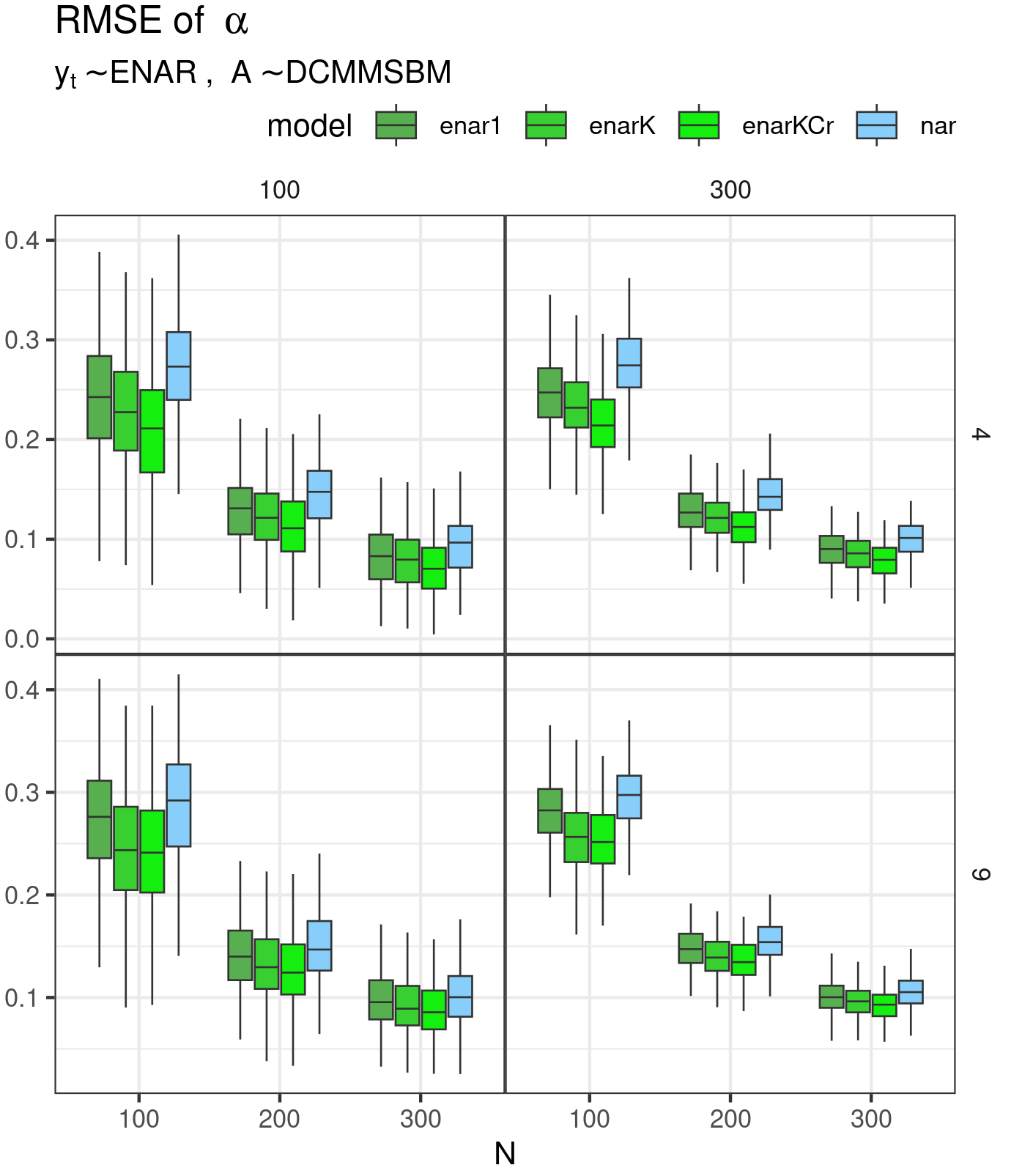}
        \includegraphics[width=.45\textwidth]{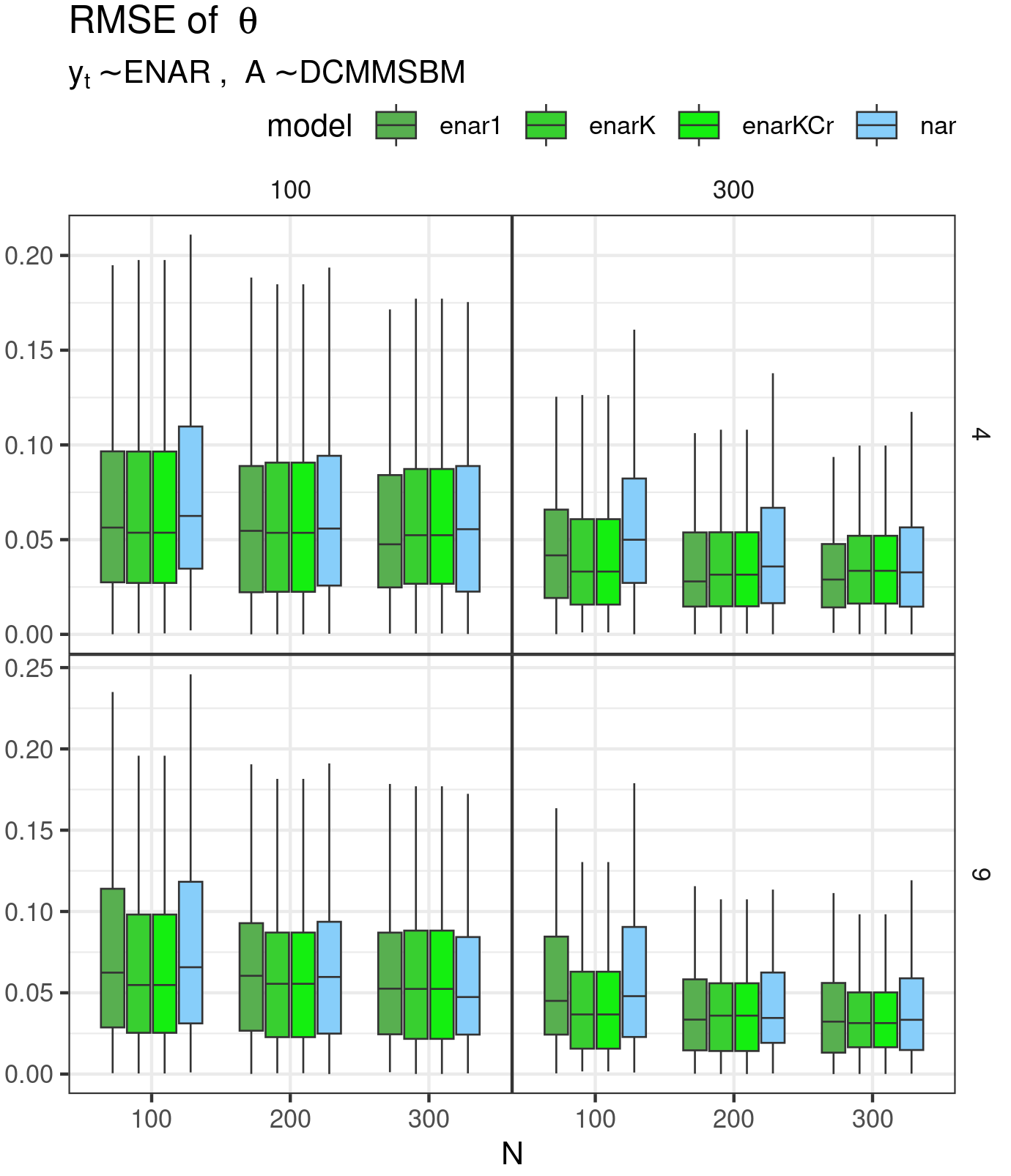}
    \end{subfigure}
    \caption{Boxplot of estimates of $\theta$ and $\alpha$ from ENAR with $\bU=\bU_{ev}$ for $K_o\in\brcs{1,K,\hat K}$, and NAR model when data is generated from ENAR model with DCMMSBM. \vspace{-5pt}}
    \label{fig-misK_dcmmsbm-ev}
\end{figure}

Next, we perform a simulation to compare the performance of the estimators when \( K \) is selected according to our criterion \( Cr(k) \), searched over \( k \in \{1, \dots, J_N = \|\bA\|_{\infty}^{1/2} \} \) using \( p_{NT}(k) = \frac{k}{N+T} \), with the estimators when the true \( K \) is used and when \( K \) is underselected to 1.
In Figure \ref{fig-misK_dcmmsbm-se}, we plot the RMSEs for estimation of $\alpha$ and $\theta$ for increasing $N$ and fixed $T=3$ for the ENAR model with $\bU=\bU_{se}$, and in \ref{fig-misK_dcmmsbm-ev}, we plot the RMSEs for growing $N,T$ for the model with $\bU-\bU_{ev}$. In all cases, the estimators with selected $K=\hat K$ perform similarly or sometimes even better than the estimators with true $K$. In contrast, the estimators with $K=1$ and $K=0$ (which corresponds to the NAR model) underperform. For both modeling setups, our method with a fully data-driven choice of the number of latent vectors provides effective control of confounding due to latent homophily.

\section{Simulations with general latent space network model}
Next, we consider simulation results for the AMNAR model. 
We consider the latent space network model (LSN) discussed in section \ref{sec-lsn} for the population distribution of $\bA$, and generate the networks similarly to \cite{ma2020universal,li2023statistical}.
First, we generate a matrix of latent vectors $\bQ\in\R^{N\times K}$ with i.i.d. entries from $\cN(0, 4)$ and a vector of nodal degree heterogeneity parameters $\bv\in\R^N$ with i.i.d. entries from $\operatorname{Uniform}(-1, 1)$. Then, we center $\bQ$ and $\bv$ to obtain $\bX=\bxs{\bQ|\bv}$. We use sigmoid function $\sigma(x)=\frac{1}{1+\exp(-x)}$ as the link function for $p_{ij}=\sigma({\bq_i}\T\bq_j+v_i+v_j)$ so that the network is generated as $a_{ij}\overset{ind.}{\sim}\operatorname{Bernoulli}(p_{ij})$. 

Recalling Theorem \ref{thm-amnar-clt-NT}, we use the growth factor ${\tilde\rho_{NT}}^{1/2}$ for $\bX$ with the choice of ${\tilde\rho_{NT}}^{1/2}=\frac{1}{\sqrt[4]{NT^2}}$.
For model parametrization, we set the parameters in the same way as ENAR except for the $\beta_m$ parameters associated with the latent effects of AMNAR, which we set as
$\lrp{1,-1/2,...,\lrp{-1}^K/(K+1)}\T\in\R^{K+1}$.
We use package `randnet' in R \citep{li2022randnet} to estimate $\bQ$ and $\bv$.

To track the model performance as its dimension grows, we take $N\in\brcs{50, 100, 200}$ and $K=3$. 
We also consider finite $T$ case where $T=3$ and growing $T$ case where $T\in\brcs{50, 100, 200}$. In all cases, we run 200 replications.

\begin{figure}[h]
    \centering
    \begin{subfigure}{0.5\textwidth}
        \includegraphics[width=1\textwidth]{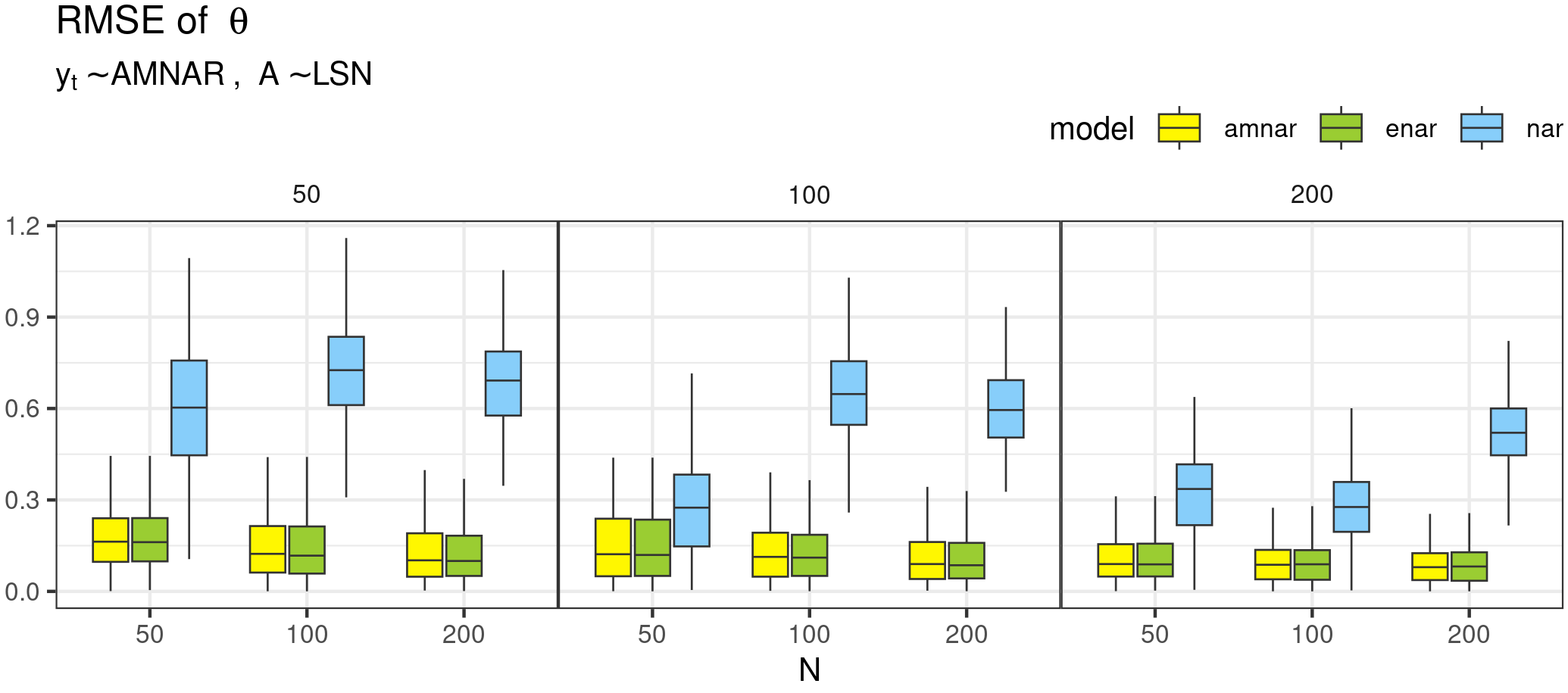}
    \end{subfigure}%
        \begin{subfigure}{0.5\textwidth}
        \includegraphics[width=1\textwidth]{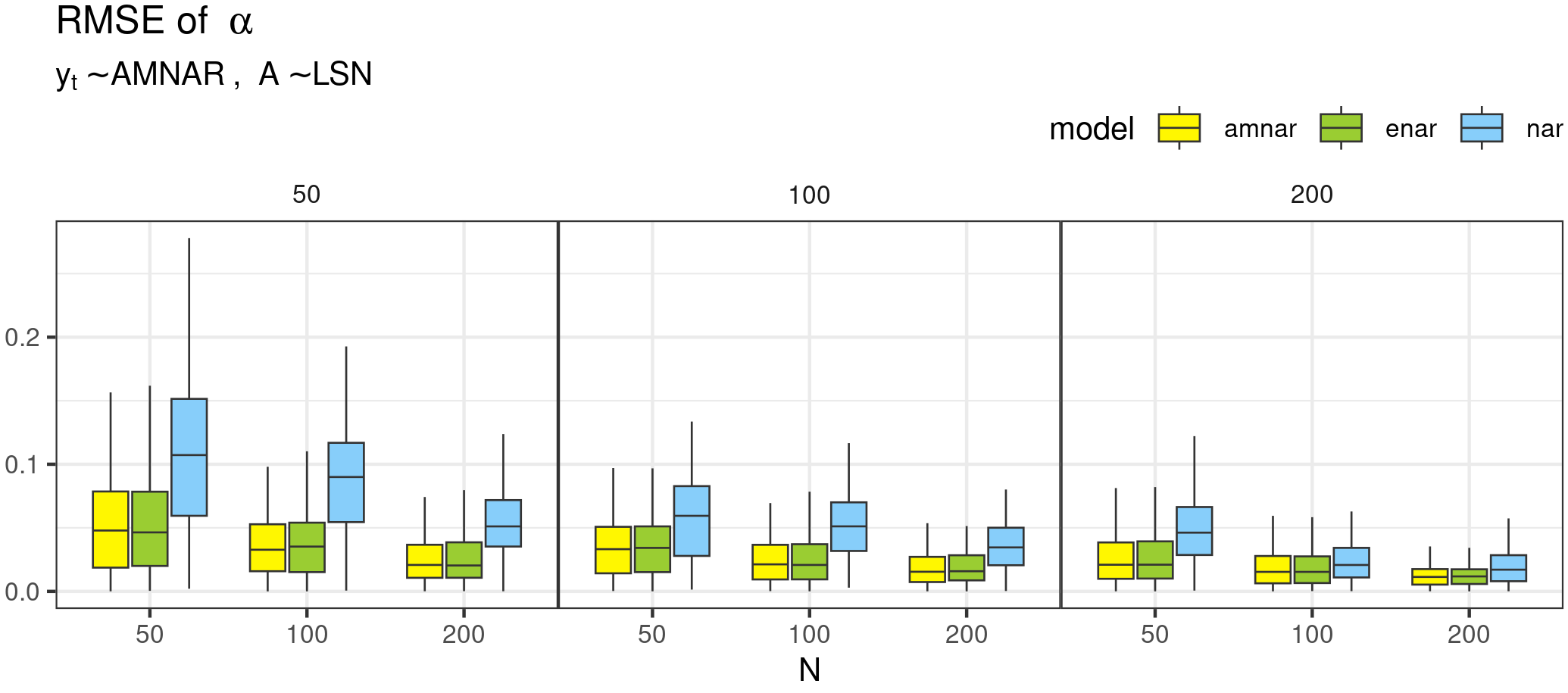}
    \end{subfigure}
    \caption{Boxplot of estimates of $\theta$ (left) and  $\alpha$ (right) from  AMNAR, ENAR, and NAR models with increasing $N$ when data is generated from AMNAR model with LSN. $K$ is set to be $3$, and the columns corresponds to $T=50, 100, 200$.
    \vspace{-5pt}}
    \label{fig-amnar_lsn}
\end{figure}

\subsection{Generating data from AMNAR model}

\begin{figure}
    \centering
         \begin{subfigure}{0.75\textwidth}
        \includegraphics[width=1\textwidth]{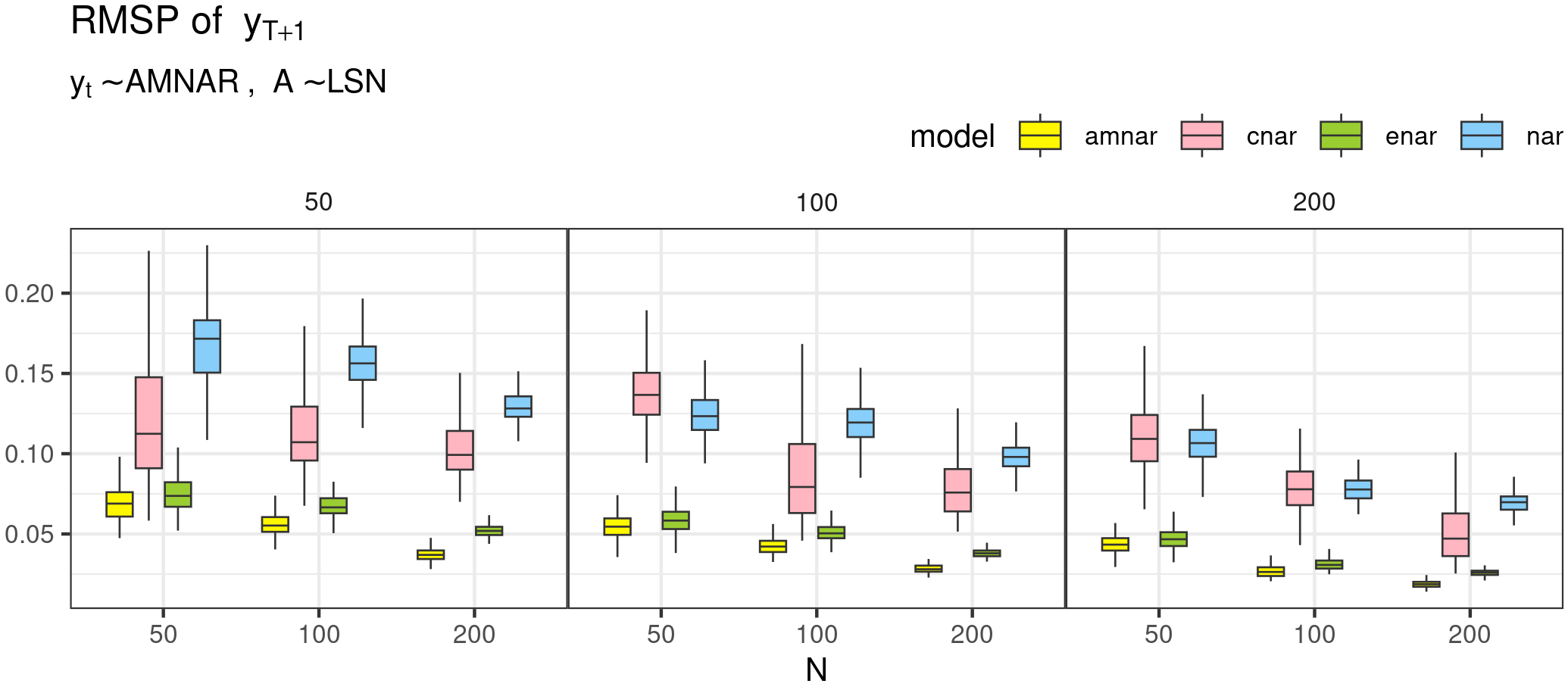}
    \end{subfigure}%
    \caption{Boxplot of the prediction error from AMNAR, ENAR, CNAR, and NAR model when data is generated from AMNAR model with LSN.
     \vspace{-5pt}}
    \label{fig-amnar_pred}
\end{figure}
The RMSE boxplots of peer influence and momentum effects are shown in Figures \ref{fig-amnar_lsn} for data generated from the AMNAR model with LSN. 
Each column and row of the facet grid corresponds to a different value of \( T \) and \( K \), respectively, while the figures within the grids are with increasing $N$. 
From Figure \ref{fig-amnar_lsn}, we observe that when the true model is AMNAR, the RMSE of its estimates of \( \alpha \) and \( \theta \) consistently decreases as \( N \) and \( T \) grows, as expected from our theorems. For estimates from the NAR model, while the RMSE for $\alpha$ parameter still decreases with increasing $N$, the RMSE for the estimation of \( \theta \)  continues to remain high even when $N$ and $T$ increase.

\begin{figure}
    \centering

    \begin{subfigure}{0.5\textwidth}
        \includegraphics[width=1\textwidth]{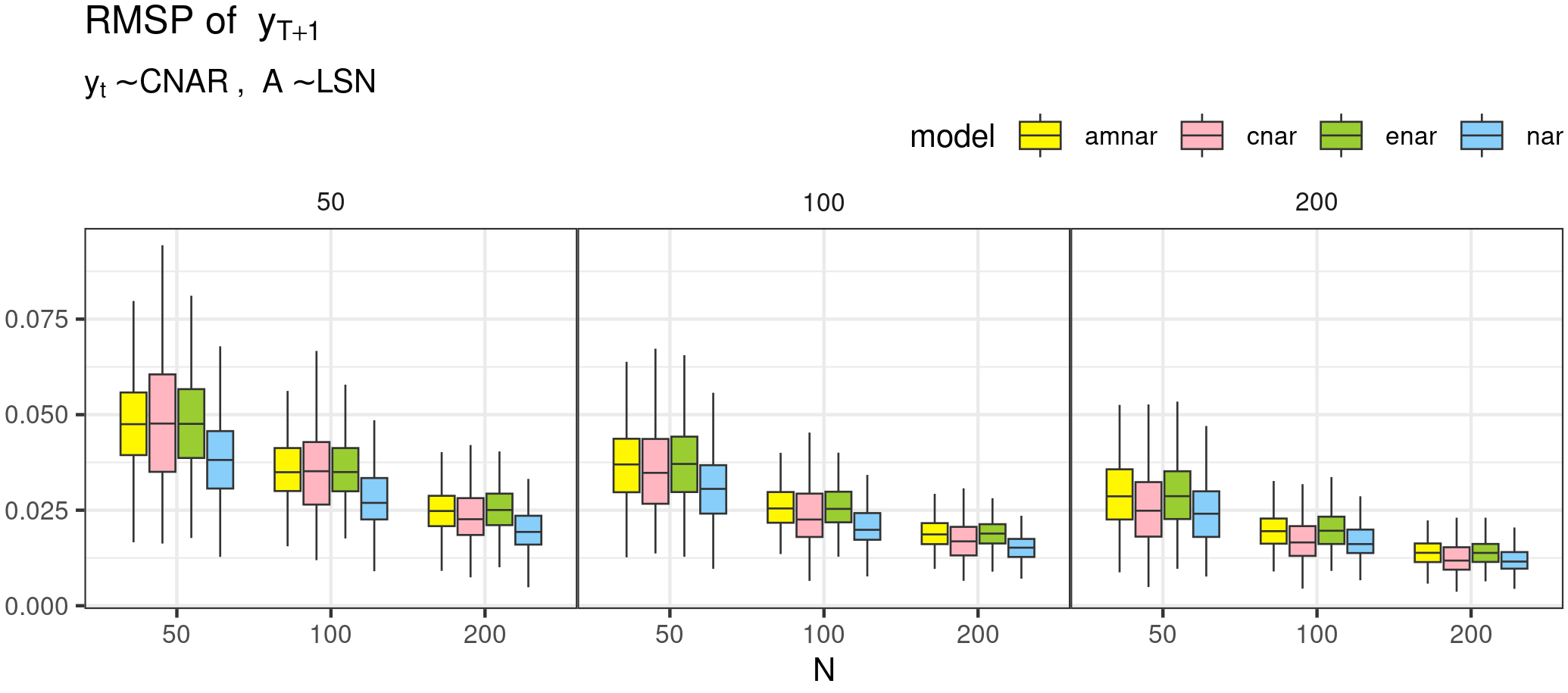}
    \end{subfigure}%
    \begin{subfigure}{0.5\textwidth}
        \includegraphics[width=1\textwidth]{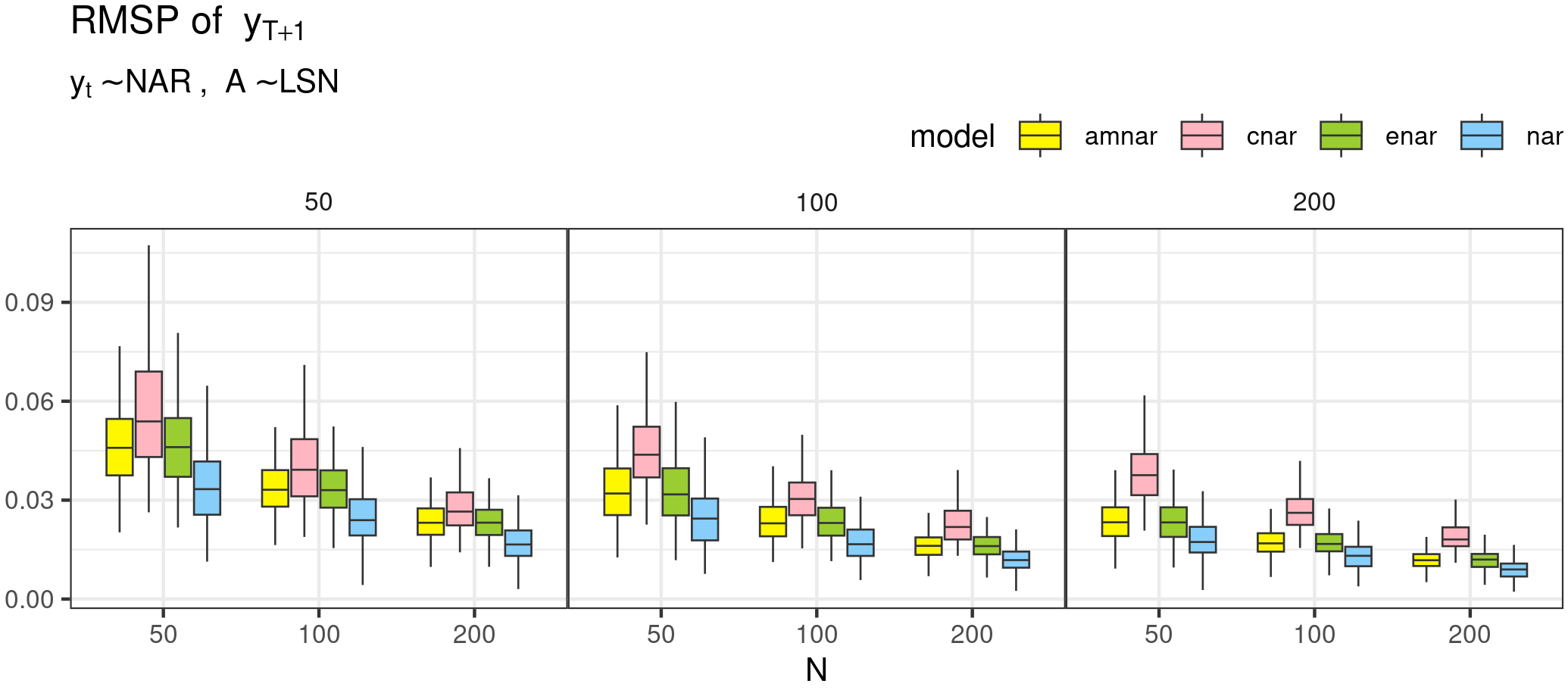}
    \end{subfigure}%
    \caption{Model misspecification: Boxplot of the prediction error from CNAR, ENAR, and NAR model when data is generated from  CNAR and NAR models respectively.
     \vspace{-5pt}}
    \label{fig-pred_all2}
\end{figure}

The predictions of AMNAR, as shown in Figure \ref{fig-amnar_pred} for graphs generated from LSN also improve with both \( N \) and \( T \). This figure shows that if AMNAR is the true data-generating model, then adjusting for the effects of latent position and degree heterogeneity improves prediction over NAR and CNAR models. This also validates the utility of the AMNAR model for the time series prediction problem. Interestingly, we note that ENAR estimators perform quite well in this scenario in comparison to both NAR and CNAR even though the data was generated from the AMNAR model.

\subsection{Comparison Under Model misspecification}

Next, we generate $\mathbf{y}_t$ from NAR and CNAR while the underlying networks were generated from the LSN model. This corresponds to model misspecification for AMNAR and ENAR models and we can compare the accuracy of prediction as well as parameter estimating in this setting. 
In Figure \ref{fig-pred_all2} (left), we see a comparable performance of both AMNAR and ENAR to CNAR even when the data is generated from the CNAR model.

When the true model is NAR, as expected, the predictive ability of NAR was the best overall for all settings of $N, T, K$ (Figure \ref{fig-pred_all2} (right)). However, in each case, the AMNAR and ENAR models came close in terms of predictive ability while the CNAR model produced large errors, especially when $N$ and $T$ were smaller.

\subsection{Finite \texorpdfstring{$T$}{T} case}

Next, we investigate the performance of AMNAR and NAR estimators in terms of accuracy of parameter estimation in the fixed $T$ case. We set $T=3$, and $K=3$ and increased $N$.
\begin{figure}
    \centering
    \begin{subfigure}{.245\textwidth}
        \includegraphics[width=1\textwidth]{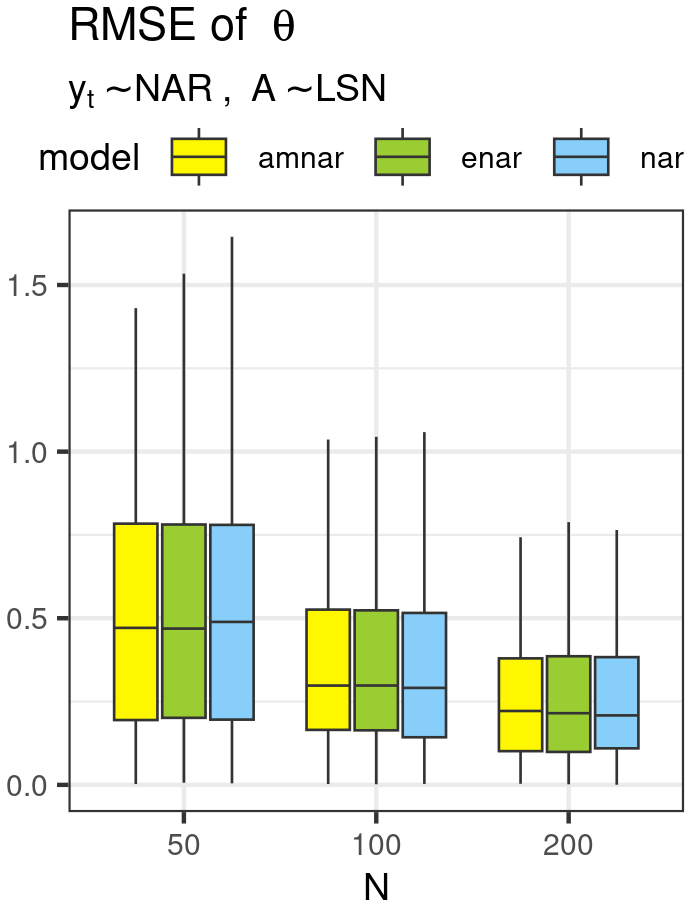}
    \end{subfigure}
    \begin{subfigure}{.245\textwidth}
        \includegraphics[width=1\textwidth]{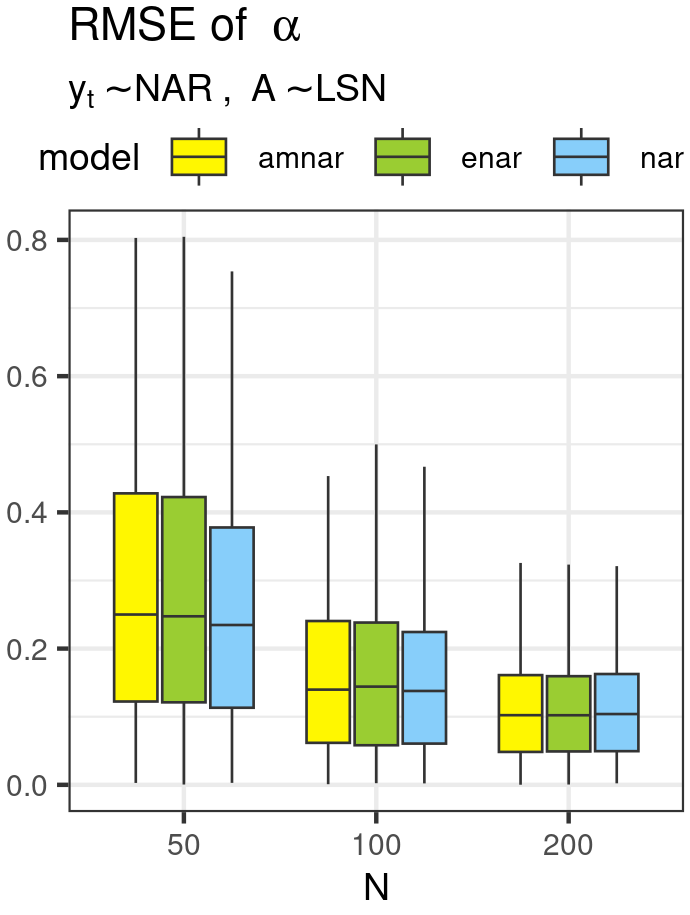}
    \end{subfigure}
    \begin{subfigure}{.245\textwidth}
        \includegraphics[width=1\textwidth]{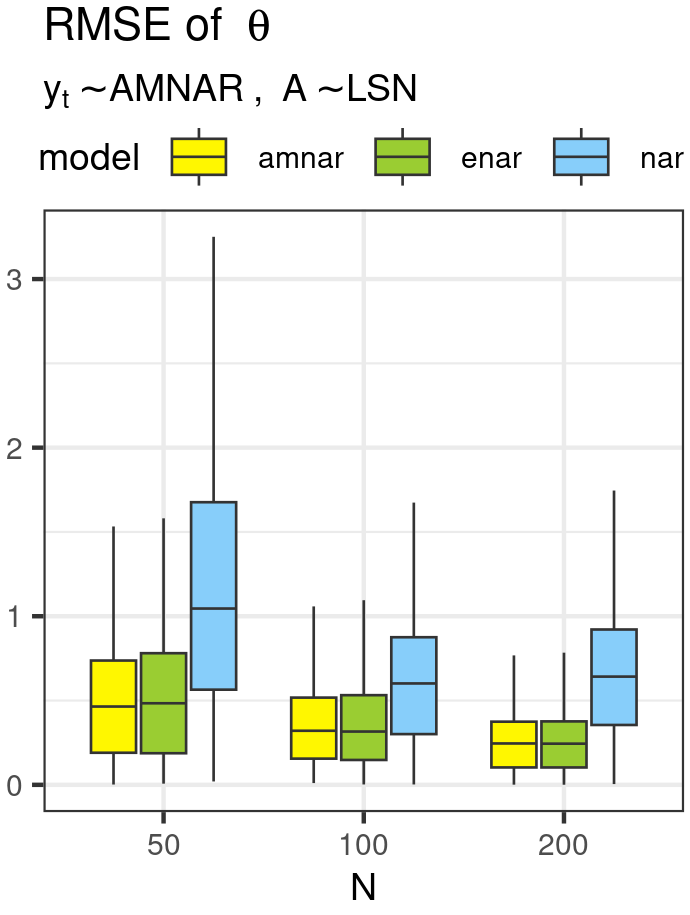}
    \end{subfigure}%
    \begin{subfigure}{.245\textwidth}
        \includegraphics[width=1\textwidth]{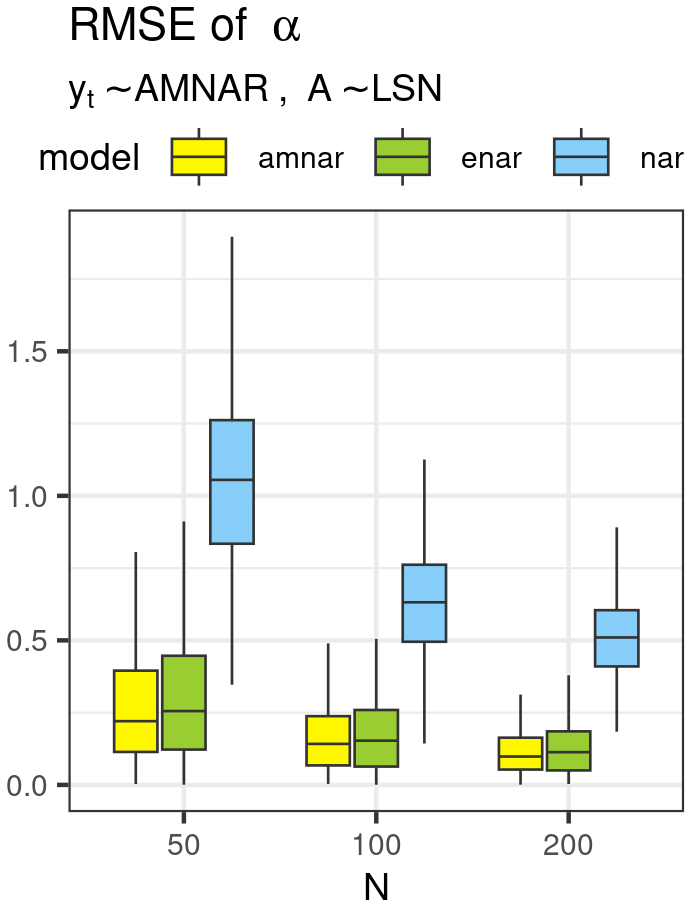}
    \end{subfigure}
    \caption{Boxplot of estimates of $\theta$ and $\alpha$ from AMNAR, ENAR, and NAR model when data is generated from AMNAR and NAR model with LSN.
     \vspace{-5pt}}
    \label{fig-finiteT_lsn}
\end{figure}
In Figure \ref{fig-finiteT_lsn}, it is again clear that AMNAR can estimate $\alpha$ and $\theta$ relatively well, even under model misspecification, as the estimation errors from AMNAR are comparable to NAR and decrease with increasing $N$. In contrast, when the data is generated from AMNAR, we see that the estimate of $\theta$ from NAR is biased and continues to show high error even when $N$ increases, while the estimation error decreases for AMNAR and ENAR.

Aligning well with our previous observations in growing $N$ and $T$ cases, AMNAR shows robust estimation performance under various model misspecifications, while achieving better prediction and parameter estimation performance when the data is generated from AMNAR in both growing and fixed model dimension cases.

\section{Real Data Example}
In this section, we will analyze two datasets. The first one is a finite-time dataset where the primary goal is to infer causal peer effects and effects of covariates, and the second one is a time series dataset where the goal is both accurate prediction and parameter estimation. For all datasets, we select $K$ using our IC based selection method with $p_{f,NT}(k)=\frac{k^2}{N+T}$, and $J_N=\sqrt{N}$.
\smallskip

\noindent \textbf{Knecht dutch students delinquency and alcohol data:}  The first dataset we analyze with this new method is the longitudinal Dutch students' friendship network and delinquency study by Andrea Knecht \cite{knecht2008friendship,knecht2010friendship,he2019multiplicative}. The dataset is taken from the R package ``xergm.common''. This longitudinal data consists of friendship networks along with responses related to delinquency and alcohol consumption and some demographic covariates measured at four time points on 26 students in one classroom. The measurements are taken 3 months apart during the first year of secondary school. The demographic information includes sex, age, religion and ethnicity. There are two response variables that we are interested in - alcohol consumption measured at waves 2,3,4 and delinquency, which is defined as a rounded average over four types of minor delinquency (stealing, vandalism, graffiti, and fighting) and measured at all 4 waves. We construct an average friendship network by taking the average of the friendship networks in waves 1 and 2. For alcohol consumption, we fit the ENAR and NAR models using 2 time periods, waves 2 and 3, and then predict the responses for Wave 4. For the delinquency response, we fit two models, one using data from waves 1 and 2, and predicting the response in wave 3, and the other using data from waves 1, 2, and 3, and predict the response in wave 4.  Using our IC method, we select $K=4$ for delinquency as the response and $K=6$ for alcohol as the response. In both cases, we also fit a linear regression model with only the demographic covariates and do not include the lagged response and lagged peer effects, which we call the OLS model. 
\begin{wraptable}{l}{0.7\textwidth}
        \centering
    \begin{tabular}{p{2.5cm}|p{2cm}|p{1.5cm}|p{1cm}|p{1cm}}
      Model   & ENAR (selected $K$) & ENAR ($K=2$) & NAR & OLS \\
      \hline
      Alcohol   & 2.546 & \textbf{1.9133} & 1.9726 & 2.5628\\
      Delinquency 3  & \textbf{0.6278} & 0.6712 & 0.6942 & 0.7375\\
      Delinquency 4  & \textbf{0.8983} & 1.2425 & 1.0644 & 1.2257\\
      \hline
    \end{tabular}
    \caption{Comparison of Mean square prediction error (MSPE) for alcohol use (wave 4) and delinquency response from 3rd and 4th wave respetively.} 
    \label{knecht_mspe}
\end{wraptable}
We compare the model fit in terms of out-of-sample prediction in Table \ref{knecht_mspe} for 4 models, that include two ENAR models with selected number of $K$ and setting $K=2$ respectively. We observe that ENAR performs quite well in terms of mean square prediction error for both responses. The boxplots in Figure \ref{knecht_boxplot} show the predicted values from the ENAR, NAR, and OLS for various levels of actual response. We see that for both alcohol and delinquency, predictions from the ENAR model are higher for higher values of the actual response indicating a good model fit to the data with strong predictive ability. Table \ref{knecht_estimates} in the Appendix shows the parameter estimates from the models. We note that both the lagged response and lagged peer response parameters are insignificant in both NAR and ENAR models. The coefficients corresponding to most predictors are also insignificant in all models.

\smallskip

\noindent \textbf{Wind speed time series data:} Next we apply the ENAR model to a multivariate time series data containing wind speed measurements over 721 time periods at 102 weather stations in England and Wales. We take this dataset from the R package GNAR \cite{knight2019generalised}. This is a data with large $T$ and large $N$. We assess the accuracy of model fits in terms of the ability to accurately predict responses in 1 time period ahead.
\begin{figure}[h]
    \centering
            \begin{subfigure}{0.4 \textwidth}
       \includegraphics[width=0.9 \textwidth]{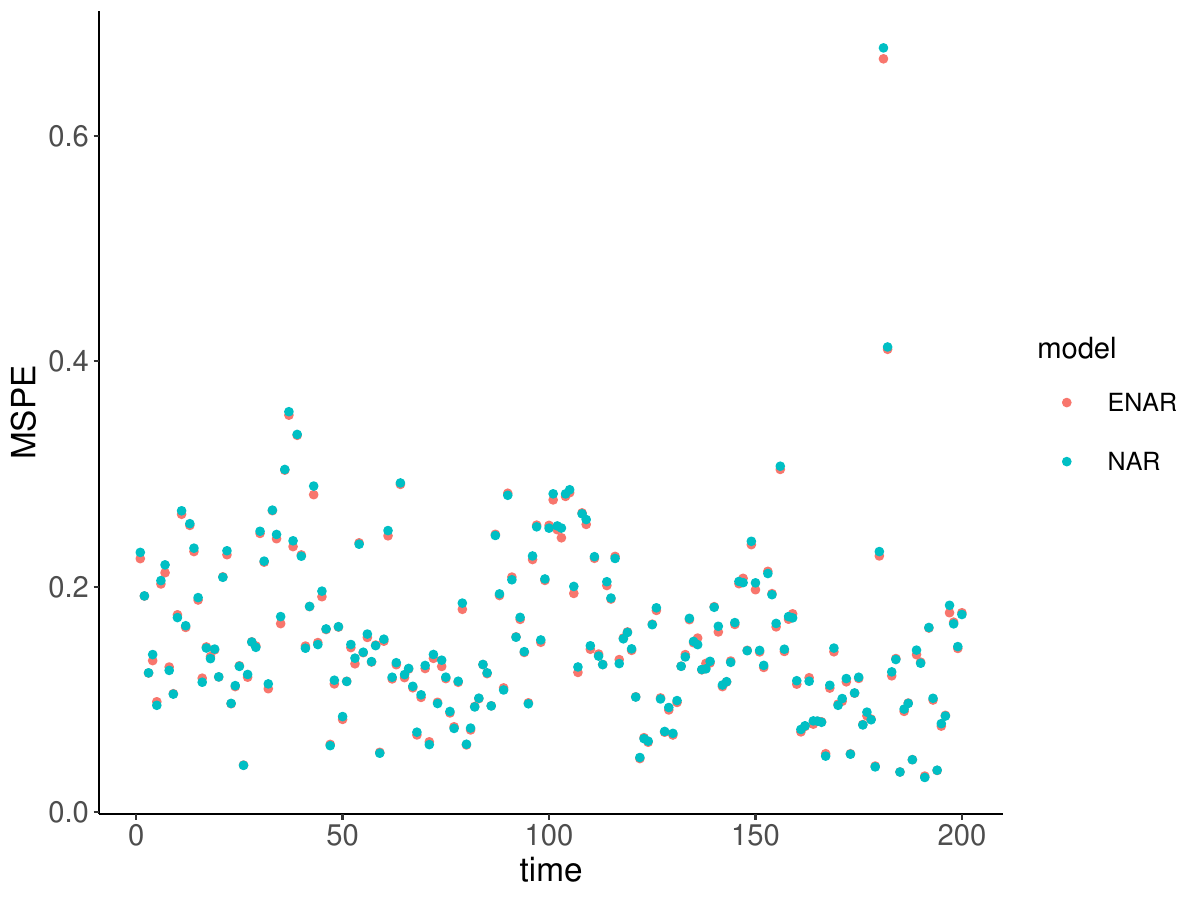}   
    \end{subfigure}%
        \begin{subfigure}{0.3 \textwidth}
       \includegraphics[width=0.9 \textwidth]{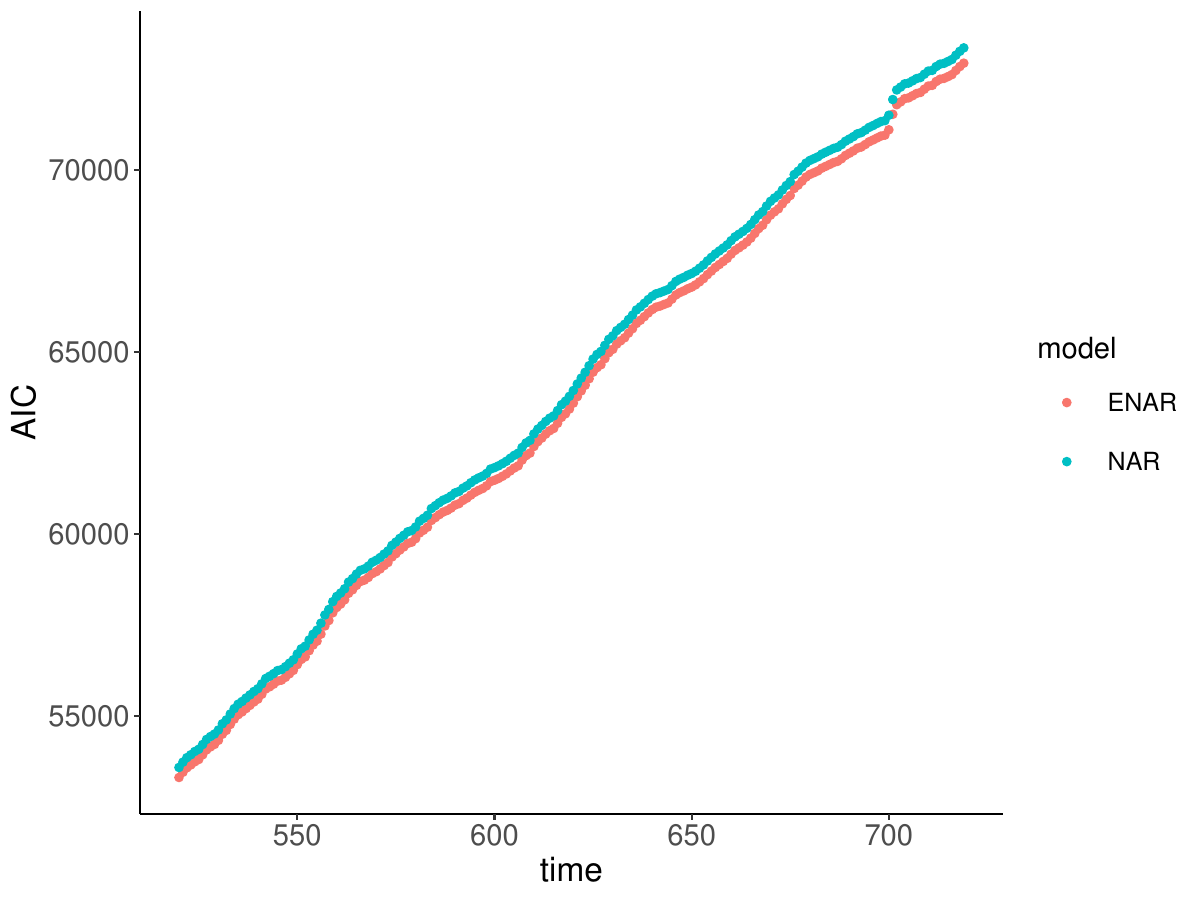}   
    \end{subfigure}%
        \begin{subfigure}{0.3 \textwidth}
       \includegraphics[width=0.9 \textwidth]{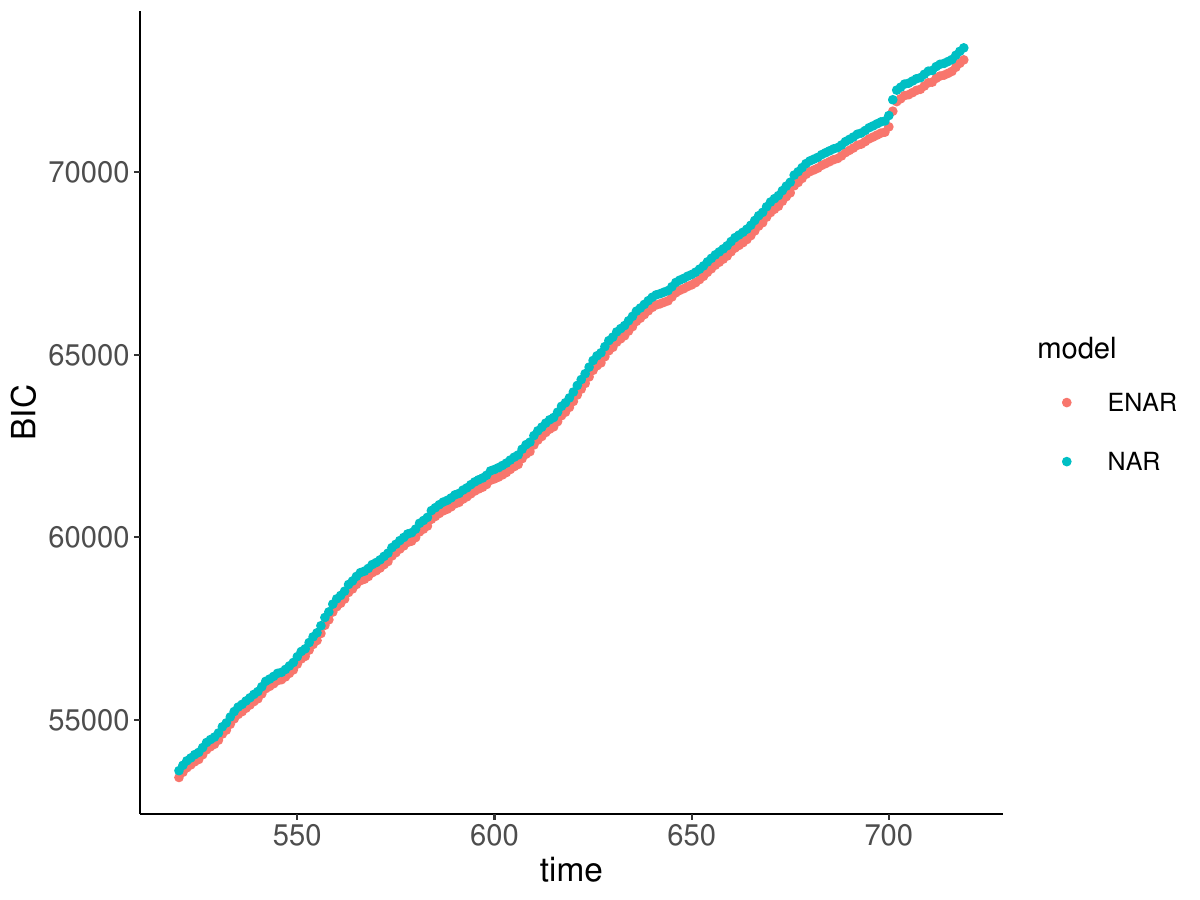}   
    \end{subfigure}
    \caption{(left) 1 step ahead mean square prediction error, (center) AIC and (right) BIC of model with increasing time for the Wind speed data.
     \vspace{-5pt}}
    \label{vswind_aic}
\end{figure}
To compare the models over a range of periods, we perform 1 step ahead predictions 200 times. In the $i$ th prediction task, we fit the NAR and ENAR models to the time series until time $519+i$ and predict the response at $520+i$ th time. We continue to increase $i$ and slide the window of training data until $i=200$. For the ENAR model we estimate the dimension of the latent variables as $K=10$ using our IC model selection criterion. 
\begin{wrapfigure}{l}{0.7 \textwidth}
    \centering
    \begin{subfigure}{0.35 \textwidth}
     \centering
     \includegraphics[width=0.8 \textwidth]{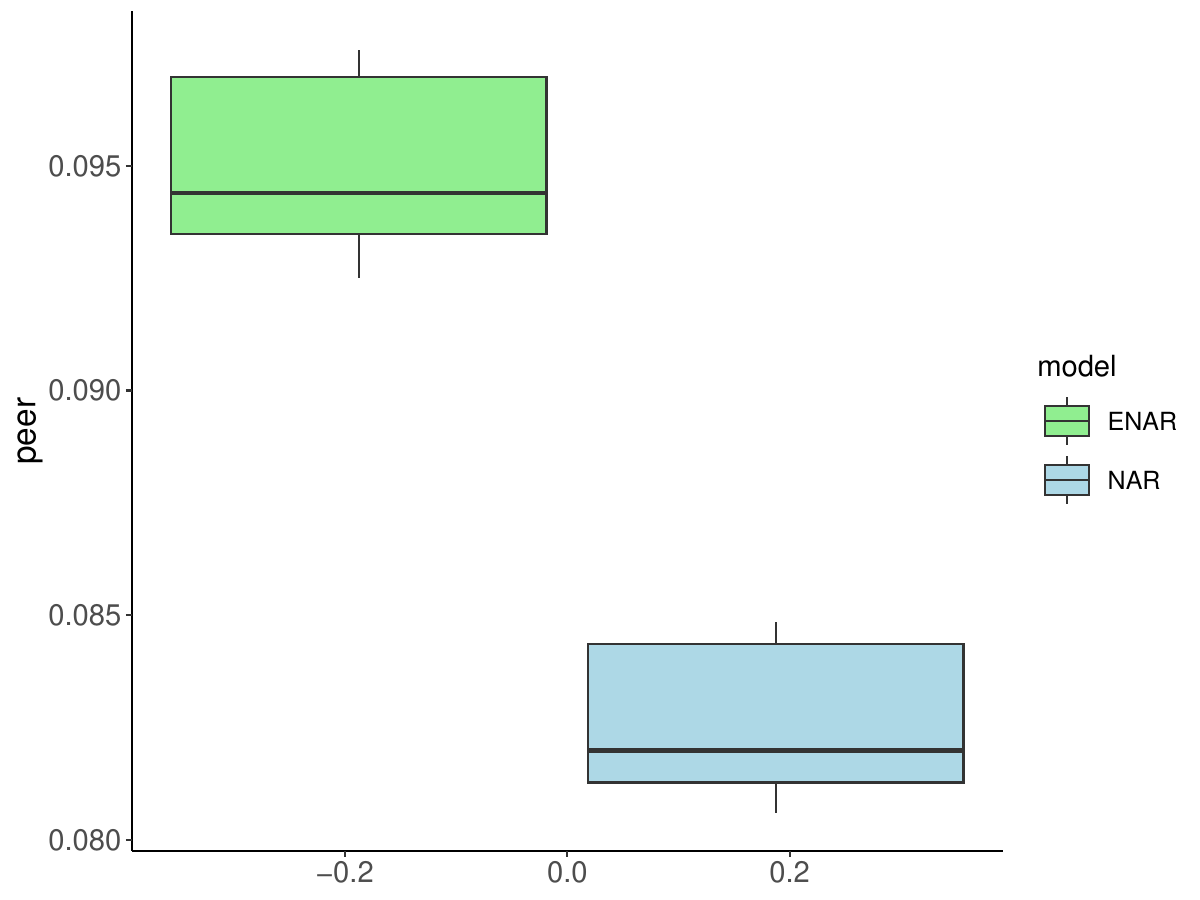}  
    \end{subfigure}%
    \begin{subfigure}{0.35 \textwidth}
       \centering
        \includegraphics[width=0.8 \textwidth]{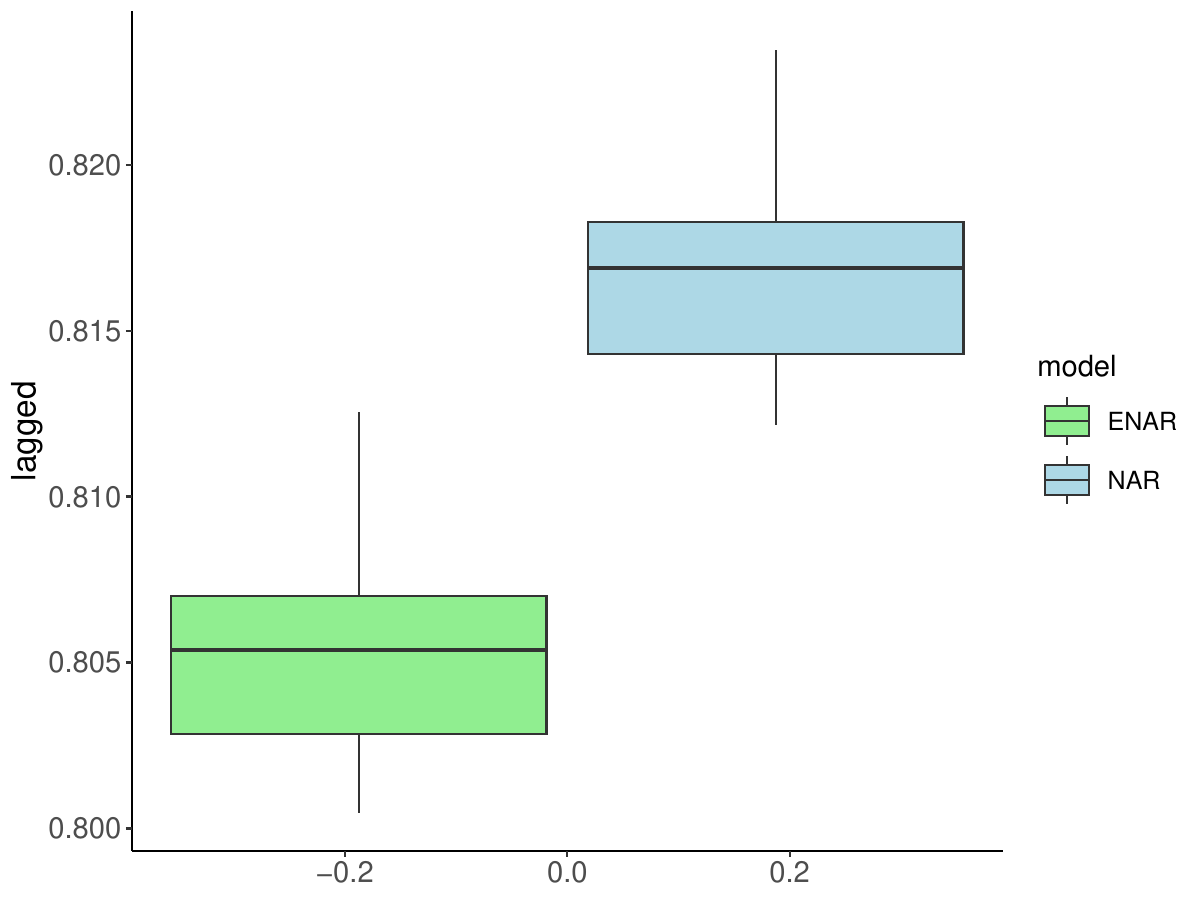}
    \end{subfigure}   
    \caption{Boxplots of (left) Peer effect parameter and (right) lagged effect parameter of $200$ estimates of ENAR and NAR model with training sample size $520$ to $720$.
    \vspace{-15pt}}
    \label{vswind_boxplot}
\end{wrapfigure}
Figure \ref{vswind_aic} (left) shows the 1 step ahead prediction error from the ENAR and NAR model. Out of these 200 test time windows, ENAR gives a smaller MSPE in 64\% of the cases. Therefore we conclude ENAR outperforms NAR in this prediction task. Both the AIC and BIC criteria in Figure \ref{vswind_aic} also point to superior model fit by the ENAR model over the NAR model. The parameter estimates of the peer effect and lagged effect (momentum effect) parameters along with their standard errors for one representative model with $T=520$ is shown in Table \ref{vswind_estimates} in the Appendix. We can see that both effects are statistically significant, with the momentum effect being roughly 10 times that of the peer effect. The parameter estimates for both of these effects are significantly different in the ENAR model from those in the NAR model, as is evident from the size of the differences of these estimates relative to their standard errors. In figure \ref{vswind_boxplot} we display the boxplots of parameter estimates corresponding to the peer effect and lagged effect parameters from the 200 ENAR and NAR models. The plot shows a clear difference in the parameter estimates from the two models.

\section{Conclusion}
The ENAR and AMNAR models can successfully address two major statistical problems. It embeds the homophily latent effects in the network auto-regressive model, tackling both the consistent estimation of causal peer effects and time series predictive performance enhancement. We proved that the estimators of AMNAR and ENAR have asymptotic normality in both long-term and finite-time cases. Our numerical study also illustrated that the estimation accuracy of the key peer effect parameter and predictive performance of the ENAR and AMNAR models are comparable to or better than that of other considered competitors. Our theoretical results provide conclusive answers to several henceforth open questions, including how to model peer effects with latent variables in a growing $N$ and $K$ setup, what are the tradeoffs between measurement error and omitted variable bias, how to select the dimension of latent variables, and for the growing $T$ case of network time series autoregressive model, is it possible to consistently estimate model parameters when latent variables are included. Therefore, this work greatly enhances currently available approaches for both causal effect estimation and prediction tasks.

\bibliographystyle{AoS/imsart-number} 
\bibliography{coevol,references,sarpeer} 

\begin{thebibliography}{45}

\bibitem{an2022causal}
\begin{barticle}[author]
\bauthor{\bsnm{An},~\bfnm{Weihua}\binits{W.}}, \bauthor{\bsnm{Beauvile},~\bfnm{Roberson}\binits{R.}} \AND \bauthor{\bsnm{Rosche},~\bfnm{Benjamin}\binits{B.}}
(\byear{2022}).
\btitle{Causal network analysis}.
\bjournal{Annual Review of Sociology}
\bvolume{48}
\bpages{23--41}.
\end{barticle}
\endbibitem

\bibitem{athreya2017statistical}
\begin{barticle}[author]
\bauthor{\bsnm{Athreya},~\bfnm{Avanti}\binits{A.}}, \bauthor{\bsnm{Fishkind},~\bfnm{Donniell~E}\binits{D.~E.}}, \bauthor{\bsnm{Tang},~\bfnm{Minh}\binits{M.}}, \bauthor{\bsnm{Priebe},~\bfnm{Carey~E}\binits{C.~E.}}, \bauthor{\bsnm{Park},~\bfnm{Youngser}\binits{Y.}}, \bauthor{\bsnm{Vogelstein},~\bfnm{Joshua~T}\binits{J.~T.}}, \bauthor{\bsnm{Levin},~\bfnm{Keith}\binits{K.}}, \bauthor{\bsnm{Lyzinski},~\bfnm{Vince}\binits{V.}} \AND \bauthor{\bsnm{Qin},~\bfnm{Yichen}\binits{Y.}}
(\byear{2017}).
\btitle{Statistical inference on random dot product graphs: a survey}.
\bjournal{The Journal of Machine Learning Research}
\bvolume{18}
\bpages{8393--8484}.
\end{barticle}
\endbibitem

\bibitem{basu2015regularized}
\begin{barticle}[author]
\bauthor{\bsnm{Basu},~\bfnm{Sumanta}\binits{S.}} \AND \bauthor{\bsnm{Michailidis},~\bfnm{George}\binits{G.}}
(\byear{2015}).
\btitle{{Regularized estimation in sparse high-dimensional time series models}}.
\bjournal{The Annals of Statistics}
\bvolume{43}
\bpages{1535 -- 1567}.
\end{barticle}
\endbibitem

\bibitem{cai2018rate}
\begin{barticle}[author]
\bauthor{\bsnm{Cai},~\bfnm{T.~Tony}\binits{T.~T.}} \AND \bauthor{\bsnm{Zhang},~\bfnm{Anru}\binits{A.}}
(\byear{2018}).
\btitle{{Rate-optimal perturbation bounds for singular subspaces with applications to high-dimensional statistics}}.
\bjournal{The Annals of Statistics}
\bvolume{46}
\bpages{60 -- 89}.
\end{barticle}
\endbibitem

\bibitem{cape2017kato}
\begin{barticle}[author]
\bauthor{\bsnm{Cape},~\bfnm{Joshua}\binits{J.}}, \bauthor{\bsnm{Tang},~\bfnm{Minh}\binits{M.}} \AND \bauthor{\bsnm{Priebe},~\bfnm{Carey~E.}\binits{C.~E.}}
(\byear{2017}).
\btitle{The Kato–Temple inequality and eigenvalue concentration with applications to graph inference}.
\bjournal{Electronic Journal of Statistics}
\bvolume{11}
\bpages{3954--3978}.
\end{barticle}
\endbibitem

\bibitem{cape2019signal}
\begin{barticle}[author]
\bauthor{\bsnm{Cape},~\bfnm{Joshua}\binits{J.}}, \bauthor{\bsnm{Tang},~\bfnm{Minh}\binits{M.}} \AND \bauthor{\bsnm{Priebe},~\bfnm{Carey~E}\binits{C.~E.}}
(\byear{2019}).
\btitle{Signal-plus-noise matrix models: eigenvector deviations and fluctuations}.
\bjournal{Biometrika}
\bvolume{106}
\bpages{243--250}.
\end{barticle}
\endbibitem

\bibitem{cape2019two}
\begin{barticle}[author]
\bauthor{\bsnm{Cape},~\bfnm{Joshua}\binits{J.}}, \bauthor{\bsnm{Tang},~\bfnm{Minh}\binits{M.}} \AND \bauthor{\bsnm{Priebe},~\bfnm{Carey~E.}\binits{C.~E.}}
(\byear{2019}).
\btitle{{The two-to-infinity norm and singular subspace geometry with applications to high-dimensional statistics}}.
\bjournal{The Annals of Statistics}
\bvolume{47}
\bpages{2405 -- 2439}.
\end{barticle}
\endbibitem

\bibitem{chen2023community}
\begin{barticle}[author]
\bauthor{\bsnm{Chen},~\bfnm{Elynn~Y}\binits{E.~Y.}}, \bauthor{\bsnm{Fan},~\bfnm{Jianqing}\binits{J.}} \AND \bauthor{\bsnm{Zhu},~\bfnm{Xuening}\binits{X.}}
(\byear{2023}).
\btitle{Community network auto-regression for high-dimensional time series}.
\bjournal{Journal of Econometrics}
\bvolume{235}
\bpages{1239--1256}.
\end{barticle}
\endbibitem

\bibitem{christakis2007spread}
\begin{barticle}[author]
\bauthor{\bsnm{Christakis},~\bfnm{Nicholas~A}\binits{N.~A.}} \AND \bauthor{\bsnm{Fowler},~\bfnm{James~H}\binits{J.~H.}}
(\byear{2007}).
\btitle{The spread of obesity in a large social network over 32 years}.
\bjournal{New England journal of medicine}
\bvolume{357}
\bpages{370--379}.
\end{barticle}
\endbibitem

\bibitem{christakis2013social}
\begin{barticle}[author]
\bauthor{\bsnm{Christakis},~\bfnm{Nicholas~A}\binits{N.~A.}} \AND \bauthor{\bsnm{Fowler},~\bfnm{James~H}\binits{J.~H.}}
(\byear{2013}).
\btitle{Social contagion theory: examining dynamic social networks and human behavior}.
\bjournal{Statistics in medicine}
\bvolume{32}
\bpages{556--577}.
\end{barticle}
\endbibitem

\bibitem{fosdick2015testing}
\begin{barticle}[author]
\bauthor{\bsnm{Fosdick},~\bfnm{Bailey~K}\binits{B.~K.}} \AND \bauthor{\bsnm{Hoff},~\bfnm{Peter~D}\binits{P.~D.}}
(\byear{2015}).
\btitle{Testing and modeling dependencies between a network and nodal attributes}.
\bjournal{Journal of the American Statistical Association}
\bvolume{110}
\bpages{1047--1056}.
\end{barticle}
\endbibitem

\bibitem{goldsmith2013social}
\begin{barticle}[author]
\bauthor{\bsnm{Goldsmith-Pinkham},~\bfnm{Paul}\binits{P.}} \AND \bauthor{\bsnm{Imbens},~\bfnm{Guido~W}\binits{G.~W.}}
(\byear{2013}).
\btitle{Social networks and the identification of peer effects}.
\bjournal{Journal of Business \& Economic Statistics}
\bvolume{31}
\bpages{253--264}.
\end{barticle}
\endbibitem

\bibitem{guan2023spectral}
\begin{barticle}[author]
\bauthor{\bsnm{Guan},~\bfnm{Yawen}\binits{Y.}}, \bauthor{\bsnm{Page},~\bfnm{Garritt~L}\binits{G.~L.}}, \bauthor{\bsnm{Reich},~\bfnm{Brian~J}\binits{B.~J.}}, \bauthor{\bsnm{Ventrucci},~\bfnm{Massimo}\binits{M.}} \AND \bauthor{\bsnm{Yang},~\bfnm{Shu}\binits{S.}}
(\byear{2023}).
\btitle{Spectral adjustment for spatial confounding}.
\bjournal{Biometrika}
\bvolume{110}
\bpages{699--719}.
\end{barticle}
\endbibitem

\bibitem{hall2014martingale}
\begin{bbook}[author]
\bauthor{\bsnm{Hall},~\bfnm{Peter}\binits{P.}} \AND \bauthor{\bsnm{Heyde},~\bfnm{Christopher~C}\binits{C.~C.}}
(\byear{2014}).
\btitle{Martingale limit theory and its application}.
\bpublisher{Academic press}.
\end{bbook}
\endbibitem

\bibitem{he2019multiplicative}
\begin{barticle}[author]
\bauthor{\bsnm{He},~\bfnm{Yanjun}\binits{Y.}} \AND \bauthor{\bsnm{Hoff},~\bfnm{Peter~D}\binits{P.~D.}}
(\byear{2019}).
\btitle{Multiplicative coevolution regression models for longitudinal networks and nodal attributes}.
\bjournal{Social Networks}
\bvolume{57}
\bpages{54--62}.
\end{barticle}
\endbibitem

\bibitem{hoff2021additive}
\begin{barticle}[author]
\bauthor{\bsnm{Hoff},~\bfnm{Peter}\binits{P.}}
(\byear{2021}).
\btitle{Additive and multiplicative effects network models}.
\bjournal{Statistical Science}.
\end{barticle}
\endbibitem

\bibitem{hoff2002latent}
\begin{barticle}[author]
\bauthor{\bsnm{Hoff},~\bfnm{Peter~D}\binits{P.~D.}}, \bauthor{\bsnm{Raftery},~\bfnm{Adrian~E}\binits{A.~E.}} \AND \bauthor{\bsnm{Handcock},~\bfnm{Mark~S}\binits{M.~S.}}
(\byear{2002}).
\btitle{Latent space approaches to social network analysis}.
\bjournal{Journal of the american Statistical association}
\bvolume{97}
\bpages{1090--1098}.
\end{barticle}
\endbibitem

\bibitem{knecht2010friendship}
\begin{barticle}[author]
\bauthor{\bsnm{Knecht},~\bfnm{Andrea}\binits{A.}}, \bauthor{\bsnm{Snijders},~\bfnm{Tom~AB}\binits{T.~A.}}, \bauthor{\bsnm{Baerveldt},~\bfnm{Chris}\binits{C.}}, \bauthor{\bsnm{Steglich},~\bfnm{Christian~EG}\binits{C.~E.}} \AND \bauthor{\bsnm{Raub},~\bfnm{Werner}\binits{W.}}
(\byear{2010}).
\btitle{Friendship and delinquency: Selection and influence processes in early adolescence}.
\bjournal{Social Development}
\bvolume{19}
\bpages{494--514}.
\end{barticle}
\endbibitem

\bibitem{knecht2008friendship}
\begin{bphdthesis}[author]
\bauthor{\bsnm{Knecht},~\bfnm{Andrea~Beate}\binits{A.~B.}}
(\byear{2008}).
\btitle{Friendship selection and friends' influence: dynamics of networks and actor attributes in early adolescence},
\btype{PhD thesis},
\bpublisher{University Utrecht}.
\end{bphdthesis}
\endbibitem

\bibitem{knight2019generalised}
\begin{barticle}[author]
\bauthor{\bsnm{Knight},~\bfnm{Marina}\binits{M.}}, \bauthor{\bsnm{Leeming},~\bfnm{Kathryn}\binits{K.}}, \bauthor{\bsnm{Nason},~\bfnm{Guy}\binits{G.}} \AND \bauthor{\bsnm{Nunes},~\bfnm{Matthew}\binits{M.}}
(\byear{2020}).
\btitle{Generalized Network Autoregressive Processes and the GNAR Package}.
\bjournal{Journal of Statistical Software}
\bvolume{96}
\bpages{1–36}.
\bdoi{10.18637/jss.v096.i05}
\end{barticle}
\endbibitem

\bibitem{le2022linear}
\begin{barticle}[author]
\bauthor{\bsnm{Le},~\bfnm{Can~M}\binits{C.~M.}} \AND \bauthor{\bsnm{Li},~\bfnm{Tianxi}\binits{T.}}
(\byear{2022}).
\btitle{Linear regression and its inference on noisy network-linked data}.
\bjournal{Journal of the Royal Statistical Society Series B: Statistical Methodology}
\bvolume{84}
\bpages{1851--1885}.
\end{barticle}
\endbibitem

\bibitem{lei2015consistency}
\begin{barticle}[author]
\bauthor{\bsnm{Lei},~\bfnm{Jing}\binits{J.}} \AND \bauthor{\bsnm{Rinaldo},~\bfnm{Alessandro}\binits{A.}}
(\byear{2015}).
\btitle{Consistency of spectral clustering in stochastic block models}.
\bjournal{The Annals of Statistics}
\bvolume{43}.
\end{barticle}
\endbibitem

\bibitem{li2023statistical}
\begin{barticle}[author]
\bauthor{\bsnm{Li},~\bfnm{Jinming}\binits{J.}}, \bauthor{\bsnm{Xu},~\bfnm{Gongjun}\binits{G.}} \AND \bauthor{\bsnm{Zhu},~\bfnm{Ji}\binits{J.}}
(\byear{2023}).
\btitle{Statistical Inference on Latent Space Models for Network Data}.
\bjournal{arXiv preprint arXiv:2312.06605}.
\end{barticle}
\endbibitem

\bibitem{li2022randnet}
\begin{barticle}[author]
\bauthor{\bsnm{Li},~\bfnm{T}\binits{T.}}, \bauthor{\bsnm{Levina},~\bfnm{E}\binits{E.}}, \bauthor{\bsnm{Zhu},~\bfnm{J}\binits{J.}} \AND \bauthor{\bsnm{Le},~\bfnm{CM}\binits{C.}}
(\byear{2022}).
\btitle{Randnet: Random Network Model Estimation, Selection and Parameter Tuning}.
\bjournal{R package version 0.5}.
\end{barticle}
\endbibitem

\bibitem{lunde2023conformal}
\begin{barticle}[author]
\bauthor{\bsnm{Lunde},~\bfnm{Robert}\binits{R.}}, \bauthor{\bsnm{Levina},~\bfnm{Elizaveta}\binits{E.}} \AND \bauthor{\bsnm{Zhu},~\bfnm{Ji}\binits{J.}}
(\byear{2023}).
\btitle{Conformal prediction for network-assisted regression}.
\bjournal{arXiv preprint arXiv:2302.10095}.
\end{barticle}
\endbibitem

\bibitem{ma2020universal}
\begin{barticle}[author]
\bauthor{\bsnm{Ma},~\bfnm{Zhuang}\binits{Z.}}, \bauthor{\bsnm{Ma},~\bfnm{Zongming}\binits{Z.}} \AND \bauthor{\bsnm{Yuan},~\bfnm{Hongsong}\binits{H.}}
(\byear{2020}).
\btitle{Universal latent space model fitting for large networks with edge covariates}.
\bjournal{Journal of Machine Learning Research}
\bvolume{21}
\bpages{1--67}.
\end{barticle}
\endbibitem

\bibitem{mcfowland2021estimating}
\begin{barticle}[author]
\bauthor{\bsnm{McFowland~III},~\bfnm{Edward}\binits{E.}} \AND \bauthor{\bsnm{Shalizi},~\bfnm{Cosma~Rohilla}\binits{C.~R.}}
(\byear{2023}).
\btitle{Estimating causal peer influence in homophilous social networks by inferring latent locations}.
\bjournal{Journal of the American Statistical Association}
\bpages{1--12}.
\end{barticle}
\endbibitem

\bibitem{nath2022identifying}
\begin{barticle}[author]
\bauthor{\bsnm{Nath},~\bfnm{Shanjukta}\binits{S.}}, \bauthor{\bsnm{Warren},~\bfnm{Keith}\binits{K.}} \AND \bauthor{\bsnm{Paul},~\bfnm{Subhadeep}\binits{S.}}
(\byear{2022}).
\btitle{Identifying Peer Influence in Therapeutic Communities Adjusting for Latent Homophily}.
\bjournal{arXiv preprint arXiv:2203.14223}.
\end{barticle}
\endbibitem

\bibitem{o2014estimating}
\begin{barticle}[author]
\bauthor{\bsnm{O'Malley},~\bfnm{A~James}\binits{A.~J.}}, \bauthor{\bsnm{Elwert},~\bfnm{Felix}\binits{F.}}, \bauthor{\bsnm{Rosenquist},~\bfnm{J~Niels}\binits{J.~N.}}, \bauthor{\bsnm{Zaslavsky},~\bfnm{Alan~M}\binits{A.~M.}} \AND \bauthor{\bsnm{Christakis},~\bfnm{Nicholas~A}\binits{N.~A.}}
(\byear{2014}).
\btitle{Estimating peer effects in longitudinal dyadic data using instrumental variables}.
\bjournal{Biometrics}
\bvolume{70}
\bpages{506--515}.
\end{barticle}
\endbibitem

\bibitem{pearl2009causality}
\begin{bbook}[author]
\bauthor{\bsnm{Pearl},~\bfnm{Judea}\binits{J.}}
(\byear{2009}).
\btitle{Causality}.
\bpublisher{Cambridge university press}.
\end{bbook}
\endbibitem

\bibitem{rohe2018note}
\begin{barticle}[author]
\bauthor{\bsnm{Rohe},~\bfnm{Karl}\binits{K.}}, \bauthor{\bsnm{Tao},~\bfnm{Jun}\binits{J.}}, \bauthor{\bsnm{Han},~\bfnm{Xintian}\binits{X.}} \AND \bauthor{\bsnm{Binkiewicz},~\bfnm{Norbert}\binits{N.}}
(\byear{2018}).
\btitle{A note on quickly sampling a sparse matrix with low rank expectation}.
\bjournal{Journal of Machine Learning Research}
\bvolume{19}
\bpages{1--13}.
\end{barticle}
\endbibitem

\bibitem{rubin2022statistical}
\begin{barticle}[author]
\bauthor{\bsnm{Rubin-Delanchy},~\bfnm{Patrick}\binits{P.}}, \bauthor{\bsnm{Cape},~\bfnm{Joshua}\binits{J.}}, \bauthor{\bsnm{Tang},~\bfnm{Minh}\binits{M.}} \AND \bauthor{\bsnm{Priebe},~\bfnm{Carey~E}\binits{C.~E.}}
(\byear{2022}).
\btitle{A statistical interpretation of spectral embedding: The generalised random dot product graph}.
\bjournal{Journal of the Royal Statistical Society Series B: Statistical Methodology}
\bvolume{84}
\bpages{1446--1473}.
\end{barticle}
\endbibitem

\bibitem{rudelson2013hanson}
\begin{barticle}[author]
\bauthor{\bsnm{Rudelson},~\bfnm{Mark}\binits{M.}} \AND \bauthor{\bsnm{Vershynin},~\bfnm{Roman}\binits{R.}}
(\byear{2013}).
\btitle{{Hanson-Wright inequality and sub-gaussian concentration}}.
\bjournal{Electronic Communications in Probability}
\bvolume{18}
\bpages{1 -- 9}.
\end{barticle}
\endbibitem

\bibitem{shalizi2011homophily}
\begin{barticle}[author]
\bauthor{\bsnm{Shalizi},~\bfnm{Cosma~Rohilla}\binits{C.~R.}} \AND \bauthor{\bsnm{Thomas},~\bfnm{Andrew~C}\binits{A.~C.}}
(\byear{2011}).
\btitle{Homophily and contagion are generically confounded in observational social network studies}.
\bjournal{Sociological methods \& research}
\bvolume{40}
\bpages{211--239}.
\end{barticle}
\endbibitem

\bibitem{tang2017semiparametric}
\begin{barticle}[author]
\bauthor{\bsnm{Tang},~\bfnm{Minh}\binits{M.}}, \bauthor{\bsnm{Athreya},~\bfnm{Avanti}\binits{A.}}, \bauthor{\bsnm{Sussman},~\bfnm{Daniel~L}\binits{D.~L.}}, \bauthor{\bsnm{Lyzinski},~\bfnm{Vince}\binits{V.}}, \bauthor{\bsnm{Park},~\bfnm{Youngser}\binits{Y.}} \AND \bauthor{\bsnm{Priebe},~\bfnm{Carey~E}\binits{C.~E.}}
(\byear{2017}).
\btitle{A semiparametric two-sample hypothesis testing problem for random graphs}.
\bjournal{Journal of Computational and Graphical Statistics}
\bvolume{26}
\bpages{344--354}.
\end{barticle}
\endbibitem

\bibitem{tang2018limit}
\begin{barticle}[author]
\bauthor{\bsnm{Tang},~\bfnm{Minh}\binits{M.}} \AND \bauthor{\bsnm{Priebe},~\bfnm{Carey~E}\binits{C.~E.}}
(\byear{2018}).
\btitle{Limit theorems for eigenvectors of the normalized Laplacian for random graphs}.
\bjournal{The Annals of Statistics}
\bvolume{46}
\bpages{2360--2415}.
\end{barticle}
\endbibitem

\bibitem{vanderweele2011sensitivity}
\begin{barticle}[author]
\bauthor{\bsnm{VanderWeele},~\bfnm{Tyler~J}\binits{T.~J.}}
(\byear{2011}).
\btitle{Sensitivity analysis for contagion effects in social networks}.
\bjournal{Sociological Methods \& Research}
\bvolume{40}
\bpages{240--255}.
\end{barticle}
\endbibitem

\bibitem{vanderweele2012and}
\begin{barticle}[author]
\bauthor{\bsnm{VanderWeele},~\bfnm{Tyler~J}\binits{T.~J.}}, \bauthor{\bsnm{Ogburn},~\bfnm{Elizabeth~L}\binits{E.~L.}} \AND \bauthor{\bsnm{Tchetgen},~\bfnm{Eric J~Tchetgen}\binits{E.~J.~T.}}
(\byear{2012}).
\btitle{Why and when" flawed" social network analyses still yield valid tests of no contagion}.
\bjournal{Statistics, Politics and Policy}
\bvolume{3}.
\end{barticle}
\endbibitem

\bibitem{vershynin2018high}
\begin{bbook}[author]
\bauthor{\bsnm{Vershynin},~\bfnm{Roman}\binits{R.}}
(\byear{2018}).
\btitle{High-dimensional probability: An introduction with applications in data science}
\bvolume{47}.
\bpublisher{Cambridge university press}.
\end{bbook}
\endbibitem

\bibitem{wu2023random}
\begin{barticle}[author]
\bauthor{\bsnm{Wu},~\bfnm{Weichi}\binits{W.}} \AND \bauthor{\bsnm{Leng},~\bfnm{Chenlei}\binits{C.}}
(\byear{2023}).
\btitle{A Random Graph-based Autoregressive Model for Networked Time Series}.
\bjournal{arXiv preprint arXiv:2309.08488}.
\end{barticle}
\endbibitem

\bibitem{xie2024entrywise}
\begin{barticle}[author]
\bauthor{\bsnm{Xie},~\bfnm{Fangzheng}\binits{F.}}
(\byear{2024}).
\btitle{Entrywise limit theorems for eigenvectors of signal-plus-noise matrix models with weak signals}.
\bjournal{Bernoulli}
\bvolume{30}
\bpages{388--418}.
\end{barticle}
\endbibitem

\bibitem{Xie2019EfficientEF}
\begin{barticle}[author]
\bauthor{\bsnm{Xie},~\bfnm{Fangzheng}\binits{F.}} \AND \bauthor{\bsnm{Xu},~\bfnm{Yanxun}\binits{Y.}}
(\byear{2023}).
\btitle{Efficient Estimation for Random Dot Product Graphs via a One-Step Procedure}.
\bjournal{Journal of the American Statistical Association}
\bvolume{118}
\bpages{651--664}.
\end{barticle}
\endbibitem

\bibitem{zhu2020grouped}
\begin{barticle}[author]
\bauthor{\bsnm{Zhu},~\bfnm{Xuening}\binits{X.}} \AND \bauthor{\bsnm{Pan},~\bfnm{Rui}\binits{R.}}
(\byear{2020}).
\btitle{Grouped network vector autoregression}.
\bjournal{Statistica Sinica}
\bvolume{30}
\bpages{1437--1462}.
\end{barticle}
\endbibitem

\bibitem{zhu2017network}
\begin{barticle}[author]
\bauthor{\bsnm{Zhu},~\bfnm{Xuening}\binits{X.}}, \bauthor{\bsnm{Pan},~\bfnm{Rui}\binits{R.}}, \bauthor{\bsnm{Li},~\bfnm{Guodong}\binits{G.}}, \bauthor{\bsnm{Liu},~\bfnm{Yuewen}\binits{Y.}} \AND \bauthor{\bsnm{Wang},~\bfnm{Hansheng}\binits{H.}}
(\byear{2017}).
\btitle{{Network vector autoregression}}.
\bjournal{The Annals of Statistics}
\bvolume{45}
\bpages{1096 -- 1123}.
\end{barticle}
\endbibitem

\bibitem{zhu2019network}
\begin{barticle}[author]
\bauthor{\bsnm{Zhu},~\bfnm{Xuening}\binits{X.}}, \bauthor{\bsnm{Wang},~\bfnm{Weining}\binits{W.}}, \bauthor{\bsnm{Wang},~\bfnm{Hansheng}\binits{H.}} \AND \bauthor{\bsnm{H{\"a}rdle},~\bfnm{Wolfgang~Karl}\binits{W.~K.}}
(\byear{2019}).
\btitle{Network quantile autoregression}.
\bjournal{Journal of econometrics}
\bvolume{212}
\bpages{345--358}.
\end{barticle}
\endbibitem

\end{thebibliography}

\newpage

\begin{appendix}
\setcounter{table}{0}
\renewcommand{\thetable}{A\arabic{table}}
\setcounter{figure}{0}
\renewcommand{\thefigure}{A\arabic{figure}}

\section{Proofs for ENAR model in Sections \ref{sec-station} \& \ref{sec-enar-consist}}
\subsection{Technical Results}
Here, we list some technical results that are useful in proving the asymptotic properties of our estimators for ENAR and AMNAR.
\begin{proposition}
    If $X_n=O(Y_n)$ whp. and $Y_n=O(Z_n)$ whp., then $X_n=O(Z_n)$ whp. (or as.) 
    If $X_n=O(Y_n)$ whp., then we have $\pr\lrp{\brcs{\vrt{X_n}\leq C\vrt{Y_n}}~ev.}=1$ by Borel-Cantelli lemma.
\end{proposition}
By the model assumptions, our responses of interest, $\by_t$, will have stationary sub-gaussian distributions.
To show the concentration of our estimators, we employ the theories of sub-Gaussian concentration as discussed in \cite{vershynin2018high}. However, since we assumed that both the observed graph $\bA$ and covariates $\bz_{it}$ are random, it necessitates investigating the asymptotic behaviors of the inner products of multiple random components. Therefore, we cannot directly apply well-known sub-Gaussian concentration results like Bernstein's inequality or Hoeffding's inequality.

For example, in the case of the Hanson-Wright inequality \citep{rudelson2013hanson}, we do not know how the tail behavior of concentrations will differ when studying the quadratic forms associated with random matrices. Under our assumptions on the random graphs and random predictors, we achieve stochastic boundedness for these elements. This fact can provide analogous concentration results.
\begin{lemma}[Hoeffding's Inequality]\label{lmm-hoeffding-whp}
    Let $\by\in\R^n$ be a sub-gaussian random vector such that $\max_i\nrm{Y_i}{\psi_2}<\infty$ with independent coordinates and zero mean. Let $\bx\in\R^n$ be a random vector such that $\nrm{\bx}{}=O(g)$ whp. for some $g>0$ and is independent with $\by$. Then, there exist some $c>0, n_0>0$ such that
    \[\pr(|\bx\T\by|\geq \upsilon)\leq2\exp\lrp{-\frac{c\upsilon^2}{g^2}}+\frac{1}{n}\]
    for all $n>n_0$ and $\upsilon>0$. If $g$ is a constant, then we have no $1/n$ on the RHS.
\end{lemma}
\begin{proof}
    Note that by general Hoeffding's inequality, with probability one we have
    \[\pr(|\bx\T\by|\geq \upsilon)=\pr\lrp{\vrt{\bx\T\by}\geq\upsilon|\bx}\leq2\exp\lrp{-\frac{c\upsilon^2}{\nrm{\bx}{}^2}}.\]
    Denote the LHS and RHS by $f_n(\bx)$ and $h_n(\bx)$, respectively.
    By assumption, there exist contants $M,n_0>0$ such that $\pr\lrp{\nrm{\bx}{}>Mg}<1/n$ for all $n>n_0$. Then, for a sequence of events $A_n\triangleq\brcs{\nrm{\bx}{}\leq Mg}$ we have
    \[f_n(\bx)\mathbb{I}_{A_n}\leq h_n(\bx)\mathbb{I}_{A_n}\leq2\exp\lrp{-\frac{c'\upsilon^2}{g^2}}\mathbb{I}_{A_n}\]
    for some constant $c'>0$. As $A_n$ is $\sigma(\bx)$-measurable, taking expectation, we get
    \[\E\bxs{\pr\lrp{A_n\cap\brcs{\vrt{\bx\T\by}\geq\upsilon}|\bx}}=\pr\lrp{A_n\cap\brcs{\vrt{\bx\T\by}\geq\upsilon}}\leq2\exp\lrp{-\frac{c'\upsilon^2}{g^2}}\pr(A_n)\]
    hence
    \begin{align*}
        \pr&\lrp{\brcs{\vrt{\bx\T\by}\geq\upsilon}\backslash {A_n}^c}\leq2\exp\lrp{-\frac{c'\upsilon^2}{g^2}} \\
        &\therefore~~\pr\lrp{\vrt{\bx\T\by}\ge\upsilon}\leq2\exp\lrp{-\frac{c'\upsilon^2}{g^2}}+\pr({A_n}^c).
    \end{align*}
    We will have no $\frac{1}{n}$ in the upper bound for almost surely bounded cases.
\end{proof}
\begin{lemma}
    \label{lmm-hoeffding-Op}
    In Lemma \ref{lmm-hoeffding-whp}, now suppose that $\nrm{\bx}{}=O_\pr(g)$. Then, $\forall\epsilon,\upsilon>0$ there exist some $c>0, n_0>0$ such that
    \[\pr(|\bx\T\by|\geq \upsilon)\leq2\exp\lrp{-\frac{c\upsilon^2}{g^2}}+\epsilon\]
    for all $n>n_0$.   
\end{lemma}
The proof of Lemma \ref{lmm-hoeffding-Op} coincides with Lemma \ref{lmm-hoeffding-whp}.
\begin{lemma} \label{thm-hwineq-whp}
    (Hanson-Wright Inequality) Let $\by\in\R^n$ be a random vector with independent coordinates and zero mean such that $\max_i\|Y_i\|_{\psi_2}<\infty$. Let $\bS\in\R^{n\times n}$ be a random matrix which is independent with $\by$ such that $\nrm{\bS}{F}^2=O(h)$ whp. and $\nrm{\bS}{}=O(k)$ whp. for some $h,k>0$. Then, there exists $n_0>0$ such that
    \[\pr\lrp{|\by\T\bS\by-\E(\by\T\bS\by)|>\upsilon}\leq
    2\exp\bxs{-c\min\lrp{\frac{\upsilon^2}{h},\frac{\upsilon}{k}}}+\frac{2}{n}\]
    for all $n>n_0$ and $\upsilon>0$. If $\nrm{\bS}{F}^2=O(h)$ as. and $\nrm{\bS}{}=O(k)$ as., then we have 
    \[\pr\lrp{|\by\T\bS\by-\E(\by\T\bS\by)|>\upsilon}\leq
    2\exp\bxs{-c\min\lrp{\frac{\upsilon^2}{h},\frac{\upsilon}{k}}}\]
    for all $n>0$.
\end{lemma}

\subsection{Concentration of \texorpdfstring{$\hat\bU-\tilde\bU$}{ASE}}\label{sec-prf-ase-cncnt}
In this section, we prove the asymptotic order of the measurement errors when approximating $\bU$ by $\hat\bU$ using either eigenvectors or spectral embedding.
\begin{proof}[Proof of Proposition \ref{prp-rdpg-cnt}]
    The spectral norm version of Lemma 5.1 of \cite{lei2015consistency} is straightforward as we have $\inf_{\bH\in\cO_K}\nrm{\bU_\bA-\bU_\bP\bH}{}=\Theta\brcs{\nrm{\sin\bC(\bU_\bP,\bU_\bA)}{}}$ and:
    \[\nrm{\sin\bC(\bU_\bP,\bU_\bA)}{}=\Theta\brcs{\nrm{\bU_\bP{\bU_\bP}\T\lrp{\bI-\bU_\bA{\bU_\bA}}\T}{}}\]
    where $\sin\bC(\bU_\bP,\bU_\bA)\triangleq\operatorname{diag}\bxs{\sin\circ\cos^{-1}s_1({\bU_\bP}\T\bU_\bA),\dots,\sin\circ\cos^{-1}s_K({\bU_\bP}\T\bU_\bA)}$ denotes the distances between the subspaces spanned by $\bU_\bP$ and $\bU_\bA$, measured in the $\sin$'s of canonical angles (Lemma 1 of \citealp{cai2018rate}).
    By Davis--Kahan Theorem we have
    \begin{align*}
        &\nrm{\bU_\bP{\bU_\bP}\T\lrp{\bI-\bU_\bA{\bU_\bA}}\T}{}\le\frac{2\nrm{\bA-\bP}{}}{s_K(\bP)}=O\lrp{\frac{1}{\sqrt{N\rho_N}}}~whp.
    \end{align*}
    so, there exists $\bH_{ev}\in\cO_K$ such that $\nrm{\hat\bU_{ev}-\bU_{ev}\bH_{ev}}{}=O\lrp{\frac{1}{\sqrt{N\rho_N}}}~whp.$, and this completes the proof for the concentration of $\hat\bU_{ev}-\bU_{ev}\bH_{ev}$.
    
    Now we proceed to the case of spectral embedding.
    Let us denote the orthogonal matrices containing left and right singular vectors of ${\bU_\bP}\T\bU_\bA$ by $\bH_1$ and $\bH_2$ respectively and define an orthogonal matrix $\bH^*\triangleq\bH_1{\bH_2}\T\in\cO_K$.
    Then, we have
    \begin{align*}
        \lrp{\bH^*|\bS_\bA|^{1/2}-{\bS_\bP}^{1/2}\bH^*}_{ij}=\lrp{\bH^*}_{ij}\frac{|s_j(\bA)|-s_i(\bP)}{|s_j(\bA)|^{1/2}+s_i(\bP)^{1/2}}\le\lrp{\bH^*}_{ij}\frac{|s_j(\bA)|-s_i(\bP)}{s_K(\bP)^{1/2}}
    \end{align*}
    hence by Kato-Temple inequality \citep{cape2017kato} and $\nrm{\bH}{F}=\sqrt{K}$,
    \begin{align*}
        \nrm{\bH^*|\bS_\bA|^{1/2}-{\bS_\bP}^{1/2}\bH^*}{}\le\nrm{\bH^*|\bS_\bA|^{1/2}-{\bS_\bP}^{1/2}\bH^*}{F}=O\lrp{\sqrt{\frac{K\log^2N}{N\rho_N}}}~whp.
    \end{align*}
    Let $\bR_1\triangleq\bU_\bP{\bU_\bP}\T\bU_\bA-\bU_\bP\bH^*$ and $\bR_2\triangleq\bH^*|\bS_\bA|^{1/2}-{\bS_\bP}^{1/2}\bH^*$.
    Note that $\bU_\bA|\bS_\bA|^{1/2}=\bA\bU_\bA|\bS_\bA|^{-1/2}$ and ${\bU_\bP}\T\bP=\bS_\bP{\bU_\bP}\T$.
    Then, 
    \begin{align*}
        \bU_\bA|\bS_\bA|^{1/2}-&\bU_\bP{\bS_\bP}^{1/2}\bH^*=\bU_\bA|\bS_\bA|^{1/2}-\bU_\bP\bH^*|\bS_\bA|^{1/2}+\bU_\bP\bR_2\\
        &=\lrp{\bU_\bA-\bU_\bP{\bU_\bP}\T\bU_\bA}|\bS_\bA|^{1/2}+\bR_1|\bS_\bA|^{1/2}+\bU_\bP\bR_2\\
        &=\lrp{\bI_N-\bU_\bP{\bU_\bP}\T}\bA\bU_\bA|\bS_\bA|^{-1/2}+\bR_1|\bS_\bA|^{1/2}+\bU_\bP\bR_2\\
        &=\lrp{\bI_N-\bU_\bP{\bU_\bP}\T}\lrp{\bA-\bP}\bU_\bA|\bS_\bA|^{-1/2}+\bR_1|\bS_\bA|^{1/2}+\bU_\bP\bR_2.
    \end{align*}
    It is easy to check that $\nrm{{\bU_\bP}\T\bU_\bA-\bH^*}{}=O\lrp{\frac{1}{N\rho_N}}$ whp. as
    \begin{align*}
        \nrm{{\bU_\bP}\T\bU_\bA-\bH^*}{}&\le\max_{i=1,\dots,K}\brcs{1-s_i({\bU_\bP}\T\bU_\bA)}\\
        &\le\max_i\vrt{1-s_i({\bU_\bP}\T\bU_\bA)^2}=\nrm{\sin\bC(\bU_\bP,\bU_\bA)}{}^2\lesssim\frac{\nrm{\bA-\bP}{}^2}{s_K(\bP)^2}.
    \end{align*}
    So, we observe that
    \begin{align*}
        \nrm{\bR_1}{}&\le\nrm{{\bU_\bP}\T\bU_\bA-\bH^*}{}=O\lrp{\frac{1}{N\rho_N}},
        \nrm{\bR_2}{}=O\lrp{\sqrt{\frac{K\log^2N}{N\rho_N}}}
        ~whp.
    \end{align*}
    and
    \begin{align*}
        \nrm{\lrp{\bI_N-\bU_\bP{\bU_\bP}\T}\lrp{\bA-\bP}\bU_\bA|\bS_\bA|^{-1/2}}{}\le\nrm{\bA-\bP}{}|s_K(\bS_\bA)|^{-1/2}=O\lrp{1}~whp.
    \end{align*}
    by $|s_K(\bA)|=\Theta\lrp{N\rho_N}~whp.$
    Therefore, we have
    \[\nrm{\bU_\bA|\bS_\bA|^{1/2}-\bU_\bP{\bS_\bP}^{1/2}\bH^*}{}=O\lrp{1+\sqrt{\frac{K\log^2N}{N\rho_N}}}~whp.\]
\end{proof}
In the following theorem, we establish the asymptotic orders of ASE concentration under under-embedding, i.e., the case where we embed $k$ eigenvectors of $\bA$ while the true latent dimension of the graph regime, $\operatorname{rank}(\bP)$, is $K > k$.

\begin{proof}[Proof of Theorem \ref{thm-rdpg-cnt2}]
    Basic steps are equal to the proof of Proposition \ref{prp-rdpg-cnt}.
    Again we have $\inf_{\bH_k\in\cO_{K,k}}\nrm{\bU_{\bA,k}-\bU_\bP\bH_k}{}=\Theta\brcs{\nrm{\sin\bC(\bU_\bP,\bU_{\bA,k})}{}}$ and:
    \[\nrm{\sin\bC(\bU_\bP,\bU_{\bA,k})}{}=\Theta\brcs{\nrm{\bU_\bP{\bU_\bP}\T\lrp{\bI-\bU_{\bA,k}{\bU_{\bA,k}}}\T}{}}.\]
    Now let $\bA_k=\bU_{\bA,k}\bS_{\bA,k}{\bU_{\bA,k}}\T$ be a truncated eigenvalue decomposition of $\bA$ where $\bS_{\bA,k}$ contains $k$ ($\le K$) largest magnitude eigenvalues of $\bA$ and $\bU_{\bA,k}\in\cO_{N,k}$ is a matrix with corresponding $k$ eigenvectors.
    By Weyl's inequality, for $\bU_\bA=\bxs{\bu_{\bA,1}\dots\bu_{\bA,K}}$ we have
    \[\nrm{\bA-\bA_k}{}=\nrm{\sum_{i=k+1}^Ks_i(\bA)\bu_{\bA,i}{\bu_{\bA,i}}\T}{}=s_{k+1}(\bA)\le s_{k+1}(\bP)+\nrm{\bA-\bP}{}.\]
    So, we have $\nrm{\bA_k-\bP}{}\le s_{k+1}(\bP)+2\nrm{\bA-\bP}{}$.
    Then, by Davis--Kahan Theorem,
    \begin{align*}
        &\nrm{\bU_\bP{\bU_\bP}\T\lrp{\bI-\bU_{\bA,k}{\bU_{\bA,k}}}\T}{}\le\frac{2\nrm{\bA_k-\bP}{}}{s_K(\bP)}\le\frac{2s_{k+1}(\bP)}{s_K(\bP)}+O\lrp{\frac{1}{\sqrt{N\rho_N}}}~whp.
    \end{align*}
    provided that $N\rho_N=\omega\lrp{\log^4N}$.
    Therefore, there exists $\bH_{sv,k}\in\cO_{K,k}$ such that $\nrm{\bU_{\bA,k}-\bU_{\bP}\bH_{sv,k}}{}=q_{k+1}+O\lrp{\frac{1}{\sqrt{N\rho_N}}}~whp.$ where $q_{k+1}\triangleq\frac{2s_{k+1}(\bP)}{s_K(\bP)}$. 
    Notice that $q_k\ge q_{k+1}$ for $k=1,\dots,K$ and $q_{k+1}=\Theta(1)$ as $s_i(\bP)=\Theta\lrp{N\rho_N}$ for all $i=1,\dots,K$ and $q_{k+1}=0$ for all $k\ge K$.
    This proves the first conclusion.
    
    Now consider the case of spectral embedding.
    Let us denote the orthogonal matrices containing left and right singular vectors of ${\bU_\bP}\T\bU_{\bA_k}$ by $\bH_{1,k}$ and $\bH_{2,k}$ respectively and define $\bH_k^*\triangleq\bH_{1,k}{\bH_{2,k}}\T\in\cO_{K,k}$. We have
    \begin{align*}
        \lrp{\bH_k^*|\bS_{\bA_k}|^{1/2}-{\bS_\bP}^{1/2}\bH_k^*}_{ij}=\lrp{\bH_k^*}_{ij}\frac{|s_j(\bA_k)|-s_i(\bP)}{|s_j(\bA_k)|^{1/2}+s_i(\bP)^{1/2}}\begin{cases}
            i=1,\dots,K \\ j=1,\dots,k
        \end{cases}
    \end{align*}
    hence by $\nrm{\bH_k^*}{F}=\sqrt{k}$ and the increased magnitudes of the elements at $i>k$,
    \begin{align*}
        \nrm{\bH_k^*|\bS_{\bA_k}|^{1/2}-{\bS_\bP}^{1/2}\bH_k^*}{}\le\nrm{\bH_k^*|\bS_{\bA_k}|^{1/2}-{\bS_\bP}^{1/2}\bH_k^*}{F}=O\lrp{\sqrt{kN\rho_N}}~whp.
    \end{align*}
    Let $\bR_1\triangleq\bU_\bP{\bU_\bP}\T\bU_{\bA_k}-\bU_\bP\bH_k^*$ and $\bR_2\triangleq\bH_k^*|\bS_{\bA_k}|^{1/2}-{\bS_\bP}^{1/2}\bH_k^*$. 
    Since $\nrm{\bU_{\bA_k}|\bS_{\bA_k}|^{1/2}}{}=\nrm{{\bA_k}\bU_{\bA_k}|\bS_{\bA_k}|^{-1/2}}{}$ and ${\bU_\bP}\T\bP=\bS_\bP{\bU_\bP}\T$, we have
    \begin{align*}
        \|\bU_{\bA_k}|\bS_{\bA_k}|^{1/2}-&\bU_\bP{\bS_\bP}^{1/2}\bH_k^*\|{}\le\nrm{\bU_{\bA_k}|\bS_{\bA_k}|^{1/2}-\bU_\bP\bH_k^*|\bS_{\bA_k}|^{1/2}}{}+\nrm{\bU_\bP\bR_2}{}\\
        &\le\nrm{\lrp{\bU_{\bA_k}-\bU_\bP{\bU_\bP}\T\bU_{\bA_k}}|\bS_{\bA_k}|^{1/2}}{}+\nrm{\bR_1|\bS_{\bA_k}|^{1/2}}{}+\nrm{\bR_2}{}\\
        &\le\nrm{\lrp{\bI_N-\bU_\bP{\bU_\bP}\T}{\bA_k}\bU_{\bA_k}|\bS_{\bA_k}|^{-1/2}}{}+\nrm{\bR_1|\bS_{\bA_k}|^{1/2}}{}+\nrm{\bR_2}{}\\
        &\le\nrm{\lrp{\bI_N-\bU_\bP{\bU_\bP}\T}\lrp{{\bA_k}-\bP}\bU_{\bA_k}|\bS_{\bA_k}|^{-1/2}}{}+\nrm{\bR_1|\bS_{\bA_k}|^{1/2}}{}+\nrm{\bR_2}{}.
    \end{align*}
    Check that $\nrm{{\bU_\bP}\T\bU_{\bA_k}-\bH_k^*}{}=O\lrp{1}$ whp., and we have
    \begin{align*}
        \nrm{\bR_1}{}&=\nrm{{\bU_\bP}\T\bU_{\bA_k}-\bH_k^*}{}=O\lrp{1},
        \nrm{\bR_2}{}=O\lrp{\sqrt{N\rho_N}}
        ~whp.
    \end{align*}
    and
    \begin{align*}
        \nrm{\lrp{\bI_N-\bU_\bP{\bU_\bP}\T}\lrp{{\bA_k}-\bP}\bU_{\bA_k}|\bS_{\bA_k}|^{-1/2}}{}\le\nrm{{\bA_k}-\bP}{}\nrm{|\bS_{\bA_k}|^{-1/2}}{}=O\lrp{\sqrt{N\rho_N}}~whp.
    \end{align*}
    Therefore, we have
    \[\nrm{\bU_{\bA_k}|\bS_{\bA_k}|^{1/2}-\bU_\bP{\bS_\bP}^{1/2}\bH_k^*}{}=O\lrp{\sqrt{N\rho_N}}~whp.\]
\end{proof}
\subsection{Stationarity}
To show stationarity, we adapt the proofs of \cite{zhu2017network} to the setup of ENAR.
\begin{proof}[Proof of Theorem \ref{thm-enar-station}]
    First, note that the maximum absolute eigenvalue of $\cL_\bA$ is strictly less than one.
    So it is straightforward that the spectral radius of $\bG$, denoted by $g$, satisfies
    \begin{align}\label{eq-g}
        g\leq|\alpha|+|\theta|<1
    \end{align}
    with probability one.
    Therefore, $\sum_{j=0}^\infty\bG^j\tilde\bcE_{t-j}$ exists as., and $\by_t$ in \ref{eq-enar-stationsol} is a strictly stationary process.
    It is straightforward that \ref{eq-enar-stationsol} satisfies \ref{eq-enar}.
    Next, assume that $\bar\by_t$ is another strictly stationary solution with $\E\|\bar\by_t\|<\infty$. Then, 
    $$\bar\by_t=\sum_{j=0}^{m-1}\bG^j(\bU\beta+\tilde\bcE_{t-j})+\bG^m\bar\by_{t-m}$$ for any positive integer $m$. Therefore, \[\E^*\|\by_t-\bar\by_t\|=\E^*\left\|\sum_{j=m}^\infty\bG^j(\bU\beta+\tilde\bcE_{t-j})-\bG^m\bar\by_t\right\|\leq Cg^m\] for a constant $C$ independent of $t$ and $m$. Growing $m$ to infinity, we get $\E\|\by_t-\bar\by_t\|=0$ hence $\by_t=\bar\by_t$ almost surely.
\end{proof}
Therefore, as $\bG^j\tilde\bcE_{t-j}\sim\cN_N\bxs{\bo,\lrp{\sigma^2+\gamma\T\Sigma_z\gamma}\bG^{2j}}$, the movinag average representation in \ref{eq-enar-stationsol} will follow a normal distribution with mean $\E^*(\by_t)=\lrp{\bI_N-\bG}^{-1}\bU\beta$ and $\operatorname{Var}^*(\by_t)=\lrp{\sigma^2+\gamma\T\Sigma_z\gamma}\sum_{j=0}^\infty\bG^{2j}$ for all $t$. After solving the Yule-Walker equation for determining $\Gamma$, we get Lemma \ref{lmm-enar-meancov}.
Next, we prove Theorem \ref{thm-enar-station2} according to the following definition.
\begin{definition}[\citealp{zhu2017network}]
    \label{def-sstation}
    Let $\{\by_t\in\R^N\}$ be an $N$--dimensional vector with $N\rightarrow\infty$. Define $\cM\triangleq\{\omega\in\R^\infty:\sum_{i=1}^\infty|w_i|<\infty\}$. For each $\omega\in\cM$, let $\bw_N=(w_1,...,w_N)\T\in\R^N$ be the truncated $N$--dimensional process. $\by_t$ is said to be strictly stationary if $\forall \omega\in\cM$
    \begin{enumerate}
        \item $\by_t^\omega=\lim_{N\rightarrow\infty}{\bw_N}\T\by_t$ exists almost surely.
        \item $\by_t^\omega$ is strictly stationary.
    \end{enumerate}
\end{definition}
\begin{proof}[Proof of Theorem \ref{thm-enar-station2}]
    To prove the existence of a stationary solution, it suffices to show that \ref{eq-enar-stationsol} is strictly stationary according to the above definition. Write $|\bA|_e$ as a matrix of absolute elements of a matrix $\bA$. Moreover, write $\bA\preccurlyeq\bB$ if $\bB$ is not less than $\bA$ elementwisely.
    Recall that $\by_t=\sum_{j=0}^{m-1}\bG^j(\bU\beta+\tilde\bcE_{t-j})+\bG^m\by_{t-m}$ hence $$\by_t=\lim_{m\rightarrow\infty}\by_t=\sum_{j=0}^\infty\bG^j(\bU\beta+\tilde\bcE_{t-j}).$$
    For the columns of $\bU=\bU_{ev}$, say $U_1,...,U_K$, We have $\bU\beta=\sum_{j=1}^KU_j\beta_j$ and its $\ell_\infty$-norm is bounded as $\nrm{\bU\beta}{\infty}\leq\sum_{j=1}^K\vrt{\beta_j}\nrm{U_j}{\infty}\leq\nrm{\beta}{1}$  hence $\vrt{\bU\beta}_e\preccurlyeq\nrm{\beta}{1}\bon_N$. 
    Next if $\bU=\bU_{se}$, then $\nrm{\bU\beta}{\infty}\lesssim\nrm{\bU_\bP}{\infty}s_1(\bP)^{1/2}\nrm{\beta}{\infty}\le\sqrt{K}\nrm{\bU_\bP}{\tti}s_1(\bP)^{1/2}\nrm{\beta}{\infty}=O\lrp{\sqrt{K\rho_N}\nrm{\beta}{\infty}}$  hence $\vrt{\bU\beta}_e\preccurlyeq \nrm{\beta}{\infty}\sqrt{K\rho_N}\bon_N$.
    Note that $K$ is fixed in this proof and $\rho_N$ is not greater than one.
    So, for both choice of $\bU$, we have $\E\vrt{\bU\beta+\tilde\bcE_{t-j}}_e\preccurlyeq C\cdot\bon_N$ for $C=\sqrt{K}\nrm{\beta}{1}+\E\vrt{\bz\T_1\gamma}+\E\vrt{\epsilon_{11}}$.
    Since $\cL_\bA$ is normal and symmetric, its spectral decomposition can be given as $\cL_\bA=\bU_\cL\bS_\cL{\bU_\cL}\T$ with orthogonal eigenvectors $\bU_\cL\in\R^{N\times N}$ and $\bS_\cL$ containing corresponding eigenvalues of $\cL_\bA$.
    So, we have $\nrm{{\cL_\bA}^j}{\infty}\leq\nrm{\bU_\cL}{\infty}\nrm{{\bU_\cL}\T}{\infty}\rho\lrp{\cL_\bA}^j\leq C'$ for a constant $C'>0$ independent of $j$.
    Therefore, $|\bG|_e^j\bon_N=(|\theta|\cL_\bA+|\alpha|\bI_N)^j\bon_N\preccurlyeq C'(|\theta|+|\alpha|)^j\bon_N$. Consequently,
    \begin{align*}
        \E^*\vrt{{\bw_N}\T\by_t}\leq\nrm{\bw_N}{1}\E^*\nrm{\by_t}{\infty}\leq\sum_{i=1}^\infty|w_i|\sum_{j=0}^\infty\E^*\nrm{\vrt{\bG}_e^j\vrt{\bU\beta+\tilde\bcE_{t-j}}_e}{\infty}\\
        \lesssim\sum_{i=1}^\infty|w_i|\sum_{j=0}^\infty(|\theta|+|\alpha|)^j
    \end{align*} 
    implying that $\lim_{N\rightarrow\infty}{\bw_N}\T\by_t$ exists almost surely. Next, assume that $\bar\by_t$ is another strictly stationary solution with a finite first moment. Then, $\E\vrt{\bar\by_t}_e\precsim\bon_N$. We have 
    \[\E^*\vrt{{\bw_N}\T(\by_t-\bar\by_t)}=\E^*\vrt{\sum_{j=m}^\infty{\bw_N}\T\bG^j(\bU\beta+\tilde\bcE_{t-j})-{\bw_N}\T\bG^m\bar\by_{t-m}}\]
    which is bounded above by the product of a constant and
    $$\sum_{i=1}^\infty |w_i|\sum_{j=m}^\infty\brcs{(|\alpha|+|\theta|)^j+(|\alpha|+|\theta|)^m}$$
    for any $\omega\in\cM$. Growing $m\rightarrow\infty$, we have ${\bw_N}\T(\by_t-\bar\by_t)=0$ as. hence $\by_t=\bar\by_t$ as.
\end{proof}

\subsection{ENAR with \texorpdfstring{$\bU_{ev}$ under growing $N,T$}{U(sv) under growing N, T}}
\subsubsection{Consistency}\label{sec-enar-consist-prf}
Here, we prove the asymptotic normality of $\hat\mu$ for both cases where $K$ is fixed and growing.
First, we clarify some notations here. Let $\Gamma_\by\triangleq\operatorname{Cov}^*(\by)$. Write ${\Phi_{ev}}\triangleq\bon_T\otimes\varphi_{ev}$ and $\tilde\by\triangleq{\Gamma_\by}^{-\frac{1}{2}}(\by-{\Phi_{ev}})$ so that $\by={\Gamma_\by}^{1/2}\tilde\by+{\Phi_{ev}}$. Note that the entries of $\tilde\by$ are independent by the property of multivariate normal distribution. 

By Proposition 2.2 and 2.3 of \cite{basu2015regularized}, with probability one, we have
\[
\|\Gamma_\by\|\leq\frac{\sigma^2}{m_{\min}(\bG)}\leq\frac{\sigma^2\nrm{\bU_\bG}{}^2\nrm{{\bU_\bG}^{-1}}{}^2}{(1-g)^2}
\]
where $g$ is the spectral radius of $\bG$ in \ref{eq-g}, $\bU_\bG$ is an orthogonal matrix that contains eigenvectors of $\bG$ and $m_{\min}(\bG)\triangleq\min_{\{z\in\mathcal{C};|z|=1\}}(\bI_N-\bG z)^*(\bI_N-\bG z)$. The last upper bound holds because $\bG$ is diagonalizable. So, we have $\nrm{\Gamma_\by}{}=O(1)$ as.

\begin{proof}[Proof of Theorem \ref{thm-enar-ev-clt-NT}]
    First, define $\tilde\bU_{ev}\triangleq\bU_{ev}\bH_{ev}$ and let 
    $\bW^H_t\triangleq\bxs{\tilde\bU_{ev} \,\vert\, \by_t, \cL_\bA\by_t \, \vert\, \bZ_{t}}$,
    $\mu_{ev,w}^H=\lrp{\frac{{\beta_{ev}}\T\bH_{ev}}{\sqrt{N}},\alpha_{ev},\theta_{ev},\gamma_{ev}\T}\T$,
    $\cW_t\triangleq\bxs{\sqrt{N}\bU_{ev} \,\vert\, \by_t, \cL_\bA\by_t \, \vert\, \bZ_{t}}$,
    $\cW^H_t\triangleq\bxs{\sqrt{N}\tilde\bU_{ev} \,\vert\, \by_t, \cL_\bA\by_t \, \vert\, \bZ_{t}}$,
    and  $\hat{\cW_t}=\bxs{\sqrt{N}\hat\bU_{ev} \,\vert\, \by_t, \cL_\bA\by_t \, \vert\, \bZ_{t}}$.
    Under this representation, we have $\hat\cW_t\bD_{ev}=\sqrt{NT}\hat\bW_t$, $\sqrt{NT}{\bD_{ev}}^{-1}\mu_{ev,w}^H=\mu_{ev}^H$, and $\bW^H_t\mu_{ev}^H=\cW_t^H\mu_{ev,w}^H$.
    Collect and bind them row-wise for $t=0,...,T-1$ to obtain $NT\times\lrp{K+p+2}$ matrices $\bW^H$, $\cW^H$, and define $\hat{\cW}$ analogously.
    For $\hat\Sigma_{ev}\triangleq\frac{1}{NT}{\hat{\cW}}\T\hat{\cW}$ and $\br_{w\epsilon}\triangleq\frac{1}{NT}{\hat{\cW}}\T\bcE$, we have 
    \begin{align*}
    \hat\mu_{ev} & = \lrp{{\hat\bW}\T\hat\bW}^{-1}{\hat\bW}\T\lrp{\bW^H\mu_{ev}^H+\bcE} \\
    & = \lrp{\frac{\bD_{ev}}{NT}{\hat{\cW}}\T\hat{\cW}\bD_{ev}}^{-1}\frac{\bD_{ev}}{\sqrt{NT}}{\hat{\cW}}\T\lrp{\cW^H\mu_{ev,w}^H+\bcE} \\
    & = \lrp{\hat\Sigma_{ev}\bD_{ev}}^{-1}\frac{1}{\sqrt{NT}}\hat{\cW}\T\brcs{\lrp{\cW^H-\hat{\cW}}\mu_{ev,w}^H+\hat{\cW}\mu_{ev,w}^H+\bcE} \\
    & = {\bD_{ev}}^{-1}{\hat\Sigma_{ev}}^{-1}\brcs{\frac{1}{\sqrt{NT}}\hat{\cW}\T\lrp{\cW^H-\hat{\cW}}\mu_{ev,w}^H+\sqrt{NT}\br_{w\epsilon}}+\sqrt{NT}{\bD_{ev}}^{-1}\mu_{ev,w}^H \\
    & \therefore \bD_{ev}\lrp{\hat\mu_{ev}-\mu_{ev}^H}=\frac{{\hat\Sigma_{ev}}^{-1}}{\sqrt{NT}}\hat{\cW}\T\lrp{\cW^H-\hat{\cW}}\mu_{ev,w}^H+\sqrt{NT}{\hat\Sigma_{ev}}^{-1}\br_{w\epsilon}.
    \end{align*}
    Therefore, we next show that the first term on the RHS is negligible and $\sqrt{NT}{\hat\Sigma_{ev}}^{-1}\br_{w\epsilon}$ is converging to a multivariate normal distribution. We start by claiming that $\hat\Sigma_{ev}$ is converging to a matrix with finite entries as $N$ and $T$ tend to infinity. This will allow us to focus on the behaviors of $\frac{1}{\sqrt{NT}}{\hat{\cW}}\T(\cW^H-\hat{\cW})\mu_{ev,w}^H$ and $\sqrt{NT}\br_{w\epsilon}$, and then apply Slutsky's Theorem.
    \begin{claim}
        $\hat\Sigma_{ev}$ converges to $\Sigma_{ev}$ in probability, i.e., ${\hat\Sigma_{ev}}\Rightarrow{\Sigma_{ev}}$. 
    \end{claim}
    \proof
    As consequences of statements 1--9 from Lemma \ref{lmm-enar-ev-nasym}, we have
        \begin{align}\label{eq:enar-ev-∑}
            \hat\Sigma_{ev}&=\frac{1}{NT}\sum_{t=0}^{T-1}\bxs{\begin{matrix}
            N\bI_K & \sqrt{N}{\hat\bU_{ev}}\T\by_t & \sqrt{N}\hat\bU_{ev}\T\cL_\bA\by_t & \sqrt{N}\hat\bU_{ev}\T\bZ_t \\
            & {\by_t}\T\by_t & {\by_t}\T\cL_\bA\by_t & {\by_t}\T\bZ_t \\
            & & {\by_t}\T{\cL_\bA}^2\by_t & {\by_t}\T\cL_\bA\bZ_t \\
            & & & {\bZ_t}\T\bZ_t
            \end{matrix}}
            \\&\Rightarrow
            \lim_{N,T\rightarrow\infty}\bxs{\begin{matrix}
            \bI_K & \frac{1}{\sqrt{N}}\tilde\bU_{ev}\T{\varphi_{ev}} & \frac{1}{\sqrt{N}}\tilde\bU_{ev}\T\cL_\bA{\varphi_{ev}} & \bO_{K\times p} \\
            & \frac{1}{N}{\varphi_{ev}}\T{\varphi_{ev}}+\tau_2 & \frac{1}{N}{\varphi_{ev}}\T\cL_\bA{\varphi_{ev}}+\tau_{23} & \bo_p\T \\
            & & \frac{1}{N}{\varphi_{ev}}\T{\cL_\bA}^2{\varphi_{ev}}+\tau_3 & \bo_p\T \\
            & & & \Sigma_z
            \end{matrix}}=\Sigma_{ev}
        \end{align}
    for $N\rho_N=\omega(\log^4N)$.
    Existence of $\tau_2,\tau_{23},$ and $\tau_3$ come by dominated convergence theorem after noting that $\tr\lrp{\cL_\bA\Gamma}\leq\nrm{\cL_\bA}{}\tr\lrp{\Gamma}=O\lrp{N}$ as. and $\tr\lrp{{\cL_\bA}^2\Gamma}=O(N)$ as.
    Also noting that $\nrm{\tilde\bU_{ev}}{}=1$ and the asymptotic order of each term found in the proof of Lemma \ref{lmm-enar-ev-nasym}, we have
    \begin{align*}
        \lim_{N\rightarrow\infty}\frac{1}{\sqrt{N}}{\bh_1}\T\tilde\bU_{ev}\T{\varphi_{ev}}&=0,
        \frac{1}{\sqrt{N}}{\bh_1}\T\tilde\bU_{ev}\T\cL_\bA{\varphi_{ev}}\Rightarrow0,
        \lim_{N\rightarrow\infty}\frac{1}{N}{\varphi_{ev}}\T{\varphi_{ev}}=0,\\
        \frac{1}{N}{\varphi_{ev}}\T\cL_\bA{\varphi_{ev}}&\Rightarrow0,
        \frac{1}{N}{\varphi_{ev}}\T{\cL_\bA}^2{\varphi_{ev}}\Rightarrow0
    \end{align*}
    for all $\bh_1\in B^{K-1}$.
    \begin{claim}
        $\frac{1}{\sqrt{NT}}{\hat{\cW}}\T(\cW^H-\hat{\cW})\mu_{ev,w}^H\Rightarrow\bo_{K+p+2}$.
    \end{claim}
    \proof
    Since $\hat{\cW}$ is different from $\cW^H$ by $\sqrt{N}\hat\bU_{ev}$ only, we have
    \begin{align*}
        \frac{1}{\sqrt{NT}}{\hat{\cW}}\T\lrp{\cW^H-\hat{\cW}}\mu_{ev,w}^H
        =\begin{bmatrix}
            \sqrt{T}\hat\bU_{ev}\T \\
            \frac{1}{\sqrt{NT}}\sum_{t=0}^{T-1}{\by_t}\T \\
            \frac{1}{\sqrt{NT}}\sum_{t=0}^{T-1}{\by_t}\T\cL_\bA \\
            \frac{1}{\sqrt{NT}}\sum_{t=0}^{T-1}{\bZ_t}\T
        \end{bmatrix}
        \lrp{\hat\bU_{ev}-\tilde\bU_{ev}}{\bH_{ev}}\T\beta_{ev}.
    \end{align*}
    Then, by the statements 10--13 of Lemma \ref{lmm-enar-ev-nasym}, for any $\bh\in B^{K+p+1}$, we have
    \[\pr\bxs{\frac{1}{\sqrt{NT}}\vrt{\bh\T{\hat{\cW}}\T\lrp{\cW^H-\hat{\cW}}\mu_{ev,w}^H}>\upsilon}<C_1/N+C_2\exp\lrp{-cN^2\rho_N\upsilon^2}\]
    for some constants $c,C_1,C_2>0$ and any $\upsilon$ such that $\upsilon\ge C'\sqrt{\frac{T}{N\rho_N}}$ for some $C'>0$. Provided that $N\rho_N=\omega(T)$, we have the conclusion by Cramér–Wold.
    \begin{claim}
        $\sqrt{NT}\br_{w\epsilon}\Rightarrow\cN(\bo_{K+p+2},\sigma^2\Sigma_{ev})$.
    \end{claim}
    \proof
    It is sufficient to show that for any $\eta\in\R^{K+p+2}$ such that $\|\eta\|\leq1$, we have $\sqrt{NT}\eta\T\br_{w\epsilon}\Rightarrow\cN(0,\sigma^2\eta\T\Sigma_{ev}\eta)$. Denoting $ \xi_{Nt}\triangleq\lrp{NT}^{-1/2}\eta\T{\hat\cW_{t-1}}\T\bcE_t$ and $\cF_{Nt}=\sigma(\bA,\epsilon_{is},\bZ_{i,(s-1)};i\leq N,-\infty<s\leq t)$, $\{\sum_{s=1}^t \xi_{Ns},\cF_{Nt}\}$ constitutes a martingale array for each $N,t\leq T$. Then, we can apply Corollary 3.1 of \cite{hall2014martingale} to $\sqrt{NT}\eta\T\br_{w\epsilon}=\sum_{t=0}^{T-1}\xi_{N,t+1}$ by checking following two conditions:
    
    \textbf{(1)} $\sum_{t}\E\lrp{\xi_{N,t+1}^21_{\brcs{\vrt{\xi_{N,t+1}}>\upsilon}}|\cF_{Nt}}=o_\pr(1)$. 
    
    \textbf{(2)} $\sum_t\E\lrp{\xi_{N,t+1}^2|\cF_{Nt}}=\eta\T\sigma^2\Sigma_{ev}\eta+o_\pr(1)$.
    \proof[Proof of (1)]
    First, we have
    \[\sum_{t=0}^{T-1}\E\lrp{\xi_{N,t+1}^21_{\brcs{\vrt{\xi_{N,t+1}}>\upsilon}}|\cF_{Nt}}\leq \upsilon^{-2}\sum_t\E\lrp{\xi^4_{N,t+1}|\cF_{Nt}}.\]
    One can easily verify that
    \[
    \E\lrp{\xi_{N,t+1}^4|\cF_{Nt}}\lesssim\sigma^4\lrp{\frac{1}{NT}\eta\T{\hat{\cW_t}}\T{\hat{\cW_t}}\eta}^2.
    \]
    So, we only need to show that $\sum_t\lrp{\frac{1}{NT}\eta\T{\hat{\cW_t}}\T{\hat{\cW_t}}\eta}^2=o_\pr(1)$. First note that 
    \begin{align*}
    \lrp{NT}^{-1}&{\hat{\cW_t}}\T\hat{\cW_t}\\
    &=\bxs{\begin{matrix}
    \frac{1}{T}\bI_K & \frac{1}{\sqrt{N}T}\hat\bU_{ev}\T\by_t & \frac{1}{\sqrt{N}T}\hat\bU_{ev}\T\cL_\bA\by_t & \frac{1}{\sqrt{N}T}\hat\bU_{ev}\T\bZ_t \\
    & \frac{1}{NT}{\by_t}\T\by_t & \frac{1}{NT}{\by_t}\T\cL_\bA\by_t & \frac{1}{NT}{\by_t}\T\bZ_t \\
    & & \frac{1}{NT}{\by_t}\T{\cL_\bA}^2\by_t & \frac{1}{NT}{\by_t}\T\cL_\bA\bZ_t \\
    & & & \frac{1}{NT}\bZ_t\T\bZ_t \\
    \end{matrix}}.
    \end{align*}
    Since similar arguments can be used to show the convergence of each entry, take 
    \[\sum_{t=0}^{T-1}\lrp{\frac{1}{\sqrt{N}T}{\eta_1}\T\hat\bU_{ev}\T\by_t}^2=\frac{1}{NT^2}\sum_t{\eta_1}\T\hat\bU_{ev}\T\by_t{\by_t}\T\hat\bU_{ev}\eta_1\]
    for example, where $\eta_1\in\R^K;\nrm{\eta_1}{}\leq1$.
    Then, we have
    \begin{align*}
        \sum_{t=0}^{T-1}{\eta_1}\T\hat\bU_{ev}\T\by_t{\by_t}\T\hat\bU_{ev}\eta_1=\by\T\brcs{\bI_T\otimes\lrp{\hat\bU_{ev}\eta_1{\eta_1}\T\hat\bU_{ev}\T}}\by.
    \end{align*}
    Denoting $\hat\bU_{ev}-\tilde\bU_{ev}$ by $\Delta_{ev}$, we get 
    \[\tilde\by\T{\Gamma_\by}^{1/2}\brcs{\bI_T\otimes\lrp{\Delta_{ev}\eta_1{\eta_1}\T\Delta_{ev}\T}}{\Gamma_\by}^{1/2}\tilde\by=\tilde\by\T\brcs{\bI_T\otimes\lrp{\Gamma^{1/2}\Delta_{ev}\eta_1{\eta_1}\T\Delta_{ev}\T\Gamma^{1/2}}}\tilde\by.\]
    Letting $\hat\bS\triangleq\bI_T\otimes\lrp{\Gamma^{1/2}\Delta_{ev}\eta_1{\eta_1}\T\Delta_{ev}\T\Gamma^{1/2}}$ and $\tilde\bS\triangleq\bI_T\otimes\lrp{\Gamma^{1/2}\tilde\bU_{ev}\eta_1{\eta_1}\T\tilde\bU_{ev}\T\Gamma^{1/2}}$, we have  $\nrm{\hat\bS}{F}^2=T\tr\brcs{\lrp{\Delta_{ev}\eta_1{\eta_1}\T\Delta_{ev}\T\Gamma}^2}\leq T\nrm{\Delta_{ev}\eta_1{\eta_1}\T\Delta_{ev}\T}{}^2\tr\lrp{\Gamma^2}=O\lrp{NT\lrp{N\rho_N}^{-2}}$ whp. and $\nrm{\hat\bS}{}=\nrm{\Gamma^{1/2}\Delta_{ev}\eta_1{\eta_1}\T\Delta_{ev}\T\Gamma^{1/2}}{}\leq\nrm{\Delta_{ev}}{}^2\nrm{\Gamma}{}=O(\lrp{N\rho_N}^{-1})$ whp.
    By Lemma \ref{thm-hwineq-whp},
    \[\pr\bxs{\frac{1}{\sqrt{NT}}\vrt{\tilde\by\T\hat\bS\tilde\by-\E\lrp{\tilde\by\T\hat\bS\tilde\by}}>\upsilon}<2\exp\lrp{-c\min\brcs{\frac{\upsilon^2}{(N\rho_N)^{-2}},\frac{\upsilon\sqrt{NT}}{\lrp{N\rho_N}^{-1}}}}+\frac{2}{N}.\]
    Note that $\E\lrp{\tilde\by\T\hat\bS\tilde\by}=T\E\tr\lrp{\Delta_{ev}\eta_1{\eta_1}\T\Delta_{ev}\T\Gamma}$ and $T\tr\lrp{\Delta_{ev}\eta_1{\eta_1}\T\Delta_{ev}\T\Gamma}\leq T\nrm{\Delta_{ev}\eta_1{\eta_1}\T\Delta_{ev}\T}{}\tr\lrp{\Gamma}$ hence is $O\lrp{T/\rho_N}$ whp. Therefore, $\lim_{N,T\rightarrow\infty}\frac{\tilde\by\T\hat\bS\tilde\by}{T/\rho_N}<\infty$ as. hence $\tilde\by\T\hat\bS\tilde\by=o_\pr(NT^2)$. 
    So we have $\E\lrp{\tilde\by\T\hat\bS\tilde\by}=o\lrp{NT^2}$ by dominated convergence theorem.
    Next, observe that $\nrm{\bG}{}\leq|\alpha_1|+|\theta|<1$ hence $\nrm{(\bI_N-\bG)^{-1}}{}\leq(1-\nrm{\bG}{})^{-1}<1/(1-|\alpha_1|-|\theta|)$. So, $\nrm{{\varphi_{ev}}}{}=O(\nrm{\bU_{ev}\beta_{ev}}{})=O(1)$ as.
    Using Lemma \ref{lmm-hoeffding-whp}, we have
    \begin{align*}
        \pr\lrp{\frac{1}{\sqrt{NT}}\vrt{2\tilde\by\T\hat\bS{\Phi_{ev}}}>\upsilon}<2\exp\lrp{-\frac{cNT\upsilon^2}{T/\lrp{N\rho_N}^2}}+\frac{1}{N} \\
        \pr\lrp{\frac{1}{\sqrt{NT}}\vrt{2\tilde\by\T\tilde\bS{\Phi_{ev}}}>\upsilon}<2\exp\lrp{-\frac{cNT\upsilon^2}{T}}+\frac{1}{N}
    \end{align*}
    because $\nrm{\hat\bS{\Phi_{ev}}}{}\leq\nrm{\hat\bS}{}\sqrt{T}\nrm{{\varphi_{ev}}}{}=O(\sqrt{T}/\lrp{N\rho_N})$ whp. and $\nrm{\tilde\bS{\Phi_{ev}}}{}=O(\sqrt{T})$ as.
    Also, $\vrt{{\Phi_{ev}}\T\tilde\bS{\Phi_{ev}}}=O(T)$ as.
    Therefore, we have $\frac{1}{NT^2}\sum_t\lrp{{\eta_1}\T\hat\bU_{ev}\T\by_t}^2=o_\pr(1)$. Noting that both $\by$ and $\bZ_t$ have finite fourth-order moments, one can show that the rest are also $o_\pr(1)$ similarly.
    
    \proof[Proof of (2)]
    Since $\sum_t\E\lrp{\xi_{N,t+1}^2|\cF_{Nt}}=\frac{\sigma^2}{NT}\eta\T{\hat{\cW}}\T{\hat{\cW}}\eta=\sigma^2\eta\T\hat\Sigma_{ev}\eta$, by Lemma \ref{lmm-enar-ev-∑nasym}, we have (2).
    
    Therefore, by three claims made, we have 
    \[\bD_{ev}\lrp{\hat\mu_{ev}-\mu_{ev}^H}\Rightarrow{\Sigma_{ev}}^{-1}\cN\lrp{\bo_{K+p+2},\sigma^2\Sigma_{ev}}\]
    which leads to the desired asymptotic normality.
\end{proof}
\begin{proof}[Proof of Theorem \ref{thm-enar-ev-clt-NTK}]
    It suffices to show that for any $\bA_K$ in Theorem \ref{thm-enar-ev-clt-NTK}, we have
    \begin{align*}
    \frac{\bA_K{\hat\Sigma_{ev}}^{-1}}{\sqrt{NT}}{\hat{\cW}}\T(\cW^H-\hat{\cW})\mu_{ev,w}^H&=o_\pr(1) \label{A1eq1}.
    \end{align*}
    First provided that $\nrm{\beta_{ev}}{}=1$ as $K\rightarrow\infty$, we have $\frac{1}{\sqrt{NT}}{\hat{\cW}}\T(\cW^H-\hat{\cW})\mu_{ev,w}^H=o_\pr(1)$ by the statements \textbf{10}-\textbf{13}. of Lemma \ref{lmm-enar-ev-nasym}.
    Since $\nrm{\bA_K{\hat\Sigma_{ev}}^{-1}}{}=\nrm{\bA_K\lrp{{\hat\Sigma_{ev}}^{-1}-{\Sigma_{ev}}^{-1}+{\Sigma_{ev}}^{-1}}}{}$, we only need to check if $\nrm{{\hat\Sigma_{ev}}^{-1}-{\Sigma_{ev}}^{-1}}{}=O_\pr\lrp{\nrm{{\Sigma_{ev}}^{-1}}{}}$. Since $\nrm{\Sigma_{ev}}{}=O(1)$ and $\nrm{\hat\Sigma_{ev}-\Sigma_{ev}}{}=o_\pr\brcs{\min_i\lambda_i(\Sigma_{ev})^{-1}}$ by Lemma \ref{lmm-enar-ev-∑nasym}, we have the first claim.
    
    Next, It is sufficient to show that for any $\eta\in\R^{m}$ such that $\|\eta\|\leq1$, we have $\eta\T\bA_K{\hat\Sigma_{ev}}^{-1}\sqrt{NT}\br_{w\epsilon}\Rightarrow\cN(0,\sigma^2\eta\T\bV_{ev}\eta)$. 
    Since 
    \begin{align*}
        \eta\T\bA_K{\hat\Sigma_{ev}}^{-1}\sqrt{NT}\br_{w\epsilon}=
        \eta\T\bA_K\lrp{{\hat\Sigma_{ev}}^{-1}-{\Sigma_{ev}}^{-1}}\sqrt{NT}\br_{w\epsilon}+\eta\T\bA_K{\Sigma_{ev}}^{-1}\sqrt{NT}\br_{w\epsilon} \\
        =o_\pr(1)+\eta\T\bA_K{\Sigma_{ev}}^{-1}O_\pr(1),
    \end{align*}
    we focus on the latter. 
    Define $ \xi_{Nt}\triangleq\lrp{NT}^{-1/2}\eta\T\bA_K{\Sigma_{ev}}^{-1}{\hat\cW_{t-1}}\T\bcE_t$ and $\cF_{Nt}=\sigma(\bA,\epsilon_{is},\bZ_{i,(s-1)};i\leq N,-\infty<s\leq t)$. Then, a set of pairs $\{\sum_{s=1}^t \xi_{Ns},\cF_{Nt}\}$ constitutes a martingale array for each $N,t\leq T$. Then, we repeat the following arguments \citep{hall2014martingale}:
    
    \textbf{(1)} $\sum_{t}\E\lrp{\xi_{N,t+1}^21_{\brcs{\vrt{\xi_{N,t+1}}>\upsilon}}|\cF_{Nt}}=o_\pr(1)$. 
    
    \textbf{(2)} $\sum_t\E\lrp{\xi_{N,t+1}^2|\cF_{Nt}}=\sigma^2\eta\T\bV_{ev}\eta+o_\pr(1)$.

    \proof[Proof of (1)]
    We only need to show that 
    \[
    \sum_t\lrp{\frac{1}{NT}\eta\T\bA_K{\Sigma_{ev}}^{-1}{\hat{\cW_t}}\T{\hat{\cW_t}}{\Sigma_{ev}}^{-1}{\bA_K}\T\eta}^2=o_\pr(1).
    \]
    Since both $\bA_K$ and $\Sigma_{ev}$ have spectral norm of $O(1)$ as $K\rightarrow\infty$, 
    this can be shown in the same manner as in the Proof of (1) of Claim 3 in Theorem \ref{thm-enar-ev-clt-NT}.

    \proof[Proof of (2)]
    Since 
    \begin{align*}
    \sum_t\E\lrp{\xi_{N,t+1}^2|\cF_{Nt}}=\eta\T\bA_K{\Sigma_{ev}}^{-1}\frac{\sigma^2}{NT}{\hat{\cW}}\T{\hat{\cW}}{\Sigma_{ev}}^{-1}{\bA_K}\T\eta=\sigma^2\eta\T\bA_K{\Sigma_{ev}}^{-1}\hat\Sigma_{ev}{\Sigma_{ev}}^{-1}{\bA_K}\T\eta,
    \end{align*}
    and noting that $\hat\Sigma_{ev}=\Sigma_{ev}+o_\pr(1)$ for each $K>0$, we have (2) provided that $\lim_{K\rightarrow\infty}\bA_K{\Sigma_{ev}}^{-1}{\bA_K}\T$ exists.
\end{proof}

\subsubsection{Consistency under misspecification of \texorpdfstring{$K$}{K}}
\begin{proof}[Proof of Theorem \ref{thm-enar-ev-overK-clt-NT}]
    Now let us augment $\bD_{ev}$ as $\bD_{ev,+}\triangleq\sqrt{T}\operatorname{diag}\lrp{\bI_{K_o},\sqrt{N}\bI_{p+2}}$ and augment $\hat\cW$ in the proof of Theorem \ref{thm-enar-ev-clt-NT} as $\hat\cW_+\triangleq\bxs{\sqrt{N}\bon_T\otimes\hat\bU_{ev,K:}\,|\,\hat\cW}\in\R^{NT\times( K_o+p+2)}$
    so that $\sqrt{NT}\hat\bW_+=\hat\cW_+\bD_{ev,+}$.
    This representation gives $\hat\cW_+\mu_{ev,w+}^H=\hat\cW\mu_{ev,w}^{H}$ and $\sqrt{NT}{\bD_{ev,+}}^{-1}\mu_{ev,w+}^{H}=\mu_{ev,+}^H$.
    Define
    \begin{align}\label{eq-ev-hSw+}
        \hat\Sigma_{ev,+}\triangleq\frac{1}{NT}{\hat\cW_+}\T\hat\cW_+
    \end{align}
    and $\br_{w\epsilon,+}\triangleq\frac{1}{NT}{\hat\cW_+}\T\bcE$.
    Then, recalling the proof in the previous section,
    \begin{align*}
        \hat\mu_{ev,+} & = \lrp{\frac{\bD_{ev,+}}{NT}{\hat\cW_+}\T\hat\cW_+\bD_{ev,+}}^{-1}\frac{\bD_{ev,+}}{\sqrt{NT}}{\hat\cW_+}\T\lrp{\cW^H\mu_{ev,w}^H+\bcE} \\
        & = \lrp{\hat\Sigma_{ev,+}\bD_{ev,+}}^{-1}\frac{1}{\sqrt{NT}}{\hat\cW_+}\T\brcs{\lrp{\cW^H-\hat\cW}\mu_{ev,w}^H+\hat\cW_+\mu_{ev,w+}^H+\bcE} \\
        & = {\bD_{ev,+}}^{-1}{\hat\Sigma_{ev,+}}^{-1}\brcs{\frac{1}{\sqrt{NT}}{\hat\cW_+}\T\lrp{\cW^H-\hat\cW}\mu_{ev,w}^H+\sqrt{NT}\br_{w\epsilon,+}}+\sqrt{NT}{\bD_{ev,+}}^{-1}\mu_{ev,w+}^{H} \\
        & \therefore \bD_{ev,+}\lrp{\hat\mu_{ev,+}-\mu_{ev,+}^{H}}=\frac{{\hat\Sigma_{ev,+}}^{-1}}{\sqrt{NT}}{\hat\cW_+}\T\lrp{\cW^H-\hat\cW}\mu_{ev,w}^H+{\hat\Sigma_{ev,+}}^{-1}\sqrt{NT}\br_{w\epsilon,+}.
    \end{align*}
    Therefore, we can use similar logic as the proof of Theorem \ref{thm-enar-ev-clt-NT}.
    \begin{claim}
        ${\hat\Sigma_{ev,+}}\Rightarrow{\Sigma_{ev,+}}$.
    \end{claim}
    \proof
    Now observe that ${\hat\bU_{ev,K:}}\T\hat\bU_{ev,K:}=\bI_{K_o-K}$ and $\nrm{\hat\bU_{ev,K:}}{}=1$ so by a similar result as item 0 of Lemma \ref{lmm-enar-ev-nasym},
    \begin{align*}
        \hat\Sigma_{ev,+}&=\bxs{\begin{matrix}
        \bI_{K_o-K}& \frac{1}{\sqrt{N}T}\sum_t{\hat\bU_{ev,K:}}\T\hat\cW_t\\
        &\hat\Sigma_{ev}
        \end{matrix}}\overset{N,T\to\infty}{\Rightarrow}\bxs{\begin{matrix}
        \bI_{K_o-K} & \bO_{(K_o-K)\times(K+p+2)}\\
        & \Sigma_{ev}
        \end{matrix}}.
    \end{align*}
    This is easy to check as ${\hat\bU_{ev,K:}}\T\hat\bU_{ev}=\bO_{(K_o-K)\times K}$ and for any $\bh\in B^{ K_o-K}$, we have $\nrm{{\Phi_{ev}}}{}=O(1)~as.$ and $\nrm{\bon_T\otimes\hat\bU_{ev,K:}}{}=O(\sqrt{T})~as.$ so
    \begin{align*}
        \frac{1}{\sqrt{N}T}\sum_t\lrp{\hat\bU_{ev,K:}\bh}\T\by_t=\frac{1}{\sqrt{N}T}\lrp{\bon_T\otimes\hat\bU_{ev,K:}\bh}\T{\Gamma_\by}^{1/2}\tilde\by+O\lrp{\frac{1}{\sqrt{NT}}}~as.,\\
        \pr\bxs{\frac{1}{\sqrt{N}T}\vrt{\lrp{\bon_T\otimes\hat\bU_{ev,K:}\bh}\T{\Gamma_\by}^{1/2}\tilde\by}>\upsilon}<2\exp\lrp{-cNT\upsilon^2}
    \end{align*}
    for example.
    \begin{claim}
        $\frac{1}{\sqrt{NT}}{\hat\cW_+}\T(\cW^H-\hat\cW)\mu_{ev,w}^H=\bo_{K_o+p+2}+o_\pr(1)$.
    \end{claim}
    \proof 
    Now observe that
    \begin{align*}
        \frac{1}{\sqrt{NT}}{\hat\cW_+}\T\lrp{\cW^H-\hat\cW}\mu_{ev,w}^H
        =-\frac{1}{\sqrt{NT}}\begin{bmatrix}
            \sqrt{N}\lrp{\bon_T\otimes\hat\bU_{ev,K:}}\T \\ {\hat\cW}\T
        \end{bmatrix}
        \bon_T\otimes\lrp{\hat\bU_{ev}-\tilde\bU_{ev}}{\bH_{ev}}\T\beta_{ev}.
    \end{align*}
    Then, our additional bias term of interest is 
    \[\sqrt{T}\lrp{\hat\bU_{ev,K:}\bh}\T\lrp{\hat\bU_{ev}-\tilde\bU_{ev}}{\bH_{ev}}\T\beta_{ev}=O\lrp{\sqrt{\frac{T}{N\rho_N}}}~whp.\]
    for any $\bh\in B^{K_o-K}$, which leads to the conclusion by Lemma \ref{lmm-enar-ev-nasym} provided that $N\rho_N=\omega(T)$. 
    \begin{claim}
        $\forall\eta\in B^{K_o+p+1}$, $\sqrt{NT}\eta\T\br_{w\epsilon,+}=O_\pr(1)$.
    \end{claim}
    \proof
    Using the same steps in the third claim in the proof of Theorem \ref{thm-enar-ev-clt-NT}, we can derive that $\sqrt{NT}\br_{w\epsilon,+}\Rightarrow\cN(\bo_{ K_o+p+2},\sigma^2\Sigma_{ev,+})$.

    Therefore, we have the conclusion.
\end{proof}
\begin{proof}[Proof of Theorem \ref{thm-enar-ev-underK-consist-NT}]
    Deinfe a diagonal matrix of convergence rates $\bD_{ev,-}\triangleq\sqrt{T}\operatorname{diag}\lrp{\bI_{K_o},\sqrt{N}\bI_{p+2}}$ and a truncated design matrix with $\hat\bU_{ev,-}$ embedded over the entire time, 
    \[\hat\cW_{-}\triangleq\bxs{\sqrt{N}\bon_T\otimes\hat\bU_{ev,-}\,|\,\by,\lrp{\bI_T\otimes\cL_\bA}\by\,|\,\bZ}\in\R^{NT\times(K_o+p+2)}.\]
    Let $\hat\cW_{:K}\in\R^{(K+p+2)\times NT}$ denote the augmented version of $\hat\cW_-$ which contains zeros between first and second block of $\hat\cW_-$ above:
    \[\hat\cW_{:K}\triangleq\bxs{\sqrt{N}\bon_T\otimes\hat\bU_{ev,-}\,|\bO_{NT\times(K-K_o)}\,|\,\by,\lrp{\bI_T\otimes\cL_\bA}\by\,|\,\bZ}\in\R^{NT\times(K+p+2)}.\]
    Define $\mu_{ev}^{H_-}\triangleq\lrp{\beta_{ev}\T\bH_{ev,-},\mu_{-\beta_{ev}}\T}\T\in\R^{K_o+p+2}$ and $\mu_{ev,w}^{H_-}\triangleq\lrp{\frac{1}{\sqrt{N}}\beta_{ev}\T\bH_{ev,-},\mu_{-\beta_{ev}}\T}\T$.
    This expression gives $\sqrt{NT}\hat\bW_-=\hat\cW_-\bD_{ev,-}$ and $\hat\cW_{:K}\mu_{ev,w}^H=\hat\cW_-\mu_{ev,w}^{H_-}$.
    $\hat\Sigma_{ev,-}$ and $\br_{w\epsilon,-}$ are defined analogously as the case of over-selection.
    Then, we can derive the following in the same way as previous Theorems:
    \begin{align*} 
        \bD_{ev,-}\lrp{\hat\mu_{ev,-}-\mu_{ev}^{H_-}}=\frac{{\hat\Sigma_{ev,-}}^{-1}}{\sqrt{NT}}\hat\cW_-\T\lrp{\cW^{H}-\hat\cW_{:K}}\mu_{ev,w}^H+\sqrt{NT}{\hat\Sigma_{ev,-}}^{-1}\br_{w\epsilon,-}.
    \end{align*}
    Let $\Sigma_{ev,-}$ be a version of $\Sigma_{ev}$ which misses $K-K_o$ rows and columns corresponding to $K_o+1,\dots,K$-th latent variables.

    Next, by Proposition \ref{prp-rdpg-cnt} there exists $\bH_{ev}\in\cO_K$ such that
    \begin{align*}
        \nrm{\hat\bU_{ev}-\bU_{ev}\bH_{ev}}{}=O\lrp{1/\sqrt{N\rho_N}}~whp.
    \end{align*}
    such that $\bH_{ev}=\bxs{\bH_{ev,-}\,|\,\bH_{ev,:K}}$.
    Therefore, there exists $\bH_{ev,K_o}\in\cO_{K,K_o}$ such that
    \[\nrm{\hat\bU_{ev,-}-\bU_{ev}\bH_{ev,-}}{}\le\nrm{\hat\bU_{ev,-}-\bU_{ev}\bH_{ev,K_o}}{}+\nrm{\bI_{K_o}-{\bH_{ev,K_o}}\T\bH_{ev,-}}{}=O(1)~whp.\]
    by Theorem \ref{thm-rdpg-cnt2}.
    
    \begin{claim}
        ${\hat\Sigma_{ev,-}}\Rightarrow{\Sigma_{ev,-}}$. 
    \end{claim}
    \proof
    Now we only need to show the following:
    \begin{enumerate}
        \item $\pr\lrp{\frac{1}{\sqrt{N}T}\vrt{{\bh_1}\T\sum_t\brcs{\hat\bU_{ev,-}\T\by_t-\lrp{\bU_{ev}\bH_{ev,-}}\T{\varphi_{ev}}}}>\upsilon}\lesssim e^{-d_2\upsilon^2}+1/N$
        \item $\pr\bxs{\frac{1}{\sqrt{N}T}\vrt{{\bh_1}\T\sum_t\brcs{\hat\bU_{ev,-}\T\cL_\bA\by_t-\lrp{\bU_{ev}\bH_{ev,-}}\T\cL_\bA{\varphi_{ev}}}}>\upsilon}\lesssim e^{-d_2\upsilon^2}+1/N$
        \item $\pr\lrp{\frac{1}{\sqrt{N}T}\vrt{\sum_t{\bh_1}\T\hat\bU_{ev,-}\T\bZ_t\bh_2}>\upsilon}\lesssim e^{-d_2\upsilon^2}+1/N$
    \end{enumerate}
    where $d_2\triangleq c_2NT$ for fixed $c_2>0$.
    These can be easily shown by the similar proof to 1--3 of Lemma \ref{lmm-enar-ev-nasym}.
    Also, we have $\lim_{N,T\to\infty}\frac{1}{\sqrt{N}}\lrp{\bU_{ev}\bH_{ev,-}}\T{\varphi_{ev}}=\bo_{K_o}$ and $\lim_{N,T\to\infty}\frac{1}{\sqrt{N}}\lrp{\bU_{ev}\bH_{ev,-}}\T\cL_\bA{\varphi_{ev}}=\bo_{K_o}$.
    If $K_o=0$, then we have the reduced asymptotic covariance as $\Sigma_{-u}$.
    
    Next, we analyze the entry-wise asymptotic orders of $\frac{1}{\sqrt{NT}}{\hat\cW_-}\T(\cW^H-\hat\cW_{:K})\mu_{ev,w}^H$.
    First note that $\bH_{ev,-}{\bH_{ev,-}}\T+\bH_{ev,:K}{\bH_{ev,:K}}\T=\bI_K$ as $\bH_{ev}\in\cO_K$ and
    \[\cW^H-\hat\cW_{:K}=\sqrt{N}\bon_T\otimes\bxs{\bU_{ev}\bH_{ev,-}-\hat\bU_{ev,-}\,|\,\bU_{ev}\bH_{ev,:K}\,|\,\bO_{N\times(p+2)}}\]
    Then, $\frac{1}{\sqrt{NT}}{\hat\cW_-}\T\lrp{\cW^H-\hat\cW_{:K}}\mu_{ev,w}^H$ equals
    \begin{align*}
        \frac{1}{\sqrt{NT}}\sum_t\hat\cW_{t,-}\T
        \brcs{\lrp{\bU_{ev}\bH_{ev,-}-\hat\bU_{ev,-}}{\bH_{ev,-}}\T+\bU_{ev}\bH_{ev,:K}{\bH_{ev,:K}}\T}\beta_{ev}\\
        =\bxs{\begin{matrix} 
        \sqrt{T}{\hat\bU_{ev,-}}\T \\
        \frac{1}{\sqrt{NT}}\sum_t{\by_t}\T \\
        \frac{1}{\sqrt{NT}}\sum_t{\by_t}\T\cL_\bA \\
        \frac{1}{\sqrt{NT}}\sum_t{\bZ_t}\T
        \end{matrix}}\lrp{\bU_{ev}-\hat\bU_{ev,-}{\bH_{ev,-}}\T}\beta_{ev}.
    \end{align*}
    Denoting the RHS by $\bq_{K_o}$ which depends on $K_o$, we have $\bq_0=\bU_{ev}\beta_{ev}$ for $K_o=0$ as $\hat\bU_{ev,-}{\bH_{ev,-}}\T$ is removed. So, $\nrm{\bq_0}{}=O(1)$.
    Otherwise, $\nrm{\bq_{K_o}}{}=O\lrp{1}~whp.$ by noting that 
    \[\nrm{\bU_{ev}-\hat\bU_{ev,-}{\bH_{ev,-}}\T}{}=\nrm{\hat\bU_{ev,-}-\bU_{ev}\bH_{ev,-}}{}.\]
    Then, we can easily show the following by similar arguments used in proving 10--13 of Lemma \ref{lmm-enar-ev-nasym}:
    \begin{enumerate}
        \item $\sqrt{T}\vrt{\lrp{\hat\bU_{ev,-}\bh_1}\T\bq_{K_o}}=O\lrp{\sqrt{T}}~whp.$
        \item $\pr\bxs{\frac{1}{\sqrt{NT}}\vrt{\sum_t\lrp{\by_t-{\varphi_{ev}}}\T\bq_o}>\upsilon}<e^{-cN\upsilon^2}$
        \item $\pr\bxs{\frac{1}{\sqrt{NT}}\vrt{\sum_t\lrp{\by_t-{\varphi_{ev}}}\T\cL_\bA\bq_o}>\upsilon}<e^{-cN\upsilon^2}$
        \item $\pr\bxs{\frac{1}{\sqrt{NT}}\vrt{\lrp{\sum_t{\bZ_t}\bh_2}\T\bq_o}>\upsilon}<e^{-cN\upsilon^2}$
    \end{enumerate}
    for any $\upsilon>0$ where $\bh_1\in B^{K_o-1},\bh_2\in B^{p-1}$.
    Now partition $\frac{1}{\sqrt{NT}}{\hat\cW_-}\T\lrp{\cW^H-\hat\cW_{:K}}\mu_{ev,w}^H$ into three sub-vectors of sizes $K_o$, $2$, and $p$ denoted by $\bb_1,\bb_2$, and $\bb_3$ respectively.
    These imply the following:
    \begin{align*}
        &b_{21}=\sqrt{\frac{T}{N}}{\varphi_{ev}}\T\bq_{K_o}+o_\pr(1),\quad
        b_{22}=\sqrt{\frac{T}{N}}{\varphi_{ev}}\T\cL_\bA\bq_{K_o}+o_\pr(1),\\
        &{\bh_2}\T\bb_3=o_\pr(1)
    \end{align*}
    These give the specified asymptotic orders of the estimation error terms.
    
    Now, using the same steps in the third claim in the proof of Theorem \ref{thm-enar-ev-clt-NT}, we can derive that $\sqrt{NT}\br_{w\epsilon,-}\Rightarrow\cN(\bo_{ K_o+p+2},\sigma^2\Sigma_{ev,-})$.
    
    Therefore, recalling the rates of $\bD_{ev,-}$, we can summarize as follows:
    \begin{align*}
        \sqrt{T}\lrp{\hat\beta_{ev,-}-\bH_{ev,-}\T\beta_{ev}}&=\bb_1+\bt_1,\quad
        \sqrt{NT}\begin{bmatrix}
            \hat\alpha_{ev,-}-\alpha_{ev}\\
            \hat\theta_{ev,-}-\theta_{ev}
        \end{bmatrix}=\bb_2+\bt_2,\\
        \sqrt{NT}\lrp{\hat\gamma_{ev,-}-\gamma_{ev}}&=\bt_3
    \end{align*}
    where
    \begin{align*}
        &{\bh_1}\T\bb_1=O\lrp{\sqrt{\frac{T}{N\rho_N}}}~whp.,\quad{\bh_2}\T\bb_2=O\lrp{\sqrt{\frac{T}{N}}}~whp.,\\
        &{\bh_1}\T\bt_1=O_\pr(1),{\bh_2}\T\bt_2=O_\pr(1),{\bh_3}\T\bt_3=O_\pr(1)
    \end{align*}
    for any $\bh_1\in B^{K_o-1},\bh_2\in B^1,\bh_3\in B^{p-1}$.
\end{proof}

\subsubsection{Asymptotic results} 
In the following lemma, we state each item for $N>N_0$ with large enough $N_0>0$.
\begin{lemma}\label{lmm-enar-ev-nasym}
    Let $\bh_i\in B^{j_i-1}$ for $j_1=K$, $j_2=p$.
    Then under the conditions of Theorem \ref{thm-enar-ev-clt-NT}, there exist $N_0,T_0>0$ and $\upsilon_0(N_0,T_0)>0$ such that the statements \textbf{1.}--\textbf{13.} hold for all $\upsilon\in\lrp{0,\upsilon_0}$ and $N>N_0,T>T_0$:
    \begin{enumerate} 
        \item $\pr\lrp{\frac{1}{\sqrt{N}T}\vrt{{\bh_1}\T\sum_t(\hat\bU_{ev}\T\by_t-\tilde\bU_{ev}\T{\varphi_{ev}})}>\upsilon}\lesssim e^{-d_1\upsilon^2}+e^{-d_2\upsilon^2}+1/N$ 
        \item $\pr\bxs{\frac{1}{\sqrt{N}T}\vrt{{\bh_1}\T\sum_t\lrp{\hat\bU_{ev}\T\cL_\bA\by_t-\tilde\bU_{ev}\T\cL_\bA{\varphi_{ev}}}}>\upsilon}\lesssim e^{-d_1\upsilon^2}+e^{-d_2\upsilon^2}+1/N$ 
        \item $\pr\lrp{\frac{1}{\sqrt{N}T}\vrt{\sum_t{\bh_1}\T\hat\bU_{ev}\T\bZ_t\bh_2}>\upsilon}\lesssim e^{-d_1\upsilon^2}+e^{-d_2\upsilon^2}+1/N$ 
        \item $\pr\bxs{\frac{1}{NT}\vrt{\sum_t\brcs{{\by_t}\T\by_t-\E\tr(\Gamma)-{\varphi_{ev}}\T\varphi_{ev}}}>\upsilon}\lesssim e^{-d_2\upsilon^2}+e^{-d_3\upsilon^2}$ 
        \item $\pr\bxs{\frac{1}{NT}\vrt{\sum_t\brcs{{\by_t}\T\cL_\bA\by_t-\E\tr(\cL_\bA\Gamma)-\varphi_{ev}\T\cL_\bA\varphi_{ev}}}>\upsilon}e^{-d_2\upsilon^2}+e^{-d_3\upsilon^2}$ 
        \item $\pr\lrp{\frac{1}{NT}\vrt{\sum_t{\by_t}\T\bZ_t\bh_2}>\upsilon}\lesssim e^{-d_4\upsilon^2}+e^{-d_3\upsilon^2}+1/N$ 
        \item $\pr\bxs{\frac{1}{NT}\vrt{\sum_t\brcs{{\by_t}\T{\cL_\bA}^2\by_t-\E\tr({\cL_\bA}^2\Gamma)-{\varphi_{ev}}\T{\cL_\bA}^2\varphi_{ev}}}>\upsilon}\lesssim e^{-d_2\upsilon^2}+e^{-d_3\upsilon^2}$ 
        \item $\pr\lrp{\frac{1}{NT}\vrt{\sum_t{\by_t}\T\cL_\bA\bZ_t\bh_2}>\upsilon}\lesssim e^{-d_4\upsilon^2}+2^{-d_3\upsilon^2}+1/N$ \item $\pr\bxs{\frac{1}{NT}\vrt{{\bh_2}\T\sum_t\lrp{{\bZ_t}\T\bZ_t-\Sigma_z}\bh_2}>\upsilon}\lesssim e^{-d_2\upsilon^2}$ 
        \item $\pr\bxs{\sqrt{T}\vrt{{\bh_1}\T\hat\bU_{ev}\T\lrp{\hat\bU_{ev}-\tilde\bU_{ev}}\beta_{ev}}>\upsilon}\lesssim \frac{1}{N}$ 
        \item $\pr\bxs{\frac{1}{\sqrt{NT}}\vrt{\sum_t{\by_t}\T\lrp{\hat\bU_{ev}-\tilde\bU_{ev}}\beta_{ev}}>\upsilon}\lesssim e^{-cN^2\rho_N\upsilon^2}+\frac{1}{N}$ 
        \item $\pr\bxs{\frac{1}{\sqrt{NT}}\vrt{\sum_t{\by_t}\T\cL_\bA\lrp{\hat\bU_{ev}-\tilde\bU_{ev}}\beta_{ev}}>\upsilon}\lesssim e^{-cN^2\rho_N\upsilon^2}+\frac{1}{N}$ 
        \item $\pr\bxs{\frac{1}{\sqrt{NT}}\vrt{{\bh_2}\T\sum_t{\bZ_t}\T\lrp{\hat\bU_{ev}-\tilde\bU_{ev}}\beta_{ev}}>\upsilon}\lesssim e^{-cN^2\rho_N\upsilon^2}+\frac{1}{N}$ 
    \end{enumerate}
    where $d_1\triangleq c_1N^2\rho_NT$, $d_2\triangleq c_2NT$, $d_3\triangleq c_3N^2T$, and $d_4\triangleq\frac{c_4NT}{p}$ and $c,c_1,\ldots,c_4,b>0$ are constants.
\end{lemma}
\begin{proof}[Proof of 1]
    Note that 
    \begin{align*}
        \sum_{t=1}^{T}{\bh_1}\T\lrp{\hat\bU_{ev}\T\by_t-\tilde\bU_{ev}\T{\varphi_{ev}}}=
    {\bh_1}\T\lrp{\bon_T\otimes\Delta_{ev}}\T{\Gamma_\by}^{1/2}\tilde\by
    +{\bh_1}\T\lrp{\bon_T\otimes\tilde\bU_{ev}}\T{\Gamma_\by}^{1/2}\tilde\by\\
    +T\lrp{\Delta_{ev}\bh_1}\T{\varphi_{ev}}
    =\lrp{\hat\bS\bh_1}\T\tilde\by
    +\lrp{\tilde\bS\bh_1}\T\tilde\by
    +T\lrp{\Delta_{ev}\bh_1}\T{\varphi_{ev}}.
    \end{align*}
    where $\hat\bS\triangleq{\Gamma_\by}^{1/2}\lrp{\bon_T\otimes\Delta_{ev}}$ and $\tilde\bS\triangleq{\Gamma_\by}^{1/2}\lrp{\bon_T\otimes\tilde\bU_{ev}}$.
    Since $\nrm{\hat\bS\bh_1}{}=O\lrp{\sqrt{\frac{T}{N\rho_N}}}~whp.$, by Lemma \ref{lmm-hoeffding-whp} we have
    \[\pr\lrp{\vrt{\lrp{\hat\bS\bh_1}\T\tilde\by}>\sqrt{N}T\upsilon}<2\exp\lrp{\frac{-cN^2\rho_NT^2\upsilon^2}{T}}+\frac{1}{N}.\]
    Likewise, since $\nrm{\tilde\bS}{}=O(\sqrt{T})$ as., by Lemma \ref{lmm-hoeffding-whp} we have $\pr\lrp{\vrt{\lrp{\tilde\bS\bh_1}\T\Tilde\by}>\sqrt{N}T\upsilon}<2\exp(-c_2NT^2\upsilon^2/T)$.
    Finally, $\vrt{T\lrp{\Delta_{ev}\bh_1}\T{\varphi_{ev}}}=O(T/\sqrt{N\rho_N})$ whp., i.e.,
    \[\pr\lrp{\vrt{\lrp{\Delta_{ev}\bh_1}\T{\varphi_{ev}}}>\frac{C}{\sqrt{N\rho_N}}}<1/N\]
    for all large enough $N$. By selecting $\upsilon$ such that $\sqrt{N}\upsilon=\omega(1/\sqrt{N\rho_N})$, i.e., $\upsilon=\omega(1/\sqrt{N^2\rho_N})$, and noting that $N\rho_N=\omega(\log^4N)$, we can reduce $\upsilon$ sufficiently small for large enough $N$. For any such $\upsilon$, we have
    \[\pr\lrp{\frac{1}{\sqrt{N}}\vrt{\lrp{\Delta_{ev}\bh_1}\T{\varphi_{ev}}}>\upsilon}\le\pr\lrp{\vrt{\lrp{\Delta_{ev}\bh_1}\T{\varphi_{ev}}}>\frac{C}{\sqrt{N\rho_N}}}<\frac{1}{N}.\]
    This completes the proof.
\end{proof}
\begin{proof}[Proof of 2]
    First note that
    \begin{align*}
        {\bh_1}\T\sum_t\lrp{\hat\bU_{ev}\T\cL_\bA\by_t-\tilde\bU_{ev}\T\cL_\bA{\varphi_{ev}}}=
    {\bh_1}\T\bxs{\bon_T\otimes\brcs{\cL_\bA\Delta_{ev}}}\T{\Gamma_\by}^{1/2}\tilde\by\\
    +{\bh_1}\T\brcs{\bon_T\otimes\cL_\bA\tilde\bU_{ev}}\T{\Gamma_\by}^{1/2}\tilde\by+T\lrp{\Delta_{ev}\bh_1}\T\cL_\bA{\varphi_{ev}}.
    \end{align*}
    For $\hat\bS\triangleq{\Gamma_\by}^{1/2}\brcs{\bon_T\otimes\lrp{\cL_\bA\Delta_{ev}}}$ and $\tilde\bS\triangleq{\Gamma_\by}^{1/2}\brcs{\bon_T\otimes\lrp{\cL_\bA\tilde\bU_{ev}}}$, we have $\nrm{\hat\bS\bh_1}{}=O\lrp{\sqrt{\frac{T}{N\rho_N}}}$ whp. because $\rho\lrp{\cL_\bA}<1$ as. hence $\pr\lrp{\vrt{\lrp{\hat\bS\bh_1}\T\tilde\by}>\sqrt{N}T\upsilon}<2\exp\lrp{-cN^2\rho_NT\upsilon^2}+\frac{1}{N}$. Since $\nrm{\tilde\bS\bh_1}{}=O\lrp{\sqrt{T}}$ as., we have $\pr\lrp{|\lrp{\tilde\bS\bh_1}\T\tilde\by|>\sqrt{N}T\upsilon}<2\exp\lrp{-cNT\upsilon^2}$. The tail probability of $\vrt{T\lrp{\Delta_{ev}\bh_1}\T\cL_\bA{\varphi_{ev}}}$ can be bounded similarly to 1.
\end{proof}
\begin{proof}[Proof of 3]
    Let $\bZ\triangleq\bxs{{\bZ_0}\T,...,{\bZ_{T-1}}\T}\T$ and $\hat\bS\triangleq\bon_T\otimes\Delta_{ev}$ and $\tilde\bS\triangleq\bon_T\otimes\tilde\bU_{ev}$. Then, we can express $\sum_t{\bh_1}\T\hat\bU_{ev}\T\bZ_t\bh_2$ as $\tr\lrp{\bZ\bh_2{\bh_1}\T{\hat\bS}\T}
    +\tr\lrp{\bZ\bh_2{\bh_1}\T{\tilde\bS}\T}$. This is equal to
    \[\vecop\lrp{\bh_2{\bh_1}\T}\T\brcs{\bI_p\otimes\lrp{\hat\bS+\tilde\bS}\T}\vecop\lrp{\bZ}.\]
    Note that $\nrm{\bI_p\otimes\hat\bS\T}{}=O\lrp{\sqrt{\frac{T}{N\rho_N}}}$ whp. and $\nrm{\bI_p\otimes\tilde\bS\T}{}=O\lrp{\sqrt{T}}$. Therefore, by Lemma \ref{lmm-hoeffding-whp}, we have
    \[\pr\bxs{\vrt{\sum_t{\bh_1}\T\hat\bU_{ev}\T\bZ_t\bh_2}>\sqrt{N}T\upsilon}\leq2\exp\lrp{-\frac{cNT\upsilon^2}{\lrp{N\rho_N}^{-1}}}+2\exp\lrp{-c'NT\upsilon^2}+\frac{1}{N}.\]
\end{proof}
\begin{proof}[Proof of 4]
    First, we have 
    \begin{align*}
        \sum_t\lrp{{\by_t}\T\by_t-\varphi_{ev}\T\varphi_{ev}}
        =\nrm{{\Gamma_\by}^{1/2}\tilde\by+\Phi_{ev}}{}^2-T\varphi_{ev}\T\varphi_{ev}
        =\tilde\by\T\Gamma_\by\tilde\by+2\Phi_{ev}\T{\Gamma_\by}^{1/2}\tilde\by.
    \end{align*}
    Note that $\nrm{\Gamma_\by}{F}^2=\sum_{i=1}^{NT}\sigma_i(\Gamma_\by)^2=O( NT)$ as. By Lemma \ref{thm-hwineq-whp}, we have \[\pr\lrp{\vrt{\tilde\by\T\Gamma_\by\tilde\by-T\tr\E\Gamma}>\upsilon}<2\exp\lrp{-c\min\brcs{\frac{\upsilon^2}{NT},\upsilon}}.\]
    By Lemma \ref{lmm-hoeffding-whp}, \[\pr\lrp{2|\Phi_{ev}\T{\Gamma_\by}^{1/2}\tilde\by|>\upsilon}<2\exp\lrp{-\frac{c'\upsilon^2}{T}}\] since $\nrm{{\Gamma_\by}^{1/2}\Phi_{ev}}{}^2\leq\nrm{\Gamma_\by}{}T\varphi_{ev}\T\varphi_{ev}=O(T)$ as.
\end{proof}
\begin{proof}[Proof of 6]
    We have 
    \begin{align*}
        \sum_t{\by_t}\T\bZ_t\bh_2=\lrp{{\Gamma_\by}^{1/2}\tilde\by+{\Phi_{ev}}}\T\bZ\bh_2=\tr\lrp{{\bh_2}\Tilde\by\T{\Gamma_\by}^{1/2}\bZ}+\tr\lrp{\bh_2{\Phi_{ev}}\T\bZ}\\
        =\vecop\lrp{\tilde\by{\bh_2}\T}\T\lrp{\bI_p\otimes{\Gamma_\by}^{1/2}}\vecop\lrp{\bZ}+\vecop\lrp{{\Phi_{ev}}{\bh_2}\T}\T\lrp{\bI_p\otimes\bI_{NT}}\vecop\lrp{\bZ}.
    \end{align*}
    Therefore, since $\nrm{\Gamma_\by}{}=O(1)$ as., $\nrm{\lrp{\bI_p\otimes{\Gamma_\by}^{1/2}}\vecop\lrp{\tilde\by{\bh_2}\T}}{}\lesssim\nrm{\vecop\lrp{\tilde\by{\bh_2}\T}}{}$ which is concentrated around $\sqrt{NTp}$ \citep{vershynin2018high} hence is $O(\sqrt{NTp})$ whp. 
    Also, we have $\nrm{\lrp{\bI_p\otimes\bI_{NT}}\vecop\lrp{{\Phi_{ev}}{\bh_2}\T}}{}=O(\sqrt{T})$ as. 
    By Lemma \ref{lmm-hoeffding-whp},
    \[\pr\lrp{\vrt{\sum_t{\by_t}\T\bZ_t\bh_2}>\upsilon}\leq2\exp\lrp{-\frac{c\upsilon^2}{NTp}}+2\exp\lrp{-\frac{c\upsilon^2}{T}}+\frac{1}{N}.\]
    Therefore, the conclusion follows after letting $NT\upsilon=\omega(\sqrt{NTp})$.
\end{proof}
\begin{proof}[Proof of 8]
    The same logic as the proof in 6. applies here. Since $\nrm{\cL_\bA}{}\leq1$, we get the same bound by letting $NT\upsilon=\omega(\sqrt{NTp})$. 
\end{proof}
\begin{proof}[Proof of 9]
    Note that 
    \begin{align*}
        \sum_t{\bh_2}\T{\bZ_t}\T\bZ_t\bh_2=\tr\lrp{\bZ\bh_2
    {\bh_2}\T\bZ\T}=\vecop\lrp{\bZ\T}\T\brcs{\bI_{NT}\otimes\lrp{\bh_2{\bh_2}\T}}\vecop\lrp{\bZ\T}.
    \end{align*}
    Note that $\E\bxs{\vecop\lrp{\bZ\T}\vecop\lrp{\bZ\T}\T}=\bI_{NT}\otimes\Sigma_z$. 
    Also, by the definition of $\bh_2$, we have $\nrm{\bI_{NT}\otimes\lrp{\bh_2\bh_2\T}}{F}^2\leq NT$ and $\nrm{\bI_{NT}\otimes\lrp{\bh_2\bh_2\T}}{}\leq1$.
    Therefore, by Hanson-Wright inequality \citep{rudelson2013hanson},
    \[\pr\bxs{\vrt{{\bh_2}\T\sum_t\bZ_t\T\bZ_t\bh_2-NT{\bh_2}\T\Sigma_z\bh_2}>\upsilon}<2\exp\bxs{-c\min\brcs{\frac{\upsilon^2}{NT},\upsilon}}.\]
\end{proof}
\begin{proof}[Proof of 10]
    Note that $\nrm{\Delta_{ev}}{}^2=O\lrp{\frac{1}{N\rho_N}}$ whp. and $\nrm{\tilde\bU_{ev}}{}=1$. Also, $\nrm{\beta_{ev}}{}=O(1)$. So,
    \[\vrt{{\bh_1}\T\hat\bU_{ev}\T\Delta_{ev}\beta_{ev}}\leq\lrp{\nrm{\Delta_{ev}}{}^2+\nrm{\Delta_{ev}}{}}\nrm{\beta_{ev}}{}=O\lrp{1/\sqrt{N\rho_N}}\]
    whp., by letting $\upsilon=\omega\lrp{\sqrt{\frac{T}{N\rho_N}}}$ and noting that $N\rho_N=\omega(T)$, we have the conclusion.
\end{proof}
\begin{proof}[Proof of 11]
    Similar to the proof of 1., taking $\hat\bS\triangleq{\Gamma_\by}^{1/2}\lrp{\bon_T\otimes\Delta_{ev}}$ which has the spectral norm of  $O\lrp{\sqrt{\frac{T}{N\rho_N}}}$ whp., we have
    \begin{align*}
        \sum_t{\by_t}\T\Delta_{ev}\beta_{ev}=
        \lrp{{\Gamma_\by}^{1/2}{\tilde\by}+{\Phi_{ev}}}\T\lrp{\bon_T\otimes\Delta_{ev}}\beta_{ev}=\tilde\by\T\hat\bS\beta_{ev}+{\Phi_{ev}}\T{\Gamma_\by}^{-\frac{1}{2}}\hat\bS\beta_{ev}.
    \end{align*}
    First noting that $\nrm{\beta_{ev}}{}=O(1)$ as $K\rightarrow\infty$, Lemma \ref{lmm-hoeffding-whp} gives
    \[\pr\lrp{\vrt{\tilde\by\T\hat\bS\beta_{ev}}\geq \sqrt{NT}\upsilon}\leq2\exp\lrp{-cN^2\rho_N\upsilon^2}+\frac{1}{N}.\]
    It is straightforward that $\vrt{{\Phi_{ev}}\T{\Gamma_\by}^{-\frac{1}{2}}\hat\bS\beta_{ev}}=O(T/\sqrt{N\rho_N})$ whp. hence by letting $\upsilon=\omega\lrp{\sqrt{\frac{T}{N^2\rho_N}}}$, we obtain the conclusion.
\end{proof}
\begin{proof}[Proof of 12]
    Let $\hat\bS\triangleq{\Gamma_\by}^{1/2}\brcs{\bon_T\otimes\lrp{\cL_\bA\Delta_{ev}}}$ which is of $O\lrp{\sqrt{\frac{T}{N\rho_N}}}$ whp. The rest of the proof coincides with the proof of 11.
\end{proof}
\begin{proof}[Proof of 13]
    Note that ${\bh_2}\T\sum_t{\bZ_t}\T\Delta_{ev}\beta_{ev}={\bh_2}\T\bZ\T\lrp{\bon_T\otimes\Delta_{ev}}\beta_{ev}$ which equals
    \[\tr\brcs{\bZ\bh_2\beta_{ev}\T\lrp{\bon_T\otimes\Delta_{ev}}\T}=\vecop\lrp{\bh_2\beta_{ev}\T}\T\brcs{\bI_p\otimes\lrp{\bon_T\otimes\Delta_{ev}}\T}\vecop\lrp{\bZ}.\]
    Since the norm of $\vecop\lrp{\bh_2\beta_{ev}\T}\T\brcs{\bI_p\otimes\lrp{\bon_T\otimes\Delta_{ev}}\T}$ is of $O\lrp{\sqrt{\frac{T}{N\rho_N}}}$ whp., we obtain the bound after applying Lemma \ref{lmm-hoeffding-whp}.
\end{proof}
\begin{lemma}\label{lmm-enar-ev-∑nasym}
    Assume that $N\rho_N=\omega\lrp{T+\log^4N}$.
    Then, there exist $N_0,T_0>0$ and $\upsilon_0(N_0,T_0)\in(0,1)$ such that for all $\upsilon\in(0,\upsilon_0)$ and $N>N_0,T>T_0$,
    \begin{align*}
        \pr\bxs{\sup_{\bh\in B^{K+p+1}}\vrt{\bh\T\lrp{\hat\Sigma_{ev}-\Sigma_{ev}}\bh}>\upsilon}\lesssim K^2\brcs{\frac{1}{N}+\exp\lrp{-\frac{cNT\upsilon^2}{K^2p}}}.
    \end{align*}
\end{lemma}
\begin{proof}
    Let $\bh\in B^{K+p+1}$ and $\be_i\in S^{K+p+1}$ be the $i^{th}$ canonical basis of $\R^{K+p+2}$, i.e., $(\be_i)_j\triangleq\mathbb{I}_{\brcs{i}}(j)$ for $j=1,\dots,K+p+2$.
    By Lemma \ref{lmm-enar-ev-nasym}, we have
    \begin{align*}
        \max_{i,j}\pr\lrp{\vrt{{\be_
        i}\T\lrp{\hat\Sigma_{ev}-\Sigma_{ev}}\be_j}>\upsilon}
        \lesssim\frac{1}{N}+\exp\lrp{-c_1\upsilon^2NT/p}+\exp\lrp{-c_2\upsilon^2TN^2\rho_N}.
    \end{align*}
    Since 
    \begin{align*}
        \sup_{\bh\in B^{K+p+1}}\vrt{\bh\T\lrp{\hat\Sigma_{ev}-\Sigma_{ev}}\bh}
        \leq\sup_\bh\nrm{\bh}{1}^2\max_{i,j}\vrt{\hat\sigma_{ev,ij}-\sigma_{ev,ij}}\lesssim K\max_{i,j}\vrt{\hat\sigma_{ev,ij}-\sigma_{ev,ij}},
    \end{align*}
    we have 
    \begin{align*}
        \pr\bxs{\sup_{\bh\in B^{K+p+1}}\vrt{\bh\T\lrp{\hat\Sigma_{ev}-\Sigma_{ev}}\bh}>\upsilon}\leq\pr\lrp{\max_{i,j}\vrt{\hat\sigma_{ev,ij}-\sigma_{ev,ij}}>\frac{ce}{K}} \\
        \lesssim K^2\brcs{\frac{1}{N}+\exp\lrp{-{c_1}'\upsilon^2\frac{NT}{K^2p}}+\exp\lrp{-{c_2}'\upsilon^2\frac{N^2\rho_NT}{K^2}}}\\
        \lesssim K^2\bxs{\frac{1}{N}+\exp\brcs{-\frac{cNT\upsilon^2}{K^2}\min\brcs{N\rho_N,\frac{1}{p}}}}.
    \end{align*}
\end{proof}
\begin{lemma}
    If $N\rho_N=\omega(T+\log^4N)$, then we have
    \[\sqrt{T}\lrp{\hat\bU_{ev}\hat\beta_{ev}-\bU_{ev}\beta_{ev}}\Rightarrow\cN\lrp{\bo_K,\sigma^2\bU_{ev}{\bU_{ev}}\T}\]
    as $\min\lrp{N,T}\rightarrow\infty$.
\end{lemma}
\begin{proof}
    Recall that $\sqrt{T}\lrp{\hat\beta_{ev}-{\bH_{ev}}\T\beta_{ev}}\Rightarrow\cN\lrp{\bo_K,\sigma^2\bI_K}$ by Theorem \ref{thm-enar-ev-clt-NT}.
    By simple algebra, we have
    \begin{align}
        \sqrt{T}\lrp{\hat\bU_{ev}\hat\beta_{ev}-\bU_{ev}\beta_{ev}}&=\sqrt{T}\lrp{\hat\bU_{ev}\hat\beta_{ev}-\tilde\bU_{ev}{\bH_{ev}}\T\beta_{ev}}\nonumber \\
        &=\Delta_{ev}\sqrt{T}\lrp{\hat\beta_{ev}-{\bH_{ev}}\T\beta_{ev}}+\sqrt{T}\Delta_{ev}{\bH_{ev}}\T\beta_{ev}+\tilde\bU_{ev}\sqrt{T}\lrp{\hat\beta_{ev}-{\bH_{ev}}\T\beta_{ev}}.\label{Uß:eq1}
    \end{align}
    By assumption, $\sqrt{T}\Delta_{ev}=O\lrp{\sqrt{\frac{T}{N\rho_N}}}$ whp. hence we have
    \[\pr\bxs{\brcs{\sqrt{T}\Delta_{ev}\leq C\sqrt{\frac{T}{N\rho_N}}}~ev.}=1\]
    for some constant $C>0$. Provided that $N\rho_N=\omega(T)$, this implies $\sqrt{T}\Delta_{ev}\rightarrow0$ as. Therefore, we have $\ref{Uß:eq1}=o_\pr(1)+\tilde\bU_{ev}\sqrt{T}\lrp{\hat\beta_{ev}-{\bH_{ev}}\T\beta_{ev}}$ and the conclusion by Slutsky.
\end{proof}

\subsection{ENAR with \texorpdfstring{$\bU_{se}$}{U(se)} under finite time}
Note that under finite time assumption, we proposed to use spectral embedding for ENAR instead of eigenvectors. i.e., $\bU_{se}=\bU_\bP{\bS_\bP}^{1/2}$ and $\hat\bU_{se}=\bU_\bA|\bS_\bA|^{1/2}$.
Here, we present theoretical results for the finite-time corollary models specified in section \ref{sec-enar-n-enr}. Recall ENR:
\[\by=\alpha_{se}\bon_N+\bU_{se}\beta_{se}+\bZ\gamma_{se}+\bcE\]
and ENAR (with finite $T$) models:
\[\by_{t+1}=\alpha_{se}\by_{t}+\theta_{se}\cL_\bA\by_{t}+\bU_{se}\beta_{se}+\bZ \gamma+\bcE_{t+1}.\]
Note that ENR can be treated as a subset model of ENAR.
Therefore, we only verify the asymptotic properties of ENAR with finite $T$.

\subsubsection{Consistency}
As we have $\bU=\bU_{se}$, $\nrm{{\varphi_{se}}}{}=O(\nrm{\bU_{se}\beta_{se}}{})$ $\pr^*$-as. (hence $\pr$-as.)
Since $\nrm{\bU_\bP{\bS_\bP}^{1/2}}{}=O(\sqrt{N\rho_N})$, we have $\nrm{{\Phi_{se}}}{}=O(\sqrt{T}\nrm{{\varphi_{se}}}{})=O(\sqrt{TN\rho_N})$ as.
Though we assume that $T$ is finite in this section, our proof of the consistency of $\hat\mu$ here keeps track of the order of $T$.
For convenience, we denote $1+\frac{K\log N}{\sqrt{N\rho_N}}$ by $\pi_{se}$ in this section so that $\nrm{\Delta_{se}}{}\triangleq\nrm{\hat\bU_{se}-\tilde\bU_{se}}{}=O\lrp{\pi_{se}}~whp.$ if $N\rho_N=\omega\lrp{\log^4N}$ by Proposition \ref{prp-rdpg-cnt}.
\begin{proof}[Proof of Theorem \ref{thm-enar-se-clt-N}]
    Here, the formulation is very similar to the case of ENAR with eigenvectors.
    First, define $\tilde\bU_{se}\triangleq\bU_{se}\bH_{se}$ and let 
    \begin{align*}
     \bW^H_t&\triangleq\bxs{\tilde\bU_{se} \,\vert\, \by_t, \cL_\bA\by_t \, \vert\, \bZ_{t}},
     &\mu_{se,w}^H\triangleq\lrp{\sqrt{\rho_N}{\beta_{se}}\T\bH_{se},\alpha_{se},\theta_{se},{\gamma_{se}}\T}\T,\\
     \cW_t&\triangleq\bxs{\frac{1}{\sqrt{\rho_N}}\bU_{se} \,\vert\, \by_t, \cL_\bA\by_t \, \vert\, \bZ_{t}},
     &\cW^H_t\triangleq\bxs{\frac{1}{\sqrt{\rho_N}}\tilde\bU_{se} \,\vert\, \by_t, \cL_\bA\by_t \, \vert\, \bZ_{t}}
    \end{align*}
    and  $\hat\cW_t\triangleq\bxs{\frac{1}{\sqrt{\rho_N}}\hat\bU_{se} \,\vert\, \by_t, \cL_\bA\by_t \, \vert\, \bZ_{t}}$.
    Under this representation, we have $\hat\cW_t\bD_{se}=\sqrt{NT}\hat\bW_t$, $\sqrt{NT}{\bD_{se}}^{-1}\mu_{se,w}^H=\mu_{se}^H$, and $\bW^H_t\mu_{se}^H=\cW_t^H\mu_{se,w}^H$.
    Collect and bind them row-wise for $t=0,...,T-1$ to obtain $NT\times\lrp{K+p+2}$ matrices $\bW^H$, $\cW^H$, and define $\hat\cW$ analogously.
    For $\hat\Sigma_{se}\triangleq\frac{1}{NT}{\hat\cW}\T\hat\cW$ and $\br_{w\epsilon}\triangleq\frac{1}{NT}{\hat\cW}\T\bcE$, we have 
    \begin{align*}
    \hat\mu_{se} & = \lrp{{\hat\bW}\T\hat\bW}^{-1}{\hat\bW}\T\lrp{\bW^H\mu_{se}^H+\bcE} \\
    & = \lrp{\frac{\bD_{se}}{NT}{\hat\cW}\T\hat\cW\bD_{se}}^{-1}\frac{\bD_{se}}{\sqrt{NT}}{\hat\cW}\T\lrp{\cW^H\mu_{se,w}^H+\bcE} \\
    & = \lrp{\hat\Sigma_{se}\bD_{se}}^{-1}\frac{1}{\sqrt{NT}}\hat\cW\T\brcs{\lrp{\cW^H-\hat\cW}\mu_{se,w}^H+\hat\cW\mu_{se,w}^H+\bcE} \\
    & = {\bD_{se}}^{-1}{\hat\Sigma_{se}}^{-1}\brcs{\frac{1}{\sqrt{NT}}\hat\cW\T\lrp{\cW^H-\hat\cW}\mu_{se,w}^H+\sqrt{NT}\br_{w\epsilon}}+\sqrt{NT}{\bD_{se}}^{-1}\mu_{se,w}^H \\
    & \therefore \bD_{se}\lrp{\hat\mu_{se}-\mu_{se}^H}=\frac{{\hat\Sigma_{se}}^{-1}}{\sqrt{NT}}\hat\cW\T\lrp{\cW^H-\hat\cW}\mu_{se,w}^H+\sqrt{NT}{\hat\Sigma_{se}}^{-1}\br_{w\epsilon}.
    \end{align*}
    Therefore, we next show that the first term on the RHS is negligible and $\sqrt{NT}{\hat\Sigma_{se}}^{-1}\br_{w\epsilon}$ is converging to a multivariate normal distribution. We start with claiming that $\hat\Sigma_{se}$ is converging to a matrix with finite entries as $N$ and $T$ tend to infinity. This will allow us to focus on the behaviors of $\frac{1}{\sqrt{NT}}{\hat\cW}\T(\cW^H-\hat\cW)\mu_{se,w}^H$ and $\sqrt{NT}\br_{w\epsilon}$, and then apply Slutsky's Theorem.
    
    \begin{claim}
        $\hat\Sigma_{se}$ converges to $\Sigma_{se}$ in probability, i.e., ${\hat\Sigma_{se}}\Rightarrow{\Sigma_{se}}$.
    \end{claim}
    \proof
    As consequences of Lemma \ref{lmm-enar-se-∑nasym}, we have
    \begin{align*}
        \hat\Sigma_{se}&=\frac{1}{NT}
        \sum_t\begin{bmatrix}
            \frac{1}{\rho_N}|\bS_\bA|& \frac{1}{\sqrt{\rho_N}}{\hat\bU_{se}}\T\by_t & \frac{1}{\sqrt{\rho_N}}{\hat\bU_{se}}\T\cL_\bA\by_t & \frac{1}{\sqrt{\rho_N}}{\hat\bU_{se}}\T\bZ_t \\
            & {\by_t}\T\by_t & {\by_t}\T\cL_\bA\by_t & {\by_t}\T\bZ_t \\
            & & {\by_t}\T{\cL_\bA}^2\by_t & {\by_t}\T\cL_\bA\bZ_t \\
            & & & {\bZ_t}\T\bZ_t
        \end{bmatrix}
        \\&\overset{N\to\infty}{\Rightarrow}
        \lim_{N\to\infty}\begin{bmatrix}
            \frac{1}{N\rho_N}\bS_\bP & \frac{1}{N\sqrt{\rho_N}}\tilde\bU_{se}\T{\varphi_{se}} & \frac{1}{N\sqrt{\rho_N}}\tilde\bU_{se}\T\cL_\bA{\varphi_{se}} & \bO_{K\times p} \\
            & \frac{1}{N}{\varphi_{se}}\T\varphi_{se}+\tau_2 & \frac{1}{N}{\varphi_{se}}\T\cL_\bA\varphi_{se}+\tau_{23} & \bo_p\T \\
            & & \frac{1}{N}{\varphi_{se}}\T{\cL_\bA}^2\varphi_{se}+\tau_3 & \bo_p\T \\
            & & & \Sigma_z
        \end{bmatrix}
    \end{align*}
    Note that $\nrm{\bS_\bP}{}=O(N\rho_N)$ and  ${\varphi_{se}}=O(\sqrt{N\rho_N})$ as.
    Existence of $\tau_2,\tau_{23},$ and $\tau_3$ come by dominated convergence theorem after noting that $\tr\lrp{\cL_\bA\Gamma}\leq\nrm{\cL_\bA}{}\tr\lrp{\Gamma}=O\lrp{N}$ as. and $\tr\lrp{{\cL_\bA}^2\Gamma}=O(N)$ as.
    Also from the assumptions and the proof of Lemma \ref{lmm-enar-se-nasym}, we have
    \begin{align*}
        \frac{1}{N\sqrt{\rho_N}}{\bh_1}\T\tilde\bU_{se}\T{\varphi_{se}}&=O(\sqrt{\rho_N}),
        \frac{1}{N\sqrt{\rho_N}}{\bh_1}\T\tilde\bU_{se}\T\cL_\bA{\varphi_{se}}=O(\sqrt{\rho_N}),\\
        \frac{1}{N}{\varphi_{se}}\T{\varphi_{se}}&=O(\rho_N),
        \frac{1}{N}{\varphi_{se}}\T\cL_\bA{\varphi_{se}}=O(\rho_N),
        \frac{1}{N}{\varphi_{se}}\T{\cL_\bA}^2{\varphi_{se}}=O(\rho_N)~as.
    \end{align*}
    for all $\bh_1\in B^{K-1}$.
    So, we obtain the asymptotic precision matrix.

    The following claim clearly shows that the parameter estimation errors except for $\hat\gamma_{se}$ are not guaranteed to be finite if $T\to\infty$.
    \begin{claim}\label{clm-enar-se-err}
        For any $\bh\in B^{K+p+1}$, $\frac{1}{\sqrt{NT}}\lrp{\hat\cW\bh}\T\lrp{\cW^H-\hat\cW}\mu_{se,w}^H=O\lrp{1}~whp.$
    \end{claim}
    \proof
    Since $\hat\cW$ is different from $\cW^H$ by latent variables only, we have
    \begin{align*}
    \frac{1}{\sqrt{NT}}{\hat\cW}\T\lrp{\cW^H-\hat\cW}\mu_{se,w}^H
    =\begin{bmatrix}
        \sqrt{\frac{T}{N\rho_N}}{\hat\bU_{se}}\T\ \\
        \frac{1}{\sqrt{NT}}\sum_t{\by_t}\T \\
        \frac{1}{\sqrt{NT}}\sum_t{\by_t}\T\cL_\bA \\
        \frac{1}{\sqrt{NT}}\sum_t{\bZ_t}\T
    \end{bmatrix}\Delta_{se}\beta_{se}.
    \end{align*}
    Then, by the statements 10--13 of Lemma \ref{lmm-enar-se-nasym}, we have $\frac{1}{\sqrt{NT}}{\hat\cW}\T\lrp{\cW^H-\hat\cW}\mu_{se,w}^H=\begin{bmatrix}
        \bb_1 \\ 
        \bb_2 \\ 
        \bb_3
        \end{bmatrix}$
    where $\bb_1\in\R^K,\bb_2\in\R^2,\bb_3\in\R^p$ are such that $\bx\T\bb_1=O\lrp{\sqrt{\pi_{se}T}}~whp.$, $\by\T\bb_2=O\lrp{\sqrt{\pi_{se}T\rho_N}}~whp.$, and $\bz\T\bb_3=O(\sqrt{\pi_{se}/N})~whp.$ for any $\bx\in B^{K-1},\by\in B^1,\bz\in B^{p-1}$. Noting that $\pi_{se}=O(1)$, we get the conclusion.
    \begin{claim}
        $\sqrt{NT}\br_{w\epsilon}\Rightarrow\cN(\bo_{K+p+2},\sigma^2\Sigma_{se})$.
    \end{claim}
    \proof
    It is sufficient to show that for any $\eta\in B^{K+p+1}$, we have $\sqrt{NT}\eta\T\br_{w\epsilon}\Rightarrow\cN(0,\sigma^2\eta\T\Sigma_{se}\eta)$. Denoting $ \xi_{Nt}\triangleq\lrp{NT}^{-1/2}\eta\T{\hat\cW_{t-1}}\T\bcE_t$ and $\cF_{Nt}=\sigma(\bA,\epsilon_{is},\bZ_{i,(s-1)};i\leq N,-\infty<s\leq t)$, $\{\sum_{s=1}^t \xi_{Ns},\cF_{Nt}\}$ constitutes a martingale array for each $N,t\leq T$. Then, we can apply Corollary 3.1 of \cite{hall2014martingale} to $\sqrt{NT}\eta\T\br_{w\epsilon}=\sum_t\xi_{N,t+1}$ by checking following two conditions, (1) and (2).
    
    \textbf{(1)} $\sum_{t}\E\lrp{\xi_{N,t+1}^21_{\brcs{\vrt{\xi_{N,t+1}}>\upsilon}}|\cF_{Nt}}=o_\pr(1)$. 
    
    \textbf{(2)} $\sum_t\E\lrp{\xi_{N,t+1}^2|\cF_{Nt}}=\eta\T\sigma^2\Sigma_{se}\eta+o_\pr(1)$.
    
    \proof[Proof of (1)]
    First, we have
    \[\sum_{t=0}^{T-1}\E\lrp{\xi_{N,t+1}^21_{\brcs{\vrt{\xi_{N,t+1}}>\upsilon}}|\cF_{Nt}}\leq \upsilon^{-2}\sum_t\E\lrp{\xi^4_{N,t+1}|\cF_{Nt}}.\]
    One can easily verify that
    \[
    \E\lrp{\xi_{N,t+1}^4|\cF_{Nt}}\lesssim\sigma^4\lrp{\frac{1}{NT}\eta\T{\hat\cW_t}\T{\hat\cW_t}\eta}^2.
    \]
    So, we only need to show that $\sum_t\lrp{\frac{1}{NT}\eta\T{\hat\cW_t}\T{\hat\cW_t}\eta}^2=o_\pr(1)$. First note that 
    \begin{align*}
    \lrp{NT}^{-1}&{\hat\cW_t}\T\hat\cW_t\\
    &=\bxs{\begin{matrix}
    \frac{1}{NT\rho_N}\bS_\bA & \frac{1}{NT\sqrt{\rho_N}}{\hat\bU_{se}}\T\by_t & \frac{1}{NT\sqrt{\rho_N}}{\hat\bU_{se}}\T\cL_\bA\by_t & \frac{1}{NT\sqrt{\rho_N}}{\hat\bU_{se}}\T\bZ_t \\
    & \frac{1}{NT}{\by_t}\T\by_t & \frac{1}{NT}{\by_t}\T\cL_\bA\by_t & \frac{1}{NT}{\by_t}\T\bZ_t \\
    & & \frac{1}{NT}{\by_t}\T{\cL_\bA}^2\by_t & \frac{1}{NT}{\by_t}\T\cL_\bA\bZ_t \\
    & & & \frac{1}{NT}\bZ_t\T\bZ_t \\
    \end{matrix}}.
    \end{align*}
    Since similar arguments can be used to show the convergence of each entry, take 
    \[\sum_t\lrp{\frac{1}{NT\sqrt{\rho_N}}{\eta_1}\T{\hat\bU_{se}}\T\by_t}^2=\frac{1}{N^2T^2\rho_N}\sum_t{\eta_1}\T{\hat\bU_{se}}\T\by_t{\by_t}\T\hat\bU_{se}\eta_1\]
    for example, where $\eta_1\in B^{K-1}$.
    Then, we have
    \begin{align*}
        \sum_t{\eta_1}\T{\hat\bU_{se}}\T\by_t{\by_t}\T\hat\bU_{se}\eta_1=\by\T\brcs{\bI_T\otimes\lrp{\hat\bU_{se}\eta_1{\eta_1}\T{\hat\bU_{se}}\T}}\by.
    \end{align*}
    First, we have
    \[\tilde\by\T{\Gamma_\by}^{1/2}\brcs{\bI_T\otimes\lrp{\hat\bU_{se}\eta_1{\eta_1}\T{\hat\bU_{se}}\T}}{\Gamma_\by}^{1/2}\tilde\by=\tilde\by\T\brcs{\bI_T\otimes\lrp{\Gamma^{1/2}\hat\bU_{se}\eta_1{\eta_1}\T{\hat\bU_{se}}\T\Gamma^{1/2}}}\tilde\by.\]
    Letting $\hat\bS\triangleq\bI_T\otimes\lrp{\Gamma^{1/2}\hat\bU_{se}\eta_1{\eta_1}\T{\hat\bU_{se}}\T\Gamma^{1/2}}$, we have  $\nrm{\hat\bS}{F}^2=T\tr\brcs{\lrp{\hat\bU_{se}\eta_1{\eta_1}\T{\hat\bU_{se}}\T\Gamma}^2}\leq T\nrm{\hat\bU_{se}\eta_1{\eta_1}\T{\hat\bU_{se}}\T}{}^2\tr\lrp{\Gamma^2}=O\lrp{NT\cdot N^2{\rho_N}^2}~whp.$ and $\nrm{\hat\bS}{}=\nrm{\Gamma^{1/2}\hat\bU_{se}\eta_1{\eta_1}\T{\hat\bU_{se}}\T\Gamma^{1/2}}{}\leq\nrm{\hat\bU_{se}}{}^2\nrm{\Gamma}{}=O(N\rho_N)~whp.$
    By Lemma \ref{thm-hwineq-whp},
    \begin{align*}
        \pr&\bxs{\frac{1}{N^2T^2\rho_N}\vrt{\tilde\by\T\hat\bS\tilde\by-\E\lrp{\tilde\by\T\hat\bS\tilde\by}}>\upsilon}\\
        &<2\exp\brcs{-c\min\lrp{\frac{\upsilon^2N^4T^4{\rho_N}^2}{N^3T{\rho_N}^2},\frac{\upsilon N^2T^2\rho_N}{N\rho_N}}}+1/N.
    \end{align*}
    Note that $\E\lrp{\tilde\by\T\hat\bS\tilde\by}=T\E\tr\lrp{\hat\bU_{se}\eta_1{\eta_1}\T{\hat\bU_{se}}\T\Gamma}$ and 
    \[T\tr\lrp{\hat\bU_{se}\eta_1{\eta_1}\T{\hat\bU_{se}}\T\Gamma}\leq T\nrm{\hat\bU_{se}\eta_1{\eta_1}\T{\hat\bU_{se}}\T}{}\tr\lrp{\Gamma}\]
    hence is $O\lrp{N^2T\rho_N}~whp.$
    Therefore, $\frac{\tilde\by\T\hat\bS\tilde\by}{N^2T\rho_N}<\infty=O(1)~whp.$ hence $\tilde\by\T\hat\bS\tilde\by=o_\pr(N^2T^2\rho_N)$. 
    So we have $\E\lrp{\tilde\by\T\hat\bS\tilde\by}=o\lrp{N^2T^2\rho_N}$ by dominated convergence theorem.
    Also, $\vrt{{\Phi_{se}}\T\hat\bS{\Phi_{se}}}=O(N^2T{\rho_N}^2)~whp.$
    Therefore, we have $\frac{1}{N^2T^2\rho_N}\sum_t\lrp{{\eta_1}\T{\hat\bU_{se}}\T\by_t}^2=o_\pr(1)$. Noting that both $\by$ and $\bZ_t$ have finite fourth-order moments, one can show that the rest are also $o_\pr(1)$ similarly.

    \proof[Proof of (2)]
    Since $\sum_t\E\lrp{\xi_{N,t+1}^2|\cF_{Nt}}=\frac{\sigma^2}{NT}\eta\T{\hat\cW}\T{\hat\cW}\eta=\sigma^2\eta\T\hat\Sigma_{se}\eta$, by Lemma \ref{lmm-enar-se-∑nasym}, we have (2).
    
    Therefore, by the three claims, we have the desired asymptotic normality after dividing $\bD_{se}\lrp{\hat\mu_{se}-\mu_{se}^H}$ by $\sqrt{T}$.
\end{proof}

\begin{proof}[Proof of Theorem \ref{thm-enar-se-clt-NK}]
    We can follow the steps in the proof of Theorem \ref{thm-enar-se-clt-N}.
    Since $N\rho_N=\omega(\log^4N)$, we have $\pi_{se}=C+o\lrp{\frac{\sqrt{K}}{\log N}}$ for some constant $C>0$.
    Therefore, by $N=\omega(K^2\log K)$ we have $N=\omega\lrp{\pi_{se}}$ and $\log K-N/K^2\to-\infty$ hence $\hat\Sigma_{se}\Rightarrow\Sigma_{se}$ by Lemma \ref{lmm-enar-se-nasym} and Lemma \ref{lmm-enar-se-∑nasym}.
    Note that an additional order of $\log K$ is required on $N=\omega(K^2)$ because $T$ is assumed to be finite.
    Check that $\bA_K\bD_{se}\lrp{\hat\mu_{se}-\mu_{se}^H}=\frac{\bA_K{\hat\Sigma_{se}}^{-1}}{\sqrt{NT}}{\hat\cW}\T\lrp{\cW^H-\hat\cW}\mu_{se,w}^H+\sqrt{NT}\bA_K{\hat\Sigma_{se}}^{-1}\br_{w\epsilon}$ and $\nrm{\bA_K}{}=O(1)$.
    For convenience, write $\bd\triangleq\frac{1}{\sqrt{NT}}{\hat\cW}\T\lrp{\cW^H-\hat\cW}\mu_{se,w}^H$. Since
    \[\nrm{\bA_K{\hat\Sigma_{se}}^{-1}\bd}{}\lesssim\nrm{\bA_K\lrp{{\hat\Sigma_{se}}^{-1}-{\Sigma_{se}}^{-1}}\bd}{}+\nrm{\bA_K{\Sigma_{se}}^{-1}\bd}{},\]
    we check if $\nrm{\bA_K\lrp{{\hat\Sigma_{se}}^{-1}-{\Sigma_{se}}^{-1}}\bd}{}=o_\pr\lrp{\nrm{\bA_K{\Sigma_{se}}^{-1}\bd}{}}$.
    It suffices to check if it is satisfied that $\nrm{{\hat\Sigma_{se}}^{-1}-{\Sigma_{se}}^{-1}}{}=o_\pr\lrp{\nrm{{\Sigma_{se}}^{-1}}{}}$. 
    First, check that
    \[\sup_{\bh\in S^{K+p+1}}\vrt{\bh\T\lrp{\hat\Sigma_{se}-\Sigma_{se}}\bh}\le\sup_{\bh\in \bB^{K+p+1}}\vrt{\bh\T\lrp{\hat\Sigma_{se}-\Sigma_{se}}\bh}\]
    hence by Lemma \ref{lmm-enar-se-∑nasym}, we confirm that this is true as
    \[\nrm{{\hat\Sigma_{se}}^{-1}-{\Sigma_{se}}^{-1}}{}\lesssim s_{K+p+2}\lrp{{\hat\Sigma_{se}}}^{-1}{}\nrm{\hat\Sigma_{se}-\Sigma_{se}}{}s_{K+p+2}(\Sigma_{se})^{-1}={s_{K+p+2}(\Sigma_{se})}^{-1}o_\pr(1).\]
    Therefore, it suffices to check the order of $\bA_K{\Sigma_{se}}^{-1}\bd$.
    First, since $N=\omega(K^2\log K)$, we have $\pi_{se}=C+o\lrp{\frac{\sqrt{K}}{\log K}}$.
    Therefore from Claim \ref{clm-enar-se-err}, we have $\bx\T\bA_K{\Sigma_{se}}^{-1}\bd=O\lrp{1}~whp.$ for any $\bx\in B^{m-1}$ as $\nrm{\bA_1}{}=O\lrp{\frac{\log K}{\sqrt{K}}}$.
    
    Next, it is sufficient to show that for any $\eta\in B^{m-1}$, we have $\eta\T\bA_K{\hat\Sigma_{se}}^{-1}\sqrt{NT}\br_{w\epsilon}\Rightarrow\cN(0,\sigma^2\eta\T\bV_{se}\eta)$. 
    Since 
    \begin{align*}
        \eta\T\bA_K{\hat\Sigma_{se}}^{-1}\sqrt{NT}\br_{w\epsilon}=
        \eta\T\bA_K\lrp{{\hat\Sigma_{se}}^{-1}-{\Sigma_{se}}^{-1}}\sqrt{NT}\br_{w\epsilon}+\eta\T\bA_K{\Sigma_{se}}^{-1}\sqrt{NT}\br_{w\epsilon} \\
        =o_\pr(1)+\eta\T\bA_K{\Sigma_{se}}^{-1}O_\pr(1),
    \end{align*}
    we focus on the latter. 
    Define $ \xi_{Nt}\triangleq\lrp{NT}^{-1/2}\eta\T\bA_K{\Sigma_{se}}^{-1}{\hat\cW_{t-1}}\T\bcE_t$ and a filtration $\cF_{Nt}=\sigma(\bA,\epsilon_{is},\bZ_{i,(s-1)};i\leq N,-\infty<s\leq t)$. Then, a set of pairs $\{\sum_{s=1}^t \xi_{Ns},\cF_{Nt}\}$ constitutes a martingale array for each $N,t\leq T$. Then, we repeat the following arguments \citep{hall2014martingale}:
    
    \textbf{(1)} $\sum_{t}\E\lrp{\xi_{N,t+1}^21_{\brcs{\vrt{\xi_{N,t+1}}>\upsilon}}|\cF_{Nt}}=o_\pr(1)$. 

    \proof
    We only need to show that 
    \[
    \sum_t\lrp{\frac{1}{NT}\eta\T\bA_K{\Sigma_{se}}^{-1}{\hat\cW_t}\T{\hat\cW_t}{\Sigma_{se}}^{-1}{\bA_K}\T\eta}^2=o_\pr(1).
    \]
    Since both $\bA_K$ and $\Sigma_{se}$ have spectral norm of $O(1)$ as $K\rightarrow\infty$, 
    this can be shown in the same manner as in the proof of (1) of the third claim in proving Theorem \ref{thm-enar-se-clt-N}.
    
    \textbf{(2)} $\sum_t\E\lrp{\xi_{N,t+1}^2|\cF_{Nt}}=\sigma^2\eta\T\bV_{se}\eta+o_\pr(1)$.
    
    \proof
    This can be shown by the arguments in the proof of (2) of the third claim in proving Theorem \ref{thm-enar-se-clt-N}.

    This completes the proof of Theorem \ref{thm-enar-se-clt-NK}.
\end{proof}

\subsubsection{Consistency under misspecification of \texorpdfstring{$K$}{K}}
\begin{proof}[Proof of Theorem \ref{thm-enar-se-overK-clt-N}]
    Now let us augment $\bD_{se}$ as $\bD_{se,+}\triangleq\sqrt{NT}\operatorname{diag}\lrp{\sqrt{\rho_N}\bI_{K_o},\bI_{p+2}}$ and augment $\hat\cW$ in the proof of Theorem \ref{thm-enar-se-clt-N} as $\hat\cW_+\triangleq\bxs{\frac{1}{\sqrt{\rho_N}}\bon_T\otimes\hat\bU_{se,+}\,|\,\hat\cW}\in\R^{NT\times( K_o+p+2)}$
    so that $\sqrt{NT}\bW_+=\hat\cW_+\bD_{se,+}$.
    Also, we write $\mu_{se,+}^H=\lrp{{\bo_{K_o-K}}\T,{\mu_{se}^H}\T}\T$ and $\mu_{w,se+}^H=\lrp{{\bo_{K_o-K}}\T,{\mu_{se,w}^H}\T}\T$.
    This representation gives $\hat\cW_+\mu_{w,se+}^H=\hat\cW\mu_{se,w}^{H}$ and $\sqrt{NT}{\bD_{se,+}}^{-1}\mu_{w,se+}^{H}=\mu_{se,+}^H$.
    Define
    \begin{align}\label{eq-se-hSw+}
        \hat\Sigma_{se,+}\triangleq\frac{1}{NT}{\hat\cW_+}\T\hat\cW_+
    \end{align}
    and $\br_{w\epsilon,+}\triangleq\frac{1}{NT}{\hat\cW_+}\T\bcE$.
    Then, recalling the proof in the previous section,
    \begin{align*}
        \hat\mu_{se,+} & = \lrp{\frac{\bD_{se,+}}{NT}{\hat\cW_+}\T\hat\cW_+\bD_{se,+}}^{-1}\frac{\bD_{se,+}}{\sqrt{NT}}{\hat\cW_+}\T\lrp{\cW^H\mu_{se,w}^H+\bcE} \\
        & = \lrp{\hat\Sigma_{se,+}\bD_{se,+}}^{-1}\frac{1}{\sqrt{NT}}{\hat\cW_+}\T\brcs{\lrp{\cW^H-\hat\cW}\mu_{se,w}^H+\hat\cW_+\mu_{w,se+}^H+\bcE} \\
        & = {\bD_{se,+}}^{-1}{\hat\Sigma_{se,+}}^{-1}\brcs{\frac{1}{\sqrt{NT}}{\hat\cW_+}\T\lrp{\cW^H-\hat\cW}\mu_{se,w}^H+\sqrt{NT}\br_{w\epsilon,+}}+\sqrt{NT}{\bD_{se,+}}^{-1}\mu_{w,se+}^{H} \\
        & \therefore \hat\Sigma_{se,+}\bD_{se,+}\lrp{\hat\mu_{se,+}-\mu_{se,+}^{H}}=\frac{1}{\sqrt{NT}}{\hat\cW_+}\T\lrp{\cW^H-\hat\cW}\mu_{se,w}^H+\sqrt{NT}\br_{w\epsilon,+}.
    \end{align*}
    Therefore, we can use similar logic as the proof of Theorem \ref{thm-enar-se-clt-N}.
    \begin{claim}
        ${\hat\Sigma_{se,+}}\Rightarrow{\Sigma_{se,+}}$.
    \end{claim}
    \proof
    Now observe that ${\hat\bU_{se,+}}\T\hat\bU_{se,+}=|\bS_{\bA,:K}|$ which is a diagonal matrix containing $K+1,..., K_o$-th leading eigenvalues of $\bA$. 
    Also, by Weyl's inequality,
    \[\nrm{\bS_{\bA,:K}}{}=\nrm{\bS_{\bA,:K}-\bS_{\bP,:K}}{}\le\nrm{\bA-\bP}{}=O(\sqrt{N\rho_N})~whp.\]
    hence $\nrm{\hat\bU_{se,+}}{}=\nrm{\bS_{\bA,:K}}{}^{1/2}=O(\sqrt[4]{N\rho_N})~whp.$ so by a similar result as item 0 of Lemma \ref{lmm-enar-se-nasym},
    \begin{align*}
        \hat\Sigma_{se,+}&=\bxs{\begin{matrix}
        \frac{1}{N\rho_N}|\bS_{\bA,:K}|& \frac{1}{NT\sqrt{\rho_N}}\sum_t{\hat\bU_{se,+}}\T\hat\cW_t\\
        &\hat\Sigma_{se}
        \end{matrix}}\overset{N\to\infty}{\Rightarrow}\bxs{\begin{matrix}
        \bO_{K_o-K} & (0)\\
        & \Sigma_{se}
        \end{matrix}}
    \end{align*}
    which is easy to check as ${\hat\bU_{se,+}}\T\hat\bU_{se}=\bO_{(K_o-K)\times K}$ and for any $\bh\in B^{ K_o-K}$, we have $\nrm{{\Phi_{se}}}{}=O(\sqrt{TN\rho_N})~as.$ and $\nrm{\bon_T\otimes\hat\bU_{se,+}}{}=O(\sqrt[4]{T^2N\rho_N})~whp.$ so
    \begin{align*}
        \frac{1}{NT\sqrt{\rho_N}}\bh\T\sum_t{\hat\bU_{se,+}}\T\by_t=\frac{1}{NT\sqrt{\rho_N}}\bh\T\lrp{\bon_T\otimes\hat\bU_{se,+}}\T{\Gamma_\by}^{1/2}\tilde\by+O\lrp{\sqrt[4]{\frac{\rho_N}{N}}}~whp.,\\
        \pr\bxs{\frac{1}{NT\sqrt{\rho_N}}\vrt{\bh\T\lrp{\bon_T\otimes\hat\bU_{se,+}}\T{\Gamma_\by}^{1/2}\tilde\by}>\upsilon}<2\exp\lrp{-cNT\sqrt{N\rho_N}\upsilon^2}
    \end{align*}
    for example.
    \begin{claim}
    For any $\bh_1\in B^{K_o-K-1},\bh_2\in B^{K-1},\bh_3\in B^{1},\bh_4\in B^{p-1}$,
        \[\frac{1}{\sqrt{NT}}\operatorname{diag}\lrp{\bh_1,\dots,\bh_4}\T{\hat\cW_+}\T(\cW^H-\hat\cW)\mu_{se,w}^H=\begin{bmatrix}
            O\lrp{\frac{\sqrt{KT}}{\sqrt[4]{N\rho_N}}}~whp. \\ 
            O\lrp{\sqrt{KT}}~whp. \\
            O\lrp{\sqrt{KT\rho_N}}~whp. \\ 
            o_\pr(1)
        \end{bmatrix}.\]
    \end{claim}
    \proof 
    Now observe that
    \begin{align*}
        \frac{1}{\sqrt{NT}}{\hat\cW_+}\T\lrp{\cW^H-\hat\cW}\mu_{se,w}^H
        =-\frac{1}{\sqrt{NT}}\begin{bmatrix}
            \frac{1}{\sqrt{\rho_N}}\lrp{\bon_T\otimes\hat\bU_{se,+}}\T \\ {\hat\cW}\T
        \end{bmatrix}
        \bon_T\otimes\lrp{\hat\bU_{se}-\tilde\bU_{se}}{\bH_{se}}\T\beta_{se}.
    \end{align*}
    Then, our additional bias term of interest is 
    \[\sqrt{\frac{T}{N\rho_N}}\lrp{\hat\bU_{se,+}\bh}\T\lrp{\hat\bU_{se}-\tilde\bU_{se}}{\bH_{se}}\T\beta_{se}=O\lrp{\sqrt[4]{\frac{K^2T^2}{N\rho_N}}}~whp.\]
    for any $\bh\in B^{K_o-K-1}$ which leads to the conclusion by the observation made in the proof of the second claim in proving Theorem \ref{thm-enar-se-clt-N}.
    \begin{claim}
        $\sqrt{NT}\br_{w\epsilon,+}\Rightarrow\cN(\bo_{ K_o+p+2},\sigma^2\Sigma_{se,+})$.
    \end{claim}
    \proof
    This is trivial from the third claim in the proof of Theorem \ref{thm-enar-se-clt-N}.

    Therefore, we have the conclusion.
\end{proof}

\begin{proof}[Proof of Theorem \ref{thm-enar-se-underK-consist-N}]
    Similarly to the proof of Theorem \ref{thm-enar-ev-underK-consist-NT}, deinfe a diagonal matrix of convergence rates $\bD_{se,-}\triangleq\sqrt{NT}\operatorname{diag}\lrp{\sqrt{\rho_N}\bI_{K_o},\bI_{p+2}}$ and a truncated design matrix with $\hat\bU_{se,-}$ embedded over the entire time, 
    \[\hat\cW_{-}\triangleq\bxs{\frac{1}{\sqrt{\rho_N}}\bon_T\otimes\hat\bU_{se,-}\,|\,\by,\lrp{\bI_T\otimes\cL_\bA}\by\,|\,\bZ}\in\R^{NT\times(K_o+p+2)}.\]
    Let $\hat\cW_{:K}\in\R^{(K+p+2)\times NT}$ denote the augmented version of $\hat\cW_-$ which contains zeros between first and second block of $\hat\cW_-$ above:
    \[\hat\cW_{:K}\triangleq\bxs{\frac{1}{\sqrt{\rho_N}}\bon_T\otimes\hat\bU_{se,-}\,|\bO_{NT\times(K-K_o)}\,|\,\by,\lrp{\bI_T\otimes\cL_\bA}\by\,|\,\bZ}\in\R^{NT\times(K+p+2)}.\]
    Define $\mu_{se}^{H_-}\triangleq\lrp{{\beta_{se}}\T\bH_{se,-},\mu_{se,-\beta}\T}\T\in\R^{K_o+p+2}$ and $\mu_{se,w}^{H_-}\triangleq\lrp{\sqrt{\rho_N}{\beta_{se}}\T\bH_{se,-},\mu_{se,-\beta}\T}\T$.
    This expression gives $\sqrt{NT}\hat\bW_-=\hat\cW_-\bD_{se,-}$ and $\hat\cW_{:K}\mu_{se,w}^H=\hat\cW_-\mu_{se,w}^{H_-}$.
    $\hat\Sigma_{se,-}$ and $\br_{w\epsilon,-}$ are defined analogously as the case of over-selection.
    Then, we can derive the following in the same way as previous Theorems:
    \begin{align*}
        \bD_{se,-}\lrp{\hat\mu_{se,-}-\mu_{se}^{H_-}}=\frac{{\hat\Sigma_{se,-}}^{-1}}{\sqrt{NT}}\hat\cW_-\T\lrp{\cW^{H}-\hat\cW_{:K}}\mu_{se,w}^H+\sqrt{NT}{\hat\Sigma_{se,-}}^{-1}\br_{w\epsilon,-}.
    \end{align*}
    Let $\Sigma_{se,-}$ be a version of $\Sigma_{se}$ which misses $K-K_o$ rows and columns corresponding to $K_o+1,\dots,K$-th latent variables.

    Next, by Proposition \ref{prp-rdpg-cnt} there exists $\bH_{se}\in\cO_K$ such that
    \begin{align*}
        \nrm{\hat\bU_{se}-\bU_{se}\bH_{se}}{}=O\lrp{1}~whp.
    \end{align*}
    such that $\bH_{se}=\bxs{\bH_{se,-}\,|\,\bH_{se,:K}}$.
    Therefore, there exists $\bH_{se,K_o}\in\cO_{K,K_o}$ such that
    \[\nrm{\hat\bU_{se,-}-\bU_{se}\bH_{se,-}}{}\le\nrm{\hat\bU_{se,-}-\bU_{se}\bH_{se,K_o}}{}+\nrm{\bI_{K_o}-{\bH_{se,K_o}}\T\bH_{se,-}}{}=O(\sqrt{N\rho_N})~whp.\]
    by Theorem \ref{thm-rdpg-cnt2}.
    \begin{claim}\label{clm-enar-se-underK-Sg}
        For any $\bh\in B^{K_o+p+1}$, $\bh\T\lrp{{\hat\Sigma_{se,-}}-{\Sigma_{se,-}}}\bh=O_\pr(1)$. 
    \end{claim}
    \proof
    Now we only need to show the following:
    \begin{enumerate}
        \item $\pr\lrp{\frac{1}{NT\sqrt{\rho_N}}\vrt{{\bh_1}\T\sum_t\brcs{\hat\bU_{se,-}\T\by_t-\lrp{\bU_{se}\bH_{se,-}}\T{\varphi_{se}}}}>\upsilon}\lesssim e^{-d_0\upsilon^2}+1/N$
        \item $\pr\bxs{\frac{1}{NT\sqrt{\rho_N}}\vrt{{\bh_1}\T\sum_t\brcs{\hat\bU_{se,-}\T\cL_\bA\by_t-\lrp{\bU_{se}\bH_{se,-}}\T\cL_\bA{\varphi_{se}}}}>\upsilon}\lesssim e^{-d_0\upsilon^2}+1/N$
        \item $\pr\lrp{\frac{1}{NT\sqrt{\rho_N}}\vrt{\sum_t{\bh_1}\T\hat\bU_{se,-}\T\bZ_t\bh_2}>\upsilon}\lesssim e^{-NT\upsilon^2}+1/N$
    \end{enumerate}
    where $d_0\triangleq c_0/\rho_N$ for a constant $c_0>0$ and $\upsilon=\omega\lrp{\sqrt{\rho_N}}$.
    These can be easily shown by the similar proof to 1--3 of Lemma \ref{lmm-enar-se-nasym}.
    Also, we have $\frac{1}{N\sqrt{\rho_N}}\lrp{\bU_{se}\bH_{se,-}\bh_1}\T{\varphi_{se}}=\Theta(\sqrt{\rho_N})~as.$ and $\frac{1}{N\sqrt{\rho_N}}\lrp{\bU_{se}\bH_{se,-}\bh_1}\T\cL_\bA{\varphi_{se}}=\Theta\lrp{\sqrt{\rho_N}}$.
    Therefore, according to the same argument as Lemma \ref{lmm-enar-se-∑nasym}, due to 1--2. we no longer obtain convergence in probability but boundedness in probability.
    If $K_o=0$, then we have the reduced asymptotic covariance as $\Sigma_{-u}$.
    
    Next, we analyze the entry-wise asymptotic orders of $\frac{1}{\sqrt{NT}}{\hat\cW_-}\T(\cW^H-\hat\cW_{:K})\mu_{se,w}^H$.
    First note that $\bH_{se,-}{\bH_{se,-}}\T+\bH_{se,:K}{\bH_{se,:K}}\T=\bI_K$ as $\bH_{se}\in\cO_K$ and
    \[\cW^H-\hat\cW_{:K}=\frac{1}{\sqrt{\rho_N}}\bon_T\otimes\bxs{\bU_{se}\bH_{se,-}-\hat\bU_{se,-}\,|\,\bU_{se}\bH_{se,:K}\,|\,\bO_{N\times(p+2)}}.\]
    Then, $\frac{1}{\sqrt{NT}}{\hat\cW_-}\T\lrp{\cW^H-\hat\cW_{:K}}\mu_{se,w}^H$ equals
    \begin{align*}
        \frac{1}{\sqrt{NT}}\sum_t\hat\cW_{t,-}\T
        \brcs{\lrp{\bU_{se}\bH_{se,-}-\hat\bU_{se,-}}{\bH_{se,-}}\T+\bU_{se}\bH_{se,:K}{\bH_{se,:K}}\T}\beta_{se}\\
        =\bxs{\begin{matrix} 
        \sqrt{\frac{T}{N\rho_N}}{\hat\bU_{se,-}}\T \\
        \frac{1}{\sqrt{NT}}\sum_t{\by_t}\T \\
        \frac{1}{\sqrt{NT}}\sum_t{\by_t}\T\cL_\bA \\
        \frac{1}{\sqrt{NT}}\sum_t{\bZ_t}\T
        \end{matrix}}\lrp{\bU_{se}-\hat\bU_{se,-}{\bH_{se,-}}\T}\beta_{se}.
    \end{align*}
    Denoting the RHS by $\bq_{K_o}$ which depends on $K_o$, we have $\bq_0=\bU_{se}\beta_{se}$ for $K_o=0$ as $\hat\bU_{se,-}{\bH_{se,-}}\T$ is removed. So, $\nrm{\bq_{K_o}}{}=O(\sqrt{N\rho_N})~whp.$ by noting that 
    \[\nrm{\bU_{se}-\hat\bU_{se,-}{\bH_{se,-}}\T}{}=\nrm{\hat\bU_{se,-}-\bU_{se}\bH_{se,-}}{}.\]
    Then, we can easily show the following by similar arguments used in proving 10--13 of Lemma \ref{lmm-enar-se-nasym}:
    \begin{enumerate}
        \item $\sqrt{\frac{T}{N\rho_N}}\vrt{\lrp{\hat\bU_{se,-}\bh_1}\T\bq_{K_o}}=O\lrp{\sqrt{TN\rho_N}}~whp.$
        \item $\frac{1}{\sqrt{NT}}\vrt{\sum_t{\by_t}\T\bq_{K_o}}=O\lrp{\sqrt{N}\rho_N}~whp.$
        \item $\frac{1}{\sqrt{NT}}\vrt{\sum_t{\by_t}\T\cL_\bA\bq_{K_o}}=O\lrp{\sqrt{N}\rho_N}~whp.$
        \item $\pr\bxs{\frac{1}{\sqrt{NT}}\vrt{\lrp{\sum_t{\bZ_t}\bh_2}\T\bq_{K_o}}>\upsilon}<e^{-c\upsilon^2/\rho_N}$
    \end{enumerate}
    for any $\upsilon>0$ where $\bh_1\in B^{K_o-1},\bh_2\in B^{p-1}$.
    Now partition $\frac{1}{\sqrt{NT}}{\hat\cW_-}\T\lrp{\cW^H-\hat\cW_{:K}}\mu_{se,w}^H$ into three sub-vectors of sizes $K_o$, $2$, and $p$ denoted by $\bb_1,\bb_2$, and $\bb_3$ respectively.
    These give the specified asymptotic orders of the estimation error terms.
    
    Now, using the same steps in the third claim in the proof of Theorem \ref{thm-enar-se-clt-N} and recalling Claim \ref{clm-enar-se-underK-Sg}, we have $\frac{1}{\sqrt{NT}}\nrm{\hat\cW_-\eta}{}=O_\pr(1)$ for any $\eta\in B^{K_o+p+1}$.
    So, we can derive that $\sqrt{NT}\eta\T\br_{w\epsilon,-}=O_\pr(1)$ as for any $\epsilon>0$ we have $N_0,T_0>0$ such that if $N>N_0,T>T_0$, by Lemma \ref{lmm-hoeffding-Op}:
    \[\pr\lrp{\frac{1}{\sqrt{NT}}\vrt{\lrp{\hat\cW_-\eta}\T\bcE}>\upsilon}<2\exp\lrp{-c\upsilon^2}+\epsilon\]
    for each $\upsilon>0$.
    Therefore, recalling the rates of $\bD_{se,-}$, we can summarize as follows:
    \begin{align*}
        \sqrt{N\rho_N}\lrp{\hat\beta_{se,-}-{\bH_{se,-}}\T\beta_{se}}&=\bb_1+\bt_1,\quad
        \sqrt{N}\begin{bmatrix}
            \hat\alpha_{se,-}-\alpha_{se}\\
            \hat\theta_{se,-}-\theta_{se}
        \end{bmatrix}=\bb_2+\bt_2,\\
        \sqrt{N}\lrp{\hat\gamma_{se,-}-\gamma_{se}}&=\bb_3+\bt_3
    \end{align*}
    where
    \begin{align*}
        &{\bh_1}\T\bb_1=O\lrp{\sqrt{N\rho_N}}~whp.,\quad{\bh_2}\T\bb_2=O\lrp{\sqrt{N}\rho_N}~whp.,\quad{\bh_3}\T\bb_3=O_\pr\lrp{\sqrt{\rho_N}},\\
        &{\bh_1}\T\bt_1=O_\pr(1),{\bh_2}\T\bt_2=O_\pr(1),{\bh_3}\T\bt_3=O_\pr(1)
    \end{align*}
    for any $\bh_1\in B^{K_o-1},\bh_2\in B^1,\bh_3\in B^{p-1}$.
\end{proof}

\subsubsection{Asymptotic results} 
This section lists analogous results to the results in the Lemma \ref{lmm-enar-ev-nasym} for the case of ENAR with spectral embedding.
Although we keep track of $T$ here, we assume that $T$ is finite when applying the following Lemma.
\begin{lemma}\label{lmm-enar-se-nasym}
    Let $\bh_i\in\R^{j_i}$ for $j_1=K$, $j_2=p$ be real vectors such that $\nrm{\bh_i}{}\leq1$.
    Then under the conditions of Theorem \ref{thm-enar-se-clt-N}, there exist $N_0>0$ and $\upsilon_0(N_0)>0$ such that the statements 0--13 hold for all $\upsilon\in\lrp{0,\upsilon_0}$ and $N>N_0$:
    \begin{enumerate}\addtocounter{enumi}{-1}
        \item $\pr\lrp{\frac{1}{N\rho_N}\vrt{{\bh_1}\T\lrp{|\bS_\bA|-\bS_\bP}\bh_1}>\upsilon}<1/N$
        \item $\pr\lrp{\frac{1}{NT\sqrt{\rho_N}}\vrt{{\bh_1}\T\sum_t(\hat\bU_{se}\T\by_t-\tilde\bU_{se}\T{\varphi_{se}})}>\upsilon}
        \lesssim e^{-d_0\upsilon^2}+1/N$
        \item $\pr\bxs{\frac{1}{NT\sqrt{\rho_N}}\vrt{{\bh_1}\T\sum_t\lrp{\hat\bU_{se}\T\cL_\bA\by_t-\tilde\bU_{se}\T\cL_\bA{\varphi_{se}}}}>\upsilon}\lesssim e^{-d_0\upsilon^2}+1/N$
        \item $\pr\lrp{\frac{1}{NT\sqrt{\rho_N}}\vrt{\sum_t{\bh_1}\T\hat\bU_{se}\T\bZ_t\bh_2}>\upsilon}\lesssim e^{-d_0\upsilon^2}+1/N$
        \item $\pr\bxs{\frac{1}{NT}\vrt{\sum_t\brcs{{\by_t}\T\by_t-\E\tr\lrp{\Gamma}-{\varphi_{se}}\T\varphi_{se}}}>\upsilon}\lesssim e^{-d_0\upsilon^2}+e^{-d_1\upsilon^2}$
        \item $\pr\bxs{\frac{1}{NT}\vrt{\sum_t\brcs{{\by_t}\T\cL_\bA\by_t-\E\tr\lrp{\cL_\bA\Gamma}-{\varphi_{se}}\T\cL_\bA\varphi_{se}}}>\upsilon}\lesssim e^{-d_0\upsilon^2}+e^{-d_1\upsilon^2}$
        \item $\pr\lrp{\frac{1}{NT}\vrt{\sum_t{\by_t}\T\bZ_t\bh_2}>\upsilon}\lesssim e^{-d_2\upsilon^2}+e^{-d_1\upsilon^2}+1/N$
        \item $\pr\bxs{\frac{1}{NT}\vrt{\sum_t\brcs{{\by_t}\T{\cL_\bA}^2\by_t-\E\tr\lrp{{\cL_\bA}^2\Gamma}-{\varphi_{se}}\T{\cL_\bA}^2\varphi_{se}}}>\upsilon}\lesssim e^{-d_0\upsilon^2}+e^{-d_1\upsilon^2}$
        \item $\pr\lrp{\frac{1}{NT}\vrt{\sum_t{\by_t}\T\cL_\bA\bZ_t\bh_2}>\upsilon}\lesssim e^{-d_2\upsilon^2}+e^{-d_1\upsilon^2}+1/N$
        \item $\pr\bxs{\frac{1}{NT}\vrt{{\bh_2}\T\sum_t\lrp{{\bZ_t}\T\bZ_t-\Sigma_z}\bh_2}>\upsilon}\lesssim e^{-d_0\upsilon^2}$
        \item $\frac{\sqrt{T}}{\sqrt{N\rho_N}}\vrt{{\bh_1}\T\hat\bU_{se}\T\lrp{\hat\bU_{se}-\tilde\bU_{se}}\beta_{se}}=O(\sqrt{T\pi_{se}})~whp.$
        \item $\frac{1}{\sqrt{NT}}\vrt{\sum_t{\by_t}\T\lrp{\hat\bU_{se}-\tilde\bU_{se}}\beta_{se}}=O\lrp{\sqrt{\pi_{se}T\rho_N}}~whp.$
        \item $\frac{1}{\sqrt{NT}}\vrt{\sum_t{\by_t}\T\cL_\bA\lrp{\hat\bU_{se}-\tilde\bU_{se}}\beta_{se}}=O\lrp{\sqrt{\pi_{se}T\rho_N}}~whp.$
        \item $\pr\bxs{\frac{1}{\sqrt{NT}}\vrt{{\bh_2}\T\sum_t{\bZ_t}\T\lrp{\hat\bU_{se}-\tilde\bU_{se}}\beta_{se}}>\upsilon}\lesssim e^{-d_3\upsilon^2}+1/N$
    \end{enumerate}
    where $d_0\triangleq c_0NT$, $d_1\triangleq \frac{c_1NT}{\rho_N}$, $d_2\triangleq\frac{c_2NT}{p}$, and $d_3=\frac{c_3N}{K}$.
\end{lemma}
\begin{proof}[Proof of 0]
    First, $\nrm{|\bS_\bA|-\bS_\bP}{}=O(\log N)~whp.$ by inhomogeneous Erdős–Rényi model singular value perturbation bounds \cite{cape2017kato} and $N\rho_N=\omega\lrp{\log^4N}$. Therefore,
    \[\pr\lrp{\frac{1}{N\rho_N}\vrt{{\bh_1}\T\lrp{|\bS_\bA|-\bS_\bP}\bh_1}>\upsilon}\le\pr\lrp{\vrt{{\bh_1}\T\lrp{|\bS_\bA|-\bS_\bP}\bh_1}>\frac{C\log N}{N\rho_N}}<1/N\]
    for some constant $C>0$ and any $\upsilon\ge\frac{C\log N}{N\rho_N}$. So we have the desired bound for large enough $N$.
\end{proof}
\begin{proof}[Proof of 1]
    Note that 
    \begin{align*}
        \sum_{t=1}^{T}{\bh_1}\T\lrp{\hat\bU_{se}\T\by_t-\tilde\bU_{se}\T{\varphi_{se}}}=
        \brcs{\lrp{\bon_T\otimes\hat\bU_{se}}\bh_1}\T{\Gamma_\by}^{1/2}\tilde\by
        +T\lrp{\Delta_{se}\bh_1}\T{\varphi_{se}}\\
        =\lrp{\hat\bS\bh_1}\T\tilde\by+T\lrp{\Delta_{se}\bh_1}\T{\varphi_{se}}
    \end{align*}
    where $\hat\bS\triangleq{\Gamma_\by}^{1/2}\lrp{\bon_T\otimes\hat\bU_{se}}$.
    Since $\nrm{\hat\bS\bh_1}{}=O\lrp{\sqrt{TN\rho_N}}$ $whp.$, by Lemma \ref{lmm-hoeffding-whp} we have
    \[\pr^*\lrp{\vrt{\lrp{\hat\bS\bh_1}\T\tilde\by}>NT\sqrt{\rho_N}\upsilon}<2\exp\lrp{\frac{-cN^2T^2\rho_N\upsilon^2}{TN\rho_N}}+1/N.\]
    Since the RHS does not depend on $\bA$, we can replace $\pr^*$ with $\pr$.
    Next, since $\nrm{{\varphi_{se}}}{}=O(\sqrt{N\rho_N})$ $\pr^*$-as. (hence is $\pr$-as.), $\vrt{\lrp{\Delta_{se}\bh_1}\T{\varphi_{se}}}=O(\sqrt{\pi_{se}N\rho_N})~whp.$, i.e., we have $\frac{T}{NT\sqrt{\rho_N}}\vrt{\lrp{\Delta_{se}\bh_1}\T{\varphi_{se}}}=O\lrp{\sqrt{\pi_{se}/N}}~whp.$ 
    We can obtain the desired bound using the same logic as 0 provided that $N=\omega(\pi_{se})$.
\end{proof}
\begin{proof}[Proof of 2]
    \begin{align*}
        {\bh_1}\T\sum_t\lrp{\hat\bU_{se}\T\cL_\bA\by_t-\tilde\bU_{se}\T\cL_\bA{\varphi_{se}}}=
        {\bh_1}\T\brcs{\bon_T\otimes\lrp{\cL_\bA\hat\bU_{se}}}\T{\Gamma_\by}^{1/2}\tilde\by+T\lrp{\Delta_{se}\bh_1}\T\cL_\bA{\varphi_{se}}\\
        ={\bh_1}\T{\hat\bS}\T\tilde\by+T\lrp{\Delta_{se}\bh_1}\T\cL_\bA{\varphi_{se}}.
    \end{align*}
    For $\hat\bS\triangleq{\Gamma_\by}^{1/2}\brcs{\bon_T\otimes\lrp{\cL_\bA\hat\bU_{se}}}$, we have $\nrm{\hat\bS\bh_1}{}=O(\sqrt{TN\rho_N})$ $whp.$ because $\rho\lrp{\cL_\bA}=O(1)$ as. The tail probability of $\vrt{T\lrp{\Delta_{se}\bh_1}\T\cL_\bA{\varphi_{se}}}$ can be bounded similarly to 1.
\end{proof}
\begin{proof}[Proof of 3]
    Let $\bZ\triangleq\bxs{{\bZ_0}\T,...,{\bZ_{T-1}}\T}\T$ and $\hat\bS\triangleq\bon_T\otimes\hat\bU_{se}$. Then, we can express $\sum_t{\bh_1}\T\hat\bU_{se}\T\bZ_t\bh_2$ as $\tr\lrp{\bZ\bh_2{\bh_1}\T{\hat\bS}\T}$. This is equal to $\vecop\lrp{\bh_2{\bh_1}\T}\T\lrp{\bI_p\otimes{\hat\bS}\T}\vecop\lrp{\bZ}$ and note that $\nrm{\bI_p\otimes\hat\bS\T}{}=O\lrp{\sqrt{TN\rho_N}}~whp.$ Therefore, by Lemma \ref{lmm-hoeffding-whp}, we have
    \[\pr\bxs{\vrt{\sum_t{\bh_1}\T\hat\bU_{se}\T\bZ_t\bh_2}>NT\sqrt{\rho_N}\upsilon}\leq2\exp\lrp{-\frac{cN^2T^2\rho_N\upsilon^2}{TN\rho_N}}+1/N.\]
\end{proof}
\begin{proof}[Proof of 4]
    Now we observe that
    \begin{align*}
        \sum_t\lrp{{\by_t}\T\by_t-\varphi_{se}\T\varphi_{se}}
        =\tilde\by\T\Gamma_\by\tilde\by+2\Phi_{se}\T{\Gamma_\by}^{1/2}\tilde\by.
    \end{align*}
    Note that $\nrm{\Gamma_\by}{F}^2=O( NT)$ as. By Lemma \ref{thm-hwineq-whp}, we have \[\pr\lrp{\vrt{\tilde\by\T\Gamma_\by\tilde\by-T\tr\E\Gamma}>\upsilon}<2\exp\lrp{-c\min\brcs{\frac{\upsilon^2}{NT},\upsilon}}.\]
    By Lemma \ref{lmm-hoeffding-whp}, 
    \[\pr\lrp{2|\Phi_{se}\T{\Gamma_\by}^{1/2}\tilde\by|>\upsilon}<2\exp\lrp{-\frac{c'\upsilon^2}{TN\rho_N}}\] 
    since $\nrm{{\Gamma_\by}^{1/2}\Phi_{se}}{}^2\leq\nrm{\Gamma_\by}{}T\varphi_{se}\T\varphi_{se}=O(TN\rho_N)$ as.
\end{proof}
\begin{proof}[Proof of 6]
    We have 
    \begin{align*}
        \sum_t{\by_t}\T\bZ_t\bh_2=\lrp{{\Gamma_\by}^{1/2}\tilde\by+{\Phi_{se}}}\T\bZ\bh_2=\tr\lrp{{\bh_2}\Tilde\by\T{\Gamma_\by}^{1/2}\bZ}+\tr\lrp{\bh_2{\Phi_{se}}\T\bZ}\\
        =\vecop\lrp{\tilde\by{\bh_2}\T}\T\lrp{\bI_p\otimes{\Gamma_\by}^{1/2}}\vecop\lrp{\bZ}+\vecop\lrp{{\Phi_{se}}{\bh_2}\T}\T\lrp{\bI_p\otimes\bI_{NT}}\vecop\lrp{\bZ}.
    \end{align*}
    Therefore, since $\nrm{\Gamma_\by}{}=O(1)$ as., $\nrm{\lrp{\bI_p\otimes{\Gamma_\by}^{1/2}}\vecop\lrp{\tilde\by{\bh_2}\T}}{}\lesssim\nrm{\vecop\lrp{\tilde\by{\bh_2}\T}}{}$ which is concentrated around $\sqrt{NTp}$ \citep{vershynin2018high} hence is $O(\sqrt{NTp})~whp.$ 
    Also, we have
    \[\nrm{\lrp{\bI_p\otimes\bI_{NT}}\vecop\lrp{{\Phi_{se}}{\bh_2}\T}}{}=O(\sqrt{TN\rho_N})~as.\]
    By Lemma \ref{lmm-hoeffding-whp},
    \[\pr\lrp{\vrt{\sum_t{\by_t}\T\bZ_t\bh_2}>\upsilon}\leq2\exp\lrp{-\frac{c\upsilon^2}{NTp}}+2\exp\lrp{-\frac{c\upsilon^2}{TN\rho_N}}+\frac{1}{N}.\]
\end{proof}
\begin{proof}[Proof of 10]
    Noting that $\nrm{\beta_{se}}{}=O(1)$, we have $\vrt{{\bh_1}\T\hat\bU_{se}\T\Delta_{se}\beta_{se}}=O\lrp{\sqrt{\pi_{se}N\rho_N}}~whp.$ which implies $\sqrt{T/(N\rho_N)}\vrt{{\bh_1}\T\hat\bU_{se}\T\Delta_{se}\beta_{se}}=O\lrp{\sqrt{T\pi_{se}}}~whp.$
\end{proof}
\begin{proof}[Proof of 11]
    Similarly to the proof of 1., taking $\hat\bS\triangleq{\Gamma_\by}^{1/2}\lrp{\bon_T\otimes\Delta_{se}}$ which has the spectral norm of  $O\lrp{\sqrt{KT}}~whp.$, we have
    \begin{align*}
        \sum_t{\by_t}\T\Delta_{se}\beta_{se}=
        \lrp{{\Gamma_\by}^{1/2}{\tilde\by}+{\Phi_{se}}}\T\lrp{\bon_T\otimes\Delta_{se}}\beta_{se}=\tilde\by\T\hat\bS\beta_{se}+{\Phi_{se}}\T{\Gamma_\by}^{-\frac{1}{2}}\hat\bS\beta_{se}.
    \end{align*}
    It is straightforward by noting that $\nrm{\beta_{se}}{}=O(1)$ as $K\rightarrow\infty$, 
    \[\pr\lrp{\frac{1}{\sqrt{NT}}\vrt{\tilde\by\T\hat\bS\beta_{se}}\geq\upsilon}\lesssim\exp\lrp{-cN\upsilon^2/K}+1/N\]
    and ${\Phi_{se}}\T{\Gamma_\by}^{-\frac{1}{2}}\hat\bS\beta_{se}=O\lrp{T\sqrt{\pi_{se}N\rho_N}}~whp.$
\end{proof}
\begin{proof}[Proof of 13]
    Note that ${\bh_2}\T\sum_t{\bZ_t}\T\Delta_{se}\beta_{se}={\bh_2}\T\bZ\T\lrp{\bon_T\otimes\Delta_{se}}\beta_{se}$ which equals
    \[\tr\brcs{\bZ\bh_2\beta_{se}\T\lrp{\bon_T\otimes\Delta_{se}}\T}=\vecop\lrp{\bh_2\beta_{se}\T}\T\brcs{\bI_p\otimes\lrp{\bon_T\otimes\Delta_{se}}\T}\vecop\lrp{\bZ}.\]
    Since the norm of $\frac{1}{\sqrt{NT}}\vecop\lrp{\bh_2\beta_{se}\T}\T\brcs{\bI_p\otimes\lrp{\bon_T\otimes\Delta_{se}}\T}$ is of $O\lrp{\sqrt{\pi_{se}/N}}~whp.$, we obtain the bound provided that $\upsilon=\omega\lrp{\sqrt{\pi_{se}/N}}$ and $N=\omega(\pi_{se})$.
\end{proof}
\begin{lemma}\label{lmm-enar-se-∑nasym}
    There exist $N_0,T_0>0$ and $\upsilon_0(N_0,T_0)\in(0,1)$ such that for all $\upsilon\in(0,\upsilon_0)$ and $N>N_0,T>T_0$,
    \begin{align*}
        \pr\bxs{\sup_{\bh\in B^{K+p+1}}\vrt{\bh\T\lrp{\hat\Sigma_{se}-\Sigma_{se}}\bh}>\upsilon}\lesssim K^2\brcs{\frac{1}{N}+\exp\lrp{-\frac{cNT\upsilon^2}{K^2p}}}.
    \end{align*}
\end{lemma}
\begin{proof}
    Let $\bh\in B^{K+p+1}$ and $\be_i\in S^{K+p+1}$ be the $i^{th}$ canonical basis of $\R^{K+p+2}$, i.e., $(\be_i)_j\triangleq\mathbb{I}_{\brcs{i}}(j)$ for $j=1,\dots,K+p+2$.
    By 0--9 of Lemma \ref{lmm-enar-se-nasym}, we have
    \begin{align*}
        \max_{i,j}\pr\lrp{\vrt{{\be_
        i}\T\lrp{\hat\Sigma_{se}-\Sigma_{se}}\be_j}>\upsilon}
        \lesssim\frac{1}{N}+\exp\lrp{-c_1\upsilon^2NT/p}.
    \end{align*}
    Since 
    \begin{align*}
        \sup_{\bh\in B^{K+p+1}}\vrt{\bh\T\lrp{\hat\Sigma_{se}-\Sigma_{se}}\bh}\lesssim K\max_{i,j}\vrt{\hat\sigma_{se,ij}-\sigma_{se,ij}},
    \end{align*}
    we have the conclusion.
\end{proof}

\subsection{Over-selection of \texorpdfstring{$K$ by $\hat K$}{K(IC) > K}}
Here, we prove that $\pr\lrp{\brcs{\hat K\ge K}~ev.}=1$.
\begin{proof}[Proof of Theorem \ref{thm-enar-overK}]
    Let $p_{f,NT}$ be a penalty function that can vary over $f\in\brcs{se,ev}$.
    First, we see that $\hat K<K$ implies $Cr_f(k)\le Cr_f(k')$ for some $k<K$ and all $k'\in\brcs{K,\dots,J_N}$, i.e.,
    \[\brcs{\hat K<K}\subseteq\bigcup_{k<K}\bigcap_{k'=K}^{J_N}\brcs{Cr_f(k)\le Cr_f(k')}\]
    so
    \begin{align*}
        \pr\lrp{\hat K<k}\le\sum_{k<K}\pr\bxs{\cap_{k'\ge K}\brcs{Cr_f(k)\le Cr_f(k')}}\le
        K\max_{k<K}\max_{k'\ge K}\pr\bxs{Cr_f(k)\le Cr_f(k')}.
    \end{align*}
    Note that for any $k$ such that $0\le k<K$ we have
    \begin{align*}
        \max_{k'\ge K}\pr\bxs{\brcs{Cr_f(k)\le Cr_f(k')}}
        &=\max_{k'\ge K}\pr\bxs{\frac{1}{T}\by\T\lrp{\hat\Pi_f^{(k')}-\hat\Pi_f^{(k)}}\by\le p_{f,NT}(k')-p_{f,NT}(k)}\\
        &\le\pr\bxs{\frac{1}{T}\by\T\lrp{\hat\Pi_f^{(K)}-\hat\Pi_f^{(k)}}\by\le p_{f,NT}(J_N)-p_{f,NT}(k)},\\
        &\max_{k<K}\pr\bxs{\frac{1}{T}\by\T\lrp{\hat\Pi_f^{(K)}-\hat\Pi_f^{(k)}}\by\le p_{f,NT}(J_N)-p_{f,NT}(k)}\\
        &\le\pr\bxs{\frac{1}{T}\by\T\lrp{\hat\Pi_f^{(K)}-\hat\Pi_f^{(K-1)}}\by\le p_{f,NT}(J_N)-p_{f,NT}(0)}
    \end{align*}
    by the fact that $\hat\Pi_f^{(i)}\succcurlyeq\hat\Pi_f^{(j)}$ for any $i\ge j$.
    Let us denote $\hat\Pi_f^{(K)}-\hat\Pi_f^{(K-1)}$ by $\hat\Pi_{f,\Delta}$. 
    $\by\T\lrp{\hat\Pi_f^{(K)}-\hat\Pi_f^{(K-1)}}\by$ can be decomposed as
    \begin{dmath*}
        \by\T\hat\Pi_{f,\Delta}\by={\tilde\by}\T{\Gamma_\by}^{1/2}\hat\Pi_{f,\Delta}{\Gamma_\by}^{1/2}{\tilde\by}+2{\tilde\by}\T{\Gamma_\by}^{1/2}\hat\Pi_{f,\Delta}{\Gamma_\by}^{1/2}\Phi_f+\Phi_f\T{\Gamma_\by}^{1/2}\hat\Pi_{f,\Delta}{\Gamma_\by}^{1/2}\Phi_f
        \ge \Phi_f\T{\Gamma_\by}^{1/2}\hat\Pi_{f,\Delta}{\Gamma_\by}^{1/2}\Phi_f-2\vrt{{\tilde\by}\T{\Gamma_\by}^{1/2}\hat\Pi_{f,\Delta}{\Gamma_\by}^{1/2}\Phi_f}.
    \end{dmath*}
    Now we realize that for $\hat\bU_{f,i:j}$ which contains eigenvectors corresponding to $i,\dots,j$-th largest magnitude eigenvalues of $\bA$, ${\hat\bU_{f,i:j}}\,\T\hat\bU_{f,i:j}=\Lambda_{f,i:j}$ where $\Lambda_{ev,i:j}=\bI_{j-i}$ and $\Lambda_{se,i:j}=|\bS_{\bA,i:j}|$ where $\bS_{\bA,i:j}$ is a diagonal matrix with $i,\dots,j$-th largest magnitude eigenvalues of $\bA$. So,
    \begin{dmath*}
        \hat\Pi_{f,\Delta}=\frac{1}{T}\bJ_T\otimes\lrp{\hat\bU_{f,1:K}\Lambda_{f,1:K}{\hat\bU_{f,1:K}}\T-\hat\bU_{f,1:(K-1)}\Lambda_{f,1:(K-1)}{\hat\bU_{f,1:(K-1)}}\T}=\frac{\lambda_{f,K}}{T}\bJ_T\otimes\lrp{\hat\bu_{f,K}{\hat\bu_{f,K}}\T},~\lambda_{f,K}{\,=}\begin{cases}
            1, & f=ev \\ s_K(\bA), & f=se
        \end{cases}
    \end{dmath*}
    where $\bJ_T\triangleq\bon_T{\bon_T}\T$.
    This is a rank-one matrix.
    So, for $\bz_f\triangleq{\Gamma_\by}^{1/2}\Phi_f$ and $\bv\in\cO_{NT,1}$ which is the eigenvector of $\hat\Pi_{f,\Delta}$ corresponding to its unit eigenvalue, we have
    \[{\bz_f}\T\hat\Pi_{f,\Delta}\bz_f=\lrp{{\Phi_f}\T\Gamma_\by^{1/2}\bv}^2\ge T\nrm{\varphi_f}{}^2s_{NT}\lrp{\Gamma_\by}\]
    by variational characterization of singular values.
    By Proposition 2.2 and 2.3 of \cite{basu2015regularized} again, we have
    \[s_{NT}(\Gamma_\by)\ge\frac{\sigma^2}{m_{\max}(\bG)}\ge\frac{\sigma^2}{\lrp{1+\nrm{\bG}{}}^2}~as.\]
    where $m_{\max}(\bG)\triangleq\max_{\brcs{z\in\cC;|z|=1}}\lrp{\bI_N-\bG z}^*\lrp{\bI_N-\bG z}$.
    For the last lower bound, we refer to (i) of the proof of Proposition 2.2 in \cite{basu2015regularized}.
    Therefore, $s_{NT}\lrp{\Gamma_\by}>\frac{\sigma^2}{4}~as.$.
    Next, noting that $\nrm{{\Gamma_\by}^{1/2}\hat\Pi_{f,\Delta}{\Gamma_\by}^{1/2}\Phi_f}{}\le\nrm{\Phi_f}{}\nrm{\Gamma_\by}{}=O\lrp{\sqrt{T}\nrm{\varphi_f}{}}~as.$, we have
    \begin{align*}
        \pr\bxs{2\vrt{{\tilde\by}\T{\Gamma_\by}^{1/2}\hat\Pi_{f,\Delta}{\Gamma_\by}^{1/2}\Phi}>\upsilon}<2\exp\lrp{-\frac{c\upsilon^2}{T\nrm{\varphi_f}{}^2}}
    \end{align*}
    for any $\upsilon>0$.
    So, denoting $p_{f,NT}(J_N)-p_{f,NT}(0)$ by $\Delta p_{f,NT}$,
    \begin{dmath*}
        K\pr\bxs{\by\T\lrp{\hat\Pi_f^{(K)}-\hat\Pi_f^{(K-1)}}\by\le T\Delta p_{f,NT}}\\
        \le K\pr\bxs{\Phi_f\T{\Gamma_\by}^{1/2}\hat\Pi_{f,\Delta}{\Gamma_\by}^{1/2}\Phi_f-T\Delta p_{f,NT}{\le}2\vrt{{\tilde\by}\T{\Gamma_\by}^{1/2}\hat\Pi_{f,\Delta}{\Gamma_\by}^{1/2}\Phi}}
        \le K\pr\bxs{T\upsilon_*{\le}2\vrt{{\tilde\by}\T{\Gamma_\by}^{1/2}\hat\Pi_{f,\Delta}{\Gamma_\by}^{1/2}\Phi}}{\le}2K\exp\lrp{\frac{-cT\upsilon_*^2}{\nrm{\varphi_f}{}^2}}
    \end{dmath*}
    for $\upsilon_*\triangleq \frac{\sigma^2}{4}\nrm{\varphi_f}{}^2-\Delta p_{f,NT}$ provided that there exist large enough $N,T$ such that $\frac{\sigma^2}{4}\nrm{\varphi_f}{}^2>\Delta p_{f,NT}$. 
    Note that $\nrm{\varphi_f}{}^2\gtrsim s_K({\bU_{ev}}\T\bU_{ev})\ge1$ and $\nrm{\varphi_f}{}^2\le\nrm{\bU_{se}}{}^2\lesssim N\rho_N$ hence
    \[\frac{\upsilon_*}{\nrm{\varphi_f}{}}=\frac{\sigma^2}{4}\nrm{\varphi_f}{}-\frac{\Delta p_{f,NT}}{\nrm{\varphi_f}{}}\gtrsim 1-\frac{p_{f,NT}(J_N)}{\sqrt{N\rho_N}}.\]
    So if $p_{f,NT}(J_N)=o(1)$, then the conclusion follows by Borel-Cantelli Lemma provided that $T=\omega\lrp{\log K}$.

\end{proof}

\section{Proofs for Sections \ref{sec-amnar-estimation} \& \ref{sec-amnar-consist}}
\label{sec-amnar-app}
Here, we present the proof for the model stationarity and the asymptotic properties of estimators for AMNAR.
First, we provide the theoretical background for the estimation of latent variables $\bX$.
Let $\tilde\bQ\triangleq\bxs{\bQ\mid\bon_N}$.
For simplicity, $(\log f)', (\log f)'', (\log f)'''$ denote the $1^{st}, 2^{nd}, 3^{rd}$-order derivatives of the log-likelihood function $\log f(a_{ij};\chi_{ij})$ with regard to $\chi_{ij}$, respectively.
Fix a constant $s\in(0,1/2)$.
Following assumptions are analogous to the assumptions made in the section 2.2 of \cite{li2023statistical}.

\begin{enumerate}[I.]
    \item $\bX$ is contained in the constrained parameter space
    \[\Xi_K\triangleq\brcs{\bX\in\R^{N\times(K+1)}\,;\,{\bQ}\T\bon_N = \bo_K, \,{\bQ}\T{\bQ} \text{ is diagonal, } \nrm{\bX}{\tti} = O(1)\text{ as }N\rightarrow \infty}.\]
    \item There exist a positive definite $K \times K$ matrix $\Omega_q$ and some constant $\nu>0$ such that $\frac{1}{N}{\bQ}\T\bQ \to \Omega_q$ as $N \to \infty$ where $\Omega_q$ is a diagonal matrix with unique eigenvalues and $\frac{1}{N}\bv\T\bv\rightarrow\nu$.
    \item $(\log f)'''(\cdot)$ exists within $\Xi_K$. Furthermore, there exists $0< b_L < b_U$ such that $b_L \leq -(\log f)''(\cdot) \leq b_U$ and $|(\log f)'''(\cdot)| \leq b_U$ within $\Xi_K$. 
    \item There exist $t > 0$ and $u > 0$ such that for all $x \geq 0$, $P(|{(\log f)'_{ij}}| > x ) \leq \exp(-(x/t)^u)$.
    \item For any $1 \leq i \leq N$, there exist $\Sigma_i$ such that \[\frac{-1}{N}\sum_{j: j\neq i} \lrp{(\log f)''\circ\sigma}\lrp{\chi_{ij}}{\tilde\bq_j}{\tilde\bq_j}\,\T \overset{p}{\to} \Sigma_{i}.\]
    For an integer $m$ and any $m$ node indices $\mathcal{I} = (i_1, i_2, ..., i_m)$, there exists $\bS_{\mathcal{I}}$ such that $\frac{1}{\sqrt{N}}\big([S(\bx)]_{i_1}, [S(\bx)]_{i_2}, ..., [S(\bx)]_{i_m} \big) \to \mathcal{N}(0, \bS_{\mathcal{I}})$, where  $S(\bx) = \frac{\partial L}{\partial \bx}|_{\bx=\bx}$ is the score vector evaluated at the true parameters $\bx\triangleq\vecop\lrp{{\bX}\T}$, and $[S(\bx)]_{i}$ denotes the subvector of  $S(\bx)$   corresponding to all latent parameters associated with node $i$, i.e., $\bx_i$.
\end{enumerate}

Assumption I is often posed for the sake of theoretical analysis \citep{ma2020universal}. 
Conditions on ${\bQ}\T\bon_N$ and ${\bQ}\T\bQ$ in Assumptions I and II ensure the identifiability of $\bv$ and give regular conditions on the asymptotic behavior of the covariance structure of latent variables $\bQ$.
However, the diagonality assumption can be relaxed as \cite{li2023statistical} noted.
Assumption III requires $\sigma(\cdot)$ to be a smooth function and $\log f$ to be a concave log-likelihood function which constitutes a widely accepted class of link functions such as logit links. 
Assumptions IV and V ensure the fast decadence of tail density and the asymptotic distributions of maximum likelihood estimator \citep{li2023statistical}.
\begin{remark}
    To facilitate the understanding of modeling ENAR, consider a simpler example where we only consider the effect of latent variables only, i.e., $\by_t=\bU\beta+\bcE_t\in\R^N$, $t=1,\dots,T$ where we assume that the columns of a full column-rank matrix $\bU\in\R^{N\times K}$ are orthogonal.
    Then, the least square estimator $\hat\beta$, which minimizes $\sum_t\nrm{\by_t-\bU\beta}{}^2$ under normal error, satisfies the estimation equation
    \[T\bU\T\bU\hat\beta=\sum_t\bU\T\by_t.\]
    Therefore, we have
    \[\sqrt{T}\bU\T\bU\lrp{\hat\beta-\beta}\sim\lrp{\bU\T\bU}^{1/2}\cN\lrp{\bo_K,\sigma^2\bI_K}.\]
    In a typical canonical regression problem, it is a general practice to assume that the limiting variance $\lim_{N\to\infty}\bU\T\bU$ finitely exists and tends to a positive definite matrix.
    In this regard, our two candidates for latent variable for ENAR, eigenvectors of $\bP$, $\bU_{ev}=\bU_\bP$, or spectral embedding (SE) of $\bP$, $\bU_{se}=\bU_\bP{\bS_\bP}^{1/2}$, can satisfy this condition as well.
    The former gives $\bU_{ev}\T\bU_{ev}=\bI_K$ hence it will guarantee $\hat\beta_{ev}=\beta_{ev}+O_\pr(1/\sqrt{T})$.
    For the latter where we have $\bU_{se}\T\bU_{se}=\bS_\bP$, note that under our RDPG model, we have $\lambda_i(\bP)=\Theta(N\rho_N)$ for all $i=1,\dots,K$ hence $\hat\beta_{se}=\beta_{se}+O_\pr(1/\sqrt{N\rho_N})$ under finite $T$.
    These observations are related to our asymptotic results as we have $\sqrt{T}$-consistency for $\hat\beta_{ev}$ in Theorem \ref{thm-enar-ev-clt-NT} while $\hat\beta_{se}$ achieves $\sqrt{N\rho_N}$-consistency in Theorem \ref{thm-enar-se-clt-N}.
    
    Similarly in AMNAR model, we obtain the following under simpler AMNAR where we consider the effect of latent variables and regress on ${\tilde\rho_{NT}}^{1/2}\bX$ only:
    \[\sqrt{TN\tilde\rho_{NT}}\lrp{\frac{1}{N}\bX\T\bX}\lrp{\hat\beta-\beta}\sim\lrp{\frac{1}{N}\bX\T\bX}^{1/2}\cdot\cN\lrp{\bo_{K+1},\sigma^2\bI_{K+1}}.\]
    We need $\frac{1}{N}\bX\T\bX$ to exist asymptotically as $N\to\infty$, so this motivated the assumption II to include $\nrm{\bv}{}^2/N=\nu+o(1)$.
\end{remark}

\subsection{Stationarity}\label{lnar-station-prf}

Note the model can be written as $\by_{t+1}={\tilde\rho_{NT}}^{1/2}\bX\beta+\bG\by_{t}+\tilde\bcE_{t+1}$, for $\bG\triangleq\alpha\bI_N+\theta\cL_\bA$ and $\tilde\bcE_{t+1}\triangleq\bZ_{t}\gamma+\bcE_{t+1}$ and we assume that $\bA$ is generated independently with the rest of random components across all $t$.

\begin{theorem}\label{thm-amnar-station}
    If $|\alpha|+|\theta|<1$, then there is a unique strictly stationary solution to \ref{eq-lnar} with a finite first moment:
    \begin{align}\label{eq-amnar-stationsol}
        \by_t={\tilde\rho_{NT}}^{1/2}(\bI_N-\bG)^{-1}\bX\beta+\sum_{j=0}^\infty\bG^j\tilde\bcE_{t-j} 
    \end{align}
\end{theorem}

\begin{lemma}
    Define $\Gamma(h)\triangleq\operatorname{Cov}^*(\by_t,\by_{t-h})$ for all $t$. 
    Upon the conditions in Theorem \ref{thm-amnar-station} and conditional on $\bA$, \ref{eq-amnar-stationsol} follows a normal distribution with the mean 
    $\psi\triangleq{\tilde\rho_{NT}}^{1/2}(\bI_N-\bG)^{-1}\bX\beta$
    and the auto-covariance function in Lemma \ref{lmm-enar-meancov}.
\end{lemma}

\begin{theorem}\label{thm-amnar-station2}
Upon the conditions in Theorem \ref{thm-amnar-station} with $N\rightarrow\infty$, \ref{eq-amnar-stationsol} is a unique strictly stationary solution with a finite first moment. i.e., $\max_i\E|y_{it}|<\infty$.
\end{theorem}

\begin{proof}[Proof of Theorem \ref{thm-amnar-station2}]
    We have $\by_t=\sum_{j=0}^{m-1}\bG^j({\tilde\rho_{NT}}^{1/2}\bX\beta+\tilde\bcE_{t-j})+\bG^m\by_{t-m}$ hence $$\by_t=\lim_{m\rightarrow\infty}\by_t=\sum_{j=0}^\infty\bG^j({\tilde\rho_{NT}}^{1/2}\bX\beta+\tilde\bcE_{t-j}).$$
    First we have $\nrm{{\tilde\rho_{NT}}^{1/2}\bX\beta}{\infty}=\max_{i=1,...,N}{\tilde\rho_{NT}}^{1/2}\vrt{{\bx_i}\T\beta}\leq {\tilde\rho_{NT}}^{1/2}\nrm{\bX}{\tti}\nrm{\beta}{}$ hence $\vrt{{\tilde\rho_{NT}}^{1/2}\bX\beta}_e\precsim {\tilde\rho_{NT}}^{1/2}\bon_N$ independently with $j$. 
    So, we have $\E\vrt{{\tilde\rho_{NT}}^{1/2}\bX\beta+\tilde\bcE_{t-j}}_e\precsim C\cdot\bon_N$ for $C=o(1)+\E\vrt{\bz\T_1\gamma}+\E\vrt{\epsilon_{11}}$.
    The rest of the proof follows the same steps in the proof of Theorem \ref{thm-enar-station2}.

\end{proof}

\subsection{Consistency}\label{sec-amnar-consist-prf}
\begin{proof}[Proof of Theorem \ref{thm-amnar-clt-NT}]
    Let $\bD_a\triangleq\sqrt{NT}\operatorname{diag}\lrp{{\tilde\rho_{NT}}^{1/2}\bI_{K+1},\bI_{p+2}}$.
    For $\hat\Omega_m\triangleq{\bD_a}^{-1}{\hat{\bM}}\T\hat{\bM}{\bD_a}^{-1}$ and $\br_{m\epsilon}\triangleq{\bD_a}^{-1}{\hat{\bM}}\T\bcE$, we have 
    \begin{align*}
    \hat\mu_m & = {\bD_a}^{-1}\lrp{\hat\Omega_m}^{-1}{\bD_a}^{-1}\hat{\bM}\T\brcs{\lrp{\bM-\hat{\bM}}\mu_m+\hat{\bM}\mu_m+\bcE} \\
    & = {\bD_a}^{-1}{\hat\Omega_m}^{-1}\brcs{{\bD_a}^{-1}\hat{\bM}\T\lrp{\bM-\hat{\bM}}\mu_m+\br_{m\epsilon}}+\mu_m
    \end{align*}
    hence
    \[{\hat\Omega_m}\bD_a(\hat\mu_m-\mu_m)=
    {\bD_a}^{-1}{\hat{\bM}}\T(\bM-\hat{\bM})\mu_m
    +\br_{m\epsilon}.\]
    \begin{claim}
        ${\hat\Omega_m}\Rightarrow{\Omega_m}$. 
    \end{claim}
    \proof
    As consequences of Lemma \ref{lmm-amnar-∑nasym}, we have
        \begin{align*}
    \hat\Omega_m&=\frac{1}{NT}\sum_t\begin{bmatrix}
    {\hat\bX}\T\hat\bX & {\hat\bX}\T\by_t & {\hat\bX}\T\cL_\bA\by_t & {\hat\bX}\T\bZ_t \\
    & {\by_t}\T\by_t & {\by_t}\T\cL_\bA\by_t & {\by_t}\T\bZ_t \\
    & & {\by_t}\T{\cL_\bA}^2\by_t & {\by_t}\T\cL_\bA\bZ_t \\
    & & & {\bZ_t}\T\bZ_t
    \end{bmatrix}
    \\&\Rightarrow\lim_{N,T\rightarrow\infty}\bxs{\begin{matrix}
    \frac{1}{N}\bX\T\bX & \frac{1}{N}\bX\T\psi & \frac{1}{N}\bX\T\cL_\bA\psi & \bO_{(K+1)\times p} \\
    & \frac{1}{N}\E(\psi\T\psi)+\tau_2 & \frac{1}{N}\E(\psi\T\cL_\bA\psi)+\tau_{23} & \bo_p\T \\
    & & \frac{1}{N}\E(\psi\T{\cL_\bA}^2\psi)+\tau_3 & \bo_p\T \\
    & & & \Sigma_z
    \end{matrix}}.
    \end{align*}
    By assumptions in section \ref{sec-amnar-app}, we have $\nrm{\bv}{}^2=O(N)$ and $\frac{1}{N}\bQ\T\bQ\rightarrow\Omega_{q}$ hence we can infer that $\nrm{\bQ\T\bQ}{}=O(N)$. 
    Therefore, $\vrt{\bh\T\bQ\T\bv}\leq\nrm{\bQ}{}\nrm{\bv}{}=O(N)$ for $\bh\in B^{K-1}$ and $\frac{1}{N}\bX\T\bX\rightarrow\Omega_x$.
    Since $\nrm{\bX}{\tti}=O\lrp{1}$,  we have $\nrm{\bX}{}=O(\sqrt{N})$ hence $\nrm{\psi}{}\leq {\tilde\rho_{NT}}^{1/2}\nrm{\lrp{\bI_N-\bG}^{-1}}{}\nrm{\bX\beta}{}\leq {\tilde\rho_{NT}}^{1/2}\nrm{\bX}{}\nrm{\beta}{}/\lrp{1-g}=O({\tilde\rho_{NT}}^{1/2}\sqrt{N})$ as.
    Therefore, $\frac{1}{N}{\bh_1}\T\bX\T\psi$ and $\frac{1}{N}{\bh_1}\T\bX\T\cL_\bA\psi$ are $O({\tilde\rho_{NT}}^{1/2})$ and
    $\frac{1}{N}\psi\T\psi$, $\frac{1}{N}\psi\T\cL_\bA\psi$, and $\frac{1}{N}\psi\T{\cL_\bA}^2\psi$ are $O(\tilde\rho_{NT})$ as. for all $\bh_1\in B^K$.
    \begin{claim}
        ${\bD_a}^{-1}{\hat{\bM}}\T(\bM-\hat{\bM})\mu_m=o_\pr(1)$.
    \end{claim}
    \proof
    $\hat{\bM}$ is different from $\bM$ by $\hat\bX$ only, so we have
    \begin{align*}
    \bh\T{\bD_a}^{-1}{\hat{\bM}}\T\lrp{\hat{\bM}-\bM}\mu_m
    =\bh\T\bxs{\begin{matrix} 
       \lrp{\frac{\tilde\rho_{NT}}{NT}}^{1/2}{\hat\bX}\T \\
        \lrp{\frac{\tilde\rho_{NT}}{NT}}^{1/2}\sum_t{\by_t}\T \\
        \lrp{\frac{\tilde\rho_{NT}}{NT}}^{1/2}\sum_t{\by_t}\T\cL_\bA \\
        \lrp{\frac{\tilde\rho_{NT}}{NT}}^{1/2}\sum_t{\bZ_t}\T
    \end{matrix}}
    [\hat\bX-\bX\,\vert\,\bO_{N\times(p+2)}]\mu_m                        
    \end{align*}
    Then, by the statements 10--13 of Lemma \ref{lmm-amnar-nasym}, we have the conclusion by Cramér–Wold.
    \begin{claim}
        $\br_{m\epsilon}\Rightarrow\cN(\bo_{K+p+3},\sigma^2\Omega_m)$.
    \end{claim}
    \proof
    We show that for any $\eta\in\R^{K+p+3}$ such that $\|\eta\|\leq1$, it holds that $\eta\T\br_{m\epsilon}\Rightarrow\cN(0,\sigma^2\eta\T\Omega_m\eta)$.
    Denoting $ \xi_{N,t+1}\triangleq\eta\T{\bD_a}^{-1}{\hat\bM_t}\T\bcE_{t+1}$ and $\cF_{Nt}=\sigma(\bA,\epsilon_{is},\bZ_{is};i\leq N,-\infty<s\leq t)$, $\{\sum_{s=1}^t \xi_{Ns},\cF_{Nt}\}$ constitutes a martingale array for each $N,t\leq T$. Applying Corollary 3.1 of \cite{hall2014martingale}, we check:
    
    \textbf{(1)} $\sum_{t}\E\lrp{\xi_{N,t+1}^21_{\brcs{\vrt{\xi_{N,t+1}}>\upsilon}}|\cF_{Nt}}=o_\pr(1)$. 
    
    \textbf{(2)} $\sum_t\E\lrp{\xi_{N,t+1}^2|\cF_{Nt}}=\sigma^2\eta\T\Omega_m\eta+o_\pr(1)$.
    \proof[Proof of (1)]
    First, we have
    \[\sum_{t=0}^{T-1}\E\lrp{\xi_{N,t+1}^21_{\brcs{\vrt{\xi_{N,t+1}}>\upsilon}}|\cF_{Nt}}\leq \upsilon^{-2}\sum_t\E\lrp{\xi^4_{N,t+1}|\cF_{Nt}}.\]
    One can easily verify that
    \[
    \E\lrp{\xi_{N,t+1}^4|\cF_{Nt}}\lesssim\sigma^4\lrp{\eta\T{\bD_a}^{-1}{{\hat\bM_t}}\T{{\hat\bM_t}}{\bD_a}^{-1}\eta}^2.
    \]
    
    Then, we have
    \begin{align*}
    {\bD_a}^{-1}&{\hat\bM_t}\T\hat\bM_t{\bD_a}^{-1}\\
    &=\frac{1}{NT}\bxs{\begin{matrix}
    {\hat\bX}\T\hat\bX & {\hat\bX}\T\by_t & {\hat\bX}\T\cL_\bA\by_t & {\hat\bX}\T\bZ_t \\
    & {\by_t}\T\by_t & {\by_t}\T\cL_\bA\by_t & {\by_t}\T\bZ_t \\
    & & {\by_t}\T{\cL_\bA}^2\by_t & {\by_t}\T\cL_\bA\bZ_t \\
    & & & \bZ_t\T\bZ_t \\
    \end{matrix}}.
    \end{align*}
    Since similar arguments can be used to show the convergence of each entry, take 
    \[\sum_t\lrp{\frac{1}{NT}{\eta_1}\T{\hat\bX}\T\by_t}^2=\frac{1}{N^2T^2}\sum_t{\eta_1}\T{\hat\bX}\T\by_t{\by_t}\T{\hat\bX}\eta_1\]
    for example, where $\eta_1\in\R^K;\nrm{\eta_1}{}\leq1$.
    Then, we have
    \begin{align*}
    \sum_t{\eta_1}\T{\hat\bX}\T\by_t{\by_t}\T{\hat\bX}\eta_1&=\by\T\brcs{\bI_T\otimes\lrp{{\hat\bX}\eta_1{\eta_1}\T{\hat\bX}\T}}\by \\
    &=\lrp{\Gamma_{\by}^{1/2}\tilde\by+\Psi}\T\bxs{\bI_T\otimes\brcs{\lrp{\Delta+\bX}\eta_1{\eta_1}\T\lrp{\Delta+\bX}\T}}\lrp{\Gamma_{\by}^{1/2}\tilde\by+\Psi}.
    \end{align*}
    Denoting $\hat\bX-\bX$ by $\Delta$, recall that $\nrm{\Delta}{}=O_\pr(1)$ and $\nrm{\Psi}{}=O(\sqrt{TN\tilde\rho_{NT}})~as.$
    Letting $\tilde\bS\triangleq\Gamma_\by^{1/2}\brcs{\bI_T\otimes\lrp{\bX\eta_1{\eta_1}\T\bX\T}}\Gamma_\by^{1/2}$, we have $\nrm{\tilde\bS}{}=\nrm{\Gamma^{1/2}\bX\eta_1{\eta_1}\T\bX\T\Gamma^{1/2}}{}\leq\nrm{\Gamma}{}\nrm{\bX}{}^2$ hence is $O(N)$ as. and $\nrm{\tilde\bS}{F}^2=T\tr\brcs{\lrp{\bX\eta_1{\eta_1}\T\bX\T\Gamma}^2}\leq T\tr\lrp{\bX\eta_1{\eta_1}\T\bX\T}^2\nrm{\Gamma}{}^2\leq T\nrm{\bX}{}^4\nrm{\Gamma}{}=O(N^2T)$ as.
    Note that 
    \[\E\lrp{\tilde\by\T\tilde\bS\tilde\by}=\E\tr\lrp{\tilde\bS\tilde\by\tilde\by\T}=\tr\brcs{\E(\tilde\bS)\E\lrp{\tilde\by\tilde\by\T}}=T\tr\lrp{\bX\eta_1{\eta_1}\T\bX\T\E\Gamma}\]
    which is bounded above by $T\tr\lrp{\bX\eta_1{\eta_1}\T\bX\T}\nrm{\E\Gamma}{}\leq T\nrm{\bX}{}^2\E\nrm{\Gamma}{}=O(T\nrm{\bX}{}^2)=O(NT)$. 
    Applying Theorem \ref{thm-hwineq-whp}, we get
    \[\pr\bxs{\vrt{\tilde\by\T\tilde\bS\tilde\by-\E\lrp{\tilde\by\T\tilde\bS\tilde\by}}>N^2T^2\upsilon}<2\exp\lrp{-c\min\brcs{\frac{N^4T^4\upsilon^2}{N^2T},\frac{N^2T^2\upsilon}{N}}}\]
    and note that $\E\lrp{\tilde\by\T\tilde\bS\tilde\by}=o\lrp{N^2T^2}$ by dominated convergence.
    Similarly, we note that $\Psi\T\brcs{\bI_T\otimes\lrp{\bX\eta_1{\eta_1}\T\bX\T}}\Psi=O(\tilde\rho_{NT} N^2T)$ as. hence  $\sum_t{\eta_1}\T{\hat\bX}\T\by_t{\by_t}\T\hat\bX\eta_1=o_\pr(N^2T^2)$ in conclusion.
    The rest of the proof uses the same logic as well so is omitted here.
    Noting that both $\by$ and $\bZ_t$ have finite fourth-order moments, one can show that the rest terms of $\sum_t\lrp{\eta\T{\bD_a}^{-1}{\hat\bM_t}\T\hat\bM_t{\bD_a}^{-1}\eta}^2$ are also $o_\pr(N^2T^2)$ similarly.
    \proof[Proof of (2)]
    Since $\hat\Omega_m=\Omega_m+o_\pr(1)$, we directly have (2).
    
    Therefore, we have the conclusion for Theorem \ref{thm-amnar-clt-NT}.
\end{proof}

\subsection{Asymptotic results} 
\begin{lemma}\label{lmm-amnar-nasym}
    Let $\bh_i\in B^{j_i-1}$ for $j_1=K+1$, $j_2=p$.
    Then under the conditions of Theorem \ref{thm-amnar-clt-NT}, there exist $N_0,T_0>0$ and $\upsilon_0(N_0,T_0)\in(0,1)$ such that for all $\upsilon\in(0,\upsilon_0)$ and $N>N_0,T>T_0$, the statements 0--13 hold.
    \begin{enumerate}\addtocounter{enumi}{-1}
        \item $\pr\bxs{\frac{1}{N}\vrt{{\bh_1}\T\lrp{{\hat\bX}\T\hat\bX-\bX\T\bX}\bh_1}>\upsilon}<\epsilon$
        \item $\pr\lrp{\frac{1}{NT}\vrt{{\bh_1}\T\sum_t({\hat\bX}\T\by_t-\bX\T\psi)}>\upsilon}<2e^{-d_1\upsilon^2}+2e^{-d_2\upsilon^2}+
        \epsilon$
        \item $\pr\bxs{\frac{1}{NT}\vrt{{\bh_1}\T\sum_t\lrp{{\hat\bX}\T\cL_\bA\by_t-\bX\T\cL_\bA\psi}}>\upsilon}<
        2e^{-d_1\upsilon^2}+2e^{-d_2\upsilon^2}+\epsilon$
        \item $\pr\lrp{\frac{1}{NT}\vrt{\sum_t{\bh_1}\T{\hat\bX}\T\bZ_t\bh_2}>\upsilon}<2e^{-d_1\upsilon^2}+
        2e^{-d_2\upsilon^2}+\epsilon$
        \item $\pr\bxs{\frac{1}{NT}\vrt{\sum_t\brcs{{\by_t}\T\by_t-\E\tr\lrp{\Gamma}-\psi\T\psi}}>\upsilon}<2e^{-d_1\upsilon^2}+2e^{-d_2\upsilon^2}$
        \item $\pr\bxs{\frac{1}{NT}\vrt{\sum_t\brcs{{\by_t}\T\cL_\bA\by_t-\E\tr\lrp{\cL_\bA\Gamma}-\psi\T\cL_\bA\psi}}>\upsilon}<2e^{-d_1\upsilon^2}+2e^{-d_2\upsilon^2}$
        \item $\pr\lrp{\frac{1}{NT}\vrt{\sum_t{\by_t}\T\bZ_t\bh_2}>\upsilon}<
        2e^{-d_2\upsilon^2/p}+2e^{-d_3\upsilon^2}+\epsilon$
        \item $\pr\bxs{\frac{1}{NT}\vrt{\sum_t\brcs{{\by_t}\T{\cL_\bA}^2\by_t-\E\tr\lrp{{\cL_\bA}^2\Gamma}-\psi\T{\cL_\bA}^2\psi}}>\upsilon}<<2e^{-d_1\upsilon^2}+2e^{-d_2\upsilon^2}$
        \item $\pr\lrp{\frac{1}{NT}\vrt{\sum_t{\by_t}\T\cL_\bA\bZ_t\bh_2}>\upsilon}<2e^{-d_2\upsilon^2/p}+2e^{-d_3\upsilon^2}+\epsilon$
        \item $\pr\bxs{\frac{1}{NT}\vrt{{\bh_2}\T\sum_t\lrp{{\bZ_t}\T\bZ_t-\Sigma_z}\bh_2}>\upsilon}<2e^{-d_2\upsilon^2}$
        \item $\lrp{\frac{\tilde\rho_{NT}}{NT}}^{1/2}\vrt{{\bh_1}\T{\hat\bX}\T\lrp{\hat\bX-\bX}\beta}=O_\pr\lrp{\sqrt{\tilde\rho_{NT}/T}}$
        \item $\lrp{\frac{\tilde\rho_{NT}}{NT}}^{1/2}\vrt{\sum_t{\by_t}\T\lrp{\hat\bX-\bX}\beta}=O_\pr\lrp{\tilde\rho_{NT}\sqrt{T}}$
        \item $\lrp{\frac{\tilde\rho_{NT}}{NT}}^{1/2}\vrt{\sum_t{\by_t}\T\cL_\bA\lrp{\hat\bX-\bX}\beta}=O_\pr\lrp{\tilde\rho_{NT}\sqrt{T}}$
        \item $\pr\bxs{\lrp{\frac{\tilde\rho_{NT}}{NT}}^{1/2}\vrt{{\bh_2}\T\sum_t{\bZ_t}\T\lrp{\hat\bX-\bX}\beta}>\upsilon}<2e^{-cN\upsilon^2/\tilde\rho_{NT}}+\epsilon$
    \end{enumerate}
    where $d_1\triangleq c_1N^2T,d_2\triangleq c_2NT,d_3\triangleq NT/\tilde\rho_{NT}$ and $c_1,\ldots,c_3>0$ are constants.
\end{lemma}
\begin{proof}[Proof of 0]
    First note that ${\hat\bX}\T\hat\bX-\bX\T\bX=\Delta\T\lrp{\Delta+\bX}+\bX\T\Delta$ and $\lrp{\Delta\bh_1}\T\Delta\bh_1\leq\nrm{\Delta}{}^2\leq\nrm{\Delta}{F}^2=O_\pr(1)$. Also, $\nrm{\Delta\T\bX}{}=O_\pr(\nrm{\bX}{})$. Therefore,
    \begin{align*}
        \pr\bxs{\frac{1}{N}\vrt{{\bh_1}\T\lrp{{\hat\bX}\T\hat\bX-\bX\T\bX}\bh_1}>\upsilon}<\epsilon
    \end{align*}
    provided that $N\upsilon=\omega\lrp{\nrm{\bX}{}}$. Since $\nrm{\bX}{}/N=O\lrp{N^{-1/2}}$ and $1/\sqrt{N}=o(1)$, we have the conclusion.
\end{proof}
\begin{proof}[Proof of 1]
    Note that 
    \begin{align*}
        \sum_{t=1}^{T}{\bh_1}\T\lrp{{\hat\bX}\T\by_t-\bX\T\psi}=
    {\bh_1}\T\lrp{\bon_T\otimes\Delta}\T{\Gamma_\by}^{1/2}\tilde\by
    +{\bh_1}\T\lrp{\bon_T\otimes\bX}\T{\Gamma_\by}^{1/2}\tilde\by\\
    +T\lrp{\Delta\bh_1}\T\psi
    =\lrp{\hat\bS\bh_1}\T\tilde\by
    +\lrp{\tilde\bS\bh_1}\T\tilde\by
    +T\lrp{\Delta\bh_1}\T\psi.
    \end{align*}
    where $\hat\bS\triangleq{\Gamma_\by}^{1/2}\lrp{\bon_T\otimes\Delta}$ and $\tilde\bS\triangleq{\Gamma_\by}^{1/2}\lrp{\bon_T\otimes\bX}$.
    Since $\nrm{\hat\bS\bh_1}{}=O_\pr\lrp{\sqrt{T}}$, by Lemma \ref{lmm-hoeffding-Op} we have
    \[\pr\lrp{\vrt{\lrp{\hat\bS\bh_1}\T\tilde\by}>NT\upsilon}<2\exp\lrp{\frac{-cN^2T^2\upsilon^2}{T}}+\epsilon.\]
    Likewise, since $\nrm{\tilde\bS}{}=O(\sqrt{T}\nrm{\bX}{})=O(\sqrt{NT})$ as., by Lemma \ref{lmm-hoeffding-whp} again we have $\pr\lrp{\vrt{\lrp{\tilde\bS\bh_1}\T\Tilde\by}>NT\upsilon}<2\exp\lrp{-c_2NT\upsilon^2}$.
    Also, $\vrt{\lrp{\Delta\bh_1}\T\psi}=O_\pr\lrp{{\tilde\rho_{NT}}^{1/2}\nrm{\bX}{}}$ hence is $O_\pr\lrp{\sqrt{N\tilde\rho_{NT}}}$.
    By selecting $\upsilon$ such that $\upsilon=\omega(\sqrt{\tilde\rho_{NT}/N})$, we have
    \[\pr\bxs{\frac{1}{N}\vrt{\lrp{\Delta\bh_1}\T\psi}>\upsilon}\le\pr\bxs{\frac{1}{N}\vrt{\lrp{\Delta\bh_1}\T\psi}>C\sqrt{\frac{\tilde\rho_{NT}}{N}}}<\epsilon\]
    hence we obtain the conclusion for all large enough $N$ and $T$.
\end{proof}
\begin{proof}[Proof of 2]
    \begin{align*}
    {\bh_1}\T\sum_t\lrp{{\hat\bX}\T\cL_\bA\by_t-\bX\T\cL_\bA\psi}=
    {\bh_1}\T\bxs{\bon_T\otimes\brcs{\cL_\bA\Delta}}\T{\Gamma_\by}^{1/2}\tilde\by\\
    +{\bh_1}\T\brcs{\bon_T\otimes\cL_\bA\bX}\T{\Gamma_\by}^{1/2}\tilde\by+T\lrp{\Delta\bh_1}\T\cL_\bA\psi.
    \end{align*}
    For $\hat\bS\triangleq{\Gamma_\by}^{1/2}\brcs{\bon_T\otimes\lrp{\cL_\bA\Delta}}$ and $\tilde\bS\triangleq{\Gamma_\by}^{1/2}\brcs{\bon_T\otimes\lrp{\cL_\bA\bX}}$, we have $\nrm{\hat\bS\bh_1}{}=O_\pr(\sqrt{T})$ because $\rho\lrp{\cL_\bA}=O(1)$ as. hence $\pr\lrp{\vrt{\lrp{\hat\bS\bh_1}\T\tilde\by}>NT\upsilon}<2\exp\lrp{\frac{-cN^2T^2\upsilon^2}{T}}+\epsilon$. Since $\nrm{\tilde\bS\bh_1}{}=O\lrp{\sqrt{T}\nrm{\bX}{}}$ as., we have $\pr\lrp{|\lrp{\tilde\bS\bh_1}\T\tilde\by|>NT\upsilon}<2\exp\brcs{-cNT\upsilon^2}$. The tail probability of $T\vrt{\lrp{\Delta\bh_1}\T\cL_\bA\psi}$ can be bounded similarly to the proof of statement 1.
\end{proof}
\begin{proof}[Proof of 3]
    Let $\bZ\triangleq\bxs{{\bZ_0}\T,...,{\bZ_{T-1}}\T}\T$ and $\hat\bS\triangleq\bon_T\otimes\Delta$ and $\tilde\bS\triangleq\bon_T\otimes\bX$. Then, we can express $\sum_t{\bh_1}\T{\hat\bX}\T\bZ_t\bh_2$ as $\tr\lrp{\bZ\bh_2{\bh_1}\T{\hat\bS}\T}
    +\tr\lrp{\bZ\bh_2{\bh_1}\T{\tilde\bS}\T}$. This is equal to
    \[\vecop\lrp{\bh_2{\bh_1}\T}\T\brcs{\bI_p\otimes\lrp{\hat\bS+\tilde\bS}\T}\vecop\lrp{\bZ}.\]
    Note that $\nrm{\bI_p\otimes\hat\bS\T}{}=O_\pr\lrp{\sqrt{T}}$ and $\nrm{\bI_p\otimes\tilde\bS\T}{}=O\lrp{\sqrt{T}\nrm{\bX}{}}$. Therefore, by Theorem \ref{lmm-hoeffding-Op}, we have
    \[\pr\bxs{\vrt{\sum_t{\bh_1}\T{\hat\bX}\T\bZ_t\bh_2}>NT\upsilon}\leq2\exp\lrp{-\frac{cN^2T^2\upsilon^2}{T}}+2\exp\lrp{-c'NT\upsilon^2}+\epsilon.\]
\end{proof}
\begin{proof}[Proof of 6]
    We have 
    \begin{align*}
        \sum_t{\by_t}\T\bZ_t\bh_2=\lrp{{\Gamma_\by}^{1/2}\tilde\by+\Psi}\T\bZ\bh_2=\tr\lrp{{\bh_2}\tilde\by\T{\Gamma_\by}^{1/2}\bZ}+\tr\lrp{\bh_2\Psi\T\bZ}\\
        =\vecop\lrp{\tilde\by{\bh_2}\T}\T\lrp{\bI_p\otimes{\Gamma_\by}^{1/2}}\vecop\lrp{\bZ}+\vecop\lrp{\Psi{\bh_2}\T}\T\lrp{\bI_p\otimes\bI_{NT}}\vecop\lrp{\bZ}.
    \end{align*}
    Therefore, since $\nrm{\Gamma_\by}{}=O(1)$ as., $\nrm{\lrp{\bI_p\otimes{\Gamma_\by}^{1/2}}\vecop\lrp{\tilde\by{\bh_2}\T}}{}\lesssim\nrm{\vecop\lrp{\tilde\by{\bh_2}\T}}{}$ which is concentrated around $\sqrt{NTp}$ \citep{vershynin2018high} hence is $O(\sqrt{NTp})$ whp. 
    Also, we have $\nrm{\lrp{\bI_p\otimes\bI_{NT}}\vecop\lrp{\Psi{\bh_2}\T}}{}=O\lrp{\sqrt{T\tilde\rho_{NT}}\nrm{\bX}{}}$ as. 
    By Lemma \ref{lmm-hoeffding-Op},
    \[\pr\lrp{\vrt{\sum_t{\by_t}\T\bZ_t\bh_2}>NT\upsilon}\leq2\exp\lrp{-\frac{c_1NT\upsilon^2}{p}}+2\exp\lrp{-c_2NT\upsilon^2/\tilde\rho_{NT}}+\epsilon\]
    provided that $NT\upsilon=\omega(\sqrt{NTp})$.
\end{proof}
\begin{proof}[Proof of 11]
    Similar to the proof of 1., taking $\hat\bS\triangleq{\Gamma_\by}^{1/2}\lrp{\bon_T\otimes\Delta}$ which has the spectral norm of  $O_\pr\lrp{\sqrt{T}}$, we have
    \begin{align*}
        \sum_t{\by_t}\T\Delta\beta=
        \lrp{{\Gamma_\by}^{1/2}{\tilde\by}+\Psi}\T\lrp{\bon_T\otimes\Delta}\beta=\tilde\by\T\hat\bS\beta+\Psi\T{\Gamma_\by}^{-\frac{1}{2}}\hat\bS\beta.
    \end{align*}
    First noting that $\nrm{\beta}{}=O(1)$ as $K\rightarrow\infty$, Lemma \ref{lmm-hoeffding-Op} gives
    \[\pr\lrp{{\tilde\rho_{NT}}^{1/2}\vrt{\tilde\by\T\hat\bS\beta}\geq \sqrt{NT}\upsilon}\leq2\exp\lrp{\frac{-cNT\upsilon^2}{T\tilde\rho_{NT}}}+\epsilon.\]
    It is straightforward that $\lrp{\frac{\tilde\rho_{NT}}{NT}}^{1/2}\vrt{\Psi\T{\Gamma_\by}^{-\frac{1}{2}}\hat\bS\beta}=O_\pr\lrp{\frac{T\tilde\rho_{NT}\sqrt{N}}{\sqrt{NT}}}=O_\pr\lrp{\tilde\rho_{NT}\sqrt{T}}$, so the conclusion follows.
\end{proof}

\begin{lemma}\label{lmm-amnar-∑nasym}
    There exist $N_0,T_0>0$ and $\upsilon_0(N_0,T_0)\in(0,1)$ such that for all $\upsilon\in(0,\upsilon_0)$ and $N>N_0,T>T_0$,
    \begin{align*}
        \pr\bxs{\sup_{\bh\in B^{K+p+1}}\vrt{\bh\T\lrp{\hat\Omega_m-\Omega_m}\bh}>\upsilon}\lesssim
        K^2\brcs{\epsilon+\exp\lrp{-c\upsilon^2\frac{NT}{pK^2}}}.
    \end{align*}
\end{lemma}

\subsection{Theories for Corollary Models}

Similarly, we verify the asymptotic properties of AMNAR with finite $T$.

\begin{proof}[Proof of Theorem \ref{thm-amnar-clt-N}]
    Let $\tilde{\bD}_a\triangleq\sqrt{N}\operatorname{diag}\lrp{ {\tilde\rho_{NT}}^{1/2}\bI_{K+1},\bI_{p+2}}$.
    Then, $\sqrt{T}\tilde{\bD}_a=\bD_a$.
    We have 
    \begin{align*}
    \hat\mu_m & = {\bD_a}^{-1}\lrp{\hat\Omega_m}^{-1}{\bD_a}^{-1}\hat{\bM}\T\brcs{\lrp{\bM-\hat{\bM}}\mu_m+\hat{\bM}\mu_m+\bcE} \\
    & = \frac{{\tilde{\bD}_a}^{-1}}{\sqrt{T}}{\hat\Omega_m}^{-1}\brcs{\bD_a^{-1}\hat{\bM}\T\lrp{\bM-\hat{\bM}}\mu_m+\br_{m\epsilon}}+\mu_m
    \end{align*}
    hence
    \[{\hat\Omega_m}\tilde{\bD}_a(\hat\mu_m-\mu_m)=
    \frac{{\bD_a}^{-1}}{\sqrt{T}}{\hat{\bM}}\T(\bM-\hat{\bM})\mu_m
    +\frac{\br_{m\epsilon}}{\sqrt{T}}.\]
    Then, the remaining steps are same to previous proofs for consistency result.
    We refer the readers to the proof of Theorem \ref{thm-amnar-clt-NT} for the details.
\end{proof}

\section{Additional tables and figures}
\subsection{More simulations}
\subsubsection{Over-selection of \texorpdfstring{$K$}{K} by IC}
\begin{figure}
    \centering
    \includegraphics[width=0.495\linewidth]{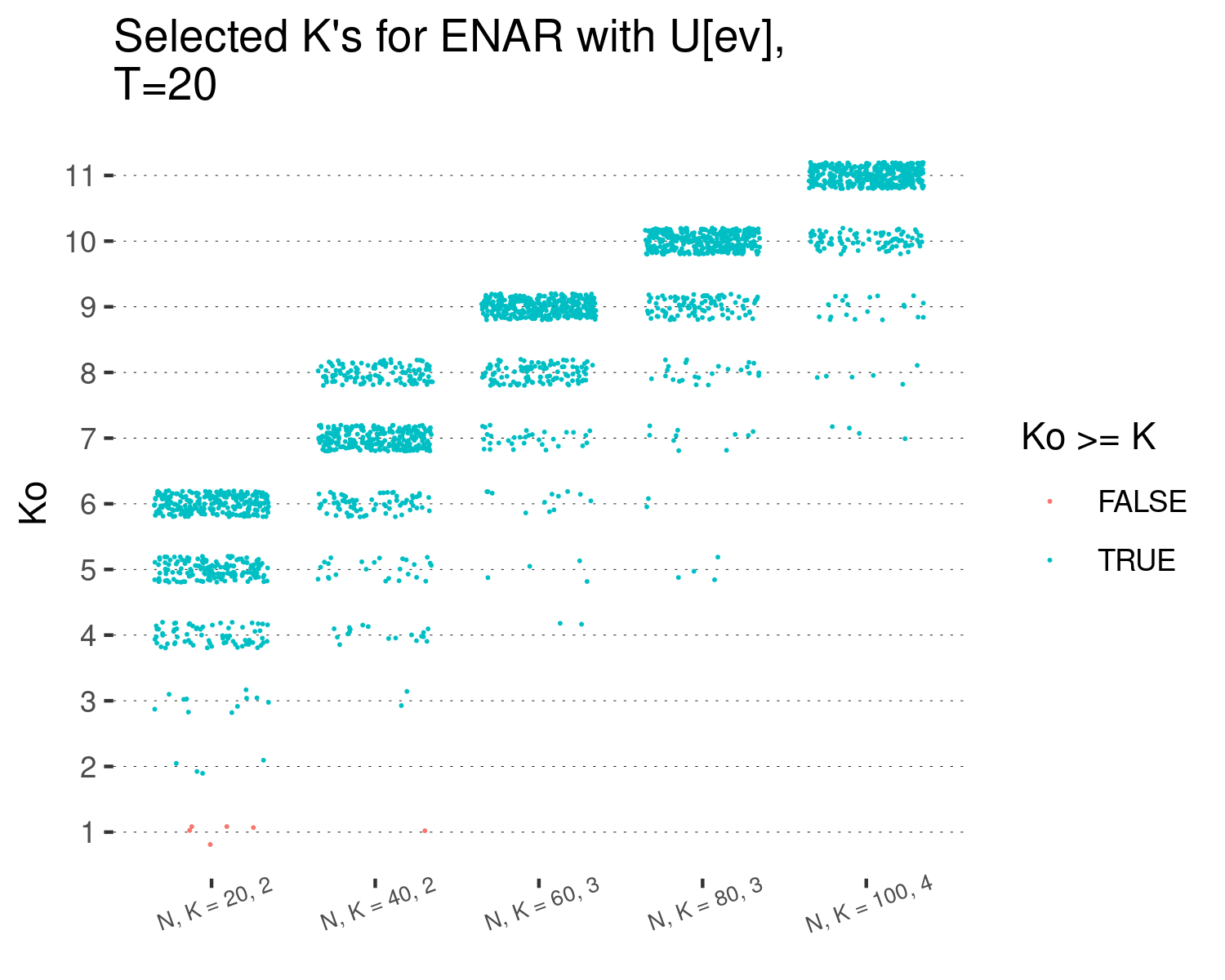}
    \includegraphics[width=0.495\linewidth]{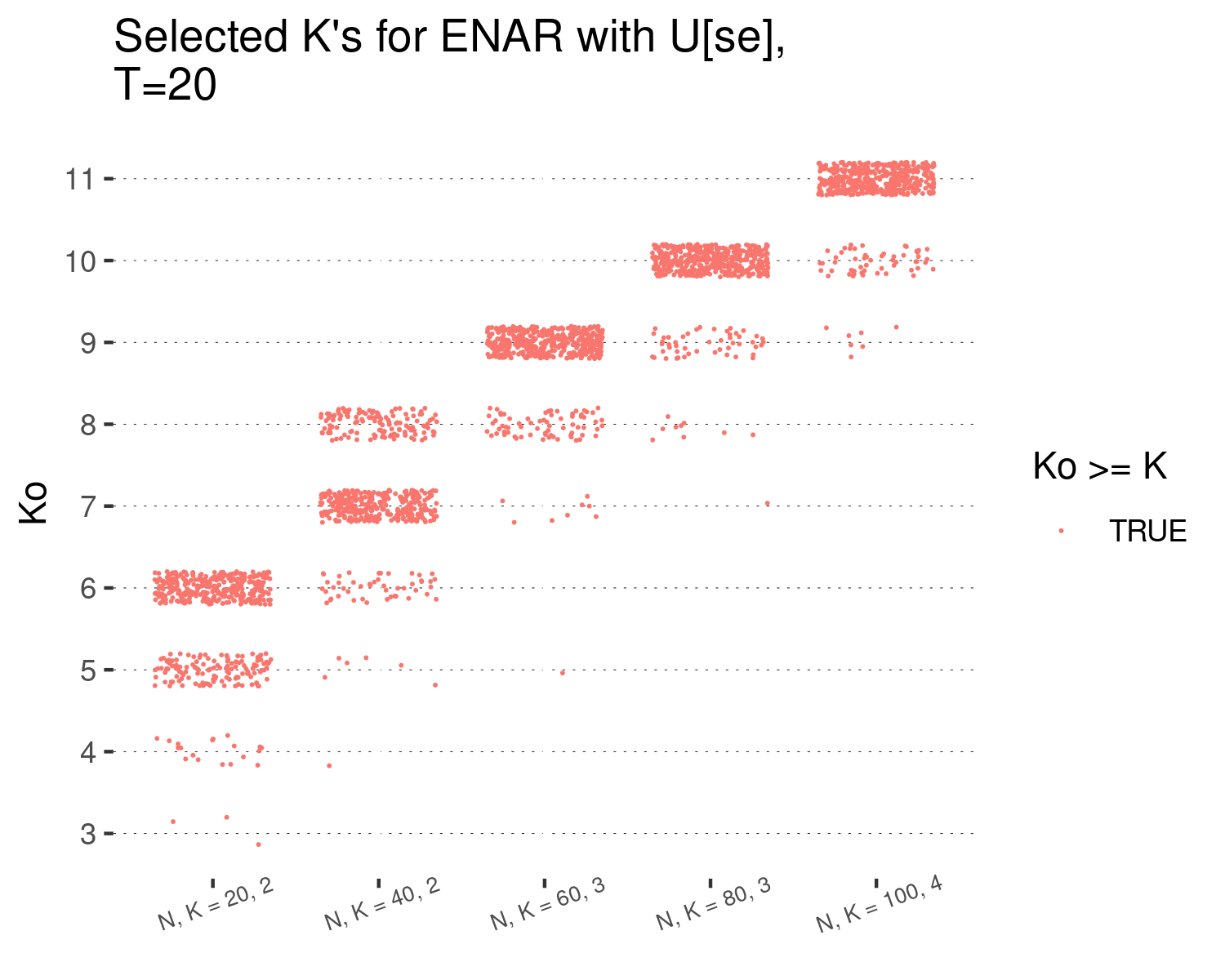}
    \caption{Jitter plots of selected $\hat K$ when $T=20$ and $N$ increases when $\bU=\bU_{ev}$ (left) and $\bU=\bU_{se}$ (right). Red dots indicate under-selections while green dots indicate $\hat K\ge K$.}
    \label{fig-enar-overK-jittplt}
\end{figure}

Here we present the simulation results for the over-selection procedure of latent dimension $K$ introduced in section \ref{sec-enar-IC}.
We pose that the network density is  $N\rho_N=\log^{2.5}N$ and choose the penalty function as $p_{NT}(k)=\frac{k}{N+T}$.
We use almost the same simulation setting we used in section \ref{sec-enar-sim}, but we only consider the case when $\bA$ is generated from DCMMSBM.
For the grid of $N,T\in\brcs{20, 40, 60, 80, 100}$, we grow $K$ to be 2, 2, 3, 3, and 4 along each of $N$.
We plot the $\frac{1}{M}\sum_{i=1}^M\mathbb{I}_{\brcs{\hat K\ge K}}$ for $M=500$ replications of Monte Carlo simulations.
When applying the criteria, we consider a grid of $k=0,1,2,\dots,J_N=\lceil\nrm{\bA}{\infty}^{1/2}\rceil$ to search for $\hat K$.
The lowest empirical probability of over-selection was 0.99 for all choices of $N,T$ and both $f\in\brcs{ev,se}$, suggesting an almost sure over-selection.
The jitter plots of selected $\hat K$ in figure \ref{fig-enar-overK-jittplt} clearly show their concentration around values greater than $K$.

\subsubsection{NAR with growing \texorpdfstring{$N,T$}{N, T}}
\begin{figure}[!ht]
    \centering
    \begin{subfigure}{0.5\textwidth}
        \includegraphics[width=1\textwidth]{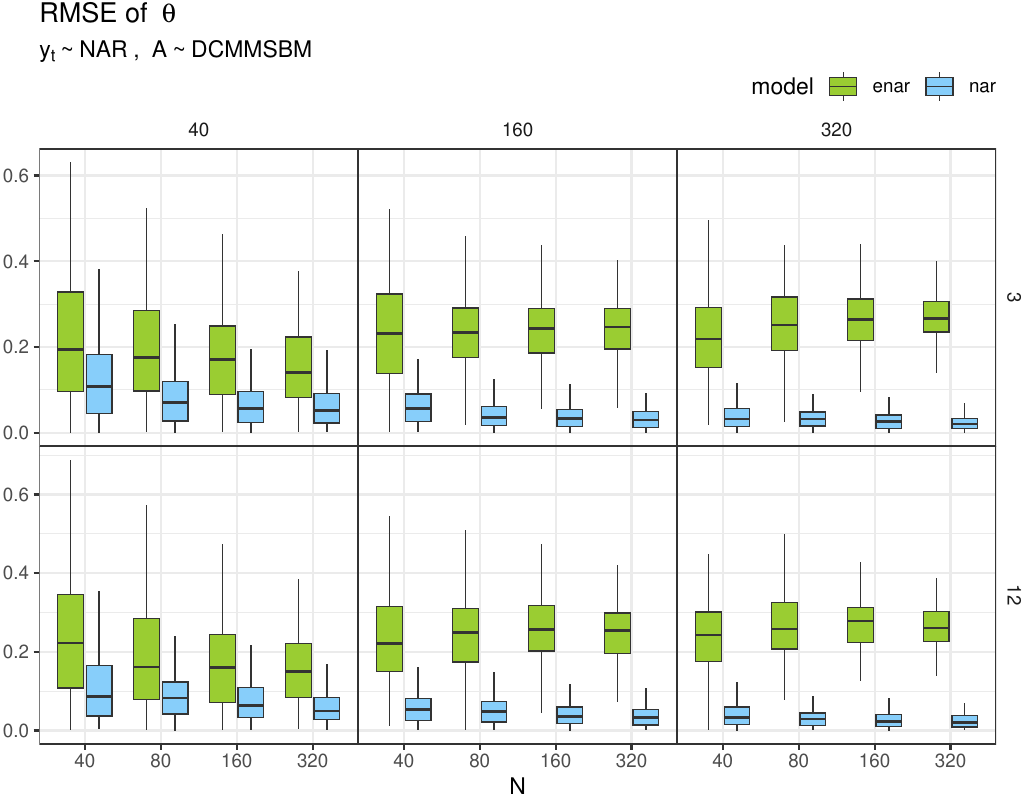}
    \end{subfigure}%
    \begin{subfigure}{0.5\textwidth}
        \includegraphics[width=1\textwidth]{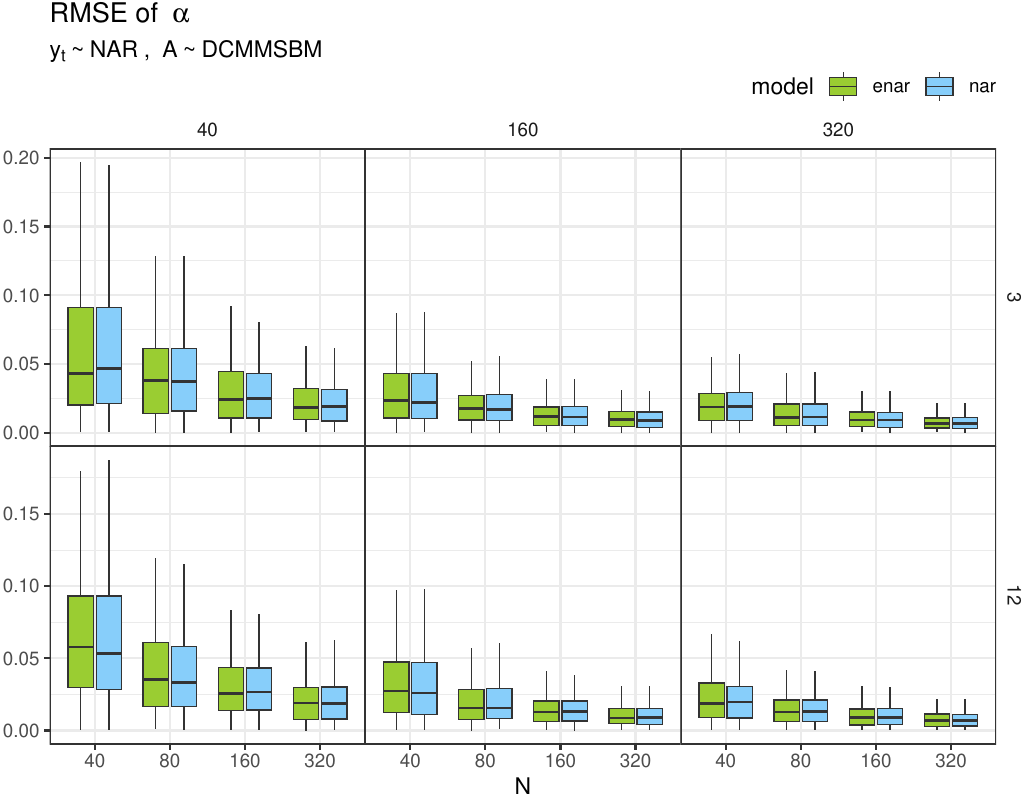}
    \end{subfigure}
    \caption{Boxplot of estimates of $\theta$ and $\alpha$ from ENAR, and NAR model when data is generated from NAR model with DCMMSBM.}
    \label{fig-nar_dcmmsbm}
\end{figure}

\begin{figure}[!ht]
    \centering
    \begin{subfigure}{0.5\textwidth}
        \includegraphics[width=1\textwidth]{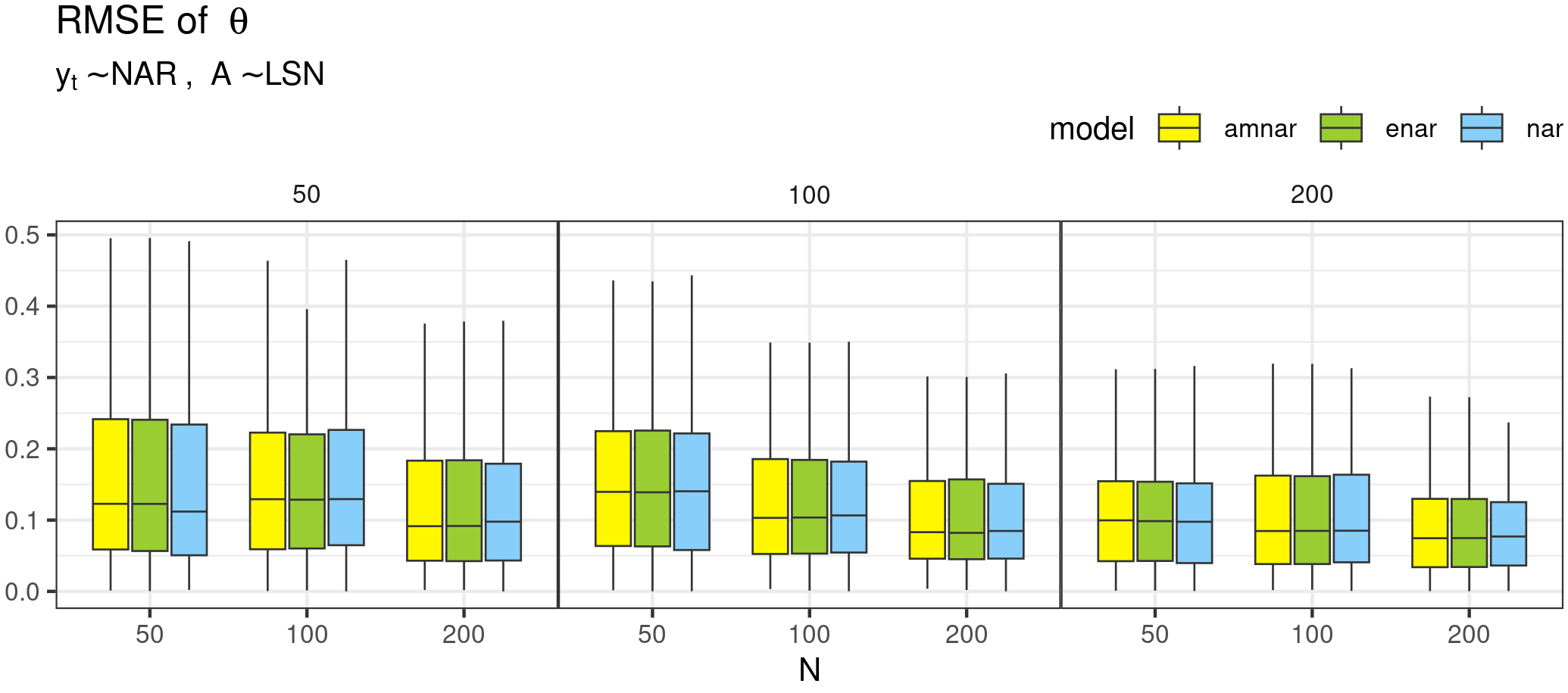}
    \end{subfigure}%
    \begin{subfigure}{0.5\textwidth}
        \includegraphics[width=1\textwidth]{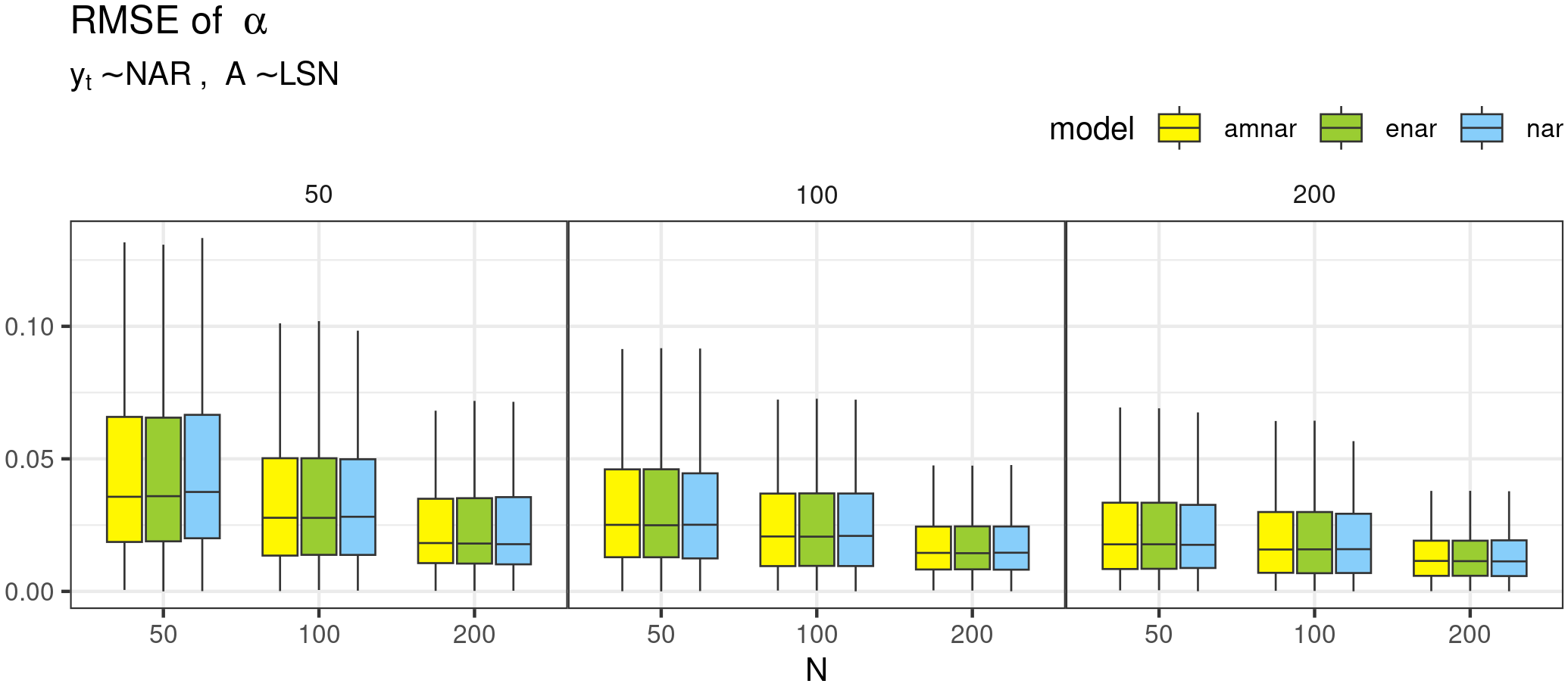}
    \end{subfigure}
    \caption{Boxplot of estimates of $\theta$ and $\alpha$ from AMNAR, ENAR, and NAR model when data is generated from NAR model with LSN.}
    \label{fig-nar_lsn}
\end{figure}

The figure \ref{fig-nar_dcmmsbm} displays the boxplots of estimates of $\theta$ and $\alpha$ parameters with growing $N$ and $T$ when the data is generated from the NAR model.
The simulation setup and the figure has been described in the main text. The figure \ref{fig-nar_lsn} shows the estimation errors when the graph is generated from LSN.

\subsection{Real data analysis}

The table \ref{knecht_estimates} below displays parameter estimates along with standard errors for the Knecht dataset. 
\begin{table}[H] 
\centering 
  \caption{Parameter estimates and standard errors for the OLS, NAR and ENAR models fitted to the alcohol consumption data and delinquency data.} 
  \label{knecht_estimates} 
  \resizebox{0.9\textwidth}{!}{
\begin{tabular}{@{\extracolsep{5pt}}lccc|ccc} 
\\[-1.8ex]\hline 
\hline \\[-1.8ex] 
 & \multicolumn{6}{c}{\textit{Dependent variable:}} \\ 
\cline{2-7} 
\\[-1.8ex] &  \multicolumn{3}{c}{Alcohol Consumption} & \multicolumn{3}{c}{Delinquency ($K=2$) }  \\ 
\\[-1.8ex] & OLS & NAR & ENAR  ($K=2$) & OLS & NAR & ENAR  ($K=2$)\\ 
\hline 
 Alc. previous &  & 0.278 & 0.372 &  & 0.101 & 0.164 \\ 
  &  & (0.288) & (0.310) & & (0.176) & (0.151) \\ 
 Peer effect &  & 0.708 & 1.788 & & $-$0.714 & $-$0.375 \\ 
  &  & (0.725) & (1.253) & & (0.583) & (0.321)  \\ 
 Sex & $-$0.533 & $-$0.328 & $-$0.669 & $-$0.061 & $-$0.369 & $-$0.083\\ 
  & (0.480) & (0.513) & (1.284) &  (0.250) & (0.661) & (0.243)  \\ 
 Age & 0.029 & 0.182 & 0.212 & $-$0.003 & $-$0.050 & 0.009 \\ 
  & (0.472) & (0.497) & (0.560) & (0.245) & (0.286) & (0.240) \\ 
 Ethnicity & 0.444 & 0.224 & 0.144 & $-$0.595 & $-$0.646 & $-$0.591 \\ 
  & (0.815) & (0.853) & (0.891) & (0.424)  & (0.459) & (0.415)  \\  
Religion & 0.664$^{**}$ & 0.540 & 0.525 & 0.277 & 0.310 & 0.294 \\ 
  & (0.315) & (0.336) & (0.356) & (0.164) & (0.189) & (0.176) \\ 
\hline \hline \\[-1.8ex] 
\end{tabular} 
}
\end{table} 

\begin{figure}[!ht]
    \centering
    \begin{subfigure}{0.5 \textwidth}
     \centering
     \includegraphics[width=0.8 \textwidth]{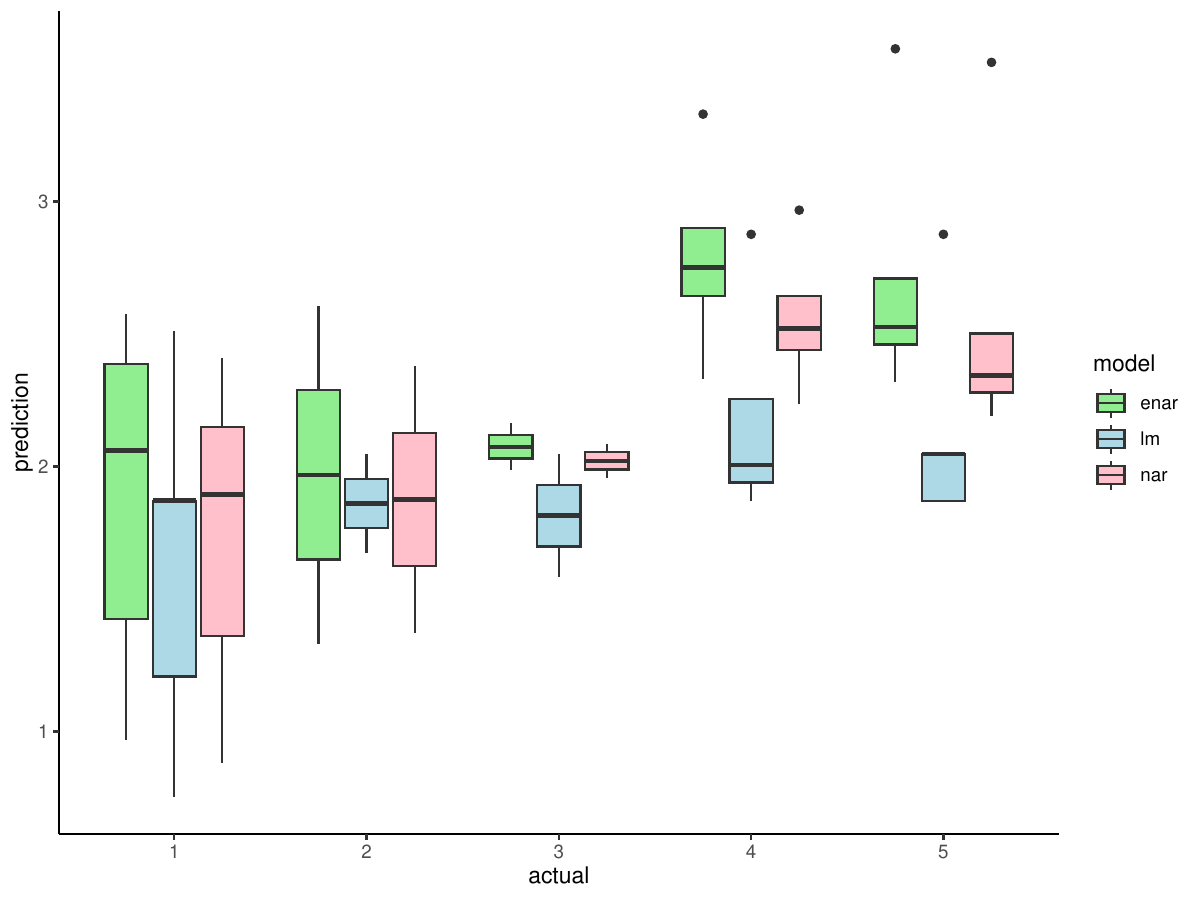}  
    \end{subfigure}%
    \begin{subfigure}{0.5 \textwidth}
       \centering
        \includegraphics[width=0.8 \textwidth]{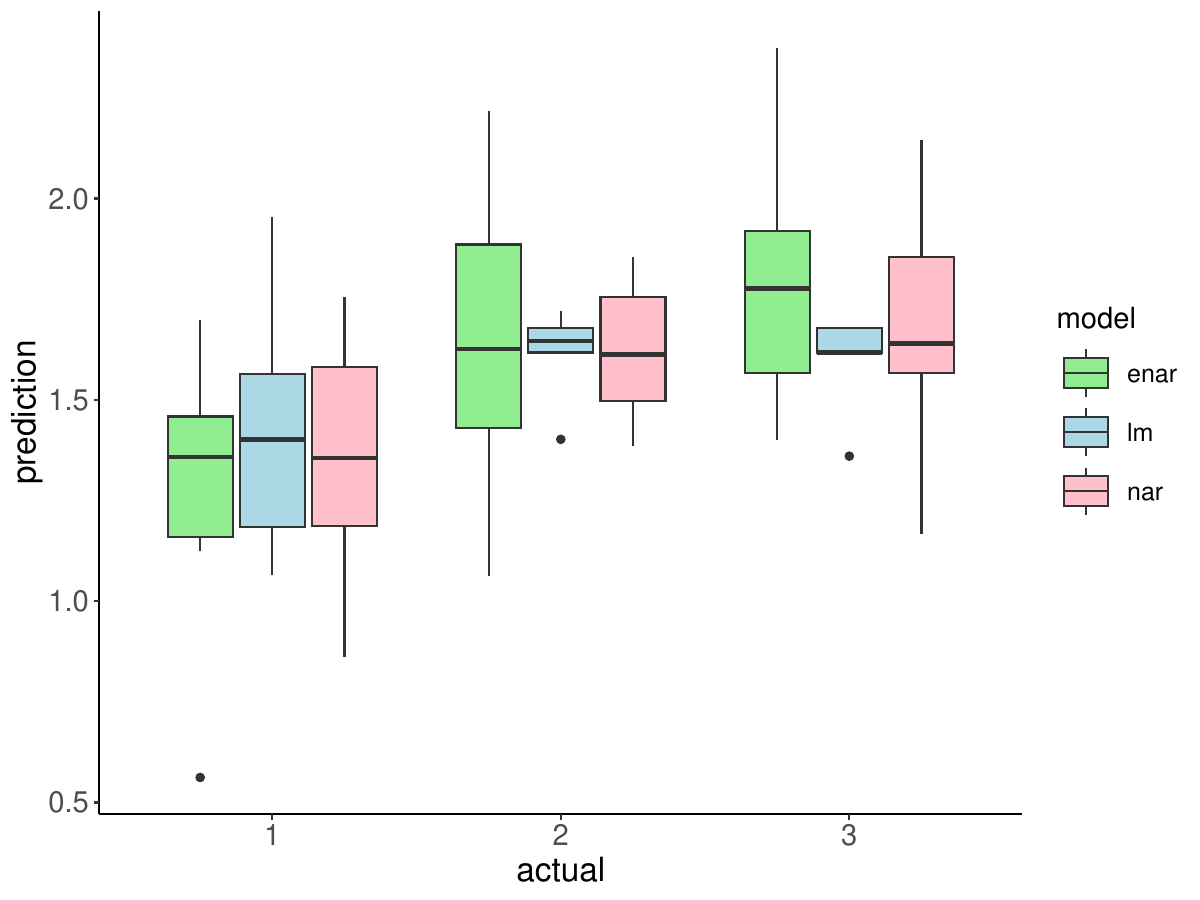}
    \end{subfigure}   
    \caption{Boxplots of predicted values from the 3 models for different labels of actual values for the response on (a) alcohol and (b) delinquency. For ENAR model, $K=2$ for alcohol and $K=4$ (selected by IC) for delinquency.}
    \label{knecht_boxplot}
\end{figure}

The figure \ref{knecht_boxplot} displays the predicted responses against the actual ones to show the quality of predictions.

Table \ref{vswind_estimates} provides estimates from ENAR and NAR for one representative model with $T=520$.

\begin{table}[H] 
\centering 
  \caption{Estimates of ENAR and NAR model for the Wind speed data where the model is fit on data between $T=[1,520]$.} 
  \label{vswind_estimates} 
\begin{tabular}{@{\extracolsep{5pt}}lcc} 
\\[-1.8ex]\hline 
\hline 
 & \multicolumn{2}{c}{\textit{Dependent variable:}} \\ 
\cline{2-3} 
 & \multicolumn{2}{c}{Wind speed} 
\\[-0ex] & ENAR & NAR \\ 
\hline 
 Ylagged & 0.810$^{***}$ & 0.823$^{***}$ \\ 
  & (0.003) & (0.002) \\ 
 LYlagged & 0.096$^{***}$ & 0.081$^{***}$ \\ 
  & (0.003) & (0.002) \\ 
\hline 
AIC & 53243.98 & 53576.56 \\ 
BIC & 53394.89 & 53612.07 \\ 
\hline 
\hline 
\textit{Note:}  & \multicolumn{2}{r}{$^{*}$p$<$0.1; $^{**}$p$<$0.05; $^{***}$p$<$0.01} \\ 
\end{tabular} 
\end{table} 
\end{appendix}

\end{document}